\documentclass{article}

\usepackage{mathrsfs}
\usepackage{tikz}
\usepackage{tikz-cd}
\usepackage[dvipsnames,svgnames]{xcolor}
\usetikzlibrary{decorations.pathmorphing}
\usetikzlibrary{decorations.markings}
\usepackage{graphicx}
\usepackage{float}
\usepackage{scrextend}
\usepackage{manfnt,pifont}
\usepackage{url}
\usepackage{lipsum}
\usepackage[font=small]{subcaption}
\usepackage[font=small]{caption}
\usepackage{tabularx}
\usepackage{array}
\usepackage{dsfont}
\usetikzlibrary{arrows.meta} 
\usepackage{subcaption}
\usepackage{appendix}
\usepackage{jheppub}

\usepackage[normalem]{ulem}

\usepackage{url}
\usepackage[outline]{contour}
\usepackage{textgreek}

\newcommand{\cL}{{\mathcal L}}
\newcommand{\bb}{\mathbf{b}}

\newcommand{\be}{\begin{equation}}
\newcommand{\ee}{\end{equation}}

\usepackage{tikz}
\usetikzlibrary{shapes}
\tikzset{line/.style={line width=0.25mm},
curve/.style={line,smooth,tension=1},
->-/.style={decoration={
  markings,
  mark=at position #1 with {\arrow[>=stealth]{>}}},postaction={decorate}},
-<-/.style={decoration={
  markings,
  mark=at position #1 with {\arrow[>=stealth]{<}}},postaction={decorate}},
}
\newcommand{\dotsol}[2]{node [circle, opacity=1, fill, inner sep=1pt, label = #1 : {#2}] {}}

\newcommand{\dotsols}[2]{node [circle, fill, inner sep=1pt, label={#1:#2}] {}}

\tikzset{
  sharp arrow/.style={
    -Stealth 
  }
}

\usepackage{CJK}
\usepackage{enumitem}
\setlist[itemize]{leftmargin=*}

\makeatletter
\DeclareFontFamily{OMX}{MnSymbolE}{}
\DeclareSymbolFont{MnLargeSymbols}{OMX}{MnSymbolE}{m}{n}
\SetSymbolFont{MnLargeSymbols}{bold}{OMX}{MnSymbolE}{b}{n}
\DeclareFontShape{OMX}{MnSymbolE}{m}{n}{
    <-6>  MnSymbolE5
   <6-7>  MnSymbolE6
   <7-8>  MnSymbolE7
   <8-9>  MnSymbolE8
   <9-10> MnSymbolE9
  <10-12> MnSymbolE10
  <12->   MnSymbolE12
}{}
\DeclareFontShape{OMX}{MnSymbolE}{b}{n}{
    <-6>  MnSymbolE-Bold5
   <6-7>  MnSymbolE-Bold6
   <7-8>  MnSymbolE-Bold7
   <8-9>  MnSymbolE-Bold8
   <9-10> MnSymbolE-Bold9
  <10-12> MnSymbolE-Bold10
  <12->   MnSymbolE-Bold12
}{}

\let\llangle\@undefined
\let\rrangle\@undefined
\DeclareMathDelimiter{\llangle}{\mathopen}%
                     {MnLargeSymbols}{'164}{MnLargeSymbols}{'164}
\DeclareMathDelimiter{\rrangle}{\mathclose}%
                     {MnLargeSymbols}{'171}{MnLargeSymbols}{'171}

\makeatletter

\def\l@subsection#1#2{}

\def\l@subsubsection#1#2{}

\makeatother

\usepackage{eso-pic}

\newcommand{\etypequiveredge}[4]{%
    \begin{scope}[shift={#1}, rotate=#2]
        \draw[#4, thick, -Stealth, bend left=25] (0.10,  0.055) to (1.40,  0.055);
        \draw[#3, thick, -Stealth, bend left=25] (0.05,  0.165) to (1.45,  0.165);
        \draw[#3, thick, -Stealth, bend left=25] (1.40, -0.055) to (0.10, -0.055);
        \draw[#4, thick, -Stealth, bend left=25] (1.45, -0.165) to (0.05, -0.165);
    \end{scope}%
}

\newcommand{\etypecenterend}[4]{%
    \begin{scope}[shift={#1}, rotate=#2]
        \draw[#4, thick, -Stealth, bend left=25] (0.10,  0.055) to (1.25,  0.055);
        \draw[#3, thick, -Stealth, bend left=25] (0.05,  0.165) to (1.30,  0.165);
        \draw[#3, thick, -Stealth, bend left=25] (1.25, -0.055) to (0.10, -0.055);
        \draw[#4, thick, -Stealth, bend left=25] (1.30, -0.165) to (0.05, -0.165);
    \end{scope}%
}

\newcommand{\etypecenterstart}[4]{%
    \begin{scope}[shift={#1}, rotate=#2]
        \draw[#4, thick, -Stealth, bend left=25] (0.25,  0.055) to (1.40,  0.055);
        \draw[#3, thick, -Stealth, bend left=25] (0.20,  0.165) to (1.45,  0.165);
        \draw[#3, thick, -Stealth, bend left=25] (1.40, -0.055) to (0.25, -0.055);
        \draw[#4, thick, -Stealth, bend left=25] (1.45, -0.165) to (0.20, -0.165);
    \end{scope}%
}

\begin{document}


\preprint{%
\begin{tabular}{l}
USTC-ICTS/PCFT-26-59\\
KIAS-P26044
\end{tabular}%
}

\title{Superstrip Algebras and Nonperturbative Spectra in Fermionic Theories}
\author{Jin Chen$^{a,b}$}
\emailAdd{zenofox@gmail.com}
\author{Zhihao Duan$^{c,d}$}
\emailAdd{xduanz@gmail.com}
\author{Qiang Jia$^{e}$}
\emailAdd{qjia1993@kaist.ac.kr}
\author{Sungjay Lee$^{f}$}
\emailAdd{sjlee@kias.re.kr}
\affiliation{$^{a}$Department of Physics, Xiamen University, Xiamen, 361005, China}
\affiliation{$^{b}$Peng Huanwu Center for Fundamental Theory, Hefei, Anhui 230026, China}
\affiliation{$^{c}$Shanghai Institute for Mathematics and Interdisciplinary Sciences (SIMIS), Shanghai 
200433, China}
\affiliation{$^{d}$Section de Math\'ematiques, Universit\'e de Gen\`eve, 1211 Gen\`eve 4, Switzerland}
\affiliation{$^{e}$Department of Physics, Korea Advanced Institute of Science \& Technology,
Daejeon 34141, Korea}
\affiliation{$^{f}$Korea Institute for Advanced Study, 85 Hoegiro, Dongdaemun-Gu, Seoul 02455, Korea}

\abstract{This is an extended version of \href{https://arxiv.org/abs/2511.22129}{2511.22129}. We develop a general construction of the superstrip algebra $\mathbf{sStr}_{\mathscr C}(\mathcal M)$, the fermionic counterpart of the strip algebra, for a gapped $(1+1)$-dimensional system with a fermionic superfusion category $\mathscr C$ and the vacuum structure described by the supermodule category $\mathcal M$. As illustrations, we work out the superstrip algebra for two simple examples: the fermionic variant of the Fibonacci fusion category and the q-type $\mathbb{Z}_2$ symmetry. We then apply the framework to the gapped IR phases of the $\mathcal N=2$ and $\mathcal N=1$ superconformal minimal models deformed by their least relevant operator and compute the Witten indices. We explain how the particle and soliton spectra, together with their fermion parity, are organized into representations of the corresponding superstrip algebras. A mixed 't Hooft anomaly between the spontaneously broken symmetry and the unbroken fermion parity $(-1)^F$ further forces each $\mathcal N =2$ soliton to carry a fractional fermion number. Finally, we set up the boundary SymTFT for fermionic systems and reproduce the results in several examples studied in this paper.}

\maketitle


\section{Introduction}

The long-distance behavior of a quantum field theory can be organized by its
symmetries and their 't~Hooft anomalies. Over the past decade this organizing
principle has been extended far beyond ordinary groups, to higher-form, higher-group,
and categorical (non-invertible) symmetries~\cite{Chang:2018iay,Thorngren:2019iar,Komargodski:2020mxz,Thorngren:2021yso,Cordova:2022ieu,Apte:2022xtu,Lin:2023uvm,Kaidi:2023maf,Zhang:2023wlu,Damia:2023ses,Cordova:2023bja,Choi:2023pdp,Antinucci:2023ezl,Bhardwaj:2024kvy,Bhardwaj:2024qiv,Cordova:2024ypu,Copetti:2024rqj,DelZotto:2024arv,Nakayama:2024msv,Cordova:2024goh,Cordova:2024nux,Chen:2023qnv,Bhardwaj:2025piv,Bhardwaj:2025jtf,Seiberg:2025bqy,Antinucci:2025fjp, Zhang:2026gqp}.
In $1+1$ dimensions a finite global symmetry is realized by topological defect
lines (TDLs) that form a fusion
category~\cite{Bhardwaj:2017xup,Aasen:2017ubm,Chang:2018iay,Chang:2022hud}, and
already the critical Ising model carries a non-invertible line, the
Kramers--Wannier duality defect. 
Such symmetries constrain strongly coupled dynamics where conventional methods
are limited.

One concrete constraint concerns the solitons of a gapped phase. It was shown in~\cite{Cordova:2024vsq,Cordova:2024iti} that the
solitons interpolating between the vacua of a spontaneously broken non-invertible
symmetry form representations of a ``strip algebra''
$\mathbf{Str}_{\mathscr C}(\mathcal M)$, a $C^\ast$-weak Hopf algebra built from
the symmetry category $\mathscr C$ and the $\mathscr C$-module category $\mathcal M$ that
describes the gapped vacua. An element of the algebra can be represented by the following ladder diagram
\begin{equation*}
\begin{gathered}
     \begin{tikzpicture}[scale=1]
        \draw[ultra thick,MidnightBlue,-<-=.55] (1,0)--(1,1);
        \draw[ultra thick,MidnightBlue,-<-=.55] (1,1)--(1,2);
        \draw[ultra thick,MidnightBlue,->-=.65] (2.5,0)--(2.5,1);
        \draw[ultra thick,MidnightBlue,->-=.7] (2.5,1)--(2.5,2);
        \draw[densely dashed,ultra thick,BurntOrange,-<-=.55] (1,0.9)--(2.5,1.15);
        \node at (0.7,.9) {$\textcolor{black}\alpha$};
        \node at (2.8,1.15) {$\textcolor{black}\beta$};
         \filldraw[black] (1,0.9) circle (2pt);
         \filldraw[black] (2.5,1.15) circle (2pt);
        \node at (1.25,0.25) {$i$};
        \node at (1.25,1.75) {$k$};
        \node at (2.25,0.25) {$j$};
        \node at (2.25,1.75) {$l$};
        \node at (1.75,0.75) {$a$};
    \end{tikzpicture}
\end{gathered}\,,
\end{equation*}
where the vertical solid lines $i,j,k,l$ stand for gapped vacua in $\mathcal{M}$, the dashed line $a$ represents the action of the symmetry operator in $\mathscr{C}$, and $\alpha,\beta$ label the junctions between $\mathscr{C}$ and the boundary module $\mathcal{M}$. The multiplication of two elements in $\mathbf{Str}_{\mathscr C}(\mathcal M)$ is defined by vertically concatenating them. Physically, the element can be understood as the symmetry operator $a$ acting on a soliton state interpolating between the vacua $i,j$ at $x\rightarrow \mp \infty$ (or a particle state if $i=j$), converting it into another soliton state interpolating between the vacua $k,l$ (or a particle state if $k=l$). The soliton/particle multiplets are then classified by the irreducible representations of the strip algebra. This analysis has so far been developed for bosonic
theories.

However, the analysis for the fermionic theories remains largely unexplored. 
A fermionic theory always carries the fermion-parity line $(-1)^F$, and its topological defect lines have $\mathbb Z_2$-graded junctions and may support a Majorana fermion. 
These features cannot be captured by an ordinary fusion category, but instead require a superfusion category~\cite{Wan:2016php,Lou:2020gfq,Hu:2021qhm,Wang:2023iqt,Ambrosino:2024ggh,Bhardwaj:2024ydc,Balasubramanian:2024nei}. It remains unclear how such intrinsically fermionic categorical symmetries constrain the non-perturbative spectrum of a given fermionic theory.

One might argue that, since any two-dimensional fermionic theory admits a bosonized description, the problem could simply be reformulated in the corresponding dual bosonic theory and analyzed using the bosonic strip algebra. This approach, however, is not satisfactory for the questions we would like to address in the present work. For example, bosonization gauges the fermion parity $(-1)^F$ and thus topological data 
associated with $(-1)^F$ as a global symmetry is no longer available 
in the bosonic description. 

The purpose of this paper is to construct the superstrip algebra
$\mathbf{sStr}_{\mathscr C}(\mathcal M)$, the fermionic counterpart of the strip
algebra, for a gapped $1+1$d theory whose non-invertible symmetry is
spontaneously broken. Our idea is to grade every junction by its fermion parity
and to track the resulting signs through the superpentagon equations, which makes
$\mathbf{sStr}_{\mathscr C}(\mathcal M)$ a $C^\ast$-weak Hopf superalgebra. Its
irreducible representations are encoded in a quiver where each node can be identified as a  vacuum and blue and red arrows count the bosonic and fermionic solitons. When $(-1)^F$
is unbroken, the type of the broken line determines the statistical nature of the vacuum: an
m-type line gives a bosonic vacuum, while a q-type line carrying a Majorana
zero mode, gives a fermionic one. The non-perturbative spectrum is consistent
only if it furnishes such a representation.

As illustrative examples, we first work out the superstrip algebras for two simple superfusion categories of the form $\mathscr{C} = \tilde{\mathscr{C}}\boxtimes \mathbb{Z}_2^{(-1)^F}$. $\mathbb{Z}_2^{(-1)^F}$ is generated by the fermion parity $(-1)^F$, and $\tilde{\mathscr{C}}$ is taken to be either the fermionic variant of the Fibonacci fusion category, where the non-invertible m-type TDL $W$ satisfies $W^2 = \mathbb{C}^{1|0} I+\mathbb{C}^{1|1}W$, or the q-type $\mathbb{Z}_2$ symmetry, whose generator $W$ is q-type and satisfies $W^2=\mathbb{C}^{1|1} I$. Here the presence of the $\mathbb{C}^{1|1}$ coefficient indicates that the corresponding trivalent junctions, $W,W,W$ and $I,W,W$ in the two cases respectively, are super vector spaces. We take the supermodule category $\mathcal{M}$ to be $\tilde{\mathscr{C}}$, which describes the gapped phase in which $\tilde{\mathscr{C}}$ is completely spontaneously broken (SSB) while $(-1)^F$ remains unbroken. We work out the superstrip algebra for the two examples, and construct the idempotes $e_i$ which project $\mathcal{A}$ onto the irreducible representations $\mathcal{A} e_i$, each of which is summarized in a quiver diagram.

We then test the framework on the $\mathcal N=2$ and $\mathcal N=1$ superconformal
minimal models with their least relevant supersymmetry-preserving deformation. 
The $\mathcal N=2$ models
are classified by the $ADE$ series, descending from the modular invariants of
$\widehat{su}(2)_k$ found by Cappelli, Itzykson and
Zuber~\cite{Cappelli:1987xt}. A generic deformation destroys this structure, but
the least relevant deformation preserves a fermionic variant of
$\widehat{su}(2)_k$, a superfusion category $\tilde{\mathscr C}_{A_{k+1}}$ that is
completely spontaneously broken in the deep infrared. Its superstrip algebra then
controls the massive spectrum: the deformed $A_{k+1}$ model has $k+1$ bosonic
vacua whose fundamental solitons fill a connected $A$-type quiver, and the
$D$- and $E$-type models follow by gauging a Frobenius algebra $\mathscr A_{D/E}$
that survives the deformation~\cite{Fuchs:2002cm,Carqueville:2012dk,Diatlyk:2023fwf}.
This is the categorical origin of the $ADE$ pattern of the soliton spectrum. A
mixed 't~Hooft anomaly between the broken line and the unbroken $(-1)^F$ further
forces each soliton to carry a fractional fermion number read off from the module $F$-symbols
in a gauge-invariant manner. All states in a
single representation are exactly mass-degenerate, in agreement with the supermultiplet structure, and the spectrum reproduces the
degenerate BPS spectra known from
integrability~\cite{Bernard:1990ti,Fendley:1991ve}. We emphasize that the
superstrip algebra uses neither supersymmetry nor integrability, and applies to
any gapped $1+1$d fermionic system with a spontaneously broken non-invertible
symmetry. For the $\mathcal N=1$ minimal models, the least relevant deformation now leads to both m- and q-type vacua. In this case, the fundamental solitons form $\mathbb Z_2$-graded or doubled $ADE$ quivers. Supersymmetry pairs the bosonic and fermionic states within each oriented soliton sector, whereas the preserved non-invertible symmetry ties different sectors on the same Dynkin graph into a single irreducible superstrip-algebra representation, thereby enforcing their mass degeneracy. The $\mathcal N=2$ application was announced in the short version of this work \cite{Chen:2025qub}. Here we provide more details and extend the analysis to $\mathcal N=1$ minimal models.

We also develop the boundary Symmetry Topological Field Theory (SymTFT) \cite{Baume:2023kkf,Bhardwaj:2023bbf,Braeger:2024jcj,Heckman:2024zdo,Cvetic:2024dzu,Choi:2024tri,Bhardwaj:2024igy} description of a 2d gapped system with superfusion category $\mathscr{C}$ and vacua described by the supermodule $\mathcal{M}$, generalizing the bosonic construction in \cite{Cordova:2024iti}. The 2d gapped vacua, labeled by $i$ as objects of the $\mathscr{C}$-module $\mathcal{M}$, are represented by the $1$d boundary of the 2d system. Under the state-operator correspondence, the particle excitation on a fixed gapped vacuum $i$ is mapped to a boundary operator $\mathcal{O}_{ii}$, while the soliton excitations interpolating between different vacua $i,j$ are mapped to boundary-changing operators $\mathcal{O}_{ij}$. The SymTFT construction expands the $(1+1)$d system into a $(2+1)$d bulk with three boundaries. To properly describe the fermionic degrees of freedom, we introduce the fermionic identity line operator $\pi$ in the SymTFT~\cite{Bhardwaj:2024ydc}, which is isomorphic to the ordinary identity line operator. The symmetry data of $\mathscr{C}$ live on the symmetry boundary $B_{\text{sym}}$, which is topological and is characterized by a \emph{fermionic Lagrangian algebra} $\mathcal{L}_{\text{sym}}$ of the bulk SymTFT. The dynamical data are encoded in the physical boundary $B_{\text{phys}}$, which is non-topological in general. The third boundary $B^*_{\text{sym}}$ is also topological, and intersects $B_{\text{sym}}$ and $B_{\text{phys}}$ along the two $1$d corners $\mathfrak{B}_{\mathcal{M}}$ and $\mathfrak{B}_{\text{phys}}$, where the corner $\mathfrak{B}_{\mathcal{M}}$ is described by the supermodule $\mathcal{M}$. The boundary operator $\mathcal{O}_{ij}$ is represented as a line operator $\rho$ living on $B^*_{\text{sym}}$, stretching between $\mathfrak{B}_{\mathcal{M}}$ and $\mathfrak{B}_{\text{phys}}$. We apply the SymTFT method to several examples, including our two illustrative examples, the fermionic Fibonacci category and the q-type $\mathbb{Z}_2$ symmetry, as well as the least relevant deformation of $\mathcal{N}=2$ $A_2$ minimal model, and find perfect agreement in all cases.

This paper is organized as follows. In Section~\ref{Sec: catandalg} we review the
superfusion category and the bosonic strip algebra, and then define the
superstrip algebra as a $C^\ast$-weak Hopf superalgebra. In
Section~\ref{sec_example_of_superstrip} we work out two illustrative examples, a
fermionic Fibonacci category and a q-type $\mathbb Z_2$ symmetry. In
Sections~\ref{sec:N_2_minimal_model} and~\ref{sec:N_1_minimal_model} we apply the
framework to the least relevantly deformed $\mathcal N=2$ and $\mathcal N=1$ minimal models, respectively. In Section~\ref{sec:fermion_number}, we provide a new perspective of the (fractional) fermion number for solitons in those deformed models from categorical symmetry. In
Section~\ref{sec:SymTFT} we recover the representation theory from a boundary
SymTFT. We conclude in Section~\ref{sec:conclusion} with discussions on future
directions. Technical details are
collected in the appendices.

\section{Superfusion category and superstrip algebra}
\label{Sec: catandalg}

In this section, we first review the (super)fusion category and the bosonic strip algebra, and then describe the algebraic properties of the superstrip algebra $\textbf{sStr}_{\mathscr{C}}(\mathcal{M})$. We focus on defining the superstrip algebra as a $C^*$-weak Hopf superalgebra, and leave more details to Appendix~\ref{app:hopf-algebra}. We work out a few examples in the following sections.

\subsection{Review of (super)fusion category}

In 1+1 dimensions, discrete 0-form global symmetries are elegantly unified and generalized through the framework of Topological Defect Lines (TDLs) \cite{Bhardwaj:2017xup,Aasen:2017ubm,Chang:2018iay,Chang:2022hud}. Traditionally, a 0-form symmetry is viewed as a unitary operator acting on the Hilbert space of states. However, in the topological framework, symmetries are elevated to codimension-one extended operators. Because our spacetime is two-dimensional, these extended operators are nothing but one-dimensional lines or loops. Being topological implies that continuous deformations of the line on the spacetime manifold do not alter the correlation functions of the theory, provided the line does not cross any local operator insertions. In other words, the topological nature is a geometric manifestation of charge conservation for group-like symmetries.

In terms of figures, a TDL $\mathcal{L}$ will be drawn as an oriented line. Its orientation reversal is then labeled by $\overline{\mathcal{L}}$. For example, if one considers an ordinary finite group symmetry $G$, a TDL, denoted as $\mathcal{L}_g$ is associated with each group element $g\in G$. Then reversing the orientation of a TDL corresponds to applying the inverse of that symmetry transformation. Mathematically, this relationship is expressed as:
\begin{equation}
    \overline{\mathcal{L}_g} = \mathcal{L}_{g^{-1}}\,.
\end{equation}
Furthermore, if one draws two such lines parallel to one another and brings them together, they fuse according to the standard group multiplication $\mathcal{L}_g \cdot \mathcal{L}_h = \mathcal{L}_{gh}$. In particular, $\overline{\mathcal{L}_g} = \mathcal{L}_{g^{-1}}$ is the inverse of $\mathcal{L}_g$. Meanwhile, the action of $\mathcal{L}_g$ on a charged operator $\mathcal{O}$ can be understood as wrapping the line around $\mathcal{O}$ and shrinking it
\begin{equation}
\begin{gathered}
\begin{tikzpicture}[scale=0.4]
\draw [line,ultra thick,dashed,MidnightBlue] (-.5,-0.75) circle (1.7) ;
\draw (-3.1,-0.75) node {$\mathcal{L}_g$};
\draw (0,1.2) node {};
\draw (-0.5,-0.75)\dotsol {right}{$\mathcal{O}$};
\draw [line,ultra thick,DarkBlue,-<-=1.0](-.47, 0.95) -- (-.46, 0.95);
\draw [line,ultra thick,DarkBlue,-<-=1.0](-.48, -2.45)--(-.49, -2.45);
\end{tikzpicture}
\end{gathered}\quad = \quad
\begin{gathered}
\begin{tikzpicture}[scale=1]
\draw (0,-0.3)\dotsol {below}{$ \widehat{\mathcal{L}}_{g}(\mathcal O)$};
\draw (0,.45) node {};
\end{tikzpicture}
\end{gathered}  
\end{equation}
Considering TDLs as symmetry generators generalizes the conventional notion of symmetry. The necessity of this generalization can be seen in perhaps the simplest non-trivial conformal field theory (CFT), i.e. the critical Ising model. The Ising CFT possesses a standard $\mathbb{Z}_2$ spin-flip symmetry, but it also contains a more exotic TDL associated with the Kramers-Wannier duality. As is well-known, this duality relates high- and low-temperature phase so it becomes a symmetry at the critical point.  If we denote the Kramers-Wannier TDL as $\mathcal{N}$ and the standard $\mathbb{Z}_2$ symmetry line as $\eta$, the fusion of two duality lines does not simply yield the identity operator $\mathcal{I}$. Instead, the fusion obeys a non-group-like algebra: 
\begin{equation}
    \mathcal{N} \cdot \mathcal{N} = I + \eta\,.
\end{equation}
This means $\mathcal{N}$ does not have an inverse, which explains why it is also called a non-invertible symmetry.

Now we present a set of axioms that characterize the non-invertible symmetries. If one restricts to bosonic theories, we can utilize the mathematical framework of a fusion category. In this framework, TDLs are treated as abstract objects, and their intersections or topological junctions are treated as morphisms between objects. However, since fermions play a crucial role in this paper, we need to upgrade the fusion category to superfusion category so as to incorporate the $\mathbb{Z}_2$ fermion parity symmetry $(-1)^F$. Below, we summarize some key properties of superfusion category and highlight those that are new compared to the fusion category:

\begin{itemize}
    \item $\mathbf{Direct\, Sum}$: Given two TDLs $\mathcal{L}_\alpha$ and $\mathcal{L}_\beta$, we can define the sum $\mathcal{L}_\alpha + \mathcal{L}_\beta$ such that the defect Hilbert space is given by a direct sum $\mathcal{H}_{\mathcal{L}_\alpha + \mathcal{L}_\beta} = \mathcal{H}_{\mathcal{L}_\alpha} \oplus \mathcal{H}_{\mathcal{L}_\beta}$. A TDL in the language of category theory corresponds to an object, and we define a \textit{simple} TDL to be a simple object in the superfusion category which cannot be further decomposed \cite{etingof2015tensor}.
        
     \item $\mathbf{Semisimplicity}$   We assume that all TDLs can be decomposed into a direct sum of simple TDLs, and the set of simple TDLs is finite. This holds true for all the cases considered in this paper.

    \item $\mathbf{Junction}$: A set of TDLs $\{\mathcal{L}_a,\mathcal{L}_b,\cdots\}$ can meet at a junction point, and we associate to it a junction vector space. In the language of category theory, it is the space of morphisms between objects. For example if the theory is conformal, the states in the defect Hilbert space at the junction $\mathcal{H}_{\mathcal{L}_a,\mathcal{L}_b,\dots}$ carry a grading given by the conformal weight. Then the junction vector space $V_{\mathcal{L}_a,\mathcal{L}_b,\dots}$ is nothing but the vacuum subspace that has zero weight. 

    \item[\textcolor{red}{\textbullet}] $\mathbf{Grading}$: In an ordinary fusion category, the junction vector spaces are finite-dimensional complex vector spaces, typically of the form $\mathbb{C}^{n}$. A TDL $\mathcal{L}$ is simple if the endomorphism junction space $V_{\mathcal{L},\overline{\mathcal{L}}}$ is isomorphic to $\mathbb{C}$. However for superfusion category, the junction vector spaces become $\mathbb{Z}_2$-graded vector spaces, and this grading dictates the bosonic or fermionic nature of the junction. 
    A typical vector space then takes the form $\mathbb{C}^{n|m}$, with $n$-dimensional bosonic subspace and $m$-dimensional fermionic subspace. 
    From a physical perspective, a TDL $\mathcal{L}$ is regarded simple if its junction vector space $V_{\mathcal{L},\overline{\mathcal{L}}}$ is isomorphic to either $\mathbb{C}^{1|0}$ or $\mathbb{C}^{1|1}$. An isomorphism to $\mathbb{C}^{1|0}$ defines an \emph{m-type} TDL, whereas an isomorphism to $\mathbb{C}^{1|1}$ designates a \emph{q-type} TDL. Ordinary bosonic TDLs within a fusion category, as well as the fermion parity TDL $(-1)^F$ in a superfusion category are examples of m-type. In contrast, a q-type TDL supports a 1D Majorana fermion on its worldline which generates the fermionic subspace in $\mathbb{C}^{1|1}$. Pictorially, we represent such 1D fermion by a red dot. Along the q-type worldline, two red dots can meet and annihilate, and exchanging their relative positions yields a minus sign.
    
\begin{equation}
    \begin{gathered}
        \begin{tikzpicture}[scale=1]
    \draw[ultra thick,dashed,MidnightBlue,-<-=.55] (3.25,-1)--(3.25,1);
    \node at (4,0) {=};
    \draw[ultra thick,dashed,MidnightBlue,-<-=.55] (4.75,-1)--(4.75,1);
    \filldraw[WildStrawberry] (4.75,-0.5) circle (2pt);
    \filldraw[WildStrawberry] (4.75,0.5) circle (2pt);
    \node at (4.5,-0.5) {$1$};
    \node at (4.5,0.5) {$2$};
    \node at (5.5,0) {$=$};
    \node at (6,0) {$-$};
    \draw[ultra thick,dashed,MidnightBlue,-<-=.55] (4.75+2,-1)--(4.75+2,1);
    \filldraw[WildStrawberry] (4.75+2,-0.5) circle (2pt);
    \filldraw[WildStrawberry] (4.75+2,0.5) circle (2pt);
    \node at (4.5+2,-0.5) {$2$};
    \node at (4.5+2,0.5) {$1$};
    \end{tikzpicture}
    \end{gathered}\qquad \,, \qquad
    \begin{gathered}
        \begin{tikzpicture}[scale=1]
    \draw[ultra thick,dashed,MidnightBlue,-<-=.55] (3.5,-1)--(3.5,1);
    \draw[ultra thick,dashed,MidnightBlue,-<-=.55] (4.5,-1)--(4.5,1);
    \filldraw[WildStrawberry] (4.5,-0.5) circle (2pt);
    \filldraw[WildStrawberry] (3.5,0.5) circle (2pt);
    \end{tikzpicture}
    \end{gathered} \quad =\quad - \quad
    \begin{gathered}
        \begin{tikzpicture}[scale=1]
    \draw[ultra thick,dashed,MidnightBlue,-<-=.55] (3.5,-1)--(3.5,1);
    \draw[ultra thick,dashed,MidnightBlue,-<-=.55] (4.5,-1)--(4.5,1);
    \filldraw[WildStrawberry] (4.5,0.5) circle (2pt);
    \filldraw[WildStrawberry] (3.5,-0.5) circle (2pt);
    \end{tikzpicture}
    \end{gathered}    
\end{equation}
    
    \item[\textcolor{red}{\textbullet}] $\mathbf{Fusion}$: Two TDLs $\mathcal{L}_a$ and $\mathcal{L}_b$ can fuse to a composite TDL, and according to the $\mathbf{Direct\, Sum}$ axiom it is further decomposed into a sum of simple TDLs. This decomposition is encoded in the so-called fusion rule,
\begin{equation}
    \mathcal{L}_a \cdot \mathcal{L}_b = \sum_{c,\text{m-type}} \mathbb{C}^{N_{ab}^{c,\,\mathbf{b}}|N_{ab}^{c,\mathbf{f}}} \cdot\mathcal{L}_c + \sum_{c,\,\text{q-type}} N_{ab}^{c} \mathcal{L}_c\,.
\end{equation}
$N_{ab}^{c,\mathbf{b}(\mathbf{f})}$ are the dimensions of the junction vector space
    \begin{equation} V_{\mathcal{L}_a,\mathcal{L}_b,\overline{\mathcal{L}}_c} = \mathbb{C}^{N_{ab}^{c,\mathbf{b}}|N_{ab}^{c,\mathbf{f}}}\,,
    \end{equation}
and we use $\mathbf{b}$ and $\mathbf{f}$ to denote bosonic and fermionic degrees of freedom respectively. When we need to emphasize their difference, we use blue dot for the bosonic subspace and red dot for the fermionic subspace respectively.
\begin{equation}
    \begin{gathered}
\begin{tikzpicture}[scale=1]
\draw [line,dashed,ultra thick,MidnightBlue,->-=.55] (0,0)  -- (0,-1) node [below=-3pt] {$\cL_c$};
\draw [line,dashed,ultra thick,BurntOrange,-<-=.55] (0,0) -- (-1,1) node [above=-3pt] {$\cL_a$};
\draw [line,dashed,ultra thick,ForestGreen,-<-=.55] (0,0) --(1,1) node [above=-3pt] {$\cL_b$};
\node at (0.5,0) {{\color{black}{$\alpha^{\mathbf{b}}$}}};
\filldraw[Cerulean] (0,0) circle (3pt);
\end{tikzpicture}
\end{gathered} \qquad
    \begin{gathered}
\begin{tikzpicture}[scale=1]
\draw [line,dashed,ultra thick,MidnightBlue,->-=.55] (0,0)  -- (0,-1) node [below=-3pt] {$\cL_c$};
\draw [line,dashed,ultra thick,BurntOrange,-<-=.55] (0,0) -- (-1,1) node [above=-3pt] {$\cL_a$};
\draw [line,dashed,ultra thick,ForestGreen,-<-=.55] (0,0) --(1,1) node [above=-3pt] {$\cL_b$};
\node at (0.5,0) {{\color{black}{$\alpha^{\mathbf{f}}$}}};
\filldraw[WildStrawberry] (0,0) circle (3pt);
\end{tikzpicture}
\end{gathered}
\end{equation}    
In the above figure, $\alpha^{\bf b}=1,\dots,N_{ab}^{c,\bf b}$ and $\alpha^{\bf f}=1,\dots,N_{ab}^{c,\bf f}$ label all possible fusion channels. For q-type $\mathcal{L}_c$, we have $N_{ab}^{c,\mathbf{b}}=N_{ab}^{c,\mathbf{f}}=N_{ab}^{c}$.

    \item[\textcolor{red}{\textbullet}] $\mathbf{Crossing}$: Given a general network of TDLs, any two topologically equivalent configurations can be consistently related to one another through a sequence of local transformations. These fundamental crossing relations, commonly known as $F$-moves, encode the associator of the TDL superfusion algebra and are depicted as
 \begin{align}
     \begin{gathered}
 \begin{tikzpicture}[scale=.75]
 \draw [line,dashed,ultra thick,MidnightBlue,->-=.55] (0,0)  -- (0,-1) node [below=-3pt] {$\cL_4$};
 \draw [line,dashed,ultra thick,BurntOrange,-<-=.55] (-1,1) -- (-2,2) node [above=-3pt] {$\cL_1$};
 \draw [line,dashed,ultra thick,ForestGreen,-<-=.55] (0,0) -- (-1,1) ;
 \draw [line,dashed,ultra thick,BurntOrange,-<-=.55] (0,0) --(2,2) node [above=-3pt] {$\cL_3$};
 \draw [line,dashed,ultra thick,MidnightBlue,-<-=.55] (-1,1) --(0,2) node [above=-3pt] {$\cL_2$};
 \draw (-.5,.5) node [ForestGreen,above right=-3pt] {$\cL_5$};
 \draw (0,-0.15) node [left]{\textcolor{black}{$\beta$}};
 \draw (-1,1) node [below]{\textcolor{black}{$\alpha$}};
 \end{tikzpicture}
 \end{gathered}
 \quad =\quad 
 \sum_{\cL_6,\delta,\gamma} \mathcal F_{\cL_4}^{\cL_1 \cL_2 \cL_3}[\cL_5,\cL_6]^{\alpha\beta}_{\delta\gamma}
 \begin{gathered}
 \begin{tikzpicture}[scale=.75]
 \draw [line,dashed,ultra thick,MidnightBlue,->-=.55] (0,0)  -- (0,-1) node [below=-4pt] {$\cL_4$};
 \draw [line,dashed,ultra thick,BurntOrange,-<-=.55] (0,0) -- (-2,2) node [above=-3pt] {$\cL_1$};
 \draw [line,dashed,ultra thick,ForestGreen,-<-=.55] (0,0) -- (1,1) ;
 \draw [line,dashed,ultra thick,BurntOrange,-<-=.55] (1,1) --(2,2) node [above=-3pt] {$\cL_3$};
 \draw [line,dashed,ultra thick,MidnightBlue,-<-=.55] (1,1) --(0,2) node [above=-3pt] {$\cL_2$};
 \draw (.5,.5) node [ForestGreen,above left=-3pt] {$\cL_6$};
 \draw (0,-0.15) node [right]{\textcolor{black}{$\delta$}};
 \draw (1,1) node [below]{\textcolor{black}{$\gamma$}};
 \end{tikzpicture}
 \end{gathered}\,,
 \end{align}
 where we use a compact notation $\alpha=(\alpha^{\bf b},\alpha^{\bf f})$, etc. Consistency yields the famous superpentagon equations,
\begin{align}
\label{eq:superpentagon}
    \sum_\varepsilon \mathcal F^{icd}_e[j,k]^{\beta\chi}_{\delta\varepsilon}\, \mathcal F^{abk}_e[i,l]^{\alpha\varepsilon}_{\phi\gamma}\ =\ (-1)^{s(\alpha) s(\delta)} \sum_{m\kappa\eta\iota} \mathcal F^{abc}_j[i,m]^{\alpha\beta}_{\eta\iota}\,\mathcal F^{amd}_e[j,l]^{\iota\chi}_{\kappa\gamma}\,\mathcal F^{bcd}_l[m,k]^{\eta\kappa}_{\delta\phi}\,,  
\end{align}
where $\cL_i$ is abbreviated to $i$ (and similarly for other indices) and $s(\alpha^{\mathbf{b}}) =0$ while $s(\alpha^{\mathbf{f}}) =1$. It turns out that these equations, together with a complete choice of gauge, severely constrain the possible values of $\mathcal F$-symbols. The number of solutions to \eqref{eq:superpentagon} is finite, and each of them is thus rigid against local deformations. For fusion category this is known as Ocneanu rigidity \cite{etingof2005fusion}, and the finiteness result for superfusion category is proved in \cite{USHER2018453}.
\end{itemize}

\subsection{Review of strip algebra}
Although our primary goal is to study the superstrip algebra, we will first review the standard strip algebra in this section to build essential physical and algebraic intuition. In the next section, we will introduce the superstrip algebra to incorporate the technical complexities associated with fermions.

In a general 2D bosonic theory with a finite global symmetry governed by a fusion category $\mathscr{C}$, the infrared (IR) phase of the symmetry is captured by a $\mathscr{C}$-module category $\mathcal{M}$ \cite{Fuchs:2012dt}. Mathematically, $\mathcal{M}$ can be regarded as a categorical generalization of a representation for a finite group. When the theory is in a gapped phase, the simple objects $v_i$ in $\mathcal{M}$ are in one-to-one correspondence with vacua of the system, denoted by $|i\rangle$. The action of the symmetry defects on these vacua is then dictated by the module structure, yielding the fusion rule:
    \begin{equation}\label{eq:fusion-module}
        \mathcal{L}_a \cdot v_i = \sum_j\widetilde{N}_{ai}^{j} v_j \,,
    \end{equation}
where $\widetilde{N}_{ai}^{j}$ are non-negative integers which counts the dimension of the junction vector space
    \begin{equation}
        V_{\mathcal{L}_a,v_i,\overline{v}_j} = \mathbb{C}^{\widetilde{N}_{ai}^{j}}\,.
    \end{equation}
This action is depicted as
    \begin{equation}
        \begin{gathered}
     \begin{tikzpicture}[scale=1]
        \filldraw[gray,opacity=0.3] (-2.5,0)--(-3.5,0)--(-3.5,2)--(-2.5,2)--(-2.5,0);
        \draw[ultra thick,MidnightBlue,->-=.55] (-2.5,2)--(-2.5,1);
        \draw[ultra thick,MidnightBlue,->-=.55] (-2.5,1)--(-2.5,0);
        \draw[densely dashed,ultra thick,BurntOrange,->-=.6] (-1.25,1)--(-2.5,1);
        \node at (-2.85,1) {$\textcolor{black}\beta$};
         \filldraw[black] (-2.5,1) circle (2pt);
        \node at (-2.9,0.3) {$v_j$};
        \node at (-2.9,1.7) {$v_i$};
        \node at (-.75,1) {$\mathcal{L}_a$};
    \end{tikzpicture}    
        \end{gathered}
    \end{equation}
with $\beta=1,\dots,\widetilde{N}_{ai}^{j}$. The gray region means trivial theory, and we use solid lines and dashed lines to represent vacua and symmetry operators separately. In the following, we will omit the grey region. 
The topological action is strictly constrained by the consistency conditions. Notably, the sequential action of $\mathcal{L}_a,\mathcal{L}_b$ on $|v_i\rangle$ must be isomorphic to the action of their fused composite. This equivalence is formalized by an associator relating $\mathcal{L}_b(\mathcal{L}_a|v_i\rangle)$ and $(\mathcal{L}_b \cdot \mathcal{L}_a) |v_i\rangle$,
whose matrix elements define the $\widetilde{F}$-move coefficients depicted as
    \begin{equation}\label{sec:Ftilde-move}
        \begin{gathered}
    
        \end{gathered}\,.
    \end{equation}
Similar to the $F$-move coefficient in $\mathscr{C}$, the $\widetilde{F}$-move coefficient should also satisfy a module pentagon equation when we consider acting three symmetry TDLs $\mathcal{L}_a,\mathcal{L}_b,\mathcal{L}_c$ in two ways on $|v_i\rangle$.

The Hilbert space of the theory can be defined by specifying the vacua at spatial infinities. For instance, $\mathcal{H}_{i,j}$ denotes 
the Hilbert space where the vacuum at left infinity 
is $|i\rangle$ while $|j\rangle$ at right infinity. If $|i\rangle=|j\rangle$ the Hilbert space $\mathcal{H}_{i,i}$ consists of the particle excitations on the vacuum $|i\rangle$. On the other hand, if $|i\rangle\neq |j\rangle$, $\mathcal{H}_{i,j}$ contains the solitonic states interpolating between two vacua. 

Non-invertible symmetries connect states across different Hilbert spaces, thereby restricting the allowed spectrum. Because these symmetries possess non-trivial kernels and cokernels, they do not act as simple bijective maps. This non-invertibility introduces ambiguities into the standard particle and soliton multiplet structures. To resolve these ambiguities in bosonic systems, recent works \cite{Cordova:2024vsq, Cordova:2024iti} introduced the \textit{strip algebra} $\textbf{Str}_{\mathscr{C}}(\mathcal{M})$ constructed from an input symmetry category $\mathscr C$ and boundary module $\mathcal M$. This algebraic framework provides a systematic method for projecting onto the proper multiplet structure, cleanly organizing the physical state degeneracies dictated by the non-invertible symmetry.

The elements in $\mathbf{Str}_{\mathscr{C}}(\mathcal{M})$ are then represented as the following diagram
\begin{equation}
\begin{gathered}
     \begin{tikzpicture}[scale=1]
        \draw[ultra thick,MidnightBlue,-<-=.55] (1,0)--(1,1);
        \draw[ultra thick,MidnightBlue,-<-=.55] (1,1)--(1,2);
        \draw[ultra thick,MidnightBlue,->-=.65] (2.5,0)--(2.5,1);
        \draw[ultra thick,MidnightBlue,->-=.7] (2.5,1)--(2.5,2);
        \draw[densely dashed,ultra thick,BurntOrange,-<-=.55] (1,0.9)--(2.5,1.15);
        \node at (0.7,.9) {$\textcolor{black}\alpha$};
        \node at (2.8,1.15) {$\textcolor{black}\beta$};
         \filldraw[black] (1,0.9) circle (2pt);
         \filldraw[black] (2.5,1.15) circle (2pt);
        \node at (1.25,0.25) {$i$};
        \node at (1.25,1.75) {$k$};
        \node at (2.25,0.25) {$j$};
        \node at (2.25,1.75) {$l$};
        \node at (1.75,0.75) {$a$};
    \end{tikzpicture}
\end{gathered}\,,
\end{equation}
\begingroup\tolerance=1000\emergencystretch=1em
where we use the subscripts $a$ for $\mathcal{L}_a$ and $i,j,k,l$ for $v_i,v_j,v_k,v_l$ in the diagram, and $\alpha = 1,\dots, \widetilde{N}_{ak}^i,\  \beta=1,\dots,\widetilde{N}_{\bar{a},j}^l$ label the fusion channel between $\mathscr{C}$ with $\mathcal{M}$. It maps elements from the Hilbert space $\mathcal{H}_{i,j}$ to $\mathcal{H}_{k,l}$.  We summarize some key properties of the strip algebra below and refer to \cite{Cordova:2024vsq, Cordova:2024iti} for more details.
\par\endgroup

\begin{itemize}
    \item \textbf{Multiplication}: For two elements in $\mathbf{Str}_{\mathscr{C}}(\mathcal{M})$, the multiplication map $\mu:\mathbf{Str}_{\mathscr{C}}(\mathcal{M})\otimes \mathbf{Str}_{\mathscr{C}}(\mathcal{M})\rightarrow \mathbf{Str}_{\mathscr{C}}(\mathcal{M})$ is given by vertically concatenating the two elements, with the left element on top of the right.
\begin{align}
\begin{gathered}

    \end{gathered}\,.
        \end{align}
It can be further simplified by applying the $\widetilde{F}$-move \eqref{sec:Ftilde-move} and its dual,
    \begin{equation}
        \begin{gathered}
            %
    \end{gathered}\,,
    \end{equation}
whose explicit expressions are given in \cite{Cordova:2024vsq, Cordova:2024iti}.
        
\item \textbf{Unit}: The unit map $\eta : \mathbb{C}\rightarrow \mathbf{Str}_{\mathscr{C}}(\mathcal{M})$ sends $1$ to the unit element of the strip algebra
\begin{equation}
    \eta :\quad 1 \quad \rightarrow \quad \sum_{i,j} \quad
\begin{gathered}
        %
    \end{gathered}\,,
\end{equation}
where $I$ is the identity TDL.

\item \textbf{Comultiplication}: The action of the comultiplication map $\Delta:\mathbf{Str}_{\mathscr{C}}(\mathcal{M}) \rightarrow \mathbf{Str}_{\mathscr{C}}(\mathcal{M}) \otimes \mathbf{Str}_{\mathscr{C}}(\mathcal{M})$ on a given element yields a tensor product defined as \cite{Cordova:2024vsq, Cordova:2024iti}
\begin{equation}
    \begin{gathered}
        %
    \end{gathered}\,,
\end{equation}
and in particular, 
    \begin{equation}
        \Delta(\eta(1)) \quad = \quad \sum_{i,j,k}\quad \begin{gathered}
        %
    \end{gathered}\,.
    \end{equation}
Although $\Delta(\eta(1))$ is not the identity element in the tensor product, it is idempotent, satisfying $\mu(\Delta(\eta(1)), \Delta(\eta(1))) = \Delta(\eta(1))$. Thus, it acts as a projector. This behavior has a natural physical motivation. Note that defining multi-particle and soliton states relies crucially on the comultiplication. Suppose we have two states $|\psi_{ij}\rangle \in \mathcal{H}_{i,j}$ and $|\psi_{kl}\rangle \in \mathcal{H}_{k,l}$ transforming under representations $R_1$ and $R_2$ of the strip algebra, with the indices indicating the asymptotic vacua at $x\rightarrow \mp \infty$. The standard tensor products for the states and representations are $|\psi_{ij}\rangle \otimes_{\mathbb{C}} |\psi_{kl}\rangle$ and $R_1 \otimes_{\mathbb{C}}R_2$, respectively. Because a composite state is only physically meaningful if the intermediate vacua align (namely, $j=k$), we must project out the incompatible configurations. By using $\Delta(\eta(1))$ as a projector, we isolate the valid composite representation as $R_1 \otimes R_2 := \Delta(\eta(1)) (R_1 \otimes_{\mathbb{C}}R_2)$. The strip algebra element $h\in \mathbf{Str}_{\mathscr{C}}(\mathcal{M})$ then acts on this physical subspace via $\Delta(h)$. As a technical remark, relaxing the requirement that $\Delta(\eta(1))$ be the identity element gives the strip algebra the structure of \textit{weak} Hopf algebra.


\item \textbf{Counit} The counit map $\epsilon:\mathbf{Str}_{\mathscr{C}}(\mathcal{M})\rightarrow \mathbb{C}$ is given by
\begin{equation}
    \epsilon:\quad
\begin{gathered}
        \begin{tikzpicture}[scale=1]
            \draw[ultra thick,MidnightBlue,->-=.55] (0,1)--(0,0);
            \draw[ultra thick,MidnightBlue,->-=.65] (0,0)--(0,-1);
            \draw[ultra thick,MidnightBlue,->-=.75] (1,0)--(1,1);
            \draw[ultra thick,MidnightBlue,->-=.65] (1,-1)--(1,0);
            \draw[densely dashed,ultra thick,BurntOrange,-<-=.6] (0,0)--(1,0.15);
            \node at (0,1.25) {$i$};
            \node at (1,1.25) {$j$};
            \node at (0,-1.25) {$k$};
            \node at (1,-1.25) {$l$};
            \node at (-0.25,0.05) {$\textcolor{black}\alpha$};
            \node at (1.25,0.15) {$\textcolor{black}\beta$};
            \node at (0.5,0.35) {$a$};
            \filldraw[black] (0,0) circle (1.5pt);
            \filldraw[black] (1,0.15) circle (1.5pt);
        \end{tikzpicture}
    \end{gathered}\quad \rightarrow \quad \delta_{ij} \delta_{kl} \delta_{\alpha \beta} \widetilde{\mathcal{F}}^{k \bar{a} i,\alpha \alpha}_{akI,11} \frac{d_a}{d_i}\,,
\end{equation}
where $d_a$ is the quantum dimension of $\mathcal{L}_a$. The counit can be viewed as the inverse of comultiplication satisfying $(\epsilon\otimes \eta)\circ \Delta = (\eta \otimes \epsilon)\circ \Delta=id$. From a physical perspective, it reduces $\mathbf{Str}_{\mathscr{C}}(\mathcal{M})\otimes \mathbf{Str}_{\mathscr{C}}(\mathcal{M})$ to $\mathbf{Str}_{\mathscr{C}}(\mathcal{M})$, allowing it to act on $R=R_1\otimes R_2$ effectively as a single reducible representation.

\item \textbf{Antipode} The antipode is a $\mathbb C$-linear map $S:\mathbf{Str}_{\mathscr{C}}(\mathcal{M})\rightarrow \mathbf{Str}_{\mathscr{C}}(\mathcal{M})$ 
\begin{equation}
    S:\quad \begin{gathered}

    \end{gathered}\,.
\end{equation}
The antipode defines an action of $h\in \mathbf{Str}_{\mathscr{C}}(\mathcal{M})$ on the dual representation defined by $R^{\vee}:=\textrm{Hom}(R,\mathbb{C})$, which is equivalently characterized by the non-degenerate pairing $(\cdot , \cdot): R^{\vee} \otimes R \rightarrow \mathbb{C}$. For any $\phi\in R^{\vee}$, the action of $h$ on $R^{\vee}$ is defined by $(h\phi,v)= (\phi ,S(h)v)$ for all $v\in R$.

\item \textbf{Conjugate} Finally, the hermitian conjugate is a $\mathbb C$-anti-linear map $\dagger : \mathbf{Str}_{\mathscr{C}}(\mathcal{M}) \rightarrow \mathbf{Str}_{\mathscr{C}}(\mathcal{M})$
\begin{equation}
\dagger :\quad 
    \begin{gathered}

    \end{gathered}\,,
\end{equation}
which is needed to define a Hilbert space.
\end{itemize}

In the next section, we discuss in detail how to construct irreducible representations. For now, we simply introduce a useful visual tool: if we are only interested in the spectrum rather than the detailed action of the strip algebra in a representation, we can make use of the so-called quiver diagram. It contains $|\mathcal{M}|$ nodes where $|\mathcal{M}|$ is the number of simple objects in $\mathcal{M}$ with a bunch of arrows:
\begin{itemize}
    \item The $|\mathcal{M}|$ nodes designate the distinct clustering vacua of the theory;
    \item Self-loops at a given node represent particle excitations above the corresponding vacuum;
    \item Arrows linking distinct nodes indicate solitonic states that interpolate between different vacua.
\end{itemize}

As an illustrative example, consider the Fibonacci fusion category $\mathscr{C}_{\text{Fib}}$, which has a single nontrivial simple object $W$ satisfying \begin{equation}
        W\times W=1+W\,.
    \end{equation}
There is a unique irreducible module $\mathcal{M}$, which coincides with the fusion category $\mathscr{C}_{\text{Fib}}$ itself. Physically, the $\mathscr{C}_{\text{Fib}}$ symmetry is spontaneously broken, resulting in two distinct vacua labeled by $|1\rangle$ and $|W\rangle$. The two irreducible representations are in one-to-one correspondence with two simple TDLs, leading to the following quiver diagrams:
\begin{itemize}
    \item Representation labeled by the identity has two degenerate particle excitations
\begin{figure}[H]
\centering
    \begin{tikzpicture}[
    dot_style/.style={circle, fill, inner sep=1.5pt},
    blue_arc/.style={Cerulean, thick, -Stealth, bend left=25},
    red_arc/.style={WildStrawberry, thick, -Stealth, bend left=25},
    label_below/.style={below=2pt}, scale=1
]

\node at (-1.5, 0) {$\mathcal R_{1}$:};

    \node[dot_style] (n0) at (0, 0) {};
    \node[dot_style] (n1) at (3, 0) {};
    \draw[Cerulean, thick, -Stealth, bend left=25] (3.2,0) arc (-70:250:0.5);
    \draw[Cerulean, thick, -Stealth, bend left=25] (.2,0) arc (-70:250:0.5);
    \node at (0,-0.75) {$|1\rangle$};
    \node at (3,-0.75) {$|W\rangle$};
\end{tikzpicture}
\end{figure}
    \item Representation labeled by $W$ has three states: a particle on the $W$-vacuum, and a pair of soliton and anti-soliton.
\begin{figure}[H]
\centering
    \begin{tikzpicture}[
    dot_style/.style={circle, fill, inner sep=1.5pt},
    blue_arc/.style={Cerulean, thick, -Stealth, bend left=25},
    red_arc/.style={WildStrawberry, thick, -Stealth, bend left=25},
    label_below/.style={below=2pt}, scale=1
]

\node at (-1.5, 0) {$\mathcal R_{W}$:};

    \node[dot_style] (n0) at (0, 0) {};
    \coordinate (n0pr) at (0.05, .15) {};
    \coordinate (n0nr) at (0.1, -.15) {};

    \node[dot_style] (n1) at (3, 0) {};
   
    \coordinate (n1pl) at (2.95, .15) {};
    \coordinate (n1nl) at (2.9, -.15) {};

    \draw[blue_arc] (n0pr) to (n1pl);
    \draw[blue_arc] (n1nl) to (n0nr);
    \draw[Cerulean, thick, -Stealth, bend left=25] (3,-0.2) arc (-160:160:0.5);
    \node at (0,-0.75) {$|1\rangle$};
    \node at (3,-0.75) {$|W\rangle$};
\end{tikzpicture}
\end{figure}
\end{itemize}

\subsection{Basic properties of superstrip algebra}
In this section, we generalize the above discussion to fermionic systems, introducing the \textit{superstrip algebra} $(\mathbf{sStr}_{\mathscr{C}}(\mathcal{M}),\mu,\eta,\Delta,\epsilon,S,\dagger)$ in order to incorporate the aforementioned fermionic nature of TDLs and junctions. Similar to the bosonic theories, we assume that the gapped phase of a 2D fermionic theory can be described by a supermodule $\mathcal{M}$ of the superfusion category $\mathscr{C}$ with the fusion rule
    \begin{equation}
        \mathcal{L}_a \cdot v_i = \sum_j\widetilde{N}_{ai}^{j,\bf b} v_j + \sum_j\widetilde{N}_{ai}^{j,\bf f} v_j \,,
    \end{equation}
where $\widetilde{N}_{ai}^{j,\bf b},\widetilde{N}_{ai}^{j,\bf f}$ are now two sets of non-negative integers reflecting the dimension of the junction vector space
    \begin{equation}
        V_{\mathcal{L}_a,v_i,\overline{v}_j} = \mathbb{C}^{\widetilde{N}_{ai}^{j,\bf b}|\widetilde{N}_{ai}^{j,\bf f}}\,,
    \end{equation}
which is depicted as
    \begin{equation}
        \begin{gathered}
     \begin{tikzpicture}[scale=1]
        \draw[ultra thick,MidnightBlue,->-=.55] (-2.5,2)--(-2.5,1);
        \draw[ultra thick,MidnightBlue,->-=.55] (-2.5,1)--(-2.5,0);
        \draw[densely dashed,ultra thick,BurntOrange,->-=.6] (-1.25,1)--(-2.5,1);
        \node at (-2.85,1) {$\textcolor{black}\beta$};
         \filldraw[black] (-2.5,1) circle (2pt);
        \node at (-2.9,0.3) {$v_j$};
        \node at (-2.9,1.7) {$v_i$};
        \node at (-.75,1) {$\mathcal{L}_a$};
    \end{tikzpicture}    
        \end{gathered}\quad {\color{black}{\beta}} = ( {\color{black}{\beta^{\bf b}}},\,{\color{black}{\beta^{\bf f}}} )
    \end{equation}
with $\beta^{\bf b}=1,\dots,\widetilde{N}_{ai}^{j,\bf b}$ and $\beta^{\bf f}=1,\dots,\widetilde{N}_{ai}^{j,\bf f}$. The basis in $\textbf{sStr}_{\mathscr{C}}(\mathcal{M})$ takes the following form
\begin{equation}\label{fig:superstrip}
\begin{gathered}
     \begin{tikzpicture}[scale=1]
        \draw[ultra thick,MidnightBlue,-<-=.55] (1,0)--(1,1);
        \draw[ultra thick,MidnightBlue,-<-=.55] (1,1)--(1,2);
        \draw[ultra thick,MidnightBlue,->-=.65] (2.5,0)--(2.5,1);
        \draw[ultra thick,MidnightBlue,->-=.7] (2.5,1)--(2.5,2);
        \draw[densely dashed,ultra thick,BurntOrange,-<-=.55] (1,0.9)--(2.5,1.15);
        \node at (0.7,.9) {$\textcolor{black}\alpha$};
        \node at (2.8,1.15) {$\textcolor{black}\beta$};
         \filldraw[black] (1,0.9) circle (2pt);
         \filldraw[black] (2.5,1.15) circle (2pt);
        \node at (1.25,0.25) {$i$};
        \node at (1.25,1.75) {$k$};
        \node at (2.25,0.25) {$j$};
        \node at (2.25,1.75) {$l$};
        \node at (1.75,0.75) {$a$};
    \end{tikzpicture}
\end{gathered}\,, \quad {\color{black}{\alpha}} = ( {\color{black}{\alpha^{\bf b}}},\,{\color{black}{\alpha^{\bf f}}} ),\quad  {\color{black}{\beta}} = ( {\color{black}{\beta^{\bf b}}},\,{\color{black}{\beta^{\bf f}}} )\,.
\end{equation}
The subscripts $a$ and $i$ refer to 
the TDL ${\cal L}_a$ and the vacuum $|i\rangle$. Compared to the strip algebra, the new ingredient is the introduction of $\alpha$ and $\beta$ dots to dictate the spin-statistics of the junction. 
Our convention is that the possible left dot $\alpha$ is by default lower than the possible right dot $\beta$. 
In the following, we will discuss the properties of superstrip algebra as a $C^*$-weak Hopf superalgebra. Because this construction closely parallels the strip algebra discussed in the previous section, we present only the core definitions here and defer further details to Appendix~\ref{app:hopf-algebra}.

\begin{itemize}
    \item $\mathbb{Z}_2$-\textbf{grading} A superalgebra means an algebra with a $\mathbb{Z}_2$ grading, which we label as $\mathbf{b}$ and $\mathbf{f}$. Namely we have a vector space decomposition $\textbf{sStr}_{\mathscr{C}}(\mathcal{M}) = \textbf{sStr}_{\mathscr{C}}(\mathcal{M})^\mathbf{b} \oplus \textbf{sStr}_{\mathscr{C}}(\mathcal{M})^\mathbf{f}$.
As the names suggest, they correspond to bosonic or fermionic degrees of freedom respectively. Given two elements $\alpha$ and $\beta$ in the superstrip algebra,
their product belongs to $\textbf{sStr}_{\mathscr{C}}(\mathcal{M})^{s(\alpha)+s(\beta)}$ where $s(\alpha), s(\beta) \in \mathbb{Z}_2$ equals 0 or 1 depending on the node being $\mathbf{b}$ or $\mathbf{f}$ respectively. 

\item \textbf{Multiplication} Given two elements in $\mathbf{sStr}_{\mathscr{C}}(\mathcal{M})$, the multiplication $\mu:\mathbf{sStr}_{\mathscr{C}}(\mathcal{M})\otimes \mathbf{sStr}_{\mathscr{C}}(\mathcal{M})\rightarrow \mathbf{sStr}_{\mathscr{C}}(\mathcal{M})$ is defined by composing the left element on top of the right, analogous to the multiplication in the strip algebra,
\begin{equation}
\begin{gathered}

    \end{gathered}\,,
        \end{equation}
Regarding the tensor product, we adopt the convention that the middle line of the left tensor factor is positioned strictly higher than that of the right factor. This is necessary in order to incorporate the Koszul sign rule. Furthermore, we can simplify the right-hand side using a sequence of module $F$-moves. We give a detailed derivation in Appendix~\ref{App:multiplication}, so we only present the final result
\begin{equation}
\begin{aligned}
\label{eq_multiplication}
   \mu: \quad \begin{gathered}
        %
    \end{gathered}\,,
    \end{aligned}
        \end{equation}
where the module $F$-move, dual module $F$-move and $\widetilde{O}$ coefficients are defined in that appendix. The sign factor $(-1)^{s({\alpha}) s({\delta})}$ arises because we need to flip the order between $\alpha$ and $\delta$ when merging the two symmetry operators. From this it is not hard to see that the $\mathbb{Z}_2$-grading structure for the product is satisfied.

\item \textbf{Unit} The unit map $\eta : \mathbb{C}\rightarrow \mathbf{sStr}_{\mathscr{C}}(\mathcal{M})$ maps $1$ to the unit element of the multiplication $\mu$, constructed the same way as the strip algebra.


\item \textbf{Comultiplication} Given an element in $\mathbf{sStr}_{\mathscr{C}}(\mathcal{M})$, the comultiplication
\[
\Delta:\mathbf{sStr}_{\mathscr{C}}(\mathcal{M}) \rightarrow \mathbf{sStr}_{\mathscr{C}}(\mathcal{M}) \otimes \mathbf{sStr}_{\mathscr{C}}(\mathcal{M})
\]
maps it to a tensor product defined by
\begin{equation}\label{eq_comultiplication}
    \Delta: \begin{gathered}
        %
\end{gathered}\,,
\end{equation}
where $\Lambda$ is a possible phase factor of moving a right node to the left \cite{Aasen:2017ubm}. In particular, 
$\Delta(\eta(1))$ is defined in the same way as the strip algebra in the previous section and serves as the projection operator. We emphasize again that not requiring $\Delta(\eta(1))$ to be identity makes the superstrip algebra a weak Hopf superalgebra only.

\item \textbf{Counit} The counit map $\epsilon:\mathbf{sStr}_{\mathscr{C}}(\mathcal{M})\rightarrow \mathbb{C}$ is given by
\begin{equation}
    \epsilon:\quad
\begin{gathered}
        \begin{tikzpicture}[scale=1]
            \draw[ultra thick,MidnightBlue,->-=.55] (0,1)--(0,0);
            \draw[ultra thick,MidnightBlue,->-=.65] (0,0)--(0,-1);
            \draw[ultra thick,MidnightBlue,->-=.75] (1,0)--(1,1);
            \draw[ultra thick,MidnightBlue,->-=.65] (1,-1)--(1,0);
            \draw[densely dashed,ultra thick,BurntOrange,-<-=.6] (0,0)--(1,0.15);
            \node at (0,1.25) {$i$};
            \node at (1,1.25) {$j$};
            \node at (0,-1.25) {$k$};
            \node at (1,-1.25) {$l$};
            \node at (-0.25,0.05) {$\textcolor{black}\alpha$};
            \node at (1.25,0.15) {$\textcolor{black}\beta$};
            \node at (0.5,0.35) {$a$};
            \filldraw[black] (0,0) circle (1.5pt);
            \filldraw[black] (1,0.15) circle (1.5pt);
        \end{tikzpicture}
    \end{gathered}\quad \rightarrow \quad \Lambda^{s(\beta) }\delta_{i,j} \delta_{k,l} \delta_{\alpha\beta} \widetilde{F}_{akI,\bb\bb}^{k\bar{a}i,\alpha\alpha} \frac{d_a}{d_i}\,.
\end{equation}
where $d_a$ is the quantum dimension of $\mathcal{L}_a$. The detailed derivation can be found in Appendix~\ref{App:multiplication}.

\item \textbf{Antipode} The antipode is a $\mathbb C$-linear map $S:\mathbf{sStr}_{\mathscr{C}}(\mathcal{M})\rightarrow \mathbf{sStr}_{\mathscr{C}}(\mathcal{M})$. Compared to the strip algebra, the extra overall factor arises because we need to exchange the relative position of the $\alpha$ and $\beta$ nodes.
\begin{equation}
    S:\quad \begin{gathered}

    \end{gathered}\,.
\end{equation}

\item \textbf{Conjugate} Finally, the hermitian conjugate is 
a $\mathbb C$-anti-linear map $\dagger : \mathbf{sStr}_{\mathscr{C}}(\mathcal{M}) \rightarrow \mathbf{sStr}_{\mathscr{C}}(\mathcal{M})$,
defined exactly as it is for the strip algebra. It is the adjoint action for operators on a Hilbert space.

\end{itemize}

In this part, we talk about the representation theory of superstrip algebra. A finite-dimensional $C^\ast$-superalgebra is semisimple, so its representation can always be decomposed into direct sums of irreducible ones. This is just a super version of the familiar property for the $C^\ast$-algebra. This makes its representation theory well-behaved and easier to study.
In fact, if we take $\mathcal M$ to be $\mathscr C$ itself, there is a systematic algorithm to determine all irreducible representations from the regular representation, which we explain now. First, we need to find a set of idempotes $e_i$ in the algebra that satisfies
\begin{equation}
    e_i e_j = \delta_{i,j} e_i\,.
\end{equation}
They form a complete set if we further have
\begin{equation}
   \text{id}=  e_1 + \cdots e_n\,.
\end{equation}
A complete set of idempotes is called minimal if the number $n$ is maximal. Given such a complete and minimal set, we can decompose the superstrip algebra $\mathcal{A}$ into a direct sum of representations
\begin{equation}\label{eq:Adecom}
    \mathcal{A} = \bigoplus_i \mathcal{A} e_i\,, 
\end{equation}
Here we regard $\mathcal{A}$ itself as a representation via its defining action and $\mathcal{A} e_i$ as a subrepresentation by acting the superstrip algebra from the left. The $\mathcal{A} e_i$ is irreducible if and only if $e_i$ cannot be written as a non-trivial sum of orthogonal idempotes. Therefore, each summand in \eqref{eq:Adecom} is irreducible because of our assumption. Two representations $\mathcal{A} e_i$ and $\mathcal{A} e_j$ are equivalent when there exists an element $\alpha$ such that
\begin{equation}
    \mathcal{A} e_i = \mathcal{A} e_j\cdot\alpha\,.
\end{equation}

We consider some explicit examples in Section~\ref{sec_example_of_superstrip} where we explicitly construct a complete set of idempotes. There is, however, one caveat: because we always assume the fermion parity $(-1)^F$ is unbroken, the supermodule category $\mathcal M$ is not precisely $\mathscr C$, but rather $\mathscr C/\mathbb{Z}_2^{(-1)^F}$. Consequently, for each irreducible representation there exists a parity-shifted one by exchanging the statistics of all excitations, which may be missed by the above constructions. A more satisfactory solution is provided by the boundary SymTFT framework, which will be discussed in Section~\ref{sec:SymTFT}. 

Finally, we also make use of the quiver diagram to encode the particle/soliton spectrum of a representation. Compared to the purely bosonic case, the only new feature is the use of blue and red arrows to denote bosonic and fermionic excitations respectively. We note that, as shown in \cite{Cordova:2024vsq}, a theory has a trivial one-form symmetry if and only if the quiver diagram for its complete set of stable solitons is connected. In Section~\ref{sec_example_of_superstrip}, we will draw the quiver diagrams for all corresponding irreducible representations.

\section{Examples of superstrip algebras}\label{sec_example_of_superstrip}

In this section, we work out two simple and concrete examples of superfusion categories, both involve a trivial TDL $I$ and a non-trivial $W$. The first example is a fermionic variant of the Fibonacci fusion category, where $W$ is an m-type non-invertible TDL satisfying $W^2 = \mathbb{C}^{1|0} I + \mathbb{C}^{1|1}W$; the second example is the q-type $\mathbb{Z}_2$ symmetry such that $W$ is a q-type TDL satisfying $W^2=\mathbb{C}^{1|1} I$. We take the supermodule category to be the superfusion category $\tilde{\mathscr C}$, which is obtained by taking the quotient of $\mathscr{C}$ by $(-1)^F$; that is, $\mathscr C=\tilde{\mathscr C}\boxtimes\mathbb Z_2^{(-1)^F}$. The choice $\mathcal M\simeq\tilde{\mathscr C}$ describes the gapped spontaneous symmetry breaking (SSB) phase in fermionic theories. We work out the associated superstrip algebra $ \mathbf{sStr}_{\mathscr{C}}(\tilde{\mathscr C})$ and their irreducible representations.

\subsection{Fibonacci superfusion category}\label{sec:superFibo}
In this case, we have a trivial TDL $I$ and an m-type TDL $W$. The fusion rules are
\begin{equation}\label{eq:fusionso3_6}
    I^2 = \mathbb{C}^{1|0}I\,,\quad W  I = I W = \mathbb{C}^{1|0} W\,, \quad W^2 = \mathbb{C}^{1|0} I + \mathbb{C}^{1|1}W\,,
\end{equation}
which are represented as 
\begin{equation}
    \begin{gathered}

\end{gathered}\,,
\end{equation} 
where $I$ and $W$ will be represented in black and green lines throughout this subsection. This superfusion category can be obtained as the fermion condensation of $\widehat{so(3)}_6$ fusion category as follows. Recall that there are four simple objects $\{I,X,Y,Z\}$ in $\widehat{so(3)}_6$ fusion category with the following non-trivial fusion rules,
\begin{equation}
    X^2 = I + X + Y,\quad X Y = X + Y + Z,\quad Y^2 = I + X + Y,\quad XZ = Y,\quad YZ = X,\quad Z^2 = I.
\end{equation}
Since $Z$ has topological spin $-1$, we can condense $Z$ using the standard fermion condensation procedure. Then $X$ and $Y$ are identified with each other and become a single m-type TDL in the condensed theory. One can further check all the fusion rules in \eqref{eq:fusionso3_6}.

\begin{figure}[!ht]
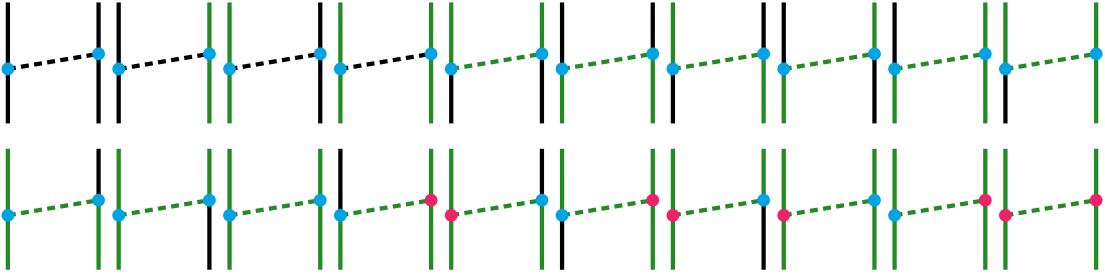

\begin{equation*}
    \begin{gathered}
    \begin{gathered}

\end{gathered}
    \end{gathered}
\end{equation*}
\caption{Generators of superstrip algebra for Fibonacci superfusion category.}
\label{graph:strip_so3_6}
\end{figure}

Now let us turn to the superstrip algebra. It has a basis of $20$ elements as shown in Figure~\ref{graph:strip_so3_6},
where we do not allow the Majorana fermion at the junction to move through the horizontal line since $W$ is m-type. As discussed in the last section, we 
need to find a maximal set of idempotes $e_i$ in the algebra satisfying
\begin{equation}
    e_i e_j = \delta_{i,j} e_i\,,\quad \text{id}=  e_1 + \cdots e_n\,.
\end{equation}
First of all, we have an obvious complete set of idempotes
\begin{equation}\label{eq:Cm11decom}
    \text{id} \quad = \quad e_1 + e_2 + e_3 + e^\prime_4 \quad = \quad    \begin{gathered}

\end{gathered}\quad\,.
\end{equation}
In order to further decompose the idempotes while maintaining orthogonality conditions, we need to seek for elements with the same left and right boundary conditions as some element in the above summand. Examining Figure~\ref{graph:strip_so3_6}, the only possibility is to decompose $e'_4$ in terms of
\begin{equation}
    \begin{gathered}
     %
\end{gathered}
\end{equation}
After trial and error, we find the decomposition
\begin{equation}
    e'_4 = e_4+e_5+e_6\,,
\end{equation}
where
\begin{equation}
\begin{split}
    &e_4 = \quad  \omega^{-1}\quad \begin{gathered}
     %
\end{gathered}\quad\,.
\end{split}
\end{equation}
One can show that together with $e_1,e_2,e_3$ in the RHS of \eqref{eq:Cm11decom}, they form a complete and minimal set. 

With a convenient choice of basis, the irreducible representations $\mathcal{A} e_i$ are generated by
\begin{equation}
\begin{gathered}
    \mathcal{A}e_1\quad : \quad \left\{ \quad \begin{gathered}

\end{gathered}\quad\right\}
\end{align}
\endgroup

Suppose we have a fermionic gapped phase where the superfusion category is spontaneously broken, then the vacua are also labeled by the TDLs. The particles/solitons are transformed under some representations of the superstrip algebra, which can be decomposed into the direct sum of the irreducible representations. The decomposition of regular representation of superstrip algebra gives us three irreducible representations $\mathcal R_1$, $\mathcal R_2$ and $\widetilde {\mathcal R}_2$. The type and statistics of the states can be read from the upper pair of vertical lines and the total number of Majorana fermions (red dots) in each ladder diagram in $\mathcal{A}e_i$. 

Let us examine them one by one. In the 2-dimensional representation $\mathcal R_1 \simeq \mathcal{A}e_1 \simeq \mathcal{A}e_5$, it consists of two particle-like excitations in the $I$-vacuum and $W$-vacuum, respectively, and both of them are bosonic since the number of Majorana fermions is even. Therefore, its quiver diagram takes the form
\begin{figure}[H]
\centering

\end{figure}
Similarly, in the 4-dimensional representation $\mathcal R_2 \simeq \mathcal{A}e_2 \simeq \mathcal{A}e_3 \simeq \mathcal{A}e_4$, it has two particle-like excitations in the $W$-vacuum, one is bosonic and the other is fermionic; it also has a pair of bosonic soliton/anti-soliton interpolating between two vacua. In other words its quiver diagram has the following form
\begin{figure}[H]
\centering
    %
\end{figure}
Finally, the other four dimensional representation $\widetilde {\mathcal R}_2 \simeq \mathcal{A}e_6$ is very similar to $\mathcal R_2$, with the only difference being the soliton/anti-soliton interpolating between two vacua are fermionic. Its quiver diagram becomes
\begin{figure}[H]
\centering
    %
\end{figure}

In fact, in Section~\ref{sec:SymTFT} when we revisit the Fibonacci superfusion category from the boundary SymTFT picture, we will see there is another representation $\widetilde{\mathcal R}_1$, which is a variant of $\mathcal R_1$ with both particle-like excitations in the $I$-vacuum and $W$-vacuum are fermionic. It cannot be seen from the decomposition of the superstrip algebra.

\subsection{\texorpdfstring{q-type $\mathbb{Z}_2$ symmetry}{q-type Z2 symmetry}}\label{sec:Z2qsuperfusion}
The next example is the q-type $\mathbb{Z}_2$-symmetry, where we have a trivial TDL $I$ and a q-type TDL $W$. The fusion rules are
\begin{equation}\label{eq:fusionfermionic_ising}
    I^2 = \mathbb{C}^{1|0}I\,,\quad I W = W I = \mathbb{C}^{1|0}W,\quad W^2 = \mathbb{C}^{1|1}I\,.
\end{equation}
which are depicted as 
\begin{equation}
    \begin{gathered}

\end{gathered}
\,,
\end{equation} 
where $I$ and $W$ are still represented in black and green lines throughout this subsection. This superfusion category can be obtained from the fermion condensation of Ising fusion category, in which we have three TDLs $\{I,\eta,\mathcal{N}\}$ with non-trivial fusion rules,
\begin{equation}
    \eta \mathcal{N} = \mathcal{N}\eta = \mathcal{N},\quad \mathcal{N}^2 = I+\eta\,.
\end{equation}
If we condense the line $\eta$, $\mathcal{N}$ becomes a q-type TDL in fermionic theory with the fusion rules given in \eqref{eq:fusionfermionic_ising}.

One can write down 25 elements for the the superstrip algebra as shown in Figure~\ref{graph:strip_fermionic_ising}.
\begin{figure}
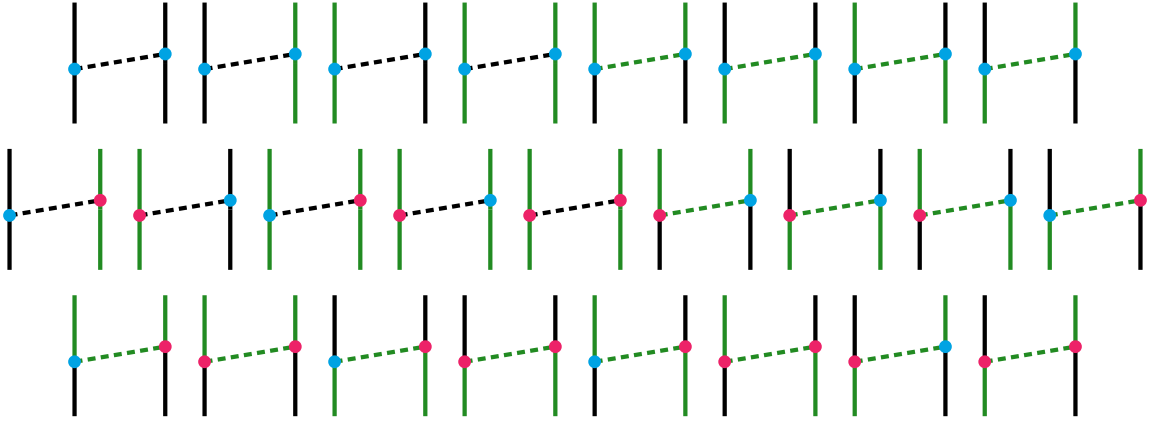

\begin{equation*}
    \begin{gathered}
    \begin{gathered}

\end{gathered}
\end{gathered}
\end{equation*}
\caption{Generators of the superstrip algebra for q-type $\mathbb{Z}_2$ symmetry.}
\label{graph:strip_fermionic_ising}
\end{figure}
However, the elements in the last line are not independent, since one can move the Majorana fermion along the green line $W$, so that they are equivalent to the elements in the second lines. Therefore, only 17 of them are independent and span a basis of the superstrip algebra.

The complete set of idempotes is
\begin{equation}
    \text{id} \quad = \quad e_1 + e_2 + e_3 + e^\prime_4 \quad = \quad    \begin{gathered}

\end{gathered}\quad\,,
\end{equation}
and $e^{\prime}_4$ is further decomposed as
\begin{equation}
   e^{\prime}_4 =  e_4 + e_5\,,
\end{equation}
in terms of
\begin{equation}
    e_4 \quad = \quad \frac{1}{2} \quad \begin{gathered}
     %
\end{gathered}\quad\,.
\end{equation}

With a convenient choices of basis, the irreducible representations $\mathcal{A} e_i$ are generated by
\begin{equation}
    \begin{gathered}
        \mathcal{A}e_1\quad : \quad     \left\{ \quad \begin{gathered}
     %
\end{gathered}\quad\right\}
    \end{gathered}
\end{equation}

In a gapped phase where the q-type TDL $W$ is spontaneously broken, the representation $R_1 \simeq \mathcal{A}e_1 \simeq \mathcal{A}e_4$ consists of three particle-like excitations. Two of them are in the $W$-vacuum and carry different statistics; the third one is a boson in the $I$-vacuum. There is a similar representation $\widetilde{R}_1 \simeq \mathcal{A}e_5$, with the only difference being that the particle in the $I$-vacuum is fermionic. Therefore their quiver diagrams are
\begin{figure}[H]
\centering
    %
\end{figure}

The representation $R_2 \simeq \mathcal{A}e_2 \simeq \mathcal{A}e_3$ consists of a boson-fermion soliton and anti-soliton pair. In terms of quiver diagram, we have
\begin{figure}[H]
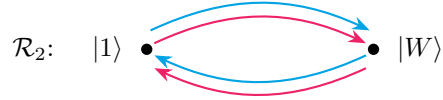

\centering
    %
\caption{The quiver diagram of the representation $\mathcal R_{2}$.}
\label{fig:Cq0quiver2}
\end{figure}
We will also revisit this example from the point of view of boundary SymTFT in Section~\ref{sec:SymTFT}.

\section{\texorpdfstring{Least relevant deformation of $\mathcal{N}=2$ minimal models}{Least relevant deformation of N=2 minimal models}}
\label{sec:N_2_minimal_model}
In this and the following section, we systematically discuss gapped phases arising from deformation of CFTs. In this section, we review and expand upon the results concerning the least relevant deformation of $\mathcal{N}=2$ minimal models, as originally presented in the short version of this work \cite{Chen:2025qub}. In the next section, we proceed to the $\mathcal{N}=1$ minimal models which require a more sophisticated analysis.

We start with some basic facts of $\mathcal{N}=2$ minimal models which can be found in standard textbooks such as \cite{Blumenhagen:2009zz}. Recall that the $\mathcal{N}=2$ superconformal algebra is an extension of Virasoro algebra, constructed by adding a $U(1)$ current $J(z)$ and two supercurrents $G^{\pm}(z)$ which are superpartners of the energy-momentum tensor $T(z)$. Depending on the choice of spin structure, the modes of $G^{\pm}(z)$ take either half-integer or integer values, corresponding to the Neveu-Schwarz (NS) and Ramond (R) sectors, respectively. Non-trivial algebraic relations are given by
\begin{equation}
\begin{aligned}
        \relax [L_m, L_n] &= (m-n) L_{m+n} +\frac{c}{12}(m^3-m)\delta_{m,-n}\,,\\
        \{G_{m}^+, G_{n}^-\} &= 2L_{m+n} +(m-n)J_{m+n} +\frac{c}{3}(m^2-\frac{1}{4})\delta_{m,-n}\,,\\
        [L_m ,G_n^{\pm}] &= (\frac{m}{2}-n) G_{m+n}^{\pm}\,,\\
        [L_m, J_n] &= -n J_{m+n}\,,\\
        [J_m, G_{n}^{\pm}] &= \pm G_{m+n}^\pm\,,\\
        [J_m,J_n] &= \frac{c}{3}m \delta_{m,-n}\,,
\end{aligned}
\end{equation}
where $\{,\}$ denotes anti-commutator and $c$ is the central charge. In particular, we learn that $L_0$ and $J_0$ commute with each other so they can label a given state in the Hilbert space simultaneously,
\begin{equation}
    L_0 |h, j\rangle = h |h, j\rangle,\quad J_0 |h, j\rangle = j |h, j\rangle\,.
\end{equation}
Regarding its representation theory, analogous to the Virasoro minimal model, there exists a discrete family of superconformal field theories governed by the $\mathcal{N}=2$ superconformal algebra. Known as the $\mathcal{N}=2$ superconformal minimal model, they are characterized by the value of $c$ labeled by a positive integer $k$, 
\begin{equation}
    c = \frac{3k}{k+2},\quad k\geq 1\,.
\end{equation}
For a given $k$, there exists a finite number of superconformal primary fields $\phi_{l,m}^s$ with the conformal weight and $U(1)$ charge,
\begin{equation}
    h_{l,m}^s = \frac{l(l+2) - (m-2s)^2}{4(k+2)} + \frac{s^2}{2},\quad j_{l,m}^s = \frac{m+ks}{k+2}\,,
\end{equation}
where $s = 0$ in the NS sector and $s = -\frac{1}{2}$ in the R sector. $l,m$ take values in the set
\begin{equation}
    (l,m) \in \{l = 0,\dots k,\ |m| \leq l,\ l =m\, (\text{mod}\ 2)\}\,.
\end{equation}

For the full superconformal theory, we need to combine the left and right moving sectors which are both representations of the $\mathcal{N}=2$ superconformal algebra. Therefore, the partition function takes the form
\begin{equation}
    Z^{(s_x,s_t)}_{\text{torus}}(q,\overline{q}) = \sum_{i,i^\prime} N^{(s_x,s_t)}_{i,i^\prime} \chi_i(q) \overline{\chi}_{i^\prime}(\overline{q}) 
\end{equation}
where $s_x, s_t$ label the NS or R sector, $\chi_i(q)$ are superconformal characters of superconformal primary fields and $N^{(s_x,s_t)}_{i,i^\prime}$ are non-negative integers known as modular pairing. Modular invariance of the torus partition function imposes strict constraints on consistent modular pairings, making a general classification highly non-trivial. However, a complete classification exists for the $\mathcal{N} = 2$ superconformal minimal models. We summarize these results in the following table \cite{Cappelli:1986ed,Gepner:1986hr,Gepner:1987qi}.
\begin{equation}
    \begin{array}{l|cccccc}
\text{type} & A_{n} & D_{n+1}^{(n \text{ odd})} & D_{n+1}^{(n \text{ even})} & E_6 & E_7 & E_8 \\
\hline
k & n-1 & 2n-2 & 2n-2 & 10 & 16 & 28\\ 
I_W^{\mathcal{N}=2} & n & n+1 & n+1 & 6 & 7 & 8 
\end{array}
\end{equation}
We also provide the Witten index $I_W^{\mathcal{N}=2}$ of each theory, which is nothing but the torus partition function with R-R spin structure $Z^{(R,R)}_{\text{torus}}(q,\overline{q})$.

Next, we deform the $\mathcal{N} = 2$ minimal models by the least relevant operator. This deformation is conveniently described using the Landau-Ginzburg (LG) formalism, which provides an explicit construction of a supersymmetric theory that flows to the desired minimal model in the infrared (IR) limit \cite{Vafa:1988uu}.

We work in 2d $\mathcal{N} = (2,2)$ superspace with coordinates $x^\mu   (\mu = 0,1)$ and Grassmann variables $\theta^\pm, \bar{\theta}^\pm$. A chiral superfield $\Phi$ is defined by the constraint $\overline{D}_{\dot{\alpha}} \Phi = (-\frac{\partial}{\partial \overline{\theta}^{\dot{\alpha}} } - i \sigma^\mu_{\alpha \dot{\alpha}}\theta^\alpha \partial_\mu)\Phi = 0$ which yields the component expansion
\begin{equation}
    \Phi(x^\mu,\theta^{\pm}) \sim \phi + \theta^+\psi_+ + \theta^- \psi_- + \theta^+ \theta^- F\,.
\end{equation}
Supersymmetric Lagrangians are easily constructed using Grassmann integration:
\begin{equation}
\mathcal{L} = \int d^4\theta \, \mathcal{K}(\Phi,\overline{\Phi}) + \left( \int d^2\theta \, \mathcal{W}(\Phi) + h.c. \right)\,,
\end{equation}
where $\mathcal{K}$ is the K\"{a}hler potential and $\mathcal{W}$ is the superpotential. This formalism extends trivially to multiple chiral superfields. Crucially, it is the superpotential $\mathcal W$ that dictates the IR fixed point of the theory.

We begin with the $A_{k+1}$-type minimal model, which can be described by a single chiral superfield $\Phi$ with the superpotential
\begin{equation}
    \mathcal{W}(\Phi) = \frac{1}{k+2} \Phi^{k+2}\,.
\end{equation}
To preserve supersymmetry, we deform the superpotential using the relevant chiral operators $\Phi^i$, $i=1,\dots,k$ which flow to the chiral primaries of the IR superconformal theory. A generic perturbation takes the form
\begin{align}
    \delta{\cal W}(\Phi) =  \sum_{j=2}^{k+1} \frac{\nu_j}{k+2-j} \Phi^{k+2-j}\, , \label{eq:N2defW}
\end{align}
where the complex parameters $\nu_j$ are chosen such that the deformed theory is completely massive with $k+1$ vacua. Since the Witten index of the $A_{k+1}$-type minimal model is $(k+1)$, we conclude that every vacuum must be bosonic.

The structure of BPS solitons interpolating the vacua is remarkably rich. As shown in \cite{Cecotti:1992rm}, the BPS spectrum is piecewise constant across the parameter space $\{\nu_j\}$, undergoing discontinuous jumps only when crossing walls of marginal stability. In this work, our focus is on the specific spectrum that emerges when the model is perturbed solely by the least relevant chiral primary field. In the LG description, it corresponds to a polynomial deformation with the leading term being $\Phi^k$ \cite{Ambrosino:2026kfu}. For simplicity, we denote this deformed theory by $A_{k+1}^{\rm def}$.

In fact, $A_{k+1}^{\rm def}$ is known to be integrable. Based on the analysis of \cite{Dijkgraaf:1990dj}, the IR effective superpotential ${\cal W}_\text{def}(\Phi)$ becomes the Chebyshev polynomial
\begin{equation}
    {\cal W}_\text{def}(\Phi = 2\cos\theta) = \frac{2}{k+2}\cos(k+2)\theta = \frac{1}{k+2}\Phi^{k+2} - \Phi^k + \cdots\,,
\end{equation}
after rescaling $\Phi$ to set the deformation parameter to one. The vacua are located at the critical points, i.e., $\partial_\Phi{\cal W}_\text{def} = 0$:
\begin{equation}
    \Phi_p = 2\cos \frac{\pi p}{k+2},\quad p=1,\dots,k+1\,.
\end{equation}
Thus, all vacua are distributed on the real line. Stable BPS states only exist between adjacent vacua. Therefore the spectrum comprises $k$ solitons and anti-solitons between $\Phi_p$ and $\Phi_{p+1}$ for $p=1,\dots,k$, whose masses are fixed by the BPS bound
\begin{equation}
    m_p = |\mathcal{W}(\Phi_p) - \mathcal{W}(\Phi_{p+1})| = \frac{4}{k+2}\,.
\end{equation}

In other words, all solitons and anti-solitons have the same mass. This exact mass degeneracy is not a mere coincidence; as will be explained below, this follows directly from the fact that they form an irreducible representation of the superstrip algebra.

First we must determine the preserved superfusion category\footnote{Since there could be emergent TDLs in the IR, our discussion is restricted to those descending from the UV.} in $A_{k+1}^{\rm def}$. To identify it, we can take a shortcut. If we bosonize the fermionic theory, it becomes a well-known coset model
\begin{equation}
    \dfrac{\mathrm{SU}(2)_k \times \mathrm{U}(1)_2}{\mathrm{U}(1)_{k+2}}\,.
\end{equation}
It contains $2(k+1)(k+2)$ bosonic primaries $\phi_{(a,c)}$ and their corresponding Verlinde lines $\mathcal L_{(a,c)}$, which furnish a fusion category denoted by $\mathscr C_b$. Here, the indices take values $a=0,1,\dots,k$ and $c\in\mathbb Z_{2k+4}$. Here we follow the notation in \cite{Gray:2008je, Cordova:2023qei}. The modular $S$-matrix is given by
\begin{equation}\label{eq:Smatrix}
     S_{(a,c),(a',c')}=\frac{1}{k+2}\sin\left(\frac{\pi(a+1)(a'+1)}{k+2}\right)\, e^{i\pi\frac{cc'}{k+2}}e^{-i\pi\frac{[a+c][a'+c']}{2}}\,,   
\end{equation}
where $[x]\equiv x\ ({\rm mod}\ 2)$. The original fermionic model $A_{k+1}$ can be recovered via fermionizing the $\mathbb{Z}_2$ symmetry TDL $\mathcal L_{(k,k+2)}$.

In $A_{k+1}$, there are $k+1$ relevant superconformal primary operators in the NS sector, denoted by $\Phi_{(l,l)}$ for $l=0,1,\dots,k$. As mentioned above, they originated from the chiral superfields $\Phi^i$ and are known as chiral primaries. They have holomorphic conformal weight equal to half of their ${\rm U}(1)_R$-charge $q_{(l,l)}$,
\begin{align}
    h_{(l,l)}=\frac{q_{(l,l)}}{2}=\frac{l}{2k+4}\,.
\end{align}
In the bosonized coset theory, $\Phi_{(l,l)}$ decomposes as
\begin{equation}
\Phi_{(l,l)}\equiv\left(\phi_{(l,l)},\,\phi_{(k-l,l+k+2)}\right)\,.
\end{equation}

Clearly $\Phi_{(k,k)}\equiv\left(\phi_{(k,k)},\,\phi_{(0,2k+2)}\right)$ is the least relevant operator. After integrating over the superspace, the deformation is controlled by the top component $\phi_{(0,2k+2)}$
\begin{equation}
\lambda\int{\rm d}^2x\,{\rm d}^2\theta\,\Phi_{(k,k)}=\lambda \int {\rm d}^2x\ \phi_{(0,2k+2)}\,.
\end{equation}

To determine the preserved TDLs under such deformation, we first compute the preserved fusion category $\tilde{\mathscr C}_b$ in the deformed bosonic theory $\mathcal B\left[A_{k+1}^{\rm def}\right]$. Performing a further fermionic anyon condensation then yields the preserved superfusion category $\tilde{\mathscr C}_{A_{k+1}}$.

The preserved TDLs that commute with $\phi_{(0,2k+2)}$ perturbation should satisfy the following condition \cite{Chang:2018iay}:
\begin{equation}
\begin{gathered}
\begin{tikzpicture}[scale=.6]
\draw [line,ultra thick,dashed,MidnightBlue] (-.5,-0.75) circle (1.125) ;
\draw (0,-2.5) node[scale=.6] {$\mathcal{L}_{(a,c)}$};
\draw (0,1.2) node {};
\draw (-0.5,-0.75)\dotsols{[scale=.6, outer sep=0pt]below=0pt}{$\phi_{(0,\,2k+2)}$};
\draw [line,ultra thick,DarkBlue,-<-=1.0](-.47, 0.375) -- (-.46, 0.375);
\draw [line,ultra thick,DarkBlue,-<-=1.0](-.48, -1.875)--(-.49, -1.875);
\end{tikzpicture}
\end{gathered}\quad = \quad
\begin{gathered}
\begin{tikzpicture}[scale=.6]
\draw [line,ultra thick,dashed,MidnightBlue] (-.5,-0.75) circle (1.125) ;
\draw (0,-2.5) node[scale=.6] {$\mathcal{L}_{(a,c)}$};
\draw (0,1.2) node {};
\draw (1.7,-0.75)\dotsols{[scale=.6, outer sep=0pt]below=0pt}{$\phi_{(0,\,2k+2)}$};
\draw [line,ultra thick,DarkBlue,-<-=1.0](-.47, 0.375) -- (-.46, 0.375);
\draw [line,ultra thick,DarkBlue,-<-=1.0](-.48, -1.875)--(-.49, -1.875);
\end{tikzpicture}
\end{gathered}
=
\left\langle\mathcal L_{(a,\,c)}\right\rangle|\phi_{(0,\,2k+2)}\rangle
\,,
\end{equation}

i.e. the charge of $\mathcal L_{(a,\,c)}$ on $\phi_{(0,\,{2k+2})}$ equals its own quantum dimension,
\begin{align}
    \left\langle\mathcal L_{(a,\,c)}\right\rangle=\frac{S_{(a,c),(0,0)}}{S_{(0,0),(0,0)}}=Q_{(a,c)}\left(\phi_{(0,2k+2)}\right)=\frac{S_{(a,c),(0,2k+2)}}{S_{(0,0),(0,2k+2)}}\,.
\end{align}
From the explicit form of the modular $S$-matrix \eqref{eq:Smatrix} we find that the preserved TDLs in $\mathcal B[A_{k+1}^{\rm def}]$ are  
\begin{align}
\tilde{\mathscr  C}_b=\{\,\mathcal L_{(a,\,0)},\ \mathcal L_{(a,\,k+2)}|a=0,1,\dots,k\}\,.
\end{align}
Their fusion rule follows from those in the bosonized $\frac{{\rm SU}(2)_k\times {\rm U}(1)_2}{{\rm U}(1)_{k+2}}$,
{\small
\begin{align}
   &\mathcal L_{(a,\,c)} \cdot \mathcal L_{(a',\,c')} = \sum_{(\alpha, \gamma)} N_{(a,c),(a',c')}^{(\alpha, \gamma)} \mathcal L_{(\alpha,\, \gamma)}\,,\\\notag
   &N_{(a,c),(a',c')}^{(\alpha, \gamma)} =
\begin{cases}
(N_{\text{SU(2)}_k})_{a,a'}^{\alpha} (N_{\text{U(1)}_{k+2}})_{c,c'}^{\gamma}\\ 
\qquad\qquad\qquad\qquad\quad\ \text{if } [a+c]\cdot[a'+c']=0 \\\\
(N_{\text{SU(2)}_k})_{a,a'}^{k-\alpha} (N_{\text{U(1)}_{k+2}})_{c,c'}^{\gamma+k+2}\\ 
\qquad\qquad\qquad\qquad\quad\  \text{if } [a+c]\cdot[a'+c']=1
\end{cases}
\end{align}
}
\!\!\!\!\! where $N_{\text{SU(2)}_k}$ and $N_{\text{U(1)}_{k+2}}$ are the fusion matrices of Kac-Moody algebra $\text{SU(2)}_k$ and $\text{U(1)}_{k+2}$ respectively. To further perform the fermionic condensation with $\mathcal{L}_{(k,k+2)}$, it is necessary to notice that
\begin{equation}
    \mathcal L_{(a,c)}\cdot\mathcal L_{(k,k+2)}=\mathcal L_{(k-a,c+k+2)}\,.
    \label{eq:Z2_on_C}
\end{equation}
\begingroup\tolerance=1000\emergencystretch=1em
Via fusion $\mathcal L_{(k,\,k+2)}$ acts as a $\mathbb{Z}_2$-symmetry on $\tilde{\mathscr C}_b$ and there is no fixed point. For the orbit $\{\mathcal L_{(a,\, 0)},\mathcal L_{(k-a,\, k+2)}\}$, we denote $\mathcal L_a\equiv \mathcal L_{(a,0)}$ as the representative.\footnote{Choosing a different representative of the orbit may change the statistics assigned to the junction space in \eqref{eq: su_2_N=2} and hence to the arrows in the quiver diagram. Physical properties, including the supermultiplet structure, remain independent of this choice.} They furnish the condensed superfusion category $\tilde{\mathscr C}_{A_{k+1}}=\{\,\mathcal L_a\,|a=0,1,2,\dots k\}$ in $A_{k+1}^{\rm def}$ with fusion rule:
\par\endgroup
\begin{equation}
    \mathcal L_a\cdot \mathcal L_{a'}=
    \begin{cases}
        \mathbb C^{1|0}\sum_{\alpha=|a-a'|\,,\,{\rm step}\ 2}^{\min(a+a',\, 2k-a-a')}  \mathcal L_\alpha\,,\quad {\rm if}\ \ a\cdot a'\ {\rm even}\,, \\[5pt]
        \mathbb C^{0|1}\sum_{\alpha=|a-a'|\,,\,{\rm step}\ 2}^{\min(a+a',\, 2k-a-a')}  \mathcal L_\alpha\,,\quad {\rm if}\ \ a\cdot a'\ {\rm odd}\,,
    \end{cases}
    \label{eq: su_2_N=2}
\end{equation}
where the fermionic junction arises whenever the fusion of the objects $\mathcal L_a$ and $\mathcal L_{a'}$ does not belong to the chosen set of representatives $\{\mathcal L_{(a,0)}\,|\,a=0,1,\dots,k\}$. For example,
\begin{align}
    \mathcal L_{(1,0)}\cdot \mathcal L_{(1,0)}&=\mathbb C^{1|0}\,\mathcal L_{(k,k+2)}+\mathbb C^{1|0}\,\mathcal L_{(k-2,k+2)}\notag\\
    &\xrightarrow[\mathcal L_{(k,k+2)}]{{\rm condense}}\mathbb C^{0|1}\,\mathcal L_{0}+\mathbb C^{0|1}\,\mathcal L_{2}\,.
\end{align}
In summary, our strategy to determine $\tilde{\mathscr C}_{A_{k+1}}$ can be visualized in Figure~\ref{fig:Commutative_diagram}.

\begin{figure}
    \centering
\begin{tikzpicture}[
    node distance=3.5cm and 4.5cm,
    every node/.style={align=center}, 
    arrow_style/.style={-latex, thick, Teal}, 
    text_style/.style={Teal}
]

\node (top_left) at (0, 2.5) {$\color{Teal}\dfrac{\mathrm{SU}(2)_k \times \mathrm{U}(1)_2}{\mathrm{U}(1)_{k+2}}$};
\node (top_right) at (6, 2.5) {$\color{Teal}\mathcal{N}=2 \ \ A_{k+1}$};
\node (bottom_left) at (0, 0) {$\color{Teal}\mathcal{B}\big[A_{k+1}^{\text{def}}\big]$};
\node (bottom_right) at (6, 0) {$\color{Teal}\mathcal{N}=2 \ \ A_{k+1}^{\text{def}}$};

\draw[sharp arrow, thick] (top_left) -- (bottom_left); 
\node at (-.5,1.5) {$\Phi_{(k,k)}^{\rm top}$};
\node at (-.6,1.167) {\rm deform.};
\draw[sharp arrow, thick] (top_right) -- (bottom_right); 
\node at (5.5,1.5) {$\Phi_{(k,k)}$};
\node at (5.4,1.167) {\rm deform.};

\draw[sharp arrow, thick] ([yshift=2pt]top_left.east) -- ([yshift=2pt]top_right.west) node[midway, above] {fermionize $\mathcal{L}_{(k,k+2)}$};
\draw[sharp arrow, thick]  ([yshift=-2pt]top_right.west) -- ([yshift=-2pt]top_left.east)  node[midway, below] {gauge $(-1)^F$};

\draw[sharp arrow, thick] ([yshift=2pt]bottom_left.east) -- ([yshift=2pt]bottom_right.west) node[midway, above] {fermionize $\mathcal{L}_{(k,k+2)}$};
\draw[sharp arrow, thick] ([yshift=-2pt]bottom_right.west) -- ([yshift=-2pt]bottom_left.east) node[midway, below] {gauge $(-1)^F$};

\node[font=\large, text_style] (c) at (0, -.917) {${\tilde{\mathscr{C}}_b}$};
\node[font=\large, text_style] (cf) at (6.1, -.917) {$\tilde{\mathscr C}_{A_{k+1}}$};

\node at (0, -.417) {$\bigcup$};
\node at (6, -.417) {$\bigcup$};

\draw[sharp arrow, thick, Teal] (c) -- (cf) node[midway, below] {fermionic condensation of $\mathcal{L}_{(k,k+2)}$};

\end{tikzpicture}
    \caption{Commutative diagram to compute the preserved superfusion category $\tilde{\mathscr C}_{A_{k+1}}$ in $A_{k+1}^{\rm def}$. }
    \label{fig:Commutative_diagram}
\end{figure}
\noindent

Once the preserved superfusion category is determined, we can apply the superstrip algebra associated with $\tilde{\mathscr C}_{A_{k+1}}$ to analyze the soliton spectra of the deformed $\mathcal N=2$ minimal models. As mentioned at the end of Section~\ref{Sec: catandalg}, we expect the IR spectrum to be encoded by a connected quiver diagram. In the present case, the simplest irreducible representation with connected diagram is generated by the element ${\mathcal L}_1\in \tilde{\mathscr C}_{A_{k+1}}$. Following the fusion rules in \eqref{eq: su_2_N=2}, the action of $\mathcal L_1$ generates an $A$-type quiver connecting all $k+1$ vacua which are all bosonic,
\begin{align}\label{fig:R_L1}
    \begin{gathered}

    \end{gathered}
\end{align}

 Here, the blue and red lines represent the bosonic and fermionic junctions, respectively, which arise from the fusion of boundaries with the ${\mathcal L}_1$-line. The quiver encodes the morphisms between Hilbert spaces defined by vacua $|r\rangle$ and $|r+1\rangle$ at $\sigma=\pm \infty$, thus capturing the boson/fermion degeneracies within those sectors (see Figure~\ref{fig:degeneracy}).

\begin{figure}[H]
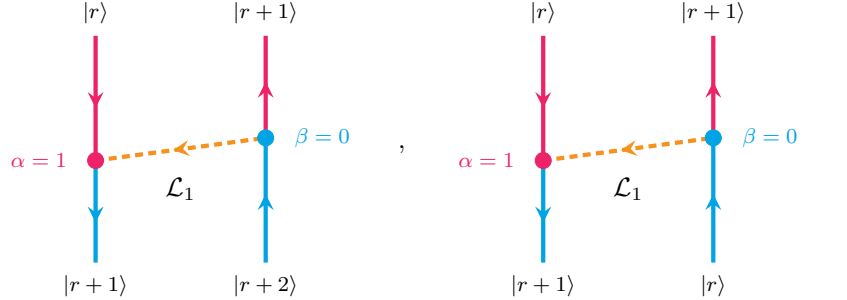
 
    \begin{subfigure}{\columnwidth}
        \centering 
    %
 
    \end{subfigure}
    \caption{Degeneracies between fermionic/bosonic states in soliton/soliton
     sectors $\mathcal H_{r,r+ 1}\leftrightarrows\mathcal H_{r+1,r+2}$, and soliton/anti-soliton ones $\mathcal H_{r,r+1}\leftrightarrows\mathcal H_{r+1,r}$ for odd $r$.}
    \label{fig:degeneracy}
\end{figure}

From this representation, we deduce the following spectral properties: 
\begin{itemize}
\item The $k+1$ vacua lack local particle excitations; all gapped states are solitons connecting adjacent vacua. 
\item Each (anti-)soliton sector $\mathcal H_{r,r+1}$ ($\mathcal H_{r+1,r}$) contains at least one bosonic or fermionic soliton state, denoted by $b_{r,r+1}$ or $f_{r,r+1}$ ($\bar b_{r+1,r}$ or $\bar f_{r+1,r}$). 

\item These states must have the same mass since they are in the same irreducible representation $\mathcal R_{{\mathcal L}_1}$, 
{
\begin{align}
\cdots=m_{\bar f_{r, r-1}}=m_{b_{r-1,r}}=m_{f_{r,r+1}}=m_{\bar b_{r+1,r}}=\cdots\,.
\end{align}
}
\end{itemize}
Due to the preserved supersymmetry for BPS states, any state (e.g. $f_{r,r+1}$) must have a super-partner ($b_{r,r+1}$ ) within the same Hilbert space sector $\mathcal H_{r,r+1}$. Together, they form an irreducible $\mathcal N=2$ massive short multiplet $s_{r,r+1} =(b_{r,r+1},f_{r,r+1})$. Therefore, the non-invertible symmetries impose constraints that are "orthogonal" to SUSY: while SUSY dictates the bosonic/fermionic pairing within each sector, the non-invertible symmetries organize the $2k$ (anti-)solitonic SUSY multiplets into a single multiplet of the superstrip algebra. Combined with supersymmetry, this result is in remarkable agreement with the spectra of the deformed $A_{k+1}^{\rm def}$ models discovered using integrability \cite{Bernard:1990ti,Fendley:1991ve}.

We can also generalize these results to the least relevant deformation of $D$ and $E$ type superconformal minimal models. At the conformal point, these $D/E$-type models are known to be generalized orbifolds of specific $A$-type theories: $D_{k+2}$ arises from $A_{2k+1}$, and $E_{6,7,8}$ from $A_{11,17,29}$, respectively. In each case, this relationship is realized by gauging the non-invertible symmetries associated with a Frobenius algebra, $\mathscr A_{D/E}$ \cite{Fuchs:2002cm, Carqueville:2012dk, Diatlyk:2023fwf}. Crucially, $\mathscr A_{D/E}$ survives the least relevant deformation and is preserved within the symmetry category $\tilde{\mathscr C}_A$ of the corresponding deformed $A$-type model (e.g., $A_{2k+1}^{\rm def}$). Hence the gapped $D/E$-type theory can be constructed by gauging the algebra object $\mathscr A_{D/E}$ directly within the deformed $A$-type theory—an operation that remains valid throughout the entire RG flow. Schematically, we have a commutative diagram in Figure~\ref{fig:preserved_DE}.

\begin{figure}
    \centering
\begin{tikzpicture}[
    node distance=3.5cm and 4.5cm,
    every node/.style={align=center}, 
    arrow_style/.style={-latex, thick, Teal}, 
    text_style/.style={Teal}
]

\node (top_left) at (0, 2.5) {$\mathcal{N}=2 \ \ A$-type};
\node (top_right) at (6, 2.5) {$\mathcal{N}=2 \ \ D/E$-type};
\node (bottom_left) at (0, 0) {$\mathcal{N}=2 \ \ A_{\ast}^{\text{def}}$};
\node (bottom_right) at (6, 0) {$\mathcal{N}=2 \ \ D_{\ast}^{\text{def}}/E_{\ast}^{\text{def}}$};

\draw[sharp arrow, thick] (top_left) -- (bottom_left); 
\node at (-.5,1.5) {$\Phi_{(k,k)}$};
\node at (-.6,1.167) {\rm deform.};
\draw[sharp arrow, thick] (top_right) -- (bottom_right); 
\node at (5.32,1.5) {least rel.};
\node at (5.4,1.167) {\rm deform.};

\draw[sharp arrow, thick] ([yshift=2pt]top_left.east) -- ([yshift=2pt]top_right.west) node[midway, above]{gauge $\mathscr A_{D/E}$};

\draw[sharp arrow, thick] ([yshift=2pt]bottom_left.east) -- ([yshift=2pt]bottom_right.west) node[midway, above] {gauge $\mathscr A_{D/E}$};

\end{tikzpicture}
    \caption{Commutative diagram to compute the preserved superfusion category in $D_\ast^{\rm def}$/$E_\ast^{\rm def}$. }
    \label{fig:preserved_DE}
\end{figure}
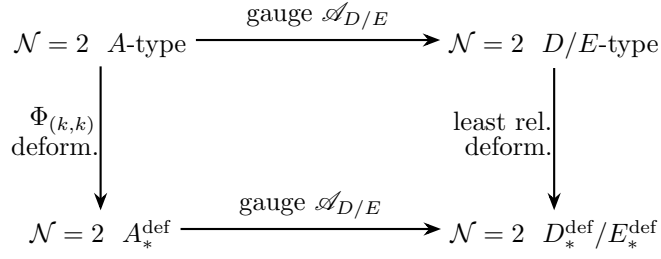

Now let us discuss the construction in detail. First, the $D_{k+2}$ minimal model is obtained from its $A_{2k+1}$ counterpart by gauging the $\mathbb Z_2$ symmetry: $\Phi\rightarrow -\Phi$. It has the following superpotential in the LG description
\begin{align}
    &\mathcal W_{A_{2k+1}}(\Phi)=\frac{1}{2k+2}\Phi^{2k+2}\notag\\
    \xrightarrow[\mathbb Z_2{\text{-gauging}}]{\Phi\rightarrow -\Phi,\,X\equiv\Phi^2}\,&\mathcal W_{D_{k+2}}(X,\,Y)=\frac{1}{2k+2}X^{k+1}+\frac{1}{2}XY^2\,.
\end{align}
Furthermore, in the $A_{2k+1}$ theory, the least relevant operator $\Phi_{(2k,2k)}=\Phi^{2k}+\cdots$ is invariant under the $\mathbb Z_2$-symmetry. Therefore it is preserved under the $\mathbb Z_2$-orbifolding and maps to the least relevant operator $\sim X^k+\cdots$. We thus conclude that the least relevantly deformed $D$-type model, $D_{k+2}^{\rm def}$, can be achieved by gauging the $\mathbb Z_2$ symmetry of $A_{2k+1}^{\rm def}$ theory. Categorically, the $\mathbb Z_2$-line to be gauged in $A_{2k+1}^{\rm def}$ is ${\mathcal L}_{2k}\in \tilde{\mathscr C}_{A_{2k+1}}$. Together with $\mathcal L_0$, it forms an algebra object in $\tilde{\mathscr C}_{A_{2k+1}}$,
\begin{align}
    \mathscr A_{D_{k+2}}\equiv\mathcal L_0\oplus\mathcal L_{2k}\,,
\end{align}
which defines a topological interface between the theories $A_{2k+1}^{\rm def}$ and $D_{k+2}^{\rm def}$.
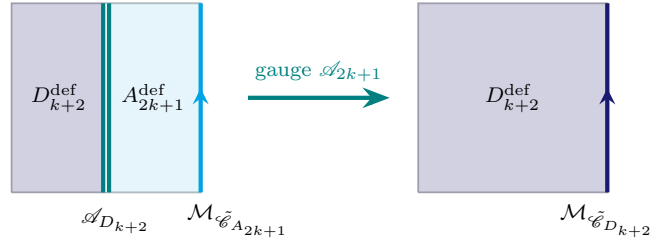
\begin{figure}[H]
    \centering
     \begin{tikzpicture}[scale=1.251]
        \draw [line,lightgray] (0,0)  -- (0,2) -- (2,2) -- (2,0) -- (0,0);
        \filldraw[MidnightBlue,opacity=0.2] (0,0)--(1,0)--(1,2)--(0,2)--(0,0);
        \filldraw[Cerulean,opacity=0.1] (1,0)--(2,0)--(2,2)--(1,2)--(1,0);
        \draw[ultra thick,Cerulean, ->-=.55] (2,0)--(2,2);
        \draw[ultra thick, double=white, draw=Teal] (1,0)--(1,2);
        \node at (.5, 1) {\footnotesize $D_{k+2}^{\rm def}$};
        \node at (1.5, 1) {\footnotesize $A_{2k+1}^{\rm def}$};
        \node at (2.4, -.3) {\footnotesize $\mathcal M_{\tilde{\mathscr C}_{A_{2k+1}}}$};
        \node at (1.1, -.3) {\footnotesize $\mathscr A_{D_{k+2}}$};
       \draw[sharp arrow, ultra thick, Teal] (2.5,1) -- (4,1) node[midway, above=2pt] {\footnotesize gauge $\mathscr A_{2k+1}$};
    \draw [line,lightgray] (4.3,0)  -- (4.3,2) -- (6.3,2) -- (6.3,0) -- (4.3,0);
    \filldraw [MidnightBlue,opacity=0.2] (4.3,0)  -- (4.3,2) -- (6.3,2) -- (6.3,0) -- (4.3,0);
        \draw[ultra thick, MidnightBlue, ->-=.55] (6.3,0)--(6.3,2);
        \node at (5.3, 1) {\footnotesize $D_{k+2}^{\rm def}$};
        \node at (6.3, -.3) {\footnotesize $\mathcal M_{\tilde{\mathscr C}_{D_{k+2}}}$};
    \end{tikzpicture}
    \caption{An object in $\mathcal M_{\tilde{\mathscr C}_{D_{k+2}}}$ obtained by fusing the interface $\mathscr A_{D_{k+2}}$ with an object in $\mathcal M_{\tilde{\mathscr C}_{A_{2k+1}}}$.} 
\end{figure}
\noindent Here $\tilde{\mathscr C}_{D_{k+2}}$ is the SSB superfusion category preserved in $D_{k+2}^{\rm def}$, and the associated module category $\mathcal M_{\tilde{\mathscr C}_{D_{k+2}}}$ encodes the vacua of this theory. The category $\tilde{\mathscr C}_{D_{k+2}}$ can be computed via a ``bosonic condensation" of $\mathscr A_{D_{k+2}}$ in $\tilde{\mathscr C}_{A_{2k+1}}$,
\begin{align}
    D_{k+2}^{\rm def}=A_{2k+1}^{\rm def}/\mathbb Z_2\,,\quad
    \tilde{\mathscr C}_{D_{k+2}}=\tilde{\mathscr C}_{A_{2k+1}}\big/{\mathscr A}_{D_{k+2}}\,.
\end{align}
To demonstrate the condensation procedure, we compute the first non-trivial example, $\tilde{\mathscr C}_{D_4}$: Within $\tilde{\mathscr C}_{A_5}$ the action of ${\mathcal L}_4$ exchanges the lines in the pairs $({\mathcal L}_0,\, {\mathcal L}_4)$, $({\mathcal L}_1,\, {\mathcal L}_3)$, while fixing the line ${\mathcal L}_2$. Therefore, after condensing ${\mathcal L}_4$, the line ${\mathcal L}_2$ splits into two simple lines, denoted by $\widetilde{\mathcal L}_{\pm}$, and the two pairs condense into two simple lines $\widetilde{\mathcal L}_0\equiv ({\mathcal L}_0,\, {\mathcal L}_4)$ and $\widetilde{\mathcal L}_1\equiv ({\mathcal L}_1,\, {\mathcal L}_3)$, respectively. From the fusion rule \eqref{eq: su_2_N=2}, one can verify that lines $\widetilde{\mathcal L}_{0,\pm}$ furnish a $\mathbb Z_3$-symmetry. Together with $\widetilde{\mathcal L}_1$, we have
\begin{align}
    &\qquad\qquad \tilde{\mathscr C}_{D_4}=\left\{\widetilde{\mathcal L}_0,\, \widetilde{\mathcal L}_+,\, \widetilde{\mathcal L}_-,\, \widetilde{\mathcal L}_1\right\}\,, \quad {\rm with}\\
    &\widetilde{\mathcal L}_{0,\pm}\cdot \widetilde{\mathcal L}_1=\mathbb C^{1|0}\,\widetilde{\mathcal L}_1\,, \quad
   \widetilde{\mathcal L}_1\cdot \widetilde{\mathcal L}_1=\mathbb C^{0|1}\left(\widetilde{\mathcal L}_0+\widetilde{\mathcal L}_++\widetilde{\mathcal L}_-\right)\,,\notag
\end{align}
which can be regarded as a variant of $TY(\mathbb Z_3)$ with non-trivial fermionic junctions for $\widetilde{\mathcal L}_1\cdot \widetilde{\mathcal L}_1$.  

Since $\tilde{\mathscr C}_{D_4}$ is completely SSB, the $D$-type deformed model has the Witten index $I_W(D_{4}^{\rm def})=4$. Repeating the superstrip algebra analysis reveals that the only connected quiver representation corresponds to the element $\widetilde{\mathcal L}_1$:
\begin{figure}[H]
\centering
    \begin{tikzpicture}[
    dot_style/.style={Cerulean, circle, fill, inner sep=1.5pt},
    blue_arc/.style={Cerulean, thick, -Stealth, bend left=25},
    red_arc/.style={WildStrawberry, thick, -Stealth, bend left=25},
    label_below/.style={below=2pt}, scale=0.80063
]

\node at (-1, 0) {$\mathcal R_{\widetilde{\mathcal L}_1}$:};

    \node[dot_style, label={[label distance=5pt]below:$0$}] (n0) at (0, 0) {};
    \coordinate (n0pr) at (0.1, .125) {};
    \coordinate (n0pl) at (-0.9, .125) {};
    
    \coordinate (n0nr) at (0.1, -.125) {};
    \coordinate (n0nl) at (-0.9, -.125) {};

    \node[dot_style, label={[label distance=5pt]right:$1$}] (n1) at (1.5, 0) {};
    
    \coordinate (n1pl) at (1.4, .125) {};
    \coordinate (n1nl) at (1.4, -.125) {};
    
    \coordinate (n1pr) at (1.442, .149) {};
    \coordinate (n1nr) at (1.658, 0.024) {};

    \coordinate (n1pr2) at (1.658, -0.024) {};
    \coordinate (n1nr2) at (1.442, -.149) {};
    \node[dot_style, label={[label distance=3pt]right:$+$}] (n2) at (2.25, 1.299) {};
    \coordinate (n2pl) at (2.092, 1.275) {};

    \coordinate (n2nl) at (2.308, 1.150) {};
    \node[dot_style, label={[label distance=3pt]right:$-$}] (n3) at (2.25, -1.299) {};
    \coordinate (n3pl) at (2.308, -1.150) {};
        
    \coordinate (n3nl) at (2.092, -1.275) {};

    \draw[blue_arc] (n0pr) to (n1pl);
    \draw[red_arc] (n1nl) to (n0nr);
    \draw[red_arc] (n1pr) to (n2pl);
    \draw[blue_arc] (n2nl) to (n1nr);
    \draw[red_arc] (n1pr2) to (n3pl);
    \draw[blue_arc]  (n3nl) to (n1nr2);
    
\end{tikzpicture}
\end{figure}
\noindent The quiver representation encodes the soliton spectra of $D_{4}^{\rm def}$ as follows: there are three degenerate sets of fundamental (anti-)solitons connecting four vacua $|0\rangle$, $|\pm\rangle$ and $|1\rangle$ in a $D_4$-quiver pattern; each of them contains a pair of bosonic/fermionic one-particle states forming an irreducible $\mathcal N=2$ massive supermultiplet. 

The generalization to $D_{k+2}^{\rm def}=A_{2k+1}^{\rm def}/\mathbb Z_2$ is straightforward: the TDL ${\mathcal L}_k\in \tilde{\mathscr C}_{A_{2k+1}}$ splits into two lines $\widetilde{\mathcal L}_{\pm}\in \tilde{\mathscr C}_{D_{k+2}}$ which is completely SSB. There are thus $k+2$ vacua, and the (anti-)soliton spectra are encoded in a $D_{k+2}$-quiver corresponding to the element $\widetilde{\mathcal L}_1\in \tilde{\mathscr C}_{D_{k+2}}$:
\begin{figure}[H]
\centering

\end{figure}
The same line of reasoning applies in parallel to the least relevant deformation of the $E$-type $\mathcal N=2$ minimal model. First, the $E_{6,7,8}$ minimal CFTs can be achieved by gauging non-invertible symmetries in $A_{11, 17, 29}$, which feature exceptional Frobenius algebras,
\begin{align}
    &\mathscr A_{E_6}=\mathcal L_0\oplus\mathcal L_6\,,\quad
    \mathscr A_{E_7}=\mathcal L_0\oplus\mathcal L_8\oplus\mathcal L_{16}\,,\notag\\
    &\mathscr A_{E_8}=\mathcal L_0\oplus\mathcal L_{10}\oplus\mathcal L_{18}\oplus\mathcal L_{28}\,,\quad
\end{align}
Because the algebras $\mathscr A_{E_{6,7,8}}$ are preserved in the corresponding least relevantly deformed model $A_{11,17,29}^{\rm def}$, we can proceed analogously to the $D$-type case. We compute the SSB superfusion categories
\begin{align}
\tilde{\mathscr C}_{E_{6,7,8}}=\tilde{\mathscr C}_{A_{11,17,29}}/\mathscr A_{E_{6,7,8}}\,,
\end{align}
preserved in $E_{6,7,8}^{\rm def}$, as well as the associated supermodule category $\mathcal M_{\tilde{\mathscr C}_{E_{6,7,8}}}$ encoding their vacua.

Every $\tilde{\mathscr C}_{E_{n}}$ (for $n=6,7,8$) contains a simple object $\widetilde{\mathcal L}_1\equiv\mathcal L_1\otimes\mathscr A_{E_n}$ that generates the entire superfusion categories. The other simple objects in $\tilde{\mathscr C}_{E_{n}}$ can be systematically determined by computing ${\rm Hom}_{\tilde{\mathscr C}_{A}}(\mathcal L_a,\,\mathcal L_a\otimes\mathscr A_{E})$ via standard procedures \cite{Fuchs:2002cm}. More details can be found in Appendix~\ref{App:algebra}.

For example, corresponding to $\widetilde{\mathcal L}_1$, we find an $E_7$-type superstrip algebra representation that encodes the vacua structure and (anti)-soliton spectrum.
\begin{figure}[H]
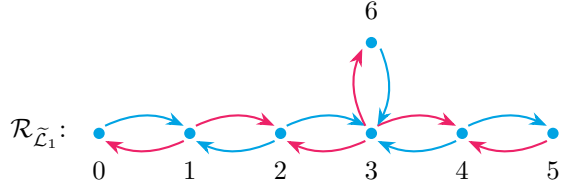

\centering

\caption{The $\mathcal R_{\widetilde{\mathcal L}_1}$ superstrip algebra representation for the least relevantly deformed $\mathcal N=2$ minimal model of $E_7$-type.}
\end{figure}

\section{\texorpdfstring{Least relevant deformation of $\mathcal N=1$ minimal models}{Least relevant deformation of N=1 minimal models}}\label{sec:N_1_minimal_model}
In this section, we examine the least relevant deformation of the $\mathcal{N}=1$ minimal models. Let us first introduce some basic results and fix our notation. The $\mathcal{N}=1$ superconformal algebra is generated by the modes of the energy-momentum tensor $T(z)$ and a single supercurrent $G(z)$. These generators satisfy the following (anti-)commutation relations:
\begin{align}
[L_m, L_n] &= (m-n)L_{m+n} + \frac{c}{12}(m^3-m)\delta_{m,-n}\,,\\
[L_m, G_n] &= \left(\tfrac{m}{2} - n\right)G_{m+n}\,, \\
\{G_m, G_n\} &= 2L_{m+n} + \frac{c}{3}\left(m^2 - \frac{1}{4}\right)\delta_{m,-n}\,,
\end{align}
where $c$ denotes the central charge. The mode indices of $G(z)$ take half-integral values for the Neveu-Schwarz (NS) sector, and integral values for the Ramond (R) sector.

Similarly, the $\mathcal{N}=1$ superconformal algebra admits a series of minimal models \cite{Friedan:1984rv,EICHENHERR198526}. For our purpose, we only discuss the unitary cases and denote them as $\mathcal {SM}_{p,p+2}$. These are characterized by their central charges and conformal dimensions:
\begin{align}
c &= \frac{3}{2}\left(1 - \frac{8}{p (p+2)}\right)\,, \\
h_{r,s} &=
\begin{cases}\label{eq:hN=1}
\frac{\big((p+2) r - p s\big)^2 - 4}{8 p (p+2)}\, , \qquad & r - s \in 2\mathbb{Z} \qquad \ (\text{NS sector})\, , \\
\frac{\big( (p+2) r - p s\big)^2 - 4}{8 p (p+2)} + \frac{1}{16}\, , \qquad & r - s \in 2\mathbb{Z} + 1 \quad (\text{R sector}) \,.
\end{cases}
\end{align}
The Kac labels $(r,s)$ are integers restricted to the domains $1\leq r \leq p-1$ and $1\leq s \leq p+1$.

Moreover, the spectrum exhibits a two-fold identification symmetry:
\begin{equation}
h_{r,s} = h_{p-r,p+2-s} \, .
\end{equation}

When $p$ is even, this symmetry yields a fixed point at $r=\frac{p}{2}$ and $s=\frac{p}{2}+1$. From \eqref{eq:hN=1}, this highest-weight state necessarily belongs to the R sector.

All possible modular invariants for $\mathcal {SM}_{p,p+2}$ are classified in \cite{Cappelli:1986ed}. This can be understood from its bosonized theory
\begin{align}\label{eq:cosetN=1}
    \mathfrak B\left[\mathcal {SM}_{p,p+2}\right]=\frac{SU(2)_{p-2}\times SU(2)_2}{SU(2)_p}\,.
\end{align}
As established in \cite{Cappelli:1987xt}, $SU(2)_k$ admits an $ADE$ classification corresponding to simply-laced simple Lie algebras. Then it is natural to expect $\mathcal {SM}_{p,p+2}$ to be classified by a pair of $ADE$ algebras. Indeed, the results derived in \cite{Cappelli:1986ed} are summarized in Table~\ref{tab:N1_ADE_Unitary}.

\begin{table}[htbp]
\centering
\renewcommand{\arraystretch}{1.8}
\begin{tabular}{llc}
\hline
\textbf{Label Pair} $(G, G')$ & \textbf{Constraint on $p$} & \textbf{Witten Index} $W$ \\
\hline
$(A_{p-1}, A_{p+1})$ & $p$ odd & $0$ \\
\hline
$(A_{p-1}, A_{p+1})$ & $p$ even & $1$ \\
\hline
$(D_{\frac{p}{2}+1}, A_{p+1})$ & $p \equiv 0 \pmod 4$ & $-1$ \\
$(D_{\frac{p}{2}+1}, A_{p+1})$ & $p \equiv 2 \pmod 4$ & $2$ \\
\hline
$(A_{p-1}, D_{\frac{p}{2}+2})$ & $p \equiv 0 \pmod 4$ & $2$ \\
$(A_{p-1}, D_{\frac{p}{2}+2})$ & $p \equiv 2 \pmod 4$ & $-1$ \\
\hline
$(A_9, E_6)$ & $p=10$ & $0$ \\
$(D_6, E_6)$ & $p=10$ & $0$ \\
$(E_6, A_{13})$ & $p=12$ & $0$ \\
$(E_6, D_8)$ & $p=12$ & $0$ \\
\hline
$(A_{15}, E_7)$ & $p=16$ & $1$ \\
$(E_7, A_{19})$ & $p=18$ & $1$ \\
\hline
$(A_{27}, E_8)$ & $p=28$ & $0$ \\
$(E_8, A_{31})$ & $p=30$ & $0$ \\
\hline
\end{tabular}
\caption{All possible modular invariants of unitary $\mathcal{N}=1$ superconformal minimal models $\mathcal{SM}_{p,p+2}$ \cite{Cappelli:1986ed}. We also provide the Witten index of each modular invariant. The overall sign is fixed by our convention of the deformed theory.}
\label{tab:N1_ADE_Unitary}
\end{table}

Clearly, $(A_{p-1},A_{p+1})$ represents the diagonal modular invariant. Let us explain, for example, the $(D,A)$ and $(A,D)$ types. They are both constructed as $\mathbb{Z}_2$ orbifolds of $(A, A)$ theory, but they have different partition functions. 

This difference can be traced back to the $ADE$ classification of $SU(2)_k$. The $D$-type modular invariant of $SU(2)_k$ is obtained by making use of a $\mathbb{Z}_2$ simple current. The precise nature is dictated by the conformal dimension of the simple current, which is determined by the affine level $k \pmod 4$:
\begin{itemize}
    \item \textbf{Type I (Chiral Extension):} If the level $k \equiv 0 \pmod 4$, the simple current possesses an integer conformal dimension. The resulting $D$-invariant corresponds to a rigorous extension of the chiral algebra.
    \item \textbf{Type II (Automorphism):} If the level $k \equiv 2 \pmod 4$, the simple current has a half-integer conformal dimension. The $D$-invariant yields an automorphism invariant (a standard $\mathbb{Z}_2$ orbifold without an extension of the chiral algebra).
\end{itemize}
Because the levels of the numerator ($p-2$) and denominator ($p$) in the coset \eqref{eq:cosetN=1} differ by exactly 2, their behavior modulo 4 is perfectly complementary. For any allowable central charge in these series, one always obtains one theory with an extended chiral symmetry and one theory governed strictly by a $\mathbb{Z}_2$ automorphism. 

We also provide the Witten index for each modular invariant in the same table. We can compute it by noting that the possible Ramond ground state corresponds to the primary field with conformal weights $(h_{p/2,p/2+1},\overline{h}_{ p/2,p/2+1 })$.

To simplify our notation, we abbreviate the $(A_{p-1},A_{p+1})$ type as $A_{p-1}$. As for $(A,D)$ or $(D,A)$ type, We restrict our discussion to the modular invariant with extended chiral symmetry, denoted collectively as $D_k$ type. In other words, for odd or even $k$, the $D_{k}$ type corresponds to $(A_{2k-3},D_{k+1})$ or $(D_k,A_{2k-1})$, respectively. Regarding exceptional modular invariants, the designation $E_k$-type refers to the $(E_k,A_\ast)$ type modular invariant.

Analogous to the $\mathcal{N} = 2$ case, the $\mathcal N=1$ minimal model also admits a Landau-Ginzburg description. We will use this formulation to study the deformation of the $A$-type modular invariant.

\subsection{\texorpdfstring{$A$-type}{A-type}}
\subsubsection{Coset construction}
We start from the coset construction of the bosonized $\mathcal N=1$ minimal model, 
\begin{align}
    \mathfrak B\left[\mathcal {SM}_{p,p+2}\right]=\frac{SU(2)_{p-2}\times SU(2)_2}{SU(2)_p}\,.
\end{align}
Any primaries in $ \mathfrak B\left[\mathcal {SM}_{p,p+2}\right]$ can be labeled by a triplet $(a,b,c)$ for $a=0,1,\dots p-2$, $b=0,1,2$ and $c=0,1,\dots, p$, where $(a,b,c)$ correspond to the integer representations of primaries in $SU(2)_{p-2}$, $SU(2)_2$ and $SU(2)_p$ respectively. The triplet $(a,b,c)$ is under the constraint
\begin{align}
    a+b+c\in 2\mathbb Z\,,
\end{align} 
and an equivalence relation,
\begin{align}
(a,b,c)\sim(p-2-a,2-b,p-c)\,.    
\end{align}
It is important to notice that, for even $p$, the equivalence relation implies a fixed point, for $a=\frac{p-2}{2}$, $b=1$ and ${c=\frac{p}{2}}$. Therefore, one has to resolve the fixed point into two primaries for the label $\left(\frac{p-2}{2},1,\frac{p}{2}\right)$. To distinguish the two primaries, it is necessary to introduce another label. Overall, we label the primaries in $ \mathfrak B\left[\mathcal {SM}_{p,p+2}\right]$ by
\begin{align}
    (a,b,c,d)=
    \begin{cases}
        (a,b,c,0)\quad{\rm for}\quad (a,b,c) \ \ {\rm not}\ \ {\rm fixed} \ \ {\rm point}\\[2ex]
        (a,b,c,\pm 1) \quad{\rm for}\quad (a,b,c) \ \ {\rm fixed} \ \ {\rm point}
    \end{cases}
\end{align}

With these preparations, we now spell out the $S$-matrix of the coset model. Basically the $S$-matrix element is assembled from the three pieces of $SU(2)_{p-2}$, $SU(2)_{2}$ and $SU(2)_{p}$, except for the fixed point $\left(\frac{p-2}{2},1,\frac{p}{2}, \pm 1\right)$. Following the discussion of \cite{Schellekens:1989am,Gaberdiel:2026sfg}, we list the $S$-matrix elements as follows:



\begin{align}
    &S_{(a,b,c,0),\,(a',b',c',0)}\notag\\
    &\quad=\sqrt{\frac{8}{p(p+2)}}\sin\left(\frac{(a+1)(a'+1)\pi}{p}\right)\sin\left(\frac{(b+1)(b'+1)\pi}{4}\right)\sin\left(\frac{(c+1)(c'+1)\pi}{p+2}\right)\notag\\[1ex]
    &S_{(a,b,c,0),\,\left(\frac{p-2}{2},1,\frac{p}{2}, \pm 1\right)}
    =S_{\left(\frac{p-2}{2},1,\frac{p}{2}, \pm 1\right),\,(a,b,c,0)}
    \notag\\
    &\quad=\sqrt{\frac{2}{p(p+2)}}\sin\left(\frac{(a+1)\pi}{2}\right)\sin\left(\frac{(b+1)\pi}{2}\right)\sin\left(\frac{(c+1)\pi}{2}\right)\notag\\[1ex]
    &S_{\left(\frac{p-2}{2},1,\frac{p}{2}, d\right),\,\left(\frac{p-2}{2},1,\frac{p}{2}, d'\right)}=
    \begin{cases}
        \frac{1}{2}\quad{\rm for}\ \ d=d'\\[2ex]
        -\frac{1}{2}\quad{\rm for}\ \ d=-d'\,.
    \end{cases}
\end{align}
The fusion rules between two primaries that are not fixed points are simply inherited from that of the $SU(2)$ Wess-Zumino-Witten model, spelled as
\begin{align}
    (a,b,c,0)\otimes(a',b',c',0)=\bigoplus_{(a'',b'',c'')}N^{(p-2)\,a''}_{a,\,a'}N^{(2)\,b''}_{b,\,b'}N^{(p)\,c''}_{c,\,c'}\,(a'',b'',c'',d)\,,
    \label{eq:fusion_N=1}
\end{align}
where 
\begin{align}
    N^{(k)\,a''}_{a,\,a'}\equiv
    \begin{cases}
    1\quad {\rm for}\ |a-a'|\leq a''\leq\min(a+a',\,2k-a-a')\quad {\rm and}\quad a+a'+a''\in 2\mathbb Z    \\[2ex]
    0 \quad {\rm otherwise}\,.
    \end{cases}
\end{align}
and label $d$ in the summand $(a'',b'',c'',d)$ on the RHS is 0 for $(a'',b'',c'')$ not a fixed point, and runs for both $\pm1$ otherwise. For the fusion of primaries containing fixed points, the fusion rules are more involved and we refer curious readers to the references \cite{Schellekens:1989am,Gaberdiel:2026sfg}.

In $ \mathfrak B\left[\mathcal {SM}_{p,p+2}\right]$, the primary $G=(0,2,0,0)$ has conformal weight $h_G=\frac{3}{2}$ and is a $\mathbb Z_2$ simple current. So one can fermionize the coset model back to the $\mathcal N=1$ minimal model. From the $S$-matrix, it is easy to verify the following $G$-charges for all primaries
\begin{align}
    Q_G(\phi_{(a,b,c,d)})=
    \begin{cases}
    +1\quad {a+c}\ \ {\rm even}\\[2ex]
    -1\quad {a+c}\ \ {\rm odd}\,.
    \end{cases}
\end{align}
This implies that, after fermionization, the field $\phi_{(a,b,c,d)}$ belongs to the NS sector if $a+c$ is even, and to the R sector if $a+c$ is odd. Furthermore, for the primaries $(a,b,c,d)$, one can also find the fusion rule as
\begin{align}
    G\cdot\phi_{(a,b,c,d)}=\begin{cases}
        \phi_{(a,2-b,c,d)}\,,\quad {\rm for}\ \ a+c\ \ {\rm even}\\[2ex]
        \phi_{(a,b,c,d)}\,,\quad {\rm for}\ \ a+c\ \ {\rm odd}\quad {\rm and}\quad d=0\\[2ex]
        \phi_{(a,b,c,-d)}\,,\quad {\rm for}\ \ a+c\ \ {\rm odd}\quad {\rm and}\quad d\neq 0\,.
    \end{cases}
    \label{eq:Gfusion_N=1}
\end{align}
Therefore the pair $\{\phi_{(a,b,c,0)},\, \phi_{(a,2-b,c,0)}\}$ with $a+c$ even furnishes an $\mathcal N=1$ superconformal multiplet in the NS sector after fermionization.

\subsubsection{The Least Relevant Deformation}
Now we consider the least relevant deformation of the $\mathcal N=1$ minimal model $\mathcal {SM}_{p,p+2}$. The deformation is induced by the NS-sector superconformal primary 
\begin{align}
    \phi^{\mathcal N=1}_{(0,2)}=\left\{\phi_{(0,2,2,0)}\,,\,\phi_{(0,0,2,0)}\right\}\,,\quad
    {\rm with}\quad \phi_{(0,0,2,0)}=G_{-1/2}\phi_{(0,2,2,0)}\,.
\end{align}
To determine the preserved TDL set, it is instructive to study the preserved TDLs in the bosonic theory $ \mathfrak B\left[\mathcal {SM}_{p,p+2}\right]$ under the deformation of operator $\phi_{(0,0,2,0)}$. Only those Verlinde lines $\mathcal L_{(a,b,c,d)}$ satisfying the condition,
\begin{align}
    \mathcal L_{(a,b,c,d)} |\phi_{(0,0,2,0)}\rangle=\frac{S_{(a,b,c,d),\,(0,0,2,0)}}{S_{(0,0,0,0),\,(0,0,2,0)}} |\phi_{(0,0,2,0)}\rangle=\langle\mathcal L_{(a,b,c,d)}\rangle|\phi_{(0,0,2,0)}\rangle=\frac{S_{(a,b,c,d),\,(0,0,0,0)}}{S_{(0,0,0,0),\,(0,0,0,0)}}|\phi_{(0,0,2,0)}\rangle\,,
\end{align}
will survive in the deformed theory. One then finds that, in the bosonic theory $ \mathfrak B\left[\mathcal {SM}_{p,p+2}\right]$, the preserved TDLs are
\begin{align}
    \mathscr C_{\mathfrak B}=\left\{\mathcal L_{(a,0,0,0)}\,,\,\mathcal L_{(a,2,0,0)}\,|\,a\ \ {\rm even}\right\}\bigcup\left\{\mathcal L_{(a,1,0,0)}\,|\,a\ \ {\rm odd}\right\}\,,\quad
    {\rm with}\ \ a=0,1,\dots,p-2\,.
\end{align}
Performing the $\mathbb Z_2$-fermionization with respect to $\mathcal L_{(0,2,0,0)}$, we obtain the $\mathcal N=1$ minimal model. Correspondingly, the superfusion category therein is obtained via fermionic condensation of the line $\mathcal L_{(0,2,0,0)}$. Notice that the fusion rule of Verlinde lines is the same as the local primaries as in eq.~\eqref{eq:Gfusion_N=1}. Therefore, one finds that the pair of lines $\mathcal L_{(a,0,0,0)}$ and $\mathcal L_{(a,2,0,0)}$ forms a 2-orbit with respect to the line $\mathcal L_{(0,2,0,0)}$, while the line $\mathcal L_{(a,1,0,0)}$ is a fixed point. Therefore, after condensation, we define the equivalence class 
\begin{align}
    \mathcal L_{a}\equiv
    \begin{cases}
        \left\{\mathcal L_{(a,0,0,0)}\,,\, \mathcal L_{(a,2,0,0)}\right\}\,,\quad{\rm for}\ \ a\ \ {\rm even}\\[2ex]
        \mathcal L_{(a,1,0,0)}\,,\quad{\rm for}\ \ a\ \ {\rm odd}
    \end{cases}
    \,,\quad
    {\rm with}\ \ a=0,1,\dots,p-2\,.
\end{align}
Clearly, the lines $\mathcal L_a$ are of m/q-type for even/odd $a$ respectively \cite{Zhou:2021ulc, Chang:2022hud}. Furthermore, their fusion rules can be determined from eq.~\eqref{eq:fusion_N=1} such that,
\begin{align}
    \mathcal L_a\cdot\mathcal L_b=
    \begin{cases}
     \sum_{c,\text{m-type}} \mathbb{C}^{N_{ab}^{c,\,\mathbf{b}}|N_{ab}^{c,\mathbf{f}}} \cdot\mathcal{L}_c \,,\quad \text{for}\ a+b\in 2\mathbb Z\\
    \\
    \sum_{c,\,\text{q-type}} N_{ab}^{c}\,\mathcal{L}_c\,,\quad \text{for}\ a+b\in 2\mathbb Z+1
    \end{cases}
\end{align}
with
\begin{align}
N_{ab}^{c,\mathbf{b}}=N_{ab}^{c,\mathbf{f}}=N_{ab}^{c}=
    \begin{cases}
    1\quad {\rm for}\ |a-b|\leq c\leq\min(a+b,\,2p-4-a-b)\quad {\rm and}\quad a+b+c\in 2\mathbb Z    \\[2ex]
    0 \quad {\rm otherwise}
    \end{cases}
    \notag
\end{align}
We denote the preserved superfusion category by $\mathscr{C}^{\mathcal N=1}_{A_{p-1}}$ for convenience.

The least relevant deformation will trigger an RG flow. Whether the flow leads to a gapless or gapped phase depends on the sign of the coupling parameter. It is instructive to introduce the LG description of the $\mathcal N=1$ minimal model. The model $\mathcal {SM}_{p,\,p+2}$ can be characterized by a single $\mathcal N=1$ real supermultiplet $\Phi=\phi+\bar\theta\psi+\frac{1}{2}\bar\theta\theta F$ and an $\mathcal N=1$ superpotential,
\begin{align}
    \mathcal W_{\mathcal N=1}(\Phi)=\frac{1}{p}\Phi^p\,,
\end{align}
which is expected to flow to the $\mathcal {SM}_{p,\,p+2}$ at the IR fixed point. A deformation of the potential will further trigger an RG flow from the above fixed point. In the case at hand, the least relevant operator corresponds to the superfield $:\Phi^{p-2}:+\cdots$, where $\cdots$ denotes some lower order correction. The deformed superpotential is thus given by
\begin{align}
    \mathcal W^{\rm def.}_{\mathcal N=1}=\frac{1}{p}\Phi^p+\lambda \Phi^{p-2}+\cdots\,.
\end{align}
Clearly, a positive $\lambda$ will lift the superpotential from $\Phi^{p}$ to $\Phi^{p-2}$ in the infra-red, indicating a massless RG flow from $\mathcal {SM}_{p,\,p+2}$ to $\mathcal {SM}_{p-2,\,p}$ for $p \geq 5$. On the other hand, we are interested in a massive RG flow from the $\mathcal {SM}_{p,\,p+2}$ to a gapped integrable model corresponding to a negative $\lambda$. For $\lambda<0$, after integration over superspace, the potential in real space is given by 
\begin{align}
V(\phi)=\left|\frac{\partial \mathcal W^{\rm def.}_{\mathcal N=1}}{\partial\Phi}\right|^2_{\Phi=\phi}\,.
\end{align}
Notice that the deformation preserves supersymmetry, and thus the supersymmetric vacua are determined by the $F$-term equation,
\begin{align}
    \frac{\partial\mathcal W^{\rm def.}_{\mathcal N=1}}{\partial\Phi}\Bigg|_{\Phi=\phi}=0\,.
\end{align}
$\frac{\partial\mathcal W^{\rm def.}_{\mathcal N=1}}{\partial\Phi}$ is a polynomial of order $p-1$. For the least relevant deformation, it has $p-1$ real roots and thus overall $p-1$ supersymmetric vacua. In fact, these vacua exactly correspond to the $p-1$ (non-invertible) lines in the spontaneously breaking superfusion category $\mathscr{C}^{\mathcal N=1}_{A_{p-1}}$. Furthermore recall that, for the $p-1$ TDLs $\mathcal L_a\in \mathscr{C}^{\mathcal N=1}_{A_{p-1}}$ with $a=0,\,1,\,\dots,\,p-2$, there are $\lfloor \frac{p-1}{2}\rfloor$ q-type ones labeled by odd $a$, and the rest are of m-type. As explained in \cite{Chen:2025qub}, the q-type SSB lines correspond to the fermionic vacua, whereas the m-type ones are for bosonic vacua. Therefore we can compute the Witten Index as
\begin{align}
    I^{\mathcal N=1}_{A_{p-1}}=N_{\text{m-type}}-N_{\text{q-type}}=
    \begin{cases}
    0\,,\quad {\rm for}\ \ p\ \ {\rm odd}\\[2ex]
    1\,,\quad {\rm for}\ \ p\ \ {\rm even}\,.
    \end{cases}
\end{align}
Since the least relevant deformation preserves supersymmetry, one finds that the Witten Index computed from the deformed theory matches perfectly with the one at the CFT critical point in the R sector, which is given in Table~\ref{tab:N1_ADE_Unitary}. In addition, the structure of SSB superfusion category $\mathscr C_{A_{p-1}}^{\mathcal N=1}$ indicates that bosonic and fermionic supersymmetric vacua are alternately and linearly aligned, as the figure below:
\begin{figure}[H]
\centering
\begin{tikzpicture}[scale=1.65]
\draw[sharp arrow, thick] (0,0) -- (0,3) node[right, yshift=0pt] {$V_{\mathcal N=1}$};
\begin{scope}[yshift=1.25cm, yscale=.5]
\draw[sharp arrow, thick] (-2.5,0) -- (2.5,0) node[below, xshift=-8pt] {$\phi_{\mathcal N=1}$};
 \draw[
    black,
   thick,
  smooth,
 samples=200,
domain=-1.83:1.83
    ] plot (\x, {(\x*\x)*(\x*\x-1)*(\x*\x-1)*(\x*\x-3)*(\x*\x-3)});
   \foreach \x in {-1.732, 0, 1.732}
     \fill[Cerulean] (\x, 0) ellipse ({1.5pt} and {3pt});
    \foreach \x in {-1, 1}
      \fill[WildStrawberry] (\x, 0) ellipse ({1.5pt} and {3pt});
\end{scope}
\end{tikzpicture}
\caption{$\mathcal {SM}^{\rm def}_{6,8}$, where the red dots denote fermionic vacua, and the blue dots for bosonic ones.}
\label{fig:N_1}
\end{figure}
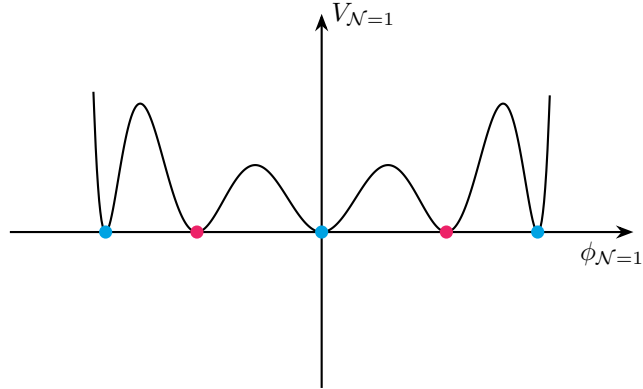

\subsubsection{\texorpdfstring{Superstrip Algebra $\mathbf{sStr}(\mathscr C_{A_{p-1}}^{\mathcal N=1})$}{Superstrip Algebra sStr(C A(p-1), N=1)}}
\begingroup\tolerance=1000\emergencystretch=1em
Now we implement the superfusion category $\mathscr C_{A_{p-1}}^{\mathcal N=1}$ data to construct its superstrip algebra $\mathbf{sStr}(\mathscr C_{A_{p-1}}^{\mathcal N=1})$. We explain in detail the simplest nontrivial case $p=3$, i.e., the tri-critical Ising model (TIM), perturbed by the least relevant term $\lambda \int \phi_{\frac{3}{5},\frac{3}{5}}$. For $\lambda >0$, the theory flows to the gapless Ising model, whereas for $\lambda <0$ it flows to a gapped phase with three vacua where the Ising category symmetry is spontaneously broken. As depicted in Figure~\ref{Tri-critical duality}, the TIM has hidden supersymmetry and can be fermionized to the $\mathcal{N}=1$ $A_2$-type minimal model.
\par\endgroup
\begin{figure}[H]
\centering
    \begin{tikzpicture}[scale=1.5]
        \node (top_left) at (0,3) {\color{Teal} Ising Model};
        \node (middle_left) at (0,1.5) {\color{Teal} TIM};
        \node (bottom_left) at (0,0) {\color{Teal} Ising SSB};
        \node (top_right) at (6,3) {\color{Teal} Majorana fermion};
        \node (middle_right) at (6,1.5) {\color{Teal} $\mathcal{N}=1$ $A_2$ MM};
        \node (bottom_right) at (6,0) {\color{Teal} $\mathbb{Z}_2$ SSB};
        \draw[sharp arrow, thick] (middle_left) -- (top_left);
        \draw[sharp arrow, thick] (middle_left) -- (bottom_left);
        \draw[sharp arrow, thick] (middle_right) -- (top_right);
        \draw[sharp arrow, thick] (middle_right) -- (bottom_right);
        \node at (-0.5,2.15) {$\lambda>0$};
        \node at (-0.5,0.85) {$\lambda<0$};
        \node at (6.5,2.15) {$\lambda>0$};
        \node at (6.5,0.85) {$\lambda<0$};
        \draw[sharp arrow, thick] ([yshift=2pt]top_left.east) -- ([yshift=2pt]top_right.west) node[midway,above] {fermionize};
        \draw[sharp arrow, thick] ([yshift=-2pt]top_right.west) -- ([yshift=-2pt]top_left.east) node[midway,below] {gauge $(-1)^F$};
        \draw[sharp arrow, thick] ([yshift=2pt]middle_left.east) -- ([yshift=2pt]middle_right.west) node[midway,above] {fermionize};
        \draw[sharp arrow, thick] ([yshift=-2pt]middle_right.west) -- ([yshift=-2pt]middle_left.east) node[midway,below] {gauge $(-1)^F$};
        \draw[sharp arrow, thick] ([yshift=2pt]bottom_left.east) -- ([yshift=2pt]bottom_right.west) node[midway,above] {fermionize};
        \draw[sharp arrow, thick] ([yshift=-2pt]bottom_right.west) -- ([yshift=-2pt]bottom_left.east) node[midway,below] {gauge $(-1)^F$};
    \end{tikzpicture}
    \caption{The duality between TIM and $\mathcal{N}=1$ $A_2$ minimal model (MM). The two columns are related by fermionization and bosonization.}
    \label{Tri-critical duality}
\end{figure}
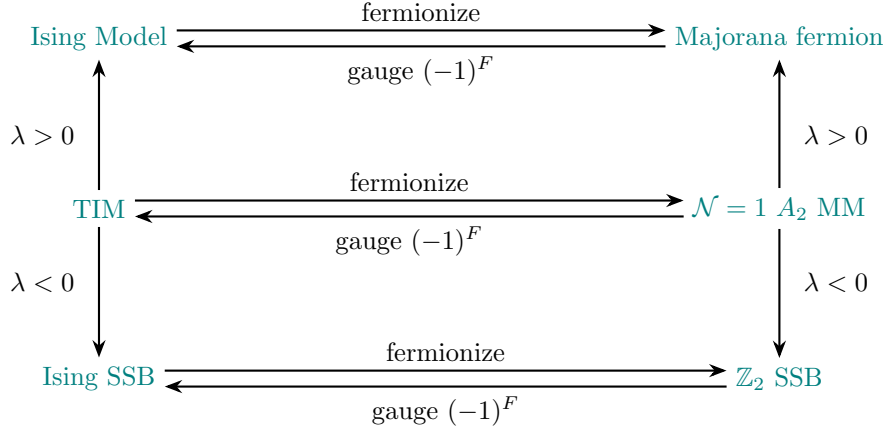
The TIM under the $\phi_{\frac{3}{5},\frac{3}{5}}$ deformation allows the LG-description with the superpotential 
$\mathcal{W}(\Phi) =\frac{1}{3} \Phi^3 + 2\lambda \Phi$. 
The potential term then becomes
\begin{equation*}
    V = \frac{1}{2}|\mathcal{W}(\phi)'|^2 + \frac{1}{2}\mathcal{W}(\phi)''\bar{\psi} \psi= \frac{1}{2} (\phi^2 + 2\lambda)^2 + \phi \bar{\psi} \psi \,. 
\end{equation*} 
The LG model has a $\mathbb{Z}^F_2 \times \mathbb{Z}_2^\eta$ global symmetry,  
the former corresponding to the fermion parity $(-1)^F$ and the latter to
the chiral $R$-symmetry 
\begin{equation}
        \eta:\quad\phi \rightarrow -\phi\,,\quad \psi\rightarrow \gamma_5 \psi\,.
        \label{eq:N=1_F_L}
\end{equation}
%
For $\lambda<0$, the potential $V$ has two minima at $\phi = \pm \sqrt{2|\lambda|}$, where the $\mathbb{Z}_2^\eta$ symmetry is spontaneously broken. At each minimum, both the scalar and fermion become massive so that the theory flows to a gapped phase. The two minima
give rise to the degenerate vacua, denoted by $|I\rangle$ and $|\eta\rangle$, which are interchanged under \eqref{eq:N=1_F_L}
	\begin{equation}
	\eta |I\rangle = |\eta\rangle\,, \quad \eta |\eta\rangle = |I\rangle\,.
	\end{equation}
The fermion parity remains unbroken. The two vacua differ by the sign of mass $m=\langle \phi \rangle$ of the Majorana fermion. One can thus
designate, without loss of generality, $|I\rangle$ as the trivial phase and $|\eta\rangle$ as the non-trivial fermionic symmetry protected topological (SPT) phase \cite{Witten:2023snr}. 

Since the $\mathbb{Z}_2^\eta$ symmetry is spontaneously broken, the $\eta$-line along the temporal direction 
can be viewed as a soliton interpolating between two vacua. One can argue that, in the presence of the soliton, 
the fermion field $\psi$ has a single real zero mode localized at the center of the soliton. As a 
consequence, $\eta$-line has to be dressed by a 1D Majorana fermion and thus becomes a 
q-type TDL obeying the fusion rule below, 
\begin{equation}
    \eta \times \eta = \mathbb C^{1|1}I\,,\quad (N_{\eta \eta}^{I,\bf b}=N_{\eta \eta}^{I,\bf f}=1)
\end{equation}
where $V_{\eta,\eta,I}=\mathbb{C}^{1|1}$, and we will use $\alpha=0,1$ to label the two channels $\mathbb{C}^{1|0}$ and $\mathbb{C}^{0|1}$. Physically, the two Majorana zero modes associated with a pair of $\eta$-lines combine into a Dirac zero mode, which spans a 2-dimensional superspace upon quantization. 

The superstrip algebra $\mathbf{sStr}(\mathscr C_{A_{2}}^{\mathcal N=1})$ consists of the building block \eqref{fig:superstrip} with $i,j,k,l,a \in \{I,\eta \}$.

Beginning with a bosonic soliton $|K_{\eta I}\rangle \in \mathcal{H}_{\eta,I}$, we can map it to a fermionic soliton $|\widetilde{K_{\eta I}}\rangle$ via
\begin{equation}
        \begin{gathered}
     %
which reverses the statistics of the soliton due to the fermionic junction $\alpha=1$. 
It can also be mapped to an anti-soliton in $\mathcal{H}_{I,\eta}$ through two inequivalent channels
\begin{equation}
\begin{gathered}
     %
 
\end{gathered}
\end{equation}
Here we set $\beta=0$, since the red dot on the $\beta$ junction can move along the $\eta$ line to the $\alpha$ junction. As a consequence, the irreducible representation of solitons consists of four states, two solitons $|K_{\eta I}\rangle,|\widetilde{K_{\eta I}}\rangle$ and two anti-solitons $|K_{I\eta}\rangle,|\widetilde{K_{I\eta}}\rangle$, with opposite statistics in each pair, represented by the quiver diagram in Figure~\ref{fig:Cq0quiver}. It agrees perfectly with the integrability results \cite{Schoutens:1990vb, Ahn:1990gn}.
\begin{figure}[H]
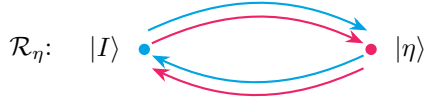

\centering

\caption{The quiver diagram of the representation $\mathcal R_{\eta}$.}
\label{fig:Cq0quiver}
\end{figure}

The discussion can be straightforwardly generalized to $\mathbf{sStr}(\mathscr C_{A_{p-1}}^{\mathcal N=1})$: There are $p-1$ bosonic and fermionic vacua arranged linearly and alternately. Between any two adjacent vacua, the excitations consist of two solitons and two anti-solitons, organized into two distinct $\mathcal N=1$ massive supermultiplets, depicted below in Figure~\ref{fig:A-type_N=1} for the case of even $p$:
\begin{figure}[H]
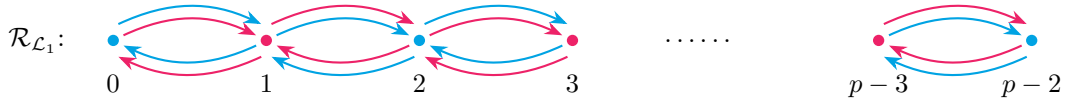

\centering
    %
\caption{The quiver diagram of the representation $\mathcal R_{{\mathcal L}_1}$.}
\label{fig:A-type_N=1}
\end{figure}

\subsection{\texorpdfstring{$D$-type}{D-type}}
Among the $A_{p-1}$-type deformed model whose SSB superfusion category is $\mathscr{C}^{\mathcal N=1}_{A_{p-1}}$, there is always a simple m-type $\mathbb Z_2$-line $\mathcal L_{p-2}$ for even $p=2k+2$, and can be gauged to obtain the deformed $D$-type models,
\begin{align}
     D_{k+2}= A_{2k+1}/\mathcal L_{2k}\,.
\end{align}
The vacua structure of $D$-type deformed model is further refined with respect to odd or even $k$. Recall that the $A_{p-1}=A_{2k+1}$-type theories in the Ramond sector display different linearly aligned vacua configurations: For odd (even) $k$ the middle vacuum is fermionic (bosonic), see Figure~\ref{fig:A_vacua} below for $k=1$ and $2$:
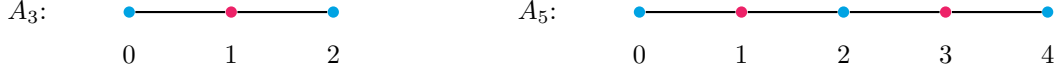
\begin{figure}[H]
\centering
    \begin{tikzpicture}[
    dot_style/.style={circle, fill, inner sep=1.5pt},
    blue_arc/.style={Cerulean, thick, -Stealth, bend left=25},
    red_arc/.style={WildStrawberry, thick, -Stealth, bend left=25},
    label_below/.style={below=2pt}, scale=1.35
]

\node at (-1, 0) {$A_3$:};

    \node[dot_style, fill=Cerulean, label={[label distance=7pt]below:$0$}] (n0) at (0, 0) {};

    \node[dot_style,fill=WildStrawberry, label={[label distance=7pt]below:$1$}] (n1) at (1, 0) {};

    \node[dot_style,fill=Cerulean, label={[label distance=7pt]below:$2$}] (n2) at (2, 0) {};

    \node[dot_style, fill=Cerulean,label={[label distance=7pt]below:$0$}] (nn0) at (5, 0) {};
   
    \node[dot_style, fill=WildStrawberry,label={[label distance=7pt]below:$1$}] (nn1) at (6, 0) {};

    \node[dot_style, fill=Cerulean,label={[label distance=7pt]below:$2$}] (nn2) at (7, 0) {};
   
    \node[dot_style, fill=WildStrawberry,label={[label distance=7pt]below:$3$}] (nn3) at (8, 0) {};

    \node[dot_style, fill=Cerulean,label={[label distance=7pt]below:$4$}] (nn4) at (9, 0) {};
    
    \draw[thick] (n0) -- (n1); 
    \draw[thick] (n1) -- (n2); 

    \draw[thick] (nn0) -- (nn1); 
    \draw[thick] (nn1) -- (nn2);
    \draw[thick] (nn2) -- (nn3);
    \draw[thick] (nn3) -- (nn4);
    
\node at (4, 0) {$A_5$:};

\end{tikzpicture}\caption{Vacua in some $A_{p-1}$-type deformed model.}
\label{fig:A_vacua}
\end{figure}
Intuitively, the $\mathbb Z_2$ orbifold is folding the $A$-type linear vacua with respect to their middle point. Given that the middle point is either fermionic or bosonic, we have two different patterns of vacua structure for the $D$-type deformed theory:
\begin{itemize}

\item 
For odd $k=2n+1$ orbifolding, i.e. $\mathcal L_{2k}=\mathcal L_{4n+2}$, the middle fermionic vacuum turns to a bosonic one, and thus the resulting vacuum pattern is again of $A$-type.

\item 
For even $k=2n$ orbifolding, i.e. $\mathcal L_{2k}=\mathcal L_{4n}$, the middle bosonic vacuum splits to two bosonic ones, and thus the resulted vacua pattern is of $D$-type;
    
\end{itemize}
The rules are demonstrated in Figure~\ref{fig:Dvacua} below,
\begin{figure}[H]
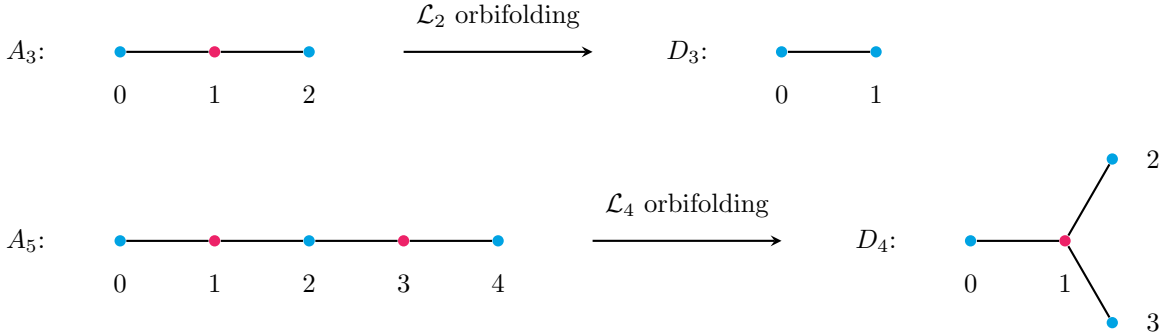

\centering
\caption{Orbifolding $A$-type deformed model to obtain $D$-type deformed model.}
\label{fig:Dvacua}
\end{figure}
\noindent 
Here we in addition remark that, for the $ D_3=A_{3}/\mathcal L_{2}$ case, the $\mathcal N=1$ $A$-type minimal model after the $\mathcal L_2$-orbifolding coincides with $\mathcal N=2$ $A_2$ theory. It has two bosonic vacua according to our rule, which is perfectly consistent with the result in Section~\ref{sec:N_2_minimal_model}. 

To see the discrepancies of the two $D$-type vacua structures, we can compute the zero temperature partition functions of the $D$-type deformed theories, which can all be obtained from their $A$-type cousins. Recall that, for $A_{2k+1}$ theory, the number of vacua for both cases can be summarized in the zero temperature partition function depending on the spin structure along spatial and temporal direction $s_x, s_t=0$ or $1$,
\begin{align}
    \mathcal Z_{s_x,\,s_t}[A_{2k+1}]=(k+1)+(-1)^{s_xs_t}k\,.
    \label{eq:pf_A}
\end{align}
Starting from \eqref{eq:pf_A}, we compute various NS/R partition functions in presence of the line $\mathcal L_{2k}$ along temporal or spatial directions. Summing them properly, one can obtain the NS/R zero temperature partition functions in $D$-type deformed theories.

\noindent
\paragraph{\textbf {$\mathcal L_{4n}$-orbifolding:}}
Let us first explain the more familiar $\mathbb Z_2$-gauging of the line $\mathcal L_{4n}$ for $k=2n$. The middle point corresponding to the SSB line $\mathcal L_{k=2n}$ is m-type, and a fixed point under the action of $\mathcal L_{4n}$. After orbifolding, it is no longer simple. Instead, it will split into two simple lines. Following eq.~\eqref{eq:pf_A}, we have the partition function in NS sector as
\begin{align}
\begin{gathered}

\end{gathered}
\quad
=
\quad\sum_{i=0}^{4n} {}_{NS}\langle i|i\rangle_{NS}=4n+1
\end{align}
positioned along the spatial direction, the line $\mathcal L_{4n}$ pairs up most of the vacua except the middle one, i.e.
\begin{align}
    \mathcal L_{4n}|i\rangle_{NS/R}=|4n-i\rangle_{NS/R}\,,\quad {\rm and}\ \ {\rm especially}\ \ \mathcal L_{4n}|2n\rangle_{NS/R}=|2n\rangle_{NS/R}\,.
    \label{eq:L4n}
\end{align}
Therefore the partition function with $\mathcal L_{4n}$ inserted horizontally is given by
\begin{align}
\begin{gathered}
%
\end{gathered}
\quad = \quad \sum_{i=0}^{4n} {}_{NS}\langle i|\mathcal L_{4n}|i\rangle_{NS}={}_{NS}\langle 2n|2n\rangle_{NS}=1\,,
\end{align}
where the black line denotes the $\mathcal L_{4n}$-line. An $S$-modular transformation will further give
\begin{align}
\begin{gathered}
%
\end{gathered}
\,\,\right)\quad=\quad 1\,.
\end{align}
However, placing $\mathcal L_{4n}$ on both spatial and temporal direction cannot be achieved via a $T$-transformation from above partition function, as it will change the boundary condition of time direction from NS to R. So instead we compute the partition function from the following path,
\begin{align}
\begin{gathered}
%
\end{gathered}
\end{align}
Applying eq.~\eqref{eq:pf_A} for $s_x=1$ and $s_t=0$, we once again have
\begin{align}
&
\begin{gathered}
%
\end{gathered}
\,\,\right)\quad=\quad 1\,
\end{align}
So overall one can compute the zero temperature partition function of the orbifolded theory in the NS sector,
\begin{align}
\begin{gathered}
%
\end{gathered}
\right)
\ =\ 2n+2\,.
\end{align}
One can similarly compute the $D$-type partition function in the Ramond sector. Noticing that, for a given Ramond vacua, the even/odd numbered ones are bosonic/fermionic, and thus we have,
\begin{align}
    (-1)^F|i\rangle_{R}=
    \begin{cases}
        +|i\rangle_R\,,\quad {\rm for}\ \ i\ \ {\rm even}\\\\
        -|i\rangle_R\,,\quad {\rm for}\ \ i\ \ {\rm odd}
    \end{cases}
\end{align}
We therefore have
\begin{align}
\begin{gathered}
%
\end{gathered}
\quad
=
\quad\sum_{i=0}^{4n} {}_{R}\langle i|(-1)^F|i\rangle_{R}=\sum_{i=0}^{4n}(-1)^i=1\,\,,
\end{align}
i.e., the Witten index of the $A$-type theory. Inserting further a $\mathcal L_{4n}$-line horizontally, and applying eq.~\eqref{eq:L4n}, one obtains
\begin{align}
\begin{gathered}
%
\end{gathered}
\quad = \quad \sum_{i=0}^{4n} {}_{R}\langle i|\mathcal L_{4n}(-1)^{F}|i\rangle_{R}=(-1)^{2n}{}_{R}\langle 2n|2n\rangle_{R}=1\,.
\end{align}
As the Ramond-Ramond sector partition functions are closed under the $S$/$T$-transformations, we also have 
\begin{align}
    \begin{gathered}
%
\end{gathered}
\ = \ 1
\end{align}
So overall we obtain the zero temperature partition function of the $D$-type theory in the Ramond sector as
\begin{align}
\begin{gathered}
%
\end{gathered}
\right)
\ =\ 2\,,
\end{align}
i.e. the total number of bosonic vacua is always two more than that of fermionic ones, which perfectly matches with the Witten index of the $D$-type minimal CFTs. One can also easily compute the partition function of NS-R, or R-NS sectors, and the overall results are finally summarized as
\begin{align}
    \mathcal Z_{s_x,\,s_t}[\mathscr D_{2n+2}]=(n+2)+(-1)^{s_xs_t}n\,.
\label{eq:pf_D1}
\end{align}

\noindent
\paragraph{\textbf {$\mathcal L_{4n+2}$-orbifolding:}}
Now we turn to the case of $k=2n+1$. The key difference here from the $k=2n$ case is that, although $\mathcal L_{4n+2}$ still shuffle the vacua from $|i\rangle$ to $|4n+2-i\rangle$ in pairs, it acts on the fixed point vacua $|2n+1\rangle$ having different eigenvalues in the NS or Ramond sector,
\begin{align}
    \mathcal L_{4n+2}|2n+1\rangle_{NS}=+|2n+1\rangle_{NS}\,,\quad{\rm and}\quad
    \mathcal L_{4n+2}|2n+1\rangle_{R}=-|2n+1\rangle_{R}\,.
    \label{eq:L4n2}
\end{align}
This additional minus sign in Ramond vacuum state will be explained soon in terms of a mixed 't Hooft anomaly between $\mathcal L_{4n+2}$ and $(-1)^F$. At this stage, let us apply eq.~\eqref{eq:L4n2} to recompute the partition functions and perform the $\mathcal L_{4n+2}$-orbifolding.

For the NS sector, we first have
\begin{align}
\begin{gathered}

\end{gathered}
\quad
=
\quad\sum_{i=0}^{4n+2} {}_{NS}\langle i|i\rangle_{NS}=4n+3
\end{align}
Placing $\mathcal L_{4n+2}$ we analogously have
\begin{align}
\begin{gathered}
%
\end{gathered}
\,\,\right)\quad=\quad 1\,.
\end{align}
Now using eq.~\eqref{eq:L4n2}, we have
\begin{align}
&
\begin{gathered}
%
\end{gathered}
\,\,\right)\quad=\quad -1\,
\end{align}
So orbifolding $\mathcal L_{4n+2}$ in NS sector gives
\begin{align}
\begin{gathered}
%
\end{gathered}
\right)\notag\\
&\quad =\quad \frac{1}{2}(4n+3+1+1-1)=2n+2\,.
\end{align}

For the Ramond sector, we have
\begin{align}
\begin{gathered}
%
\end{gathered}
\quad
=
\quad\sum_{i=0}^{4n+2} {}_{R}\langle i|(-1)^F|i\rangle_{R}=\sum_{i=0}^{4n+2}(-1)^i=1\,\,,
\end{align}
Inserting the $\mathcal L_{4n+2}$-line horizontally, and applying eq.~\eqref{eq:L4n2}, one obtains
\begin{align}
\begin{gathered}
%
\end{gathered}
\ = \ 1
\end{align}
So overall we obtain the zero temperature partition function of the $D$-type theory in the Ramond sector as
\begin{align}
\begin{gathered}
%
\end{gathered}
\right)
\ =\ 2\,,
\end{align}
i.e. the total number of bosonic vacua is once again two more than that of fermionic ones, perfectly matching with the Witten index of the $D$-type minimal CFTs. The NS-Ramond, or Ramond-NS sectors partition function are similarly computable, and the overall results can be summarized as
\begin{align}
    \mathcal Z_{s_x,\,s_t}[\mathscr D_{2n+3}]=(n+2)+(-1)^{s_xs_t}\,n\,.
    \label{eq:pf_D2}
\end{align}

\subsubsection{\texorpdfstring{Superstrip Algebra $\mathbf{sStr}(\mathscr C_{D_{k}}^{\mathcal N=1})$}{Superstrip Algebra sStr(C Dk, N=1)}}
Now we turn to study the superstrip algebra
$\mathbf{sStr}(\mathscr C_{D_{k}}^{\mathcal N=1})$ of the deformed
$D$-type theory, and establish the consistent one-particle/soliton
spectra therein. We will use the notation of the previous subsection:
the $D$-type theory is obtained from the $A$-type theory by
\[
        D_{k+2}^{\rm def}
        =
        A_{2k+1}^{\rm def}/\mathcal L_{2k}\, .
\]
Thus the index $k$ in the following formulas labels the fixed point
$\mathcal L_k$ of the parent $A_{2k+1}$ theory, while the resulting
minimal model is $D_{k+2}$. The symmetry line to be gauged is the
bosonic m-type invertible line $\mathcal L_{2k}$. Together with the
identity it defines the separable Frobenius algebra
\begin{equation}
        \mathcal A_D=\mathcal L_0\oplus \mathcal L_{2k}
        \subset \mathscr C_{A_{2k+1}}^{\mathcal N=1}\, .
        \label{eq:N1-D-Frobenius-algebra}
\end{equation}
Equivalently, $\mathcal A_D$ is the topological interface between
$A_{2k+1}^{\rm def}$ and $D_{k+2}^{\rm def}$. If this interface is pushed
to the boundary of the $D$-type theory, then the $D$-type vacua are not
labelled directly by the objects of
$\mathscr C_{A_{2k+1}}^{\mathcal N=1}$, but by the simple left
$\mathcal A_D$-modules. This is the same mechanism as in the completely
spontaneously broken $A$-type case, where the trivial algebra
$\mathcal A=\mathcal L_0$ has left modules given by the original
$A$-type objects themselves.

The computation of these left modules is reviewed in Appendix~\ref{App:algebra}; let us
summarize the part needed for the superstrip representation. In the
parent $A$-type category the gauged line acts by
\begin{equation}
        \mathcal L_{2k}\cdot \mathcal L_a=\mathcal L_{2k-a}\,,
        \qquad a=0,1,\ldots,2k\, .
        \label{eq:N1-D-Z2-action}
\end{equation}
Therefore the induced left module is
\begin{equation}
        \operatorname{Ind}_{D}(\mathcal L_a)
        =
        \mathcal A_D\otimes_{\mathcal A_f}\mathcal L_a
        =
        \mathcal L_a\oplus \mathcal L_{2k-a}\, ,
        \label{eq:N1-D-induced-module}
\end{equation}
where $\mathcal A_f$ is the fermionization algebra already condensed in
the construction of the $A$-type superfusion category. For
$a=0,\ldots,k-1$ the orbit has length two, and
\begin{equation}
        \mathcal D_a \equiv \operatorname{Ind}_{D}(\mathcal L_a),
        \qquad a=0,\ldots,k-1,
        \label{eq:N1-D-simple-orbit-modules}
\end{equation}
are simple left $\mathcal A_D$-modules. Their type is inherited from the
$A$-type object:
\begin{equation}
        \mathcal D_a
        \text{ is }
        \begin{cases}
        m\text{-type}, & a\in 2\mathbb Z,\\
        q\text{-type}, & a\in 2\mathbb Z+1.
        \end{cases}
        \label{eq:N1-D-module-parity}
\end{equation}
The only subtle point is the fixed point $\mathcal L_k$. Frobenius
reciprocity gives
\begin{equation}
        \left\langle
        \operatorname{Ind}_{D}(\mathcal L_a),
        \operatorname{Ind}_{D}(\mathcal L_a)
        \right\rangle_{\mathcal A_D}
        =
        \left\langle
        \mathcal L_a,\mathcal L_a
        \right\rangle_{\mathcal A_f},
        \qquad a=0,\ldots,k-1,
        \label{eq:N1-D-induced-hom-nonfixed}
\end{equation}
while at the fixed point
\begin{equation}
        \left\langle
        \operatorname{Ind}_{D}(\mathcal L_k),
        \operatorname{Ind}_{D}(\mathcal L_k)
        \right\rangle_{\mathcal A_D}
        =
        2\left\langle
        \mathcal L_k,\mathcal L_k
        \right\rangle_{\mathcal A_f}.
        \label{eq:N1-D-induced-hom-fixed}
\end{equation}
Since $\mathcal L_k$ is m-type for even $k$ and q-type for odd $k$,
the fixed point is resolved as
\begin{equation}
        \operatorname{Ind}_{D}(\mathcal L_k)
        =
        \begin{cases}
        \mathcal D_k^+\oplus \mathcal D_k^-,
        & k\in 2\mathbb Z,\\[3pt]
        (1\oplus \pi)\mathcal D_k,
        & k\in 2\mathbb Z+1.
        \end{cases}
        \label{eq:N1-D-fixed-point-resolution}
\end{equation}
Here $\pi$ denotes the odd one-dimensional line, so
$1\oplus\pi=\mathbb C^{1|1}$. Thus for even $k$ the middle bosonic
vacuum of the parent $A$-type chain splits into two bosonic vacua, while
for odd $k$ the middle fermionic vacuum becomes a single bosonic
m-type vacuum after the odd multiplicity in
\eqref{eq:N1-D-fixed-point-resolution} is divided out. The resulting
supermodule categories are
\begin{equation}
        \mathcal M_{D_{2n+2}}
        =
        \{\mathcal D_0,\mathcal D_1,\ldots,\mathcal D_{2n-1},
        \mathcal D_{2n}^+,\mathcal D_{2n}^-\},
        \qquad (k=2n),
        \label{eq:N1-D-even-module-category}
\end{equation}
and
\begin{equation}
        \mathcal M_{D_{2n+3}}
        =
        \{\mathcal D_0,\mathcal D_1,\ldots,\mathcal D_{2n+1}\},
        \qquad (k=2n+1).
        \label{eq:N1-D-odd-module-category}
\end{equation}
As a check, in both cases the number of m-type vacua exceeds the
number of q-type vacua by two:
\begin{equation}
        I_W(D_{k+2}^{\rm def})
        =
        \#\{m\text{-type vacua}\}
        -
        \#\{q\text{-type vacua}\}
        =2,
        \label{eq:N1-D-Witten-check}
\end{equation}
in agreement with the zero-temperature partition functions
\eqref{eq:pf_D1} and \eqref{eq:pf_D2}.

We now determine the particle/soliton spectra. Strictly speaking, the
full symmetry category of the gauged theory is the category of
$\mathcal A_D$-$\mathcal A_D$ bimodules in
$\mathscr C_{A_{2k+1}}^{\mathcal N=1}$. However, to obtain the
fundamental superstrip-algebra representation we do not need to
construct all of these bimodules. The simple line
\begin{equation}
        \widetilde{\mathcal L}_1
        \equiv
        \operatorname{Ind}_{D}(\mathcal L_1)
        =
        \mathcal A_D\otimes_{\mathcal A_f}\mathcal L_1
        \label{eq:N1-D-L1-induced}
\end{equation}
survives in the $D$-type theory and plays the role of the generator of
the connected fundamental representation. Acting with
$\widetilde{\mathcal L}_1$ on the left modules gives the junction vector
space
\begin{equation}
        \widetilde{\mathcal O}_{ab}
        \in
        \operatorname{Hom}_{\mathcal A_D}
        \left(
        \operatorname{Ind}_{D}(\mathcal L_b),
        \operatorname{Ind}_{D}
        (\mathcal L_1\otimes_{\mathcal A_f}\mathcal L_a)
        \right).
        \label{eq:N1-D-junction-space-AD}
\end{equation}
Using Frobenius reciprocity, this is equivalently
\begin{equation}
        \widetilde{\mathcal O}_{ab}
        \in
        \operatorname{Hom}_{\mathcal A_f}
        \left(
        \mathcal L_b,
        \mathcal A_D\otimes_{\mathcal A_f}
        \mathcal L_1\otimes_{\mathcal A_f}\mathcal L_a
        \right).
        \label{eq:N1-D-junction-space-Af}
\end{equation}
Thus one may compute the action in the parent $A$-type category and then
decompose the result into the simple left modules
\eqref{eq:N1-D-simple-orbit-modules} and
\eqref{eq:N1-D-fixed-point-resolution}. From the $A$-type fusion rule,
$\mathcal L_1$ only changes the label by one. Hence, for the non-fixed
modules,
\begin{equation}
        \widetilde{\mathcal O}_{ab}
        =
        \begin{cases}
        \mathbb C^{1|1}, & |a-b|=1,\\
        0, & \text{otherwise},
        \end{cases}
        \qquad 0\leq a,b\leq k-1.
        \label{eq:N1-D-nonfixed-junctions}
\end{equation}
At the fixed point one first obtains
\begin{equation}
        \widetilde{\mathcal O}_{k-1,k}
        \in
        \mathbb C^{2|2}
        =
        2(1\oplus\pi).
        \label{eq:N1-D-fixed-junction-before-resolution}
\end{equation}
For even $k$, the fixed induced module splits into
$\mathcal D_k^+\oplus\mathcal D_k^-$, so each branch carries one copy of
$\mathbb C^{1|1}$:
\begin{equation}
        \widetilde{\mathcal O}_{k-1,k^+},
        \widetilde{\mathcal O}_{k^+,k-1},
        \widetilde{\mathcal O}_{k-1,k^-},
        \widetilde{\mathcal O}_{k^-,k-1}
        \in
        \mathbb C^{1|1},
        \qquad (k\in 2\mathbb Z).
        \label{eq:N1-D-even-fixed-junctions}
\end{equation}
For odd $k$, the fixed induced module is
$(1\oplus\pi)\mathcal D_k$, and after removing this internal
$\mathbb C^{1|1}$ multiplicity one obtains
\begin{equation}
        \widetilde{\mathcal O}_{k-1,k},
        \widetilde{\mathcal O}_{k,k-1}
        \in
        \mathbb C^{1|1},
        \qquad (k\in 2\mathbb Z+1).
        \label{eq:N1-D-odd-fixed-junctions}
\end{equation}
There are no diagonal entries in this representation. Therefore the
fundamental states are solitons and anti-solitons between adjacent
vacua, rather than local particle excitations above a single vacuum.

It is useful to package the result as a quiver. An undirected edge below
means that both orientations are present and that each oriented sector
has junction space $\mathbb C^{1|1}$, i.e. one bosonic and one fermionic
state. More precisely, the edge sets are
\begin{equation}
        E(\Gamma_{D_{2n+3}})
        =
        \{(\mathcal D_a,\mathcal D_{a+1})\mid a=0,\ldots,2n\},
        \qquad (k=2n+1),
        \label{eq:N1-D-odd-edge-set}
\end{equation}
and
\begin{equation}
        E(\Gamma_{D_{2n+2}})
        =
        \{(\mathcal D_a,\mathcal D_{a+1})\mid a=0,\ldots,2n-2\}
        \cup
        \{(\mathcal D_{2n-1},\mathcal D_{2n}^{+}),
        (\mathcal D_{2n-1},\mathcal D_{2n}^{-})\},
        \qquad (k=2n).
        \label{eq:N1-D-even-edge-set}
\end{equation}
For odd $k=2n+1$ the fixed point does not split, and the last two vacua are both m-type, which is the algebraic expression of the statement that the middle fermionic $A$-type vacuum becomes
bosonic after gauging. On the other hand, for even $k=2n$, the fixed point splits. The two terminal vacua $\mathcal D_{2n}^{\pm}$ are both m-type and are not directly connected to each other in this fundamental representation; each is connected to $\mathcal D_{2n-1}$ by its own
copy of $\mathbb C^{1|1}$. The drawings below are schematic; in the lowest-rank cases the
intermediate nodes near the ellipsis are simply omitted:
\begin{figure}[H]
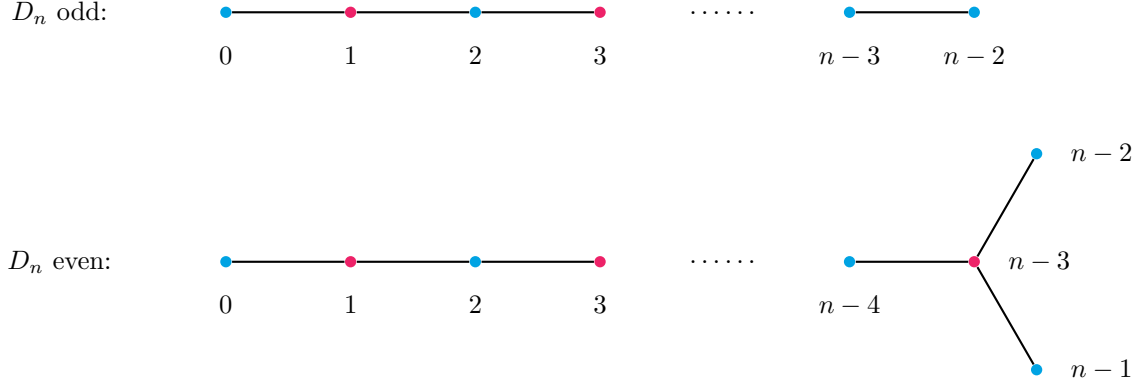

\centering
\caption{Vacua in some $D$-type deformed model.}
\label{fig:D_vacua}
\end{figure}
Consequently, every edge $M-N$ in \eqref{fig:D_vacua} gives four
fundamental one-particle states:
\begin{equation}
        K^{\rm b}_{MN},\quad K^{\rm f}_{MN},\quad
        K^{\rm b}_{NM},\quad K^{\rm f}_{NM}.
        \label{eq:N1-D-edge-states}
\end{equation}
The first two are solitons in $\mathcal H_{M,N}$ and the last two are
anti-solitons in $\mathcal H_{N,M}$. The pair
$(K^{\rm b}_{MN},K^{\rm f}_{MN})$ forms an $\mathcal N=1$ massive
supermultiplet, and similarly for the opposite orientation. Since all
edges belong to the same irreducible
$\mathcal R_{\widetilde{\mathcal L}_1}$ representation of the
superstrip algebra, the corresponding fundamental soliton multiplets
are degenerate in mass. This is the $D$-type counterpart of the
$A$-type statement above: the superstrip algebra determines which
vacuum-changing sectors are tied together by the preserved
non-invertible topological lines, while supersymmetry pairs the bosonic
and fermionic states inside each oriented soliton sector. Possible
heavier bound-state multiplets, if present, would be generated by other
objects of the $D$-type bimodule category; the fundamental spectrum
discussed here is the one generated by $\widetilde{\mathcal L}_1$.
\begin{figure}[!htbp]
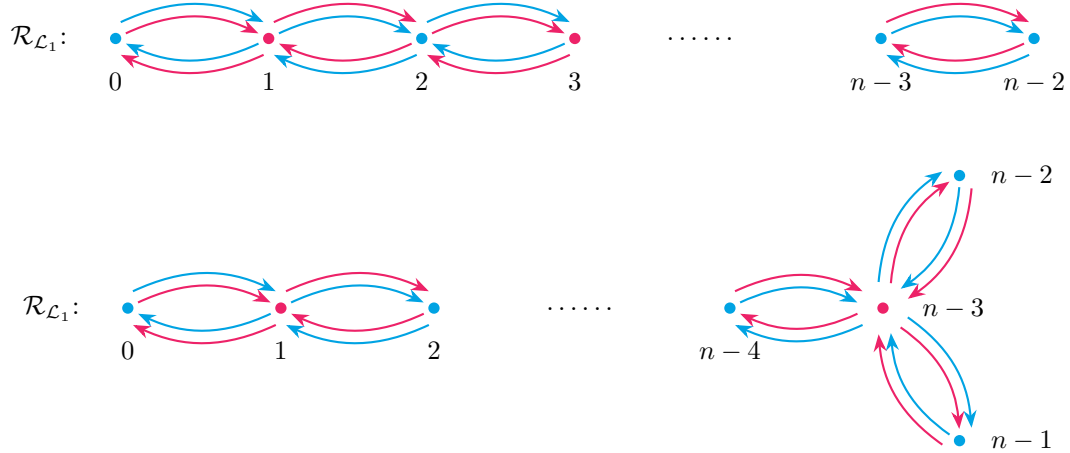

\centering

\caption{$\mathcal{N}=1$, $D_{n\,{\rm odd}}$ and $D_{n\,{\rm even}}$-type respectively}
\label{fig:Dodd-type_N=1}
\end{figure}

\subsection{\texorpdfstring{$E$-type}{E-type}}

We now turn to the exceptional $E$-type deformed models. The logic is
the same as in the $D$-type construction, but the algebra to be gauged is
no longer generated by an invertible $\mathbb Z_2$ line. Thus there is
no literal folding picture of an $A$-type chain. Instead, the vacuum
structure must be read from the simple left modules of the exceptional
Frobenius algebra. Let
\[
        \kappa_{E_6}=10,\qquad
        \kappa_{E_7}=16,\qquad
        \kappa_{E_8}=28 .
\]
The corresponding $E$-type theories are obtained by gauging the
following algebra objects in the already fermionized $A$-type categories
$\mathscr C_{A_{\kappa_G}}^{\mathcal N=1}$:
\begin{equation}
        \begin{gathered}
        \mathcal A_{E_6}
        =
        \mathcal L_0\oplus \mathcal L_6
        \subset \mathscr C_{A_{11}}^{\mathcal N=1},
        \\[3pt]
        \mathcal A_{E_7}
        =
        \mathcal L_0\oplus \mathcal L_8\oplus \mathcal L_{16}
        \subset \mathscr C_{A_{17}}^{\mathcal N=1},
        \\[3pt]
        \mathcal A_{E_8}
        =
        \mathcal L_0\oplus \mathcal L_{10}
        \oplus \mathcal L_{18}\oplus \mathcal L_{28}
        \subset \mathscr C_{A_{29}}^{\mathcal N=1}.
        \end{gathered}
        \label{eq:N1-E-Frobenius-algebras}
\end{equation}
Physically, $\mathcal A_G$ is the topological interface between the
parent $A_{\kappa_G}^{\rm def}$ theory and the exceptional
$G^{\rm def}$ theory. Pushing this interface to the boundary of the
exceptional theory gives
\[
        G^{\rm def}
        =
        A_{\kappa_G}^{\rm def}/\mathcal A_G,
        \qquad
        G=E_6,E_7,E_8,
\]
and the supersymmetric vacua are described by simple left
$\mathcal A_G$-modules. We write
\begin{equation}
        \operatorname{Ind}_G(\mathcal L_a)
        =
        \mathcal A_G\otimes_{\mathcal A_f}\mathcal L_a
        \label{eq:N1-E-induced-module}
\end{equation}
for the induced left module, where $\mathcal A_f$ is the
fermionization algebra already used to obtain
$\mathscr C_{A_{\kappa_G}}^{\mathcal N=1}$.

The simple modules are determined by Frobenius reciprocity, as worked
out in Appendix~\ref{App:algebra}. For $E_6$ one obtains
\begin{equation}
        \begin{gathered}
        \mathcal E_0=\operatorname{Ind}_{E_6}(\mathcal L_0),
        \qquad
        \mathcal E_1=\operatorname{Ind}_{E_6}(\mathcal L_1),
        \qquad
        \mathcal E_2=\operatorname{Ind}_{E_6}(\mathcal L_2)
        =
        \operatorname{Ind}_{E_6}(\mathcal L_8),
        \\[3pt]
        \mathcal E_3=\operatorname{Ind}_{E_6}(\mathcal L_9),
        \qquad
        \mathcal E_4=\operatorname{Ind}_{E_6}(\mathcal L_{10}),
        \qquad
        \mathcal E_5=
        \operatorname{Ind}_{E_6}(\mathcal L_3)
        -
        \operatorname{Ind}_{E_6}(\mathcal L_9).
        \end{gathered}
        \label{eq:N1-E6-simple-modules}
\end{equation}
Here and below, a difference such as
$\operatorname{Ind}(\mathcal L_3)-\operatorname{Ind}(\mathcal L_9)$ is
a shorthand in the Grothendieck group for the complementary simple
summand of the reducible induced module; it is not a negative physical
object. For $E_7$ the simple modules are
\begin{equation}
        \begin{gathered}
        \mathcal E_0=\operatorname{Ind}_{E_7}(\mathcal L_0),
        \qquad
        \mathcal E_1=\operatorname{Ind}_{E_7}(\mathcal L_1),
        \qquad
        \mathcal E_2=\operatorname{Ind}_{E_7}(\mathcal L_2),
        \qquad
        \mathcal E_3=\operatorname{Ind}_{E_7}(\mathcal L_3),
        \\[3pt]
        \mathcal E_4=
        \operatorname{Ind}_{E_7}(\mathcal L_6)
        -
        \operatorname{Ind}_{E_7}(\mathcal L_2),
        \qquad
        \mathcal E_5=
        \operatorname{Ind}_{E_7}(\mathcal L_5)
        -
        \operatorname{Ind}_{E_7}(\mathcal L_3),
        \\[3pt]
        \mathcal E_6=
        \operatorname{Ind}_{E_7}(\mathcal L_2)
        +
        \operatorname{Ind}_{E_7}(\mathcal L_4)
        -
        \operatorname{Ind}_{E_7}(\mathcal L_6).
        \end{gathered}
        \label{eq:N1-E7-simple-modules}
\end{equation}
For $E_8$ one similarly finds
\begin{equation}
        \begin{gathered}
        \mathcal E_0=\operatorname{Ind}_{E_8}(\mathcal L_0),
        \qquad
        \mathcal E_1=\operatorname{Ind}_{E_8}(\mathcal L_1),
        \qquad
        \mathcal E_2=\operatorname{Ind}_{E_8}(\mathcal L_2),
        \\[3pt]
        \mathcal E_3=\operatorname{Ind}_{E_8}(\mathcal L_3),
        \qquad
        \mathcal E_4=\operatorname{Ind}_{E_8}(\mathcal L_4),
        \qquad
        \mathcal E_5=
        \operatorname{Ind}_{E_8}(\mathcal L_7)
        -
        \operatorname{Ind}_{E_8}(\mathcal L_3),
        \\[3pt]
        \mathcal E_6=
        \operatorname{Ind}_{E_8}(\mathcal L_6)
        -
        \operatorname{Ind}_{E_8}(\mathcal L_4),
        \qquad
        \mathcal E_7=
        \operatorname{Ind}_{E_8}(\mathcal L_5)
        +
        \operatorname{Ind}_{E_8}(\mathcal L_3)
        -
        \operatorname{Ind}_{E_8}(\mathcal L_7).
        \end{gathered}
        \label{eq:N1-E8-simple-modules}
\end{equation}

The $m/q$ type of these modules gives the bosonic/fermionic grading of
the vacua:
\begin{equation}
        \begin{array}{c|c|c|c|c}
        G & \text{simple left modules} & m\text{-type vacua}
        & q\text{-type vacua} & I_W
        \\ \hline
        E_6
        & \mathcal E_0,\ldots,\mathcal E_5
        & \mathcal E_0,\mathcal E_2,\mathcal E_4
        & \mathcal E_1,\mathcal E_3,\mathcal E_5
        & 0
        \\
        E_7
        & \mathcal E_0,\ldots,\mathcal E_6
        & \mathcal E_0,\mathcal E_2,\mathcal E_4,\mathcal E_6
        & \mathcal E_1,\mathcal E_3,\mathcal E_5
        & 1
        \\
        E_8
        & \mathcal E_0,\ldots,\mathcal E_7
        & \mathcal E_0,\mathcal E_2,\mathcal E_4,\mathcal E_6
        & \mathcal E_1,\mathcal E_3,\mathcal E_5,\mathcal E_7
        & 0
        \end{array}
        \label{eq:N1-E-vacua-table}
\end{equation}
Thus the exceptional supermodule categories reproduce the expected number of
vacua,
\[
        |\mathcal M_{E_6}|=6,\qquad
        |\mathcal M_{E_7}|=7,\qquad
        |\mathcal M_{E_8}|=8,
\]
with Witten index computed as the difference of numbers of bosonic and fermionic vacua, $\#m-\#q$.

\subsubsection{\texorpdfstring{Superstrip Algebra $\mathbf{sStr}(\mathscr C_{E}^{\mathcal N=1})$}{Superstrip Algebra sStr(C E, N=1)}}

The fundamental superstrip-algebra representation is again generated by
the image of the $A$-type line $\mathcal L_1$:
\begin{equation}
        \widetilde{\mathcal L}_1^{(G)}
        =
        \operatorname{Ind}_{G}(\mathcal L_1)
        =
        \mathcal A_G\otimes_{\mathcal A_f}\mathcal L_1,
        \qquad
        G=E_6,E_7,E_8 .
        \label{eq:N1-E-L1-generator}
\end{equation}
As in the $D$-type case, we do not need the full
$\mathcal A_G$-$\mathcal A_G$ bimodule category to read off the
fundamental particle/soliton spectrum. We only need the action of
$\widetilde{\mathcal L}_1^{(G)}$ on the left modules. For each simple
module $\mathcal E_a$, choose a representative $X_a^{(G)}$ in the
parent $A$-type category, in the Grothendieck sense described above,
such that
\[
        \operatorname{Ind}_G(X_a^{(G)})=\mathcal E_a .
\]
The relevant junction vector space is
\begin{equation}
        \widetilde{\mathcal O}^{(G)}_{ab}
        \in
        \operatorname{Hom}_{\mathcal A_G}
        \left(
        \mathcal E_b,
        \operatorname{Ind}_{G}
        (\mathcal L_1\otimes_{\mathcal A_f}X_a^{(G)})
        \right),
        \label{eq:N1-E-junction-space-AG}
\end{equation}
or, equivalently by Frobenius reciprocity,
\begin{equation}
        \widetilde{\mathcal O}^{(G)}_{ab}
        \in
        \operatorname{Hom}_{\mathcal A_f}
        \left(
        X_b^{(G)},
        \mathcal A_G\otimes_{\mathcal A_f}
        \mathcal L_1\otimes_{\mathcal A_f}X_a^{(G)}
        \right),
        \label{eq:N1-E-junction-space-Af}
\end{equation}
The second expression is the practical computation: evaluate the
representatives $X_a^{(G)}$ in the parent $A$-type category, then
decompose the resulting induced module into the simple
$\mathcal A_G$-modules listed in
\eqref{eq:N1-E6-simple-modules}--\eqref{eq:N1-E8-simple-modules}.

In the ordered bases displayed above, Appendix~\ref{App:algebra} gives the following
matrices for the junction spaces:
\begin{equation}
        \left\langle
        \widetilde{\mathcal O}^{(E_6)}
        (\mathcal E_a\rightarrow \mathcal E_b)
        \right\rangle
        =
        (1+\pi)
        \left(
\begin{array}{cccccc}
        0 & 1 & 0 & 0 & 0 & 0 \\
        1 & 0 & 1 & 0 & 0 & 0 \\
        0 & 1 & 0 & 1 & 0 & 1 \\
        0 & 0 & 1 & 0 & 1 & 0 \\
        0 & 0 & 0 & 1 & 0 & 0 \\
        0 & 0 & 1 & 0 & 0 & 0
        \end{array}
        \right),
        \label{eq:N1-E6-junction-matrix}
\end{equation}
\begin{equation}
        \left\langle
        \widetilde{\mathcal O}^{(E_7)}
        (\mathcal E_a\rightarrow \mathcal E_b)
        \right\rangle
        =
        (1+\pi)
        \left(
        \begin{array}{ccccccc}
        0 & 1 & 0 & 0 & 0 & 0 & 0 \\
        1 & 0 & 1 & 0 & 0 & 0 & 0 \\
        0 & 1 & 0 & 1 & 0 & 0 & 0 \\
        0 & 0 & 1 & 0 & 1 & 0 & 1 \\
        0 & 0 & 0 & 1 & 0 & 1 & 0 \\
        0 & 0 & 0 & 0 & 1 & 0 & 0 \\
        0 & 0 & 0 & 1 & 0 & 0 & 0
        \end{array}
        \right),
        \label{eq:N1-E7-junction-matrix}
\end{equation}
and
\begin{equation}
        \left\langle
        \widetilde{\mathcal O}^{(E_8)}
        (\mathcal E_a\rightarrow \mathcal E_b)
        \right\rangle
        =
        (1+\pi)
        \left(
        \begin{array}{cccccccc}
        0 & 1 & 0 & 0 & 0 & 0 & 0 & 0 \\
        1 & 0 & 1 & 0 & 0 & 0 & 0 & 0 \\
        0 & 1 & 0 & 1 & 0 & 0 & 0 & 0 \\
        0 & 0 & 1 & 0 & 1 & 0 & 0 & 0 \\
        0 & 0 & 0 & 1 & 0 & 1 & 0 & 1 \\
        0 & 0 & 0 & 0 & 1 & 0 & 1 & 0 \\
        0 & 0 & 0 & 0 & 0 & 1 & 0 & 0 \\
        0 & 0 & 0 & 0 & 1 & 0 & 0 & 0
        \end{array}
        \right).
        \label{eq:N1-E8-junction-matrix}
\end{equation}
Equivalently, the nonzero entries are precisely the adjacency matrices
of the $E_6$, $E_7$, and $E_8$ Dynkin graphs, with every oriented edge
carrying the superspace $\mathbb C^{1|1}$. There are no diagonal
entries in these matrices, so the fundamental representation contains
only vacuum-changing soliton sectors.

The corresponding quivers are
\begin{figure}[H]
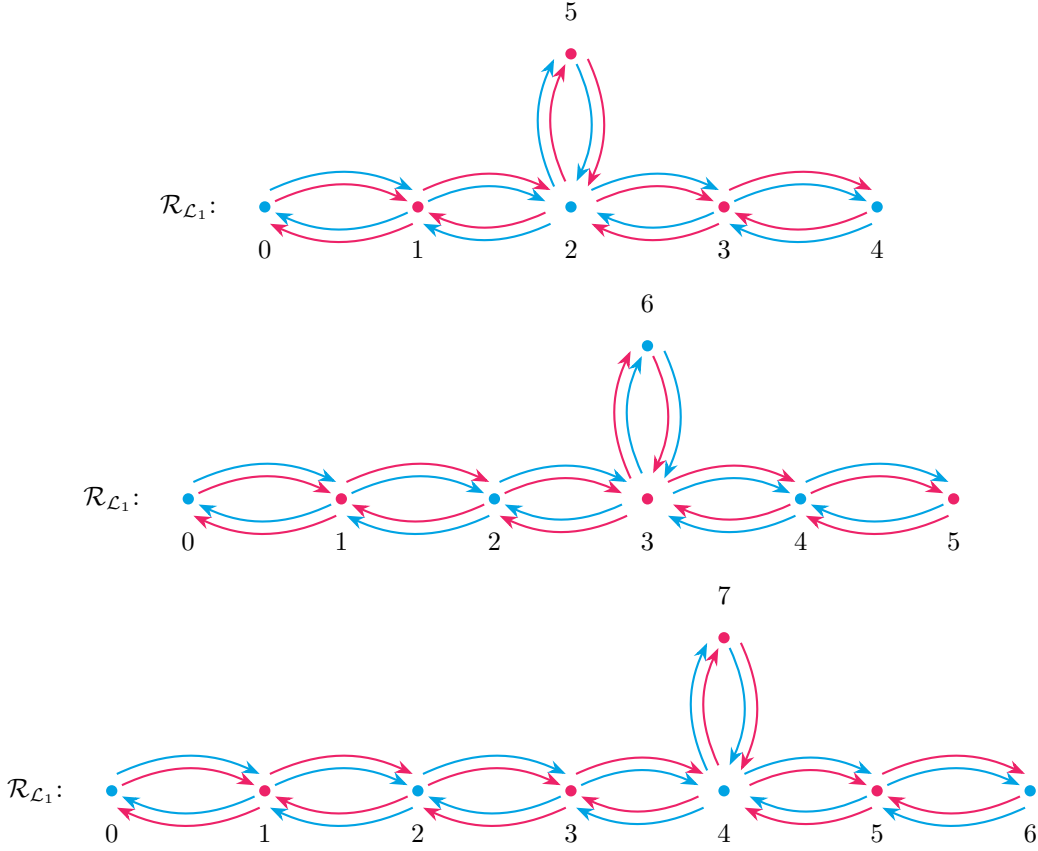

\centering
%
\caption{$\mathcal{N}=1$, $E_6$, $E_7$, and $E_8$-type}
\label{fig:E6-type_N=1}
\label{fig:E7-type_N=1}
\label{fig:E8-type_N=1}
\end{figure}
\noindent 
where blue dots denote m-type, hence bosonic, vacua and red dots
denote q-type, hence fermionic, vacua. As in the $D$-type case, an
undirected edge should be read as two oriented soliton sectors, and each
oriented sector carries one bosonic and one fermionic state:
\begin{equation}
        \mathcal H_{\mathcal E_a,\mathcal E_b}
        :
        \quad
        K^{\rm b}_{ab},\ K^{\rm f}_{ab},
        \qquad
        \mathcal H_{\mathcal E_b,\mathcal E_a}
        :
        \quad
        K^{\rm b}_{ba},\ K^{\rm f}_{ba},
        \qquad
        (\mathcal E_a,\mathcal E_b)\in E(\Gamma_G).
        \label{eq:N1-E-edge-states}
\end{equation}
The boson/fermion pair in each oriented Hilbert space forms an
$\mathcal N=1$ massive multiplet. The exceptional non-invertible
symmetry then ties all fundamental soliton multiplets on the same
$E_G$ Dynkin graph into a single irreducible representation
$\mathcal R_{\widetilde{\mathcal L}_1}^{G}$ of the superstrip algebra.
Consequently, within this fundamental representation the soliton
multiplets on all edges of a fixed $E_G$ graph are degenerate in mass.
Other representations of the full
$\mathcal A_G$-$\mathcal A_G$ bimodule category may describe heavier
particle or bound-state sectors; the universal fundamental spectrum is
the one generated by $\widetilde{\mathcal L}_1^{(G)}$.

\section{\texorpdfstring{Fermion number in $\mathcal N=1$ and $\mathcal N=2$ solitons}{Fermion number in N=1 and N=2 solitons}}\label{sec:fermion_number}
Following the discussion of the solitons in the least relevant deformed $\mathcal N=1$ and $\mathcal N=2$ minimal models, we, in this section, will use the full module $F$-symbols data to compute the fermion numbers of these soliton solutions. Especially, we will explicitly show that the fermion number of the (anti-)soliton between adjacent vacua are fractional $\pm\frac12$ in $A_2^{\mathcal N=2}$, and integral in $A_2^{\mathcal N=1}$ respectively.

\subsection{\texorpdfstring{Fermion number in $A_2^{\mathcal N=2}$}{Fermion number in A2 (N=2)}}
\label{subsec:fermion_number_N=2}
We start from the case of $A_2^{\mathcal N=2}$, where the full superfusion category is 
\begin{align}
    \mathscr C_{A_2^{\mathcal N=2}}=\mathbb Z_2^m\boxtimes \mathbb Z_2^{(-1)^F}\,,
\end{align}
where the $\mathbb Z_2^m=\mathscr {\tilde C}_{A_2}$ is the spontaneous broken symmetry with an m-type generator $\mathcal L_1$. Recall in \eqref{eq: su_2_N=2}, its fusion rules satisfy
\begin{align}
    \mathcal L_1\cdot \mathcal L_1=\mathbb C^{0|1} \mathcal L_0
\end{align}
The fractional fermion number in a soliton is due to the 't Hooft anomaly between $\mathcal L_1$ and another m-type symmetry, the $(-1)^F$, that is not spontaneously broken. Therefore one has to study the module of $\mathscr C_{A_2^{\mathcal N=2}}$,
\begin{align}
    \mathcal M_{A_2^{\mathcal N=2}}=\{m_0,\,m_1\}\,,
\end{align}
where $m_i$ are two simple objects in $\mathcal M_{A_2^{\mathcal N=2}}$, satisfying the module fusion rules:
\begin{align}
    (-1)^F\triangleright m_i=\mathbb C^{1|0}\,m_i\,,\qquad \mathcal L_1\triangleright m_0=\mathbb C^{1|0}\,m_1\,,\qquad \mathcal L_1\triangleright m_{1}=\mathbb C^{0|1}\,m_0\,.
\end{align}
The two $m_i$ characterize the two bosonic vacua of the deformed $A_2^{\mathcal N=2}$ model. Therefore a solitonic object can be diagrammatically depicted as the following configuration
\begin{align}
|{\rm sol}_{\,0,1}\rangle \quad=\quad
    \begin{gathered}

    \end{gathered}
\end{align}
To extract the necessary topological data, one has to first solve the superpentagons to the category $F$-symbols in $\mathscr C_{A_2^{\mathcal N=2}}$, then further solve module superpentagons for the module $F$-symbols $\mathcal F_M$ in $\mathcal M_{A_2^{\mathcal N=2}}$:
\begin{align}
\begin{gathered}
%
  
    \end{gathered}
 \end{align}
With the module $F$-symbols, consider the following inner product between soliton states. For better visualization, we compactify the lower and upper part of the superstrip algebra and rotate it clockwise by 90 degrees, 
\begin{align}
\langle {\rm sol}|(-1)^{2F}|{\rm sol}\rangle =
\begin{gathered}
%
 
\end{gathered}
\end{align}
where the purple, yellow and green dashed lines denote $(-1)^F$, $\eta$ and $\eta\,(-1)^F$. We apply the module $F$-moves in the following way: 
\begin{align}
\label{eq:Fmoves_in_sphere}
\begin{gathered}
%
\end{gathered}
\,,
\end{align}
where, in the above graphs, the purple bubbles have been absorbed because the quantum dimension of $(-1)^F$ is unity. The sequence of module $F$-moves picks up a phase
\begin{align}
    \kappa=\frac{\mathcal F_{M,m_1}^{m_2,\eta,(-1)^F}[m_1,\eta(-1)^F]^{10}_{01}\,\,\mathcal F_{M,m_1}^{m_2,(-1)^F,\eta(-1)^F}[m_2,\eta]^{01}_{01}}{\mathcal F_{M,m_2}^{m_1,(-1)^F,\eta(-1)^F}[m_1,\eta]^{00}_{00}\,\,\mathcal F_{M,m_2}^{m_1,\eta,(-1)^F}[m_2,\eta(-1)^F]^{00}_{00}}=-1\,.
\end{align}
Notice that the phase $\kappa$ is \emph{gauge invariant} under rescaling the module $F$-symbols,
\begin{align}
    F_{M\,\,m_2}^{m_1,\mathcal L_1,\mathcal L_2}[m_3,\mathcal L_3]^{\alpha\beta}_{\gamma\delta}\quad\longrightarrow
    \quad F_{M\,\,m_2}^{m_1,\mathcal L_1,\mathcal L_2}[m_3,\mathcal L_3]^{\alpha\beta}_{\gamma\delta}\cdot\frac{g(\mathcal L_1,\mathcal L_2,\mathcal L_3;\,\gamma)\,g(\mathcal L_3,m_1,m_2;\,\delta)}{g(\mathcal L_1,m_1,m_3;\,\alpha)\,g(\mathcal L_2,m_3,m_2;\,\beta)}\,,
\end{align}
where $g(\mathcal L_i,\mathcal L_j,\mathcal L_k;\,\alpha)$ and $g(\mathcal L_i,m_j,m_k;\,\alpha)$ are gauge transformations of the space ${\rm Hom}_{\mathbb C}(\mathcal L_i\otimes\mathcal L_j,\mathcal L_k)$ and ${\rm Hom}_{\mathbb C}(\mathcal L_i\cdot m_j, m_k)$ respectively, and $\alpha$ labels the fermionic or bosonic type of the trivalent junctions. Therefore, we conclude that
\begin{align}
\langle {\rm sol}|(-1)^{2F}|{\rm sol}\rangle=-1\,,
\end{align}
i.e. the soliton fermion number $F=\pm\frac{1}{2}$ is fractional.

Along the same line, for the case of $A_{3}^{\rm def.}$, one has full superfusion category,
\begin{align}
    \mathscr C_{A_3^{\rm def.}}=\tilde{\mathscr C}_{A_3}\boxtimes\mathbb Z_2^{(-1)^F}\,.
\end{align}
Only $\tilde{\mathscr C}_{A_3}=\{\mathcal L_0,\,\mathcal L_1,\,\mathcal L_2\}$ is SSB and results in three vacua $m_0$, $m_1$, and $m_2$ linearly aligned, satisfying
\begin{equation}
\begin{aligned}
    \mathcal L_1|m_0\rangle&=|m_1\rangle\,, &
    \mathcal L_1|m_1\rangle&=|m_0\rangle+|m_2\rangle\,, &
    \mathcal L_1|m_2\rangle&=|m_1\rangle\,,\\[1ex]
    \mathcal L_2|m_0\rangle&=|m_2\rangle\,, &
    \mathcal L_2|m_1\rangle&=|m_1\rangle\,, &
    \mathcal L_2|m_2\rangle&=|m_0\rangle\,.
\end{aligned}
\end{equation}
Using the module $F$-symbols involving lines $\mathcal L_1$ and $\mathcal L_2$, one can honestly compute
\begin{align}
    \langle {\rm sol}_{i,j}|(-1)^{2F}|{\rm sol}_{i,j}\rangle=
    \begin{cases}
        -1\,,\quad i,j\ \ {\rm adjacent}\\
        +1\,,\quad {\rm else}
    \end{cases}
    \,,
    \label{eq:sol_amp}
\end{align}
where $|{\rm sol}_{i,j}\rangle$ denotes a soliton state interpolating vacua $v_i$ and $v_j$. Eq. \eqref{eq:sol_amp} implies that only soliton states in adjacent vacua bear fractional fermion numbers, matching perfectly with the result in \cite{Fendley:1991ve}, see also discussion in Section~\ref{subsec:U1_anomaly}

\subsection{\texorpdfstring{Fermion number in $\mathcal N=2$}{Fermion number in N=2}}
\label{subsec:U1_anomaly}
For an $\mathcal N=2$ conformal theory, the above mixing anomalies between $(-1)^F$ and spontaneously broken m-type line $\mathcal L_1$ can be studied in terms of continuous $U(1)$-symmetries. Recall that, in a conformal fixed point, the $\mathcal N=2$ theory will admit two non-anomalous $U(1)_{V}$ and $U(1)_A$ symmetries corresponding to currents $J_{V}$ and $J_A$. With respect to them, one can define two conserved charges $Q_V$ and $Q_A$. When the theory is off-critical, only one of them is preserved, the other one will be broken to a subgroup $\mathbb Z_{N_f}\subset U(1)_{\rm broken}$ where $N_f$ is a model dependent integer. In an ordinary gauge theory, 2D QED with $N_f$ Dirac fermions for example, one chooses to couple gauge field $A$ to $J_V$. Then $J_V$ is guaranteed to be conserved, while $J_A$ inevitably suffers from a chiral anomaly, given by
\begin{align}
    \partial_\mu J_A^\mu=\frac{N_f}{2\pi}\epsilon^{\mu\nu}F_{\mu\nu}\,,
\end{align}
and thus the $U(1)_A$ breaks down to $Z_{N_f}$. Here $F_{\mu\nu}$ is the field strength of $A_\mu$. On the other hand, if one instead couples a gauge field $B$ to the axial current $J_A$, $J_A$ must be anomaly free, which result in an anomaly on the vector current $J_V$, similarly given by
\begin{align}
    \partial_\mu J_V^\mu=\frac{N_f}{2\pi}\epsilon^{\mu\nu}B_{\mu\nu}\,,
\end{align}
where $B_{\mu\nu}$ is the field strength of $B_\mu$ instead. 

It is important to notice that, although $Q_V$ and $Q_A$ assign different charges to fermionic excitation states depending on their chirality, they are \emph{the same} in the sense of modulo 2. Therefore, for a given system with either conservation of $J_V$ or $J_A$, one can uniquely define the fermion parity operator $(-1)^F$,
\begin{align}
    (-1)^F=\begin{cases}
        (-1)^{Q_V}\,,\ \ J_V\ {\rm \ is\ \ conserved}\\\\
        (-1)^{Q_A}\,,\ \ J_A\ {\rm \ is\ \ conserved}
    \end{cases}
\end{align}
On the other hand, the symmetry $Z_{N_f}$ from the non-preserved $U(1)$ will be further spontaneously broken. 

Back to our $\mathcal N=2$ case, for a given LG model with superpotential $\mathcal W(\Phi)$. One can write down the Lagrangian in components
\begin{align}
    \mathcal L=\mathcal L_{kin}+\left|\mathcal W'(\phi)\right|^2+\mathcal W''(\phi)\,\psi^2+h.c.\,,
\end{align}
where $\psi^2\equiv\psi_\alpha\psi^\alpha=-2\psi_R\psi_L$. Therefore, for this model, one can see that the Yukawa term explicitly breaks the $U(1)_V$-symmetry to a $Z_2$-symmetry:
\begin{align}
    \mathbb Z_2^V:\quad
    \psi_{R,L}\longrightarrow -\psi_{R,L}
\end{align}
In contrast, a chiral rotation symmetry, $U(1)_A$,
\begin{align}
    \psi_R\longrightarrow e^{i\theta}\psi_R\,,\quad 
    \psi_L\longrightarrow e^{-i\theta}\psi_L\,,
\end{align}
is preserved. Thus one defines the fermion number by
\begin{align}
    (-1)^F\equiv (-1)^{Q_A}\,,
\end{align}
as the $\mathbb Z_2$ subgroup of $U(1)_A$.

Now we can apply the standard approach in \cite{Goldstone:1981kk} to study the fractional fermion number in the soliton background. First we introduce a background gauge field $B_\mu$ to couple the axial current 
\begin{align}
    J_A^\mu=\bar\psi\gamma^\mu\gamma_5\psi\,.
\end{align}
We further adiabatically employ a vector rotation to offset the ${\rm Arg}\left(\mathcal W''(\phi)\right)$, i.e.
\begin{align}
    &\psi_{R,\, L}\longrightarrow e^{i\theta}\psi_{R,\, L}\,,\ \ {\rm with}\ \ \theta=-\frac{1}{2}\mathfrak{Im}\log\mathcal W''(\phi)\,.
\end{align}
Under this transformation, $\mathcal W''(\phi)$ is fixed to be real. However such a rotation will result in a non-unity Jacobian for the measure of the fermion fields
\begin{align}
    \mathcal D\bar\psi\mathcal D\psi\xrightarrow{\psi\rightarrow e^{i\theta}\psi}\det\mathcal J\cdot \mathcal D\bar\psi\mathcal D\psi=\mathcal D\bar\psi\mathcal D\psi\, \exp\left(\int d^2x\ \frac{\theta(x)}{2\pi}\epsilon^{\mu\nu}B_{\mu\nu}\right)\,.
\end{align}
Therefore, varying $B_0$ we have 
\begin{align}
    \left\langle J^0_A\right\rangle=\frac{\delta\Gamma_{\rm eff.}}{\delta B_0}=-\frac{1}{\pi}\partial_x\theta=\frac{1}{2\pi}\partial_x\left(\mathfrak{Im}\log\mathcal W''(\phi)\right)\,.
\end{align}
Finally, we compute a net $Q_A$-charge in a solitonic background $\phi(-\infty)=a$ and $\phi(+\infty)=b$,
\begin{align}
    \Delta Q_A=\int_{-\infty}^{+\infty} \hspace{-.3cm} dx\,\, \left\langle J^0_A\right\rangle=\frac{1}{2\pi}\mathfrak{Im}\log\mathcal W''(\phi){\Bigg \vert}^{\phi(+\infty)=b}_{\phi(-\infty)=a}
\end{align}

\subsection{\texorpdfstring{Fermion number in $A_2^{\mathcal N=1}$}{Fermion number in A2 (N=1)}}
Now let us return to the fermion number for $\mathcal N=1$ solitons. Different from the case of $\mathcal N=2$, there is no continuous $R$-symmetry to embed $(-1)^F$. So the fermion number here cannot be probed from a continuous $U(1)$ anomaly as in its $\mathcal N=2$ counterpart. However, the mixing anomaly between the $(-1)^F$ and the spontaneously broken $\mathcal L_1$ is similarly manifested in terms of the $F$-symbols as we have seen in Section~\ref{subsec:fermion_number_N=2}. The main difference is that now $\mathcal L_1$ is a q-type object, satisfying
\begin{align}
    \mathcal L_1\cdot\mathcal L_1=\mathbb C^{1|1} \mathcal L_0\,,
\end{align}
and generating the q-type $\mathbb Z_2^q$-symmetry. The full superfusion category of $A_2^{\mathcal N=1}$ is thus
\begin{align}
    \mathscr C_{A_2^{\mathcal N=1}}=\mathbb Z_2^q\boxtimes \mathbb Z_2^{(-1)^F}\,,
\end{align}
When $\mathcal L_1$ is SSB, the corresponding boundary is also q-type. The module of $\mathscr C_{A_2^{\mathcal N=1}}$ is given by
\begin{align}
    \mathcal M_{A_2^{\mathcal N=1}}=\{m_0,\,m_1\}\,,
\end{align}
but now the actions of $\mathscr C_{A_2^{\mathcal N=1}}$ on $\mathcal M_{A_2^{\mathcal N=1}}$ are
\begin{align}
    (-1)^F\triangleright m_i=\mathbb C^{1|0}\,m_i\,,\qquad \mathcal L_1\triangleright m_0=m_1\,,\qquad \mathcal L_1\triangleright m_{1}=\mathbb C^{1|1}\,m_0\,.
\end{align}
Since the q-type feature allows a Majorana zero mode to freely move, a zero mode located at one junction can move from another through either the q-type line or boundary, and pick up a phase, e.g.
\begin{align}
    \begin{gathered}

    \end{gathered}
\end{align}
where the factors $\lambda_{\alpha,\beta}$ and $\lambda_{\beta,\gamma}$ arise when the Majorana zero mode (the red dot) moves along the q-type lines $\mathcal L_1$ (yellow) and $(-1)^F\mathcal L_1$ (green). These $\lambda$-factors intrinsically depend on the junction data and are completely determined by the $F$-symbols of superfusion category $\mathscr C_{A_2^{\mathcal N=1}}$ and its $\mathcal M_{A_2^{\mathcal N=1}}$. For a through discussion on the solutions to the $F$-symbols, readers are referred to \cite{Chang:2022hud, Chang:2026pdu}. 

The $\lambda$ factors turn out to be important when one tries to move the lines to compute the VEV of $(-1)^{2F}$ in soliton backgrounds as before in $\mathcal N=2$ case, 
\begin{align}
\langle {\rm sol}|(-1)^{2F}|{\rm sol}\rangle =
\begin{gathered}
 
\end{gathered}
\end{align}
Notice now in the diagram below,
\begin{align}
\begin{gathered}
%
  
    \end{gathered}
 \end{align}
the $F$-symbol $\mathcal F_{M\,\,m_2}^{m_1,\mathcal L_1,\mathcal L_2}[m_3,\mathcal L_3]^{\alpha\beta}_{\gamma\delta}$ will not be a number, but rather a $4\times 4$ matrix for $\alpha,\,\beta,\,\gamma,\,\delta=0,1$. The matrix only has rank $2$:
\begin{align}
    &(\alpha,\beta),\ (\gamma,\delta)=(0,0)\,\ {\rm or}\ (1,1)\,,\quad {\rm for\ \ bosonic\ \ channel}\,;\notag\\
    &(\alpha,\beta),\ (\gamma,\delta)=(0,1)\,\ {\rm or}\ (1,0)\,,\quad {\rm for\ \ bosonic\ \ channel}\,.
\end{align}
The channel mixing bosonic and fermionic sectors is forbidden due to the fermionic parity in $F$-symbols,
\begin{align}
    \mathcal F_{M\,\,m_2}^{m_1,\mathcal L_1,\mathcal L_2}[m_3,\mathcal L_3]^{\alpha\beta}_{\gamma\delta}=0\,\quad {\rm unless}\quad \alpha+\beta\equiv\gamma +\delta \,(\!\!\!\!\!\!\mod 2)\,.
\end{align}
Notice that the channels, for example, from $(\alpha,\beta)=(0,1)$ to $(\gamma,\delta)=(0,1)$ or $(\gamma,\delta)=(1,0)$ are connected by moving the zero mode along the q-type line $\mathcal L_3$. Bearing this property in mind, one can apply the $F$-moves as in eq.~\eqref{eq:Fmoves_in_sphere}, and compute the factors as below:
\begin{align}
\begin{gathered}

\end{gathered}
\,,
\end{align}
where the coefficients above the arrows are obtained by applying the corresponding $F$-move as well as accompanied $\lambda$-factors to move Majorana zero modes. So overall, we pick up an accumulated phase:
\begin{align}
    \kappa=\frac{\mathcal F_{M,m_1}^{m_2,\eta,(-1)^F}[m_1,\eta(-1)^F]^{10}_{01}\,\,\mathcal F_{M,m_1}^{m_2,(-1)^F,\eta(-1)^F}[m_2,\eta]^{01}_{01}}{\mathcal F_{M,m_2}^{m_1,(-1)^F,\eta(-1)^F}[m_1,\eta]^{00}_{00}\,\,\mathcal F_{M,m_2}^{m_1,\eta,(-1)^F}[m_2,\eta(-1)^F]^{00}_{00}}=\frac12\times \frac12\times \frac12\times 8=+1\,.
\end{align}
indicating that
\begin{align}
\langle {\rm sol}|(-1)^{2F}|{\rm sol}\rangle=+1\,,
\end{align}
Therefore, in contrast to the $\mathcal N=2$ solitons, the $\mathcal N=1$ solitons from the least relevant deformation have well-defined integer fermion numbers.

\section{Boundary SymTFT and representations of superstrip algebra}\label{sec:SymTFT}

Suppose we have a soliton state $|\mathcal{O}_{ij}\rangle \in \mathcal{H}_{i,j}$, and let $\mathcal{L}_a : \mathcal{H}_{i,j} \rightarrow \mathcal{H}_{k,l}$ denote the action of a TDL. Under the state-operator correspondence, the state is mapped to a boundary operator $\mathcal{O}_{ij}$ with the action of $\mathcal{L}_a$ realized by a semi-circle wrapping $\mathcal{O}_{ij}$, as shown in Figure~\ref{Fig-State-Operator}.
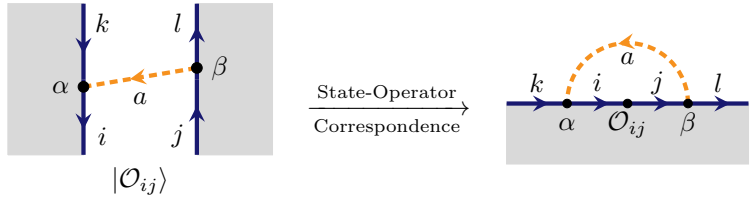
\begin{figure}[!ht]
\begin{equation*}
\begin{gathered}
     \begin{tikzpicture}
        \filldraw[gray,opacity=0.3] (0,0)--(1,0)--(1,2)--(0,2)--(0,0);
        \filldraw[gray,opacity=0.3] (2.5,0)--(3.5,0)--(3.5,2)--(2.5,2)--(2.5,0);
        \draw[ultra thick,MidnightBlue,-<-=.55] (1,0)--(1,1);
        \draw[ultra thick,MidnightBlue,-<-=.55] (1,1)--(1,2);
        \draw[ultra thick,MidnightBlue,->-=.65] (2.5,0)--(2.5,1);
        \draw[ultra thick,MidnightBlue,->-=.7] (2.5,1)--(2.5,2);
        \draw[densely dashed,ultra thick,BurntOrange,-<-=.55] (1,0.9)--(2.5,1.15);
        \node at (0.7,.9) {$\textcolor{black}\alpha$};
        \node at (2.8,1.15) {$\textcolor{black}\beta$};
         \filldraw[black] (1,0.9) circle (2pt);
         \filldraw[black] (2.5,1.15) circle (2pt);
        \node at (1.25,0.25) {$i$};
        \node at (1.25,1.75) {$k$};
        \node at (2.25,0.25) {$j$};
        \node at (2.25,1.75) {$l$};
        \node at (1.75,0.75) {$a$};
        \node at (1.75,-0.35) {$|\mathcal{O}_{ij}\rangle$};
    \end{tikzpicture}
\end{gathered} \quad \xrightarrow[\textrm{Correspondence}]{\textrm{State-Operator}} \quad 
\begin{gathered}
    \begin{tikzpicture}[scale=0.8]
    \filldraw[gray,opacity=0.3] (0,0)--(4,0)--(4,-1)--(0,-1)--(0,0);
    \draw[ultra thick,MidnightBlue,->-=.7] (0,0)--(1,0);
    \draw[ultra thick,MidnightBlue,->-=.7] (1,0)--(2,0);
    \draw[ultra thick,MidnightBlue,->-=.7] (2,0)--(3,0);
    \draw[ultra thick,MidnightBlue,->-=.7] (3,0)--(4,0);
    \draw[densely dashed,ultra thick,BurntOrange,->-=.55] (3,0) arc (0:180:1);
    \filldraw[black] (1,0) circle (2pt);
    \filldraw[black] (2,0) circle (2pt);
    \filldraw[black] (3,0) circle (2pt);
    \node at (0.5,0.35) {$k$};
    \node at (1.5,0.35) {$i$};
    \node at (2.5,0.35) {$j$};
    \node at (3.5,0.35) {$l$};
    \node at (2,0.75) {$a$};
    \node at (1,-0.35) {$\alpha$};
    \node at (3,-0.35) {$\beta$};
    \node at (2,-0.35) {$\mathcal{O}_{ij}$};
    \end{tikzpicture}
\end{gathered}
\end{equation*}
\caption{State-Operator Correspondence}
\label{Fig-State-Operator}
\end{figure}
The boundary operator $\mathcal{O}_{ij}$ furnishes a representation of the strip algebra $\textbf{Str}_{\mathscr{C}}(\mathcal{M})$
    \begin{equation}\label{eq:rep-and-dual-category}
        \textrm{Rep}(\textbf{Str}_{\mathscr{C}}(\mathcal{M})) = \mathscr{C}^*_{\mathcal{M}}\,,
    \end{equation}
where $\mathscr{C}^*_{\mathcal{M}}$ is the dual category of $\mathscr{C}$ with respect to $\mathcal{M}$. This category can be understood as the dual category obtained by gauging the symmetry associated with $\mathcal{M}$. In this section we review the SymTFT description for 2D theory with boundary based on \cite{Bhardwaj:2024igy,Copetti:2024onh}, which provides a convenient framework to describe the boundary operators. We then generalize the construction to 2D fermionic theory, motivated by \cite{Bhardwaj:2024ydc}.

\subsection*{Boundary SymTFT}

The SymTFT construction of a 2d theory $\mathfrak{T}$ with symmetry category $\mathscr{C}$, defined on the manifold $M_2$ with a boundary $\partial M_2 =\mathfrak{B}_1$, is illustrated in Figure~\ref{fig:boundary-SymTFT}.
\begin{figure}[!ht]
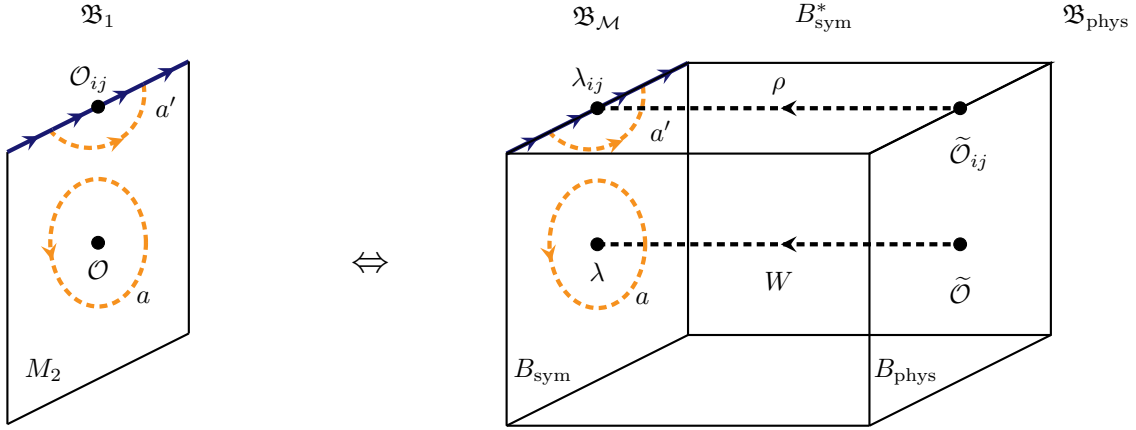

    \begin{equation*}
            \begin{gathered}

    \end{gathered} 
    \end{equation*}
    \caption{An illustrative picture of Boundary SymTFT.}
    \label{fig:boundary-SymTFT}
\end{figure}
The bulk theory living on $M_2 \times [0,1]$ is a TQFT $\mathcal{Z}(\mathscr{C})$, which depends on the symmetry category $\mathscr{C}$.
The physical boundary $B_{\rm phys}$ at $M_2\times\{1\}$ encodes the detailed dynamics of the theory, while the symmetry data associated with $\mathscr{C}$ are stored on the symmetry boundary $B_{\rm sym}$ at $M_2 \times \{0\}$. In particular, the entire symmetry category $\mathscr{C}$ lives on $B_{\rm sym}$. 

A local operator $\mathcal{O}$ in $\mathfrak{T}$ is extended into a bulk line operator $W$ stretching between two boundaries, with endpoints $\widetilde{\mathcal{O}}$ and $\lambda$ on $B_{\rm phys}$ and $B_{\rm sym}$. The action of a TDL $\mathcal{L}_a$ on $\mathcal{O}$ is realized by the linking between $\mathcal{L}$ with $W$ as shown in the figure. Notably, the symmetry acts only at the endpoint $\lambda$ and will not modify the bulk line operator $W$ itself. In fact, simple line operators $W$ in the bulk are in one-to-one correspondence with the irreducible representations $\rho_W$ of the symmetry $\mathscr{C}$, including the twist sectors~\cite{Izumi:2000qa,Evans:2010yr,MUGER2003159,Lin:2022dhv}, and the endpoint $\lambda$ on $B_{\rm sym}$ transforms as a vector in the representation $\rho_W$. For example, if $\mathscr{C}=G$ is an ordinary group, then the action of a group element $g\in G$ on $\lambda$ is given by
    \begin{equation}
        (g \circ \lambda)^i = \sum_j \lambda^j \rho_W(g)_{ji}  \  \quad (i,j=1,\dots,\dim(\rho_W))\,.
    \end{equation}

The symmetry boundary $B_{\rm sym}$ is topological and is specified by the collection of bulk line operators $W$ that can simultaneously end on the boundary. These line operators are encoded in an algebraic object known as \emph{Lagrangian algebra}\cite{Lan:2014uaa,Cong:2016ayp,Davydov:2010kfz} , which can be written as
    \begin{equation}
        \mathcal{L}_{\rm sym} = \bigoplus_{\mu} n_{\mu} W_{\mu}\,,
    \end{equation}
where $W_{\mu}$ are line operators in the SymTFT, and $n_{\mu}\in \mathbb{Z}_{\geq0}$ denotes the dimension of the representation of $\mathscr{C}$ associated with $W_{\mu}$ after shrinking the interval. Other line operators $W\notin \mathcal{L}_{\rm sym}$ cannot end on the symmetry boundary $B_{\rm sym}$, instead, they are converted to symmetry operators in $\mathscr{C}$ supported on $B_{\rm sym}$ and are responsible for the twist operators in $\mathfrak{T}$. An important property of a Lagrangian algebra is that all the line operators contained in $\mathcal{L}$ must be bosonic; in other words, their topological spin of $\theta(W)$ must be 1.

The boundary $\mathfrak{B}_1$ is expanded into an additional topological boundary $B^*_{\rm sym}$, which interpolates between $\mathfrak{B}_{\mathcal{M}}=\partial B_{\rm sym}$ and $\mathfrak{B}_{\rm phys}=\partial B_{\rm phys}$. A boundary operator $\mathcal{O}_{ij}$ is then extended as a TDL $\mathcal{L}_{\rho}$ living on the side boundary $B^*_{\rm sym}$, stretching between $\mathfrak{B}_{\rm sym}$ and $\mathfrak{B}_{\rm phys}$, with endpoints $\lambda_{ij}$ and $\widetilde{\mathcal{O}}_{ij}$, respectively. The action of the TDL $\mathcal{L}_{a'}$ is represented by the semicircular line linking with $\mathcal{L}_{\rho}$. 

Similar to the local operator $\mathcal{O}$, the irreducible representation carried by $\mathcal{O}_{ij}$ is naturally associated with the simple TDL supported on the side boundary $B^*_{\rm sym}$. According to the relation \eqref{eq:rep-and-dual-category} and the discussion therein, the corner $\mathfrak{B}_{\mathcal{M}}$ is determined by the $\mathscr{C}$-module $\mathcal{M}$, implying the dual category $\mathscr{C}^*$ lives on the side boundary $B^*_{\rm sym}$. Correspondingly, one can associate $B^*_{\rm sym}$ with another Lagrangian algebra
    \begin{equation}
        \mathcal{L}^*_{\rm sym} = \bigoplus_{\mu} n^*_{\mu} W_{\mu}\,,
    \end{equation}
and bulk line operators that cannot end on $B^*_{\rm sym}$ will transit to TDL in $\mathscr{C}^*$ instead. Conversely, one may start by specifying $B^*_{\rm sym}$ through a choice of Lagrangian algebra $\mathcal{L}^*_{\rm sym}$. Then the corner $\mathfrak{B}_{\mathcal{M}}$ is the interface between $B_{\rm sym}$ and $B^*_{\rm sym}$, from which the module category $\mathcal{M}$ can be derived.


\subsection*{Interface between topological boundaries}

The interface $\mathfrak{B}_{\mathcal{M}}$ between $B_{\rm sym}$ and $B^*_{\rm sym}$ encodes the data of module category $\mathcal{M}$. The number of simple objects in $\mathcal{M}$ is determined by the homomorphism
\begin{equation}\label{eq:dim-of-module}
|\mathcal{M}|=\dim_{\mathbb{C}} \textrm{Hom}_{\textrm{SymTFT}}(\mathcal{L_{\textrm{sym}}},\mathcal{L}^*_{\rm sym}) = \sum_{\mu} n_{\mu} n^*_{\mu}\,,
    \end{equation}
which counts the bulk line operators that can end on both $B_{\rm sym}$ and $B^*_{\rm sym}$. In the SymTFT picture, those line operators stretch between $B_{\rm sym}$ and $B^*_{\rm sym}$, and are localized at the corner $\mathfrak{B}_{\mathcal{M}}$, as depicted in Figure~\ref{fig:SymTFT-corner-operator}. After shrinking the interval, they give rise to a set of topological operators $V_{i}$ on the boundary $\mathcal{B}_1=\partial M_2$. One can perform a change of basis such that the algebra takes the form
    \begin{equation}
        \pi_{i} \times \pi_{j} = \delta_{i,j} \pi_{i} \quad (i=1,\dots, |\mathcal{M}|)\,,
    \end{equation}
and $\pi_i$ projects a generic element $v\in \mathcal{M}$ to its $v_i$ component.

\begin{figure}[H]
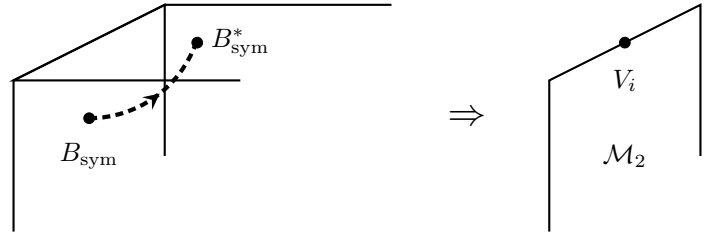

\begin{equation*}
            \begin{gathered}

    \end{gathered}
    \end{equation*}
    \caption{The bulk line operators stretching between $B_{\rm sym}$ and $B^*_{\rm sym}$ give rise to topological operators $V_i$ after shrinking the interval. }
    \label{fig:SymTFT-corner-operator}
\end{figure}

Consider a TDL supported on either $B_{\rm sym}$ or $B^*_{\rm sym}$ going across the interface $\mathfrak{B}_{\mathcal{M}}$, there are four distinct possibilities as shown in Figure~\ref{fig:lines-crossing-interface}, which we summarize below.
\begin{figure}[H]
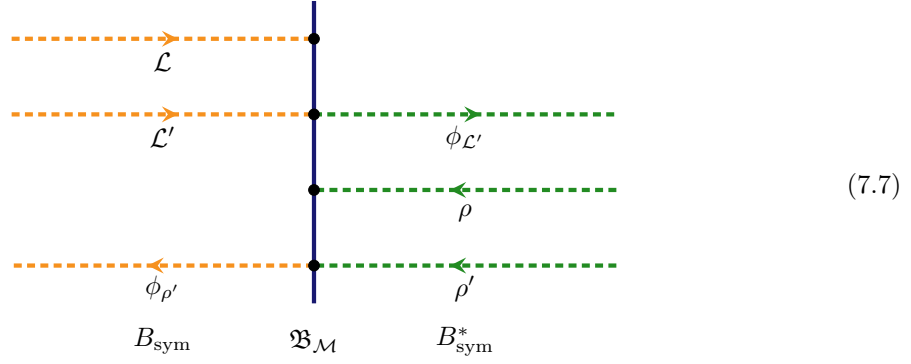

\begin{equation}
    \begin{gathered}
        %
    \end{gathered}
\end{equation}
\caption{Four possibilities when crossing the interface : 1. a symmetry TDL $L\in \mathscr{C}$ ending on the interface. 2. a symmetry TDL $\mathcal L'$ is converted into $\phi_{\mathcal{L}'}$ after crossing the interface. 3. a representation TDL $\rho \in \mathscr{C}^\ast$ ending on the interface. 4. a representation TDL $\rho'$ is converted into $\phi_{\rho'}$ after crossing the interface. }
\label{fig:lines-crossing-interface}
\end{figure}

\begin{itemize}
    \item A symmetry operator $\mathcal{L}\in\mathscr{C}$ disappears after crossing $\mathfrak{B}_{\mathcal{M}}$. Then the symmetry generated by $\mathcal{L}$ is preserved on the interface, and one has
    \begin{equation}
        \begin{gathered}
     \begin{tikzpicture}[scale=1]
        \draw[ultra thick,MidnightBlue,-<-=.55] (-2.5,2)--(-2.5,1);
        \draw[ultra thick,MidnightBlue,-<-=.55] (-2.5,1)--(-2.5,0);
        \draw[densely dashed,ultra thick,BurntOrange,->-=.6] (-4,1)--(-2.5,1);
        \node at (-2,1) {$\textcolor{black}\alpha$};
         \filldraw[black] (-2.5,1) circle (2pt);
        \node at (-2.9,0.3) {$v_i$};
        \node at (-2.9,1.7) {$v_i$};
        \node at (-4.5,1) {$\mathcal{L}$};
    \end{tikzpicture}    
        \end{gathered}\quad ,
    \end{equation}    
where the module $v_i \in \mathcal{M}$ is invariant under the action of $\mathcal{L}$.  Suppose $\mathscr{C}^*$ is the dual symmetry obtained as $\mathscr{C}/\mathscr{A}$ by gauging a subsymmetry associated with the Frobenius algebra $\mathscr{A}$, then $\mathcal{L}$ must be a component of $\mathscr{A}$ satisfying $\textrm{Hom}_{\mathscr{C}}(\mathcal{L},\mathscr{A})\neq 0$ and is absorbed by the condensation of $\mathscr{A}$ on $B^*_{\rm sym}$.
\item A symmetry operator $\mathcal{L}^\prime \in \mathscr{C}$ is converted into $\phi_{\mathcal{L}'}\in \mathscr{C}^*$ after crossing $\mathfrak{B}_{\mathcal{M}}$. Then the symmetry generated by $\mathcal{L}^\prime$ is broken by the interface. One can bend the image $\phi_{\mathcal{L}'}$ to merge it with the interface and get
    \begin{equation}
        \begin{gathered}
    
        \end{gathered} \quad \quad (v_j \in v_i\circ \phi_{\mathcal{L}'})\quad ,
    \end{equation}   
where the module $v_i$ is mapped into $v_j$ under the action of $\mathcal{L}'$.
\item A representation operator $\rho \in \mathscr{C}^*$ disappears after crossing $\mathfrak{B}_{\mathcal{M}}$. Then it gives rise to a boundary operator $\mathcal{O}_{i,i}$ which does not change the type of the module $v_i$
    \begin{equation}
        \begin{gathered}
     %
    
        \end{gathered}\quad .
    \end{equation} 
Under the state-operator correspondence, they represent the particle excitation $|\mathcal{O}_{i,i}\rangle$ on a fixed vacuum $|v_i\rangle$
\item A representation operator $\rho' \in \mathscr{C}^*$ is converted into $\phi_{\rho'} \in \mathscr{C}$ after crossing $\mathfrak{B}_{\mathcal{M}}$. We can similarly bend the image $\phi_{\rho'}$ to merge the interface, and it gives rise to the boundary changing operator $\mathcal{O}_{i,j}$
    \begin{equation}
        \begin{gathered}
     %
    
        \end{gathered}\quad \quad (v_j \in \phi_{\rho'} \circ v_i)\quad .
    \end{equation} 
    Under the state-operator correspondence, they are responsible for the soliton excitations $|\mathcal{O}_{i,j}\rangle$ interpolating between $|v_i\rangle$ and $|v_j\rangle$.
\end{itemize}

\subsection*{Example : Finite group-like symmetry $G$}

As a toy example, let us consider the case where the symmetry category is group-like, $\mathscr{C}=G$ with $G$ a finite group. The SymTFT $\mathcal{Z}(G)$ is the 3D Dijkgraaf-Witten theory whose simple line operators $W_{([g],\rho)}$ are labeled by a pair $([g],\rho_g)$. Here $[g] \in \textrm{Cl}(G)$ labels the conjugacy class of $G$ represented by an element $g\in G$, and $\rho_g$ labels the irreducible representation of the centralizer $C(g)$
    \begin{equation}
        C(g) = \{h\in G| hg=gh \}\,,
    \end{equation}
which is the subgroup of $G$ consisting of all elements that commute with $g$. In particular, when $g=e$ the centralizer $C(e)=G$ is the entire group.

The symmetry boundary $B_{\rm sym}$ is specified by the Lagrangian algebra 
    \begin{equation}
        \mathcal{L}_{\rm sym} =  \bigoplus_{\rho \in \textrm{Rep}(G)} (\dim \rho) W_{([e],\rho)}\,,
    \end{equation}
which includes all irreducible representations of $G$. In the SymTFT construction, a line operator $W_{([e],\rho)}$ that starts from an operator $\widetilde{\mathcal{O}}$ on the physical boundary $B_{\rm phys}$ and ends on the symmetry boundary $B_{\rm sym}$, gives rise a local operator $\mathcal{O}_{\rho}$ upon shrinking the interval. This operator transforms in the representation $\rho$ under the action of $G$. 

There are multiple choices for the other topological boundary $B^*_{\rm sym}$. We first choose the corresponding Lagrangian algebra to be
    \begin{equation}
        \mathcal{L}^*_{\rm sym} = \bigoplus_{[g]\in \textrm{Cl}(G)} W_{([g],\rho_{\rm id})}\,,
    \end{equation}
where $\rho_{\rm id}$ denotes the trivial representation. In this case, the summation is over all conjugacy classes $[g]\in \textrm{Cl}(G)$. The dual category living on $B^*_{\rm sym}$ is thus the representation category $\mathscr{C}^*=\textrm{Rep}(G)$. This is the dual symmetry obtained by gauging the whole $G$-symmetry, realized by condensing the Frobenius algebra $\mathscr{A}=\oplus_g g$. According to \eqref{eq:dim-of-module}, we find
    \begin{equation}
        |\mathcal{M}| = \dim_{\mathbb{C}} \textrm{Hom}_{\rm \mathcal{Z}(G)} (\mathcal{L}_{\rm sym},\mathcal{L}^*_{\rm sym})=1\,,
    \end{equation}
where the only contribution is from $\textrm{Hom}_{\rm \mathcal{Z}(G)} (W_{([e],\rho_{\rm id})},W_{([e],\rho_{\rm id})})=\mathbb{C}$, which implies that the module category is trivial, $\mathcal{M}=\textrm{Vec}_{\mathbb{C}}$. Correspondingly, all symmetry operators $g\in G$ become trivial after crossing the interface $\mathfrak{B}_{\mathcal{M}}$, since $g$ is condensed on $B^*_{\rm sym}$. As a result, the interface preserves the full symmetry $G$. Conversely, any representation operator $\rho\in \textrm{Rep}(G)$ living on $B^*_{\rm sym}$ also becomes trivial after crossing the interface, since one can restore the $G$-symmetry by gauging $\textrm{Rep}(G)$. After shrinking the interval, we have a 2D theory $\mathfrak{T}_G$ with symmetry $G$ living on $M_2$ with a symmetric boundary $\mathfrak{B}_1$, and the boundary operators $\mathcal{O}_{\rho}$ are classified by $\textrm{Rep}(G)$, see Figure~\ref{fig:symmetry-boundary}.
\begin{figure}[H]
    \begin{equation}
        \begin{gathered}
            \begin{tikzpicture}
            \draw[ultra thick,MidnightBlue,->-=.65] (4,0.5)--(4,1);
            \draw[ultra thick,MidnightBlue,->-=.65] (4,1)--(4,1.5);
            \draw[ultra thick,MidnightBlue,->-=.65] (4,1.5)--(4,2);
            \draw[ultra thick,MidnightBlue,->-=.65] (4,2)--(4,2.5);
            \draw[densely dashed,ultra thick,BurntOrange,->-=.55] (4,2) arc (90:270:0.5);
            \draw[densely dashed,ultra thick,ForestGreen,->-=.55] (6,1.5)--(4,1.5);
            \node at (3.5,2) {$g$};
            \node at (6.5,1.5) {$\rho$};
            \node at (3.75,1.5) {$\lambda_{i}$};
            \filldraw[black] (4,2) circle (2pt);
            \filldraw[black] (4,1.5) circle (2pt);
            \filldraw[black] (4,1) circle (2pt);
            \end{tikzpicture}
        \end{gathered} \quad = \quad
        \begin{gathered}
            \begin{tikzpicture}
            \draw[ultra thick,MidnightBlue,->-=.65] (4,0.5)--(4,1.5);
            \draw[ultra thick,MidnightBlue,->-=.65] (4,1.5)--(4,2.5);
            \draw[densely dashed,ultra thick,ForestGreen,->-=.55] (6,1.5)--(4,1.5);
            \node at (6.5,1.5) {$\rho$};
            \node at (3.2,1.5) {$(\rho \circ \lambda)_{i}$};
            \filldraw[black] (4,1.5) circle (2pt);
            \end{tikzpicture}
        \end{gathered}
    \end{equation}
    \caption{The interface $\mathfrak{B}_{\mathcal{M}}$ preserves the whole $G$-symmetry, and the representation operators are $\rho \in \textrm{Rep}(G)$. The junction point of $\rho$ with the interface is a vector which transforms under the representation $\rho$ upon action of $g\in G$.}
    \label{fig:symmetry-boundary}
\end{figure}

Another possible choice of $B^*_{\rm sym}$ is to take $\mathcal{L}^*_{\rm sym} = \mathcal{L}_{\rm sym}$, so that $\mathscr{C}=\mathscr{C}^*=G$. The equation \eqref{eq:dim-of-module} yields
    \begin{equation}
        |\mathcal{M}| = \dim_{\mathbb{C}} \textrm{Hom}_{\rm \mathcal{Z}(G)} (\mathcal{L}_{\rm sym},\mathcal{L}^*_{\rm sym})=\sum_{\rho \in \textrm{Rep}(G)} (\dim \rho)^2 = |G|\,.
    \end{equation}
In fact, in this case the module category is $\mathcal{M}=G$ as well. The interface $\mathfrak{B}_{\mathcal{M}}$ breaks the entire symmetry and one has
    \begin{equation}
        \begin{gathered}
     \begin{tikzpicture}[scale=1]
        \draw[ultra thick,MidnightBlue,-<-=.55] (-2.5,2)--(-2.5,1);
        \draw[ultra thick,MidnightBlue,-<-=.55] (-2.5,1)--(-2.5,0);
        \draw[densely dashed,ultra thick,BurntOrange,->-=.6] (-4,1)--(-2.5,1);
         \filldraw[black] (-2.5,1) circle (2pt);
        \node at (-2.9,0.3) {$h$};
        \node at (-2.9,1.7) {$gh$};
        \node at (-4.5,1) {$g$};
    \end{tikzpicture}    
        \end{gathered}\qquad \qquad 
        \begin{gathered}
     \begin{tikzpicture}[scale=1]
        \draw[ultra thick,MidnightBlue,-<-=.55] (-2.5,2)--(-2.5,1);
        \draw[ultra thick,MidnightBlue,-<-=.55] (-2.5,1)--(-2.5,0);
        \draw[densely dashed,ultra thick,ForestGreen,->-=.6] (-1,1)--(-2.5,1);
         \filldraw[black] (-2.5,1) circle (2pt);
        \node at (-2.9,0.3) {$h$};
        \node at (-2.9,1.7) {$hg$};
        \node at (-0.5,1) {$g$};
    \end{tikzpicture}    
        \end{gathered}\quad .
    \end{equation}
After shrinking the interval, we obtain a 2D theory $\mathfrak{T}_G$ living on $M_2$, whose boundary $\mathfrak{B}_1$ breaks the entire $G$ symmetry. The representation operator $g'\in \mathscr{C}^*=G$ gives rise to a boundary changing operator $\mathcal{O}_{h,hg'}$, which interpolates between $h$ and $hg'$ in the module category $\mathcal{M}=G$. Under the action of the symmetry $g\in \mathscr{C}$, the operator $\mathcal{O}_{h,hg'}$ is mapped to $\mathcal{O}_{gh,ghg'}$. 

We then apply the operator-state correspondence to the boundary operators to recover the particle/soliton interpretation. Depending on the choices of $\mathcal{L}^*_{\rm sym}$, one has two distinct cases:
\begin{itemize}
    \item If we choose $\mathcal{L}^*_{\rm sym} = \bigoplus_{[g]\in \textrm{Cl}(G)}  W_{([g],\rho_{\rm id})}$, we have a unique vacuum preserving the symmetry $G$ and the excitations on the vacuum transform under the representation of $G$, as expected.
    \item If we choose $\mathcal{L}^*_{\rm sym} =  \bigoplus_{\rho \in \textrm{Rep}(G)} (\dim \rho) W_{([e],\rho)}$, the symmetry $G$ is spontaneously broken and we have $|G|$ vacua $|h\rangle$ transforming according to
        \begin{equation}
            g |h\rangle = |gh\rangle\,.
        \end{equation}
    There are also $|G|$ representations of particles/solitons $|h,hg'\rangle$ classified by a group element $g'$, where $h$ and $hg'$ label the asymptotic vacuum at $\sigma\rightarrow \mp \infty$, and the degeneracy for each multiplet is also $|G|$. The action of $G$-symmetry is
        \begin{equation}
            g |h,hg'\rangle = |gh,ghg'\rangle\,.
        \end{equation}
\end{itemize}
We recover the same results in the discussion of the strip algebra of the SSB phase of $G$ discussed in \cite{Cordova:2024iti}.

There also exist other Lagrangian algebras corresponding to partial symmetry breaking of $G$, which will not be discussed here.

\subsection*{$\mathbb{Z}_2^{\pi}$-symmetry and fermionic Lagrangian algebra}

To generalize the above discussion to fermionic theories, we need to study the symmetry boundary $B_{\rm sym}$ that supports a superfusion category, together with the associated \emph{fermionic Lagrangian algebra}. To achieve this, we follow \cite{Bhardwaj:2024ydc} and first introduce the $\mathbb{Z}_2^{\pi}$ symmetry generated by $\pi \equiv 1_f$.

To motivate the introduction of the $\mathbb{Z}_2^{\pi}$ symmetry, consider a fermionic gapped phase with a mass gap $\Delta$, and a neutral massive fermion $\Psi$ propagating along the time direction, whose worldline is denoted by $W_{\Psi}$. In the deep IR, the massive fermion decouples from the low-energy dynamics, and one may expect that $W_{\Psi}\sim 1$, corresponding to ordinary identity TDL. However, the fermionic statistics of $\Psi$ should persist even in the IR, and we need to introduce the \emph{fermionic identity line defect} $\pi \equiv 1_f$ and identify $ W_{\Psi} \sim \pi \equiv 1_f$ in the IR. Furthermore, a pair of massive fermions form a composite particle with bosonic statistics, which is identified as the ordinary identity line operator. That implies $\pi$ generates a $\mathbb{Z}_2$ symmetry $\pi^2=1$.

Therefore, inserting a $\pi$-line along the time direction can be interpreted as introducing a defect corresponding to the worldline of a neutral, very massive fermion $\Psi$. Equivalently, the insertion of a $\pi$-line amounts to stacking a $(0+1)$d fermionic invertible topological field theory, which is a quantum mechanical system with a single fermionic vacuum. 

On the other hand, we can insert a $\pi$-line along the spatial direction as a symmetry operator. When the space is a circle, in the path-integral language, we can understand the $\pi$-line as a virtual fermion propagating along the spatial direction and winding around the circle. As a result, it probes the spin structure along the spatial circle and acts as $-1$ on the Ramond sector $\pi |\textrm{R}\rangle = -|\textrm{R}\rangle$.

The fermionic identity line operator $\pi$ is isomorphic to the ordinary identity line operator $1$
    \begin{equation}\label{eq:pi-identity-iso}
        \textrm{Hom}(\pi,1) = \mathbb{C}^{0|1}\,,
    \end{equation}
which says the $\pi$-line can have a fermionic topological endpoint $\mathcal{O}_{\pi}\in \textrm{Hom}(\pi,1)$
\begin{equation}
    \begin{gathered}

    \end{gathered}\quad .
\end{equation}
Recall that we have another $\mathbb{Z}_2^F$ symmetry generated by the fermion parity $(-1)^F$. Since the endpoint of $\pi$ is a fermionic operator, one must have
    \begin{equation}
        \begin{gathered}
        %
            
        \end{gathered}
    \end{equation}
Furthermore, if we switch $\pi$ and $(-1)^F$, one should also have
    \begin{equation}
        \begin{gathered}
        %
            
        \end{gathered}
    \end{equation}    
where $\mathcal{O}_R$ is a twist operator attached to $(-1)^F$. By the state-operator correspondence, this is equivalent to the statement that all states in the Ramond sector carry eigenvalue $-1$ under the action of the $\pi$ operator.

We may define an alternative fermion parity operator by stacking the $\pi$ operator
    \begin{equation}
        (-1)^F \rightarrow \pi (-1)^F\,.
    \end{equation}
Since $\pi$ flips the sign of states in the Ramond sector, this redefinition is equivalent to stacking the fermionic SPT phases given by the Arf-invariant (or the IR phase of the Kitaev chain \cite{Kitaev:2000nmw}) to the theory $\mathfrak{T}$
    \begin{equation}
        \mathfrak{T}\rightarrow \mathfrak{T} \boxtimes \textrm{Arf}\,.
    \end{equation}

To incorporate the $\mathbb{Z}_2^{\pi}$ symmetry, we stack the SymTFT with a 3D fermionic TQFT $\mathbf{sVec}$, whose only non-trivial line operator is a spin-$\frac{1}{2}$ line $\pi$
    \begin{equation}
        \textrm{SymTFT} \rightarrow \textrm{SymTFT} \boxtimes \mathbf{sVec}\,.
    \end{equation}
With the inclusion of $\mathbf{sVec}$, we can enlarge the Lagrangian algebra to include spin-$\frac{1}{2}$ (fermionic) line operators in the bulk, which takes the form
    \begin{equation}
        \mathcal{L}_f = \left(\bigoplus_b n_b W_b\right) \oplus \left(\bigoplus_f n_f (\pi W_f) \right)\,,
    \end{equation}
where the topological spins are $\theta(W_b)=1$ and $\theta(W_f)=-1$. Here each fermionic line operator $W_f$ is decorated with a $\pi$-line, which reverses its statistic so that we can treat the composite line operator $\pi W_f$ as if it is bosonic. 



In the following, we choose the Lagrangian algebra $\mathcal{L}_{\rm sym}$ to be of fermionic type, so that the symmetry boundary $B_{\rm sym}$ supports a superfusion category $\mathscr{C}$ as the symmetry. The presence of $\pi$-line allows us to distinguish bosonic and fermionic degrees of freedoms within the SymTFT picture. For example, consider a line operator $W_f$ with $\pi W_f \in \mathcal{L}_{\rm sym}$. Stretching $W_f$ between the symmetry boundary $B_{\rm sym}$ and physical boundary $B_{\rm phys}$ produces, upon shrinking the interval, an operator $\mathcal{O}_f$ attached to the $\pi$-line. Since the $\pi$-line is isomorphic to the identity line, as mentioned in \eqref{eq:pi-identity-iso}, we can cut the $\pi$-line and introduce a fermionic topological endpoint $\mathcal{O}_{\pi}$. Bringing $\mathcal{O}_{\pi}$ together with $\mathcal{O}_f$ yields a fermionic local operator $\mathcal{O}_f$
\begin{equation}
    \begin{gathered}
        \begin{tikzpicture}
            \draw[densely dotted,ultra thick,Red,->-=.55] (0,0)--(3,0);
            \filldraw[black] (3,0) circle (2pt);
            \node at (2,0.25) {$\pi$};
            \node at (3.5,0) {$\mathcal{O}_{f}$};
        \end{tikzpicture}
    \end{gathered} \quad = \quad 
    \begin{gathered}
        \begin{tikzpicture}
            \draw[densely dotted,ultra thick,Red,->-=.55] (1.5,0)--(3,0);
            \filldraw[black] (3,0) circle (2pt);
            \filldraw[WildStrawberry] (1.5,0) circle (2pt);
            \node at (2.25,0.25) {$\pi$};
            \node at (3.5,0) {$\mathcal{O}_{f}$};
            \node at (1,0) {$\mathcal{O}_{\pi}$};
        \end{tikzpicture}
    \end{gathered}\quad = \quad
    \begin{gathered}
        \begin{tikzpicture}
            \filldraw[WildStrawberry] (3,0) circle (2pt);
            \node at (3.5,0) {$\mathcal{O}_{f}$};
            \node at (3,0.25) {\ };
        \end{tikzpicture}
    \end{gathered}    \,.
\end{equation}

\subsection*{A toy example : $\mathbb{Z}_2$ symmetry}

Let us begin with a trivial fermionic theory whose only symmetry generator is the fermion parity $(-1)^F$. The associated SymTFT is the 3D $\mathbb{Z}_2$ Dijkgraaf-Witten theory $\mathcal{Z}(\mathbb{Z}_2)$ stacked with $\textbf{sVec}$.

For an abelian group $G$, each group element forms its own conjugacy class, and the corresponding centralizer group is the entire group $G$. Therefore, when $G=\mathbb{Z}_2$, the bulk line operators $W_{(g,q)}$ are labeled by the pairs $(g,q)$ where $g=\{1,\eta\}$ denotes the group elements and $q=\{0,1\}$ labels the $\mathbb{Z}_2$ charge. We will denote the four line operators as
    \begin{equation}
        1\equiv W_{(1,0)}\,,\quad e\equiv W_{(\eta,0)}\,,\quad m\equiv W_{(1,1)}\,, \quad f=em\equiv W_{(\eta,1)}\,,
    \end{equation}
with fusion rules $e^2=m^2=f^2=1$ and the topological spin $\theta(e)=\theta(m)=1,\ \theta(f)=-1$.

There exist two bosonic Lagrangian algebras
    \begin{equation}
        \mathcal{L}_{\rm Dir} = 1\oplus e\,, \quad \mathcal{L}_{\rm Neu} = 1\oplus m\,,
    \end{equation}
and they are related by the exchange of electric and magnetic line operators $e\leftrightarrow m$. Without loss of generality, we can fix $\mathcal{L}_{\rm sym} = \mathcal{L}_{\rm Dir}$, and the symmetry boundary $B_{\rm sym}$ supports an ordinary $\mathbb{Z}_2$ symmetry. In this case, the $e$-line can be absorbed into $B_{\rm sym}$, whereas the $m$-line is projected to the $\mathbb{Z}_2$ generator $\eta$. Explicitly, one has
    \begin{equation}
        \Pi_{\text{Dir}}(1)=\Pi_{\text{Dir}}(e) =1\,,\quad \Pi_{\text{Dir}}(m) = \Pi_{\text{Dir}}(f)=\eta\,,
    \end{equation}
where $\Pi_{\text{Dir}}$ denotes the projection onto the symmetry boundary $B_{\rm sym}$. After shrinking the interval, the four bulk line operators will give rise to following 2D local/twist operators
    \begin{itemize}
        \item (neutral, local) Identity line operator $1$ ends on the $B_{\textrm{sym}}$
        \item (charged, local) Electric line operator $e$ ends on the $B_{\textrm{sym}}$
        \item (neutral, twist) Magnetic line operator $m$ attached to $\eta$ on the $B_{\textrm{sym}}$
        \item (charged, twist) fermionic line operator $f$ attached to $\eta$ on the $B_{\textrm{sym}}$
    \end{itemize}

After stacking $\mathbf{sVec}$ to $\mathcal{Z}(\mathbb{Z}_2)$, we have an additional fermionic Lagrangian algebra
    \begin{equation}
        \mathcal{L}_f=1\oplus \pi f\,.
    \end{equation}
If we choose $\mathcal{L}_{\rm sym}=\mathcal{L}_f$, the projection of the four line operators onto the symmetry boundary $B_{\rm sym}$ are given by
    \begin{equation}
        \Pi_f(1)=1\,,\quad \Pi_f(e)=\pi P\,,\quad \Pi_f(m)=P\,, \quad \Pi_f(f)=\pi\,,
    \end{equation}
where $P$ denotes a $\mathbb{Z}_2$ generator. If we further identify $P$ as the fermion parity $(-1)^F$, then the four bulk line operators will give rise to
    \begin{itemize}
        \item (boson, NS sector) Identity line operator ends on the boundary
        \item (boson, R sector) Magnetic line operator $m$ attached to $P$ on the boundary
        \item (fermion, R sector) Electric line operator $e$ attached to $\pi P$ on the boundary
        \item (fermion,NS sector) Fermionic line operator $f$ attached to $\pi$ on the boundary
    \end{itemize}
Here, the operators attached to $\pi$-lines are fermionic, since one can cut the $\pi$-line and introduce a fermionic topological endpoint $\mathcal{O}_{\pi}$, as discussed above. On the other hand, if we choose $\pi P$ as the fermion parity operator, then fermionic and bosonic states are exchanged in the Ramond-sector, which is equivalent to stacking the fermionic SPT phase given by the Arf-invariant.

We now turn to the discussion of boundary operators in the SymTFT picture. For the bosonic case where $\mathcal{L}_{\rm sym} = \mathcal{L}_{\rm Dir}$, the choices of $\mathcal{L}^*_{\rm sym} = \mathcal{L}_{\rm Neu}$ and $\mathcal{L}^*_{\rm sym} = \mathcal{L}_{\rm Dir}$ correspond, respectively, to a boundary $\mathfrak{B}_1=\partial M_2$ on which the $\mathbb{Z}_2$ symmetry is preserved or spontaneously broken. On the other hand, if we fix $\mathcal{L}_{\rm sym} = \mathcal{L}_f$ so that the symmetry boundary $B_{\rm sym}$ supports the fermion parity symmetry $(-1)^F$, the two choices $\mathcal{L}^*_{\rm sym}=\mathcal{L}_{\rm Dir}$ and $\mathcal{L}^*_{\rm sym}=\mathcal{L}_{\rm Neu}$ are equivalent, because $\mathcal{L}_f$ is invariant under exchange $e\leftrightarrow m$.

Let us choose $\mathcal{L}^*_{\rm sym}=\mathcal{L}_{\rm Dir}=1\oplus e$. From eq.~\eqref{eq:dim-of-module}, we find
    \begin{equation}
        \textrm{Hom}_{\mathcal{Z}(\mathbb{Z}_2)\boxtimes \mathbf{sVec}}(\mathcal{L}_{\rm sym},\mathcal{L}^*_{\rm sym}) = \textrm{Hom}_{\mathcal{Z}(\mathbb{Z}_2)\boxtimes \mathbf{sVec}}(1,1) =\mathbb{C}\,,
    \end{equation}
which indicates that there is a unique boundary condition and the fermion parity symmetry $(-1)^F$ is preserved. To verify this, let us choose $\Pi(m)=P=(-1)^F$ as the fermion parity. On the symmetry boundary $B_{\rm sym}$, $m$ is equivalent to $m\sim m\times \pi f=\pi e$. After crossing the interface, the $e$-component is absorbed into the boundary $B^*_{\rm sym}$, leaving a $\pi$-line, which is isomorphic to identity. One can represent this process as
\begin{equation}
    \begin{gathered}

    \end{gathered}\quad .
\end{equation}
Let us then consider the representation operators $1$ and $m$ on $B^*_{\rm sym}$. The identity line operator $1$ gives a bosonic boundary operator. For the $m$-line, it is equivalent to $m \times e=f$ on $B^*_{\rm sym}$. After crossing the interface $B_{\mathcal{M}}$, it is absorbed and leaves a $\pi$-line on $B_{\rm sym}$, which implies the boundary operator is fermionic
    \begin{equation}
        \begin{gathered}
            %
        \end{gathered}\quad .
    \end{equation}
Therefore we have two kinds of boundary operators, bosonic and fermionic. It provides a trivial example for the SymTFT picture of boundary operators in fermionic theory.

\subsection*{Example : superfusion category $\mathscr{C}_q^{(0)}$}

We now turn to another example, namely the superfusion category $\mathscr{C}_q^{(0)}=\{1,\eta\}$, where $\eta$ is a q-type $\mathbb{Z}_2$ generator satisfying $\eta^2 = 1 + \pi$. Its SymTFT is realized by
    \begin{equation}
        \textrm{Ising}\boxtimes \overline{\textrm{Ising}}\boxtimes \mathbf{sVec}\,,
    \end{equation}
where Ising denotes 3D modular tensor category with simple line operators $1,\sigma,\psi$, carrying topological spin $1,e^{\pi i/8},-1$. These line operators are in one-to-one correspondence with the chiral primaries of the Ising CFT, and satisfy the fusion rules
\begin{equation}
    \psi\times \psi =1\,, \quad \psi \times \sigma =\sigma \times \psi = \sigma\,, \quad \sigma \times \sigma = 1+\psi\,.
\end{equation}
The category $\overline{\rm Ising}$ is the orientation-reversal counterpart. Its simple line operators, denoted as $1,\overline{\sigma},\overline{\psi}$, carry inverse topological spins. There are two kinds of Lagrangian algebras, the bosonic one
    \begin{equation}
        \mathcal{L}_{b} = 1 \oplus \psi\overline{\psi}  \oplus  \sigma\overline{\sigma}\,,
    \end{equation}
which supports the Ising symmetry $\{1,\eta,N\}$, and the fermionic one
    \begin{equation}
        \mathcal{L}_{f}=1\oplus \pi \psi \oplus \pi \overline{\psi} \oplus  \psi\overline{\psi}\,.
    \end{equation}

To realize the superfusion category $\mathscr{C}_q^{(0)}$, let us choose $\mathcal{L}_{\textrm{sym}}=\mathcal{L}_f$, which leads to the projections
    \begin{equation}
        \Pi_f (1)=\Pi_f(\psi \overline{\psi})=1\,,\quad \Pi_f(\psi)=\Pi_f(\overline{\psi})=\pi\,.
    \end{equation}
Since $\sigma$ and $\overline{\sigma}$ cannot terminate on the symmetry boundary $B_{\rm sym}$, they must instead project to symmetry generators supported on $B_{\rm sym}$. We therefore assume
    \begin{equation}\label{eq:sigma_projection}
        \Pi_f(\sigma) = \eta\,,\quad \Pi_f(\overline{\sigma})=\eta'\,.
    \end{equation}
Using the fusion rules $\sigma \times \sigma = 1+\psi$ and $\overline \sigma \times \overline \sigma = 1+\overline \psi$, one finds $\eta$ and $\eta'$ satisfy the q-type fusion rules
    \begin{equation}
        \eta \times \eta = \eta'\times \eta'=1 + \pi\,.
    \end{equation}
In fact, one may identify $\eta$ and $\eta'$ as the left- and right-moving fermion parity $(-1)^{F_L}$ and $(-1)^{F_R}$, respectively. Then $\Pi_f(\sigma \overline{\sigma})$ must be related to the total fermion parity $(-1)^F$, and we have
    \begin{equation}
        \Pi_f (\sigma \overline{\sigma}) = (1+\pi) (-1)^F\,.
    \end{equation}
Therefore, we see that the superfusion category supported on $B_{\rm sym}$ is $\mathscr{C}_q^{(0)}$.
    

To study the boundary operators, we fix $\mathcal{L}^*_{\rm sym}=\mathcal{L}_{b}$ and eq.~\eqref{eq:dim-of-module} gives
    \begin{equation}
        \textrm{Hom}_{\textrm{Ising}\boxtimes \overline{\textrm{Ising}}\boxtimes \mathbf{sVec}}(\mathcal{L}_{\textrm{sym}},\mathcal{L}^*_{\textrm{sym}}) = \mathbb{C}\oplus \mathbb{C}\,,
    \end{equation}
where the two $\mathbb{C}$-factors arise from $\textrm{Hom}(1,1)$ and $\textrm{Hom}(\psi \overline{\psi},\psi \overline{\psi})$, respectively. It indicates the existence of two simple objects in $\mathcal{M}$, which we denote by $v^{\pm}$. Since $\overline{\sigma} \sigma$ is condensed on $B^*_{\rm sym}$, the fermion parity operator $(-1)^F$ can end on the interface $B_{\mathcal{M}}$. On the other hand, the $\mathbb{Z}_2$ generator $\eta$ (and $\eta'$) can pass through the interface. That implies that $(-1)^F$ is preserved, whereas $\eta$ (and $\eta'$) is broken by the interface $B_{\mathcal{M}}$. We can also label the two kinds of interfaces $v^{\pm}$ using the broken symmetry as
    \begin{equation}
        v^+ \rightarrow 1\,,\quad v^- \rightarrow \eta\,.
    \end{equation}

Now let us consider the representation operators on $B^*_{\rm sym}$ given by generators of Ising symmetry $\{1,\widetilde{\eta},N\}$ satisfying the fusion rules
    \begin{equation}
        \widetilde{\eta} \times \widetilde{\eta} = 1\,,\quad \widetilde{\eta} \times N = N\times \widetilde{\eta}=N\,,\quad N\times N=1+\widetilde{\eta}\,.
    \end{equation}
and they are the projections of $1,\psi,\sigma$ on $B^*_{\rm sym}$, respectively. For the identity operator, when the interface is also labeled by identity $1$, we simply have
    \begin{equation}
        \begin{gathered}
            \begin{tikzpicture}
            
            \draw[ultra thick,dashed,ForestGreen,-<-=.55] (2,1)--(4,1);
            \draw[ultra thick,MidnightBlue,->-=.65] (2,0)--(2,1);
            \draw[ultra thick,MidnightBlue,->-=.65] (2,1)--(2,2);
            \filldraw[Cerulean] (2,1) circle (2pt);
            \node at (3,1.25) {$1$};
            \node at (1.75,0.5) {$1$};
            \end{tikzpicture}
        \end{gathered}\quad ,
    \end{equation}
which gives a trivial bosonic boundary operator. On the other hand, if the boundary is characterized by $\eta$, we have two boundary operators, one is bosonic and the other is fermionic
    \begin{equation}
        \begin{gathered}
            \begin{tikzpicture}
            
            \draw[ultra thick,dashed,ForestGreen,-<-=.55] (2,1)--(4,1);
            \draw[ultra thick,MidnightBlue,->-=.65] (2,0)--(2,1);
            \draw[ultra thick,MidnightBlue,->-=.65] (2,1)--(2,2);
            \filldraw[Cerulean] (2,1) circle (2pt);
            \node at (3,1.25) {$1$};
            \node at (1.75,0.5) {$\eta$};
            \end{tikzpicture}
        \end{gathered}\quad \quad
        \begin{gathered}
            \begin{tikzpicture}
            
            \draw[ultra thick,dashed,ForestGreen,-<-=.55] (2,1)--(4,1);
            \draw[ultra thick,MidnightBlue,->-=.65] (2,0)--(2,1);
            \draw[ultra thick,MidnightBlue,->-=.65] (2,1)--(2,2);
            \filldraw[WildStrawberry] (2,1) circle (2pt);
            \node at (3,1.25) {$1$};
            \node at (1.75,0.5) {$\eta$};
            \end{tikzpicture}
        \end{gathered}\quad ,        
    \end{equation}
where we used the fact that a $\pi$-line can end on $\eta$ due to $\eta \times \eta = 1 + \pi$. In other words, $\eta$ line can support a Majorana fermion. As a result, we see that the multiplet consists of three boundary operators. This coincides with the representation $\mathcal{R}_1$ in Section~\ref{sec:Z2qsuperfusion}.  

For the $\widetilde{\eta}$-operator, since $\pi \psi$ is condensed on the left symmetry boundary $B_{\rm sym}$, one expects that $\widetilde{\eta}$, being the projection of $\psi$, becomes a $\pi$-line after crossing the interface. If the boundary is labeled by the identity $1$, we have
    \begin{equation}
        \begin{gathered}

        \end{gathered}\quad .
    \end{equation}
If the boundary is labeled by $\eta$, we similarly have
    \begin{equation}
        \begin{gathered}
            %
        \end{gathered}\quad .        
    \end{equation}
Compared to the previous case, the difference here is that the boundary operator for the identity interface is fermionic rather than bosonic. This is the representation $\widetilde{\mathcal R}_1$ discussed in Section~\ref{sec:Z2qsuperfusion}.

For the $N$-operator, since $N$ is the projection of $\sigma$ on $B^\ast_{\rm sym}$, using eq.~\eqref{eq:sigma_projection} the $N$-line becomes an $\eta$-line after crossing the interface. We can further bend and merge it with the interface, e.g.,
\begin{equation}
        \begin{gathered}
            %
        \end{gathered}\quad .
\end{equation}
Altogether we have four kinds of boundary operators
    \begin{equation}
        \begin{gathered}
            %
        \end{gathered}\quad ,  
    \end{equation}
which is nothing but the representation $\mathcal R_2$ discussed at the end of Section~\ref{sec:Z2qsuperfusion}.

\subsection*{Example : superfusion category $\tilde{\mathscr{C}}_{A_2}$}

Another simple example is the superfusion category $\tilde{\mathscr{C}}_{A_2} = \{1,\eta\}$, where $\eta$ is the m-type $\mathbb{Z}_2$ generator satisfying $\eta^2=\pi$. It is the symmetry of the gapped phase of the deformed $\mathcal{N}=2$ $A_2$-type minimal model $A_{2}^{\rm def}$. In the bosonic coset model $\frac{\text{SU(2)}_1 \times \text{U(1)}_2}{\text{U(1)}_3}$, the preserved TDLs in $\mathcal B[A_{2}^{\rm def}]$ are  
\begin{equation}
\tilde{\mathscr C}_b=\{L_{(0,0)},L_{(1,0)},L_{(1,3)},L_{(0,3)}\}\,,
\end{equation}
which is a $\mathbb{Z}_4$ group generated by $L_{(1,0)}$. Moreover, since the $\mathbb{Z}_2$ line $L_{(1,3)}$ has spin-half, the $\mathbb{Z}_4$ symmetry is twisted by $\omega=2 \in H^3(\mathbb{Z}_4,U(1))= \mathbb{Z}_4$, as discussed in \cite{Bhardwaj:2024ydc}.

The associated SymTFT is $\mathcal{Z}_{\mathbb{Z}_4^{\omega}} \boxtimes \textbf{sVec}$, where $\mathcal{Z}_{\mathbb{Z}_4^{\omega}}$ is the $\mathbb{Z}^{\omega}_4$ Dijkgraaf-Witten theory with the twist $\omega=2 \in H^3(\mathbb{Z}_4,U(1))= \mathbb{Z}_4$. The line operators in $\mathcal{Z}_{\mathbb{Z}_4^{\omega}}$ are labeled by
\begin{equation}
    \{e^a m^b | a,b=0,1,2,3\}\,,
\end{equation}
with topological spins
\begin{equation}
    \theta(e^a m^b) = \exp \left(\frac{i\pi b (2a+b)}{4} \right)\,.
\end{equation}
We will mainly focus on two Lagrangian algebras, the bosonic one
    \begin{equation}
        \mathcal{L}_b=1\oplus e\oplus e^2\oplus e^3\,,
    \end{equation}
which supports the anomalous $\mathbb{Z}^{\omega}_4$ symmetry generated by $\Pi(m)$, and the fermionic one
    \begin{equation}
        \mathcal{L}_f=1\oplus \pi m^2\oplus e^2\oplus \pi e^2 m^2\,,
    \end{equation}
which supports the $\mathbb{Z}_2^F\times \mathbb{Z}_2$ symmetry via the identification
    \begin{equation}
        \Pi_f(e)=(-1)^F\,,\quad \Pi_f(m) = \eta\,.
    \end{equation}
Since $m^2\sim \pi$, the fusion rule is $\eta \times \eta = \pi$.

We choose $\mathcal{L}_{\textrm{sym}}=\mathcal{L}_f$ and $\mathcal{L}^*_{\rm sym}=\mathcal{L}_b$, then eq.~\eqref{eq:dim-of-module} gives
    \begin{equation}
        \textrm{Hom}_{\mathcal{Z}(\mathbb{Z}_4^{\omega})\boxtimes \textbf{sVec}}(\mathcal{L}_{\textrm{sym}},\mathcal{L}^*_{\textrm{sym}}) = \mathbb{C}\oplus \mathbb{C}\,,
    \end{equation}
where the two $\mathbb{C}$-factors arise from $\textrm{Hom}(1,1)$ and $\textrm{Hom}(e^2,e^2)$. It indicates that there are also two simple objects in $\mathcal{M}$, denoted by $v^{\pm}$. The fermion parity operator $(-1)^F$ can end on the interface, since $e$-line is condensed on $B^*_{\rm sym}$, whereas $\eta$ can go through the interface. Therefore we conclude that $(-1)^F$ is preserved and $\eta$ is broken by the interface, and the two kinds of interfaces $v^{\pm}$ can be labeled by the broken symmetries as
    \begin{equation}
        v^+ \rightarrow 1\,,\quad v^- \rightarrow \eta\,.
    \end{equation}

There are four representations labeled by $1,m,m^2,m^3$ on $B^*_{\rm sym}$. Beginning with $1$, we have two boundary operators
    \begin{equation}
        \begin{gathered}

        \end{gathered}\quad .        
    \end{equation}
For $m^2$ we have two more
    \begin{equation}
        \begin{gathered}
            %
        \end{gathered}\quad ,        
    \end{equation}
where the boundary operators are fermionic since $m^2$ is converted to $\pi$ on $B_{\rm sym}$. The m-line can cross the interface and become the $\mathbb{Z}_2$ generator $\eta$, and one can bend the $\eta$ line to merge the interface and obtain
    \begin{equation}
        \begin{gathered}
            %
        \end{gathered}\quad ,        
    \end{equation}
where we used $m^2 \sim \pi$ on $B_{\rm sym}$. Finally, for $m^3$ one has similarly
    \begin{equation}
        \begin{gathered}
            %
        \end{gathered}\quad .        
    \end{equation}
The representation labeled by the $m$-line exactly matches $\mathcal{R}_{\mathcal L_1}$ for $A_2^{\text{def}}$, whose quiver diagram is Figure~\ref{fig:R_L1} in Section~\ref{sec:N_2_minimal_model}.

\subsection*{Example : Fibonacci superfusion category}

As the last example, we consider the SymTFT $\mathcal{Z}(\widehat{so(3)}_6)$ of the $\widehat{so(3)}_6$ fusion category. However, since $\widehat{so(3)}_6$ is not a modular tensor category, its SymTFT is not simply $\widehat{so(3)}_6 \boxtimes \widehat{so(3)}_{-6}$. In this paper, we propose the SymTFT for $\widehat{so(3)}_6$ to be the modular extension of $\widehat{so(3)}_6 \boxtimes \widehat{so(3)}_{-6}$ given by
    \begin{equation}
        \mathcal{Z}(\widehat{so(3)}_6) = \left(\widehat{su(2)_{6}} \boxtimes \widehat{su(2)_{-6}}\right)/\mathcal{A}\,,\quad \text{with} \quad \mathcal{A} = (0,0) \oplus (6,6)\,,
    \end{equation}
where $\mathcal{A}$ is a condensable bosonic algebra in $\widehat{su(2)_6} \boxtimes \widehat{su(2)_{-6}}$, as we will review soon.

We label the seven Verlinde lines in $\widehat{su(2)_6}$ as $a=0,\dots,6$, with the quantum dimensions
\begin{equation}
    d_0=d_6=1\,, \quad d_1=d_5= \sqrt{2+\sqrt{2}}\,, \quad d_2=d_4= 1 + \sqrt{2}\,, \quad d_3=\sqrt{4+2\sqrt{2}}\,.
\end{equation}
The $\widehat{so(3)_6}$ subcategory consists of $\{0,2,4,6\}$. The anyon lines in $\widehat{su(2)_{6}} \boxtimes \widehat{su(2)_{-6}}$ are then labeled by the pair $(a,b)$ with $a,b=0,\dots,6$. We consider condensing the algebra $\mathcal{A}=(0,0)\oplus (6,6)$, with $(6,6)$ an abelian anyon satisfying the fusion rule
    \begin{equation}
        (6,6) \times (a,b) = (6-a,6-b)\,.
    \end{equation}
The surviving anyon should have trivial monodromy with $(6,6)$, and the monodromy matrix $M_{(a,b),(6,6)}$ is \cite{Barkeshli:2014cna}
    \begin{equation}
        M_{(a,b),(6,6)} = \frac{S^*_{(a,b),(6,6)} S_{(0,0),(0,0)}}{S_{(0,0),(6,6)} S_{(0,0),(a,b)}} = (-1)^{a+b}\,,
    \end{equation}
where $S_{(a,b),(c,d)}$ is the modular $S$-matrix. Thus only the anyons $(a,b)$ with $a+b=0\ \text{mod}\ 2$ survive the condensation. Modulo the identification $(a,b)\sim (6-a,6-b)$ upon fusing $(a,b)$ with $(6,6)$, the surviving simple objects are \cite{Ji:2026yfj}
\begin{equation}
\begin{gathered}
    \relax[0,0]\,,\quad [2,0]\,,\quad [4,0]\,, \quad [6,0]\,, \quad [0,2]\,,\quad [2,2]\,, \quad [4,2]\,, \quad [6,2]\\
    [1,1]\,,\quad [1,3]\,,\quad [1,5]\,,\quad [3,1]\,,\quad [3,3]_+\,,\quad [3,3]_-\,,
\end{gathered}
\end{equation}
where $[a,b]$ is the equivalence class of $(a,b)\sim (6-a,6-b)$, and the $\mathbb{Z}_2$ fixed point $(3,3)$ splits into two simple anyons $[3,3]_+$ and $[3,3]_-$ after condensation. The quantum dimensions for those anyons are
\begin{equation}
\begin{gathered}
    d_{[0,0]}=d_{[6,0]}=1\,,\quad d_{[2,0]}=d_{[4,0]}=d_{[0,2]}=d_{[6,2]}=1+\sqrt{2}\,,\\ d_{[1,1]}=d_{[1,5]}=d_{[3,3]_+}=d_{[3,3]_-}=2+\sqrt{2}\,,\\ d_{[3,1]}=d_{[1,3]}=2+2\sqrt{2}\,,\quad
    d_{[2,2]}=d_{[4,2]}=3+2\sqrt{2}\,.
\end{gathered}
\end{equation}
and the total quantum dimension $D=\sqrt{\sum_a d_a^2} = 8+4\sqrt{2}$. The topological spin for each anyon is
    \begin{equation}
        \begin{gathered}
            \theta_{[0,0]}=\theta_{[1,1]}=\theta_{[2,2]}=\theta_{[3,3]_+}=\theta_{[3,3]_-}=\theta_{[1,5]}=1\,,\quad \theta_{[6,0]}=\theta_{[4,2]}=-1\,,\\
            \theta_{[2,0]}=\theta_{[6,2]}=i\,,\quad \theta_{[0,2]}=\theta_{[4,0]}=-i\,,\quad \theta_{[1,3]}=e^{-\frac{3\pi i}{4}}\,, \quad\theta_{[3,1]}=e^{\frac{3\pi i}{4}}\,.
        \end{gathered}
    \end{equation}

In the following, we will focus on the two Lagrangian algebras, the bosonic one is
    \begin{equation}
        \mathcal{L}_b = [0,0]\oplus [1,1]\oplus [2,2]\oplus [3,3]_+\,,
    \end{equation}
which is diagonal and supports the fusion category $\mathscr{C}_{SO(3)_6}=\{I,X,Y,Z \}$ identified by
    \begin{equation}
        \Pi_b([0,0])=I\,,\quad \Pi_b([2,0])=X\,,\quad \Pi_b([4,0])=Y\,, \quad \Pi_b([6,0])=Z\,.
    \end{equation}
The fermionic one is
    \begin{equation}
        \mathcal{L}_f = [0,0] \oplus \pi [6,0] \oplus [2,2]\oplus \pi [4,2]\,,
    \end{equation}
and we can identify
    \begin{equation}
        \Pi_f([0,0])=I\,, \quad \Pi_f([2,0])=W\,.
    \end{equation}
Notice that $[2,0]\times [2,0] = [0,0]+[2,0]+[4,0]$ and $[4,0]=[2,0]\times [6,0]$. Acting the projection $\Pi_f$ on both sides, we recover the fusion rule
    \begin{equation}
        W\times W = 1+(1+\pi) W\,.
    \end{equation}
However, the symmetry operator $(-1)^F$ cannot be identified as simple objects in $\mathcal{Z}(\widehat{so(3)_6})$, the latter project to the combinations between $(-1)^F$ and $W(-1)^F$.

We choose $\mathcal{L}_{\text{sym}}=\mathcal{L}_f$ and $\mathcal{L}_{\text{sym}}^*=\mathcal{L}_b$, then eq.~\eqref{eq:dim-of-module} gives
    \begin{equation}
        \text{Hom}_{\mathcal{Z}(\widehat{so(3)_6})\boxtimes \textbf{sVec}}(\mathcal{L}_{\text{sym}},\mathcal{L}^*_{\text{sym}})=\mathbb{C}\oplus \mathbb{C}\,,
    \end{equation}
which implies two simple objects in $\mathcal{M}$ due to the spontaneous breaking of $W$, and we will label two kinds of interfaces as $1$ and $W$.

There are four representation operators represented by $[0,0],[2,0],[4,0],[6,0]$ on $B_{\text{sym}}^*$. Beginning with $[0,0]$, we have two boundary operators
    \begin{equation}
        \begin{gathered}

        \end{gathered}\quad .        
    \end{equation}
This gives the representation $\mathcal{R}_1$ in Section~\ref{sec:superFibo}. For $[6,0]$ we have another two
    \begin{equation}
        \begin{gathered}
            %
        \end{gathered}\quad ,        
    \end{equation}
where the boundary operators are fermionic since $[6,0]$ is converted to $\pi$ on $B_{\text{sym}}$. This is the $\widetilde{\mathcal R}_1$ representation mentioned in Section~\ref{sec:superFibo}. For $[2,0]$, it can cross the interface and become $W$ on the LHS, and one can bend and merge it with the interface to obtain the multiplets
    \begin{equation}
        \begin{gathered}
            %
        \end{gathered}\quad ,  
    \end{equation}
which becomes the representation $\mathcal{R}_2$ discussed in Section~\ref{sec:superFibo}. For $[4,0]=[2,0]\times [6,0]$, we have
    \begin{equation}
        \begin{gathered}
            %
        \end{gathered}\quad ,  
    \end{equation}
which becomes the representation $\widetilde{\mathcal{R}}_2$ discussed in Section~\ref{sec:superFibo}.

\section{Conclusion}
\label{sec:conclusion}

In this paper, we establish a comprehensive and mathematically rigorous framework for analyzing categorical symmetries in fermionic quantum field theories. We began by extending the standard bosonic strip algebra into the fermionic regime, formally defining the superstrip algebra as a $C^\ast$-weak Hopf superalgebra. In particular, the superfusion categories and their modules precisely capture Grassmann-odd topological defects and spin-structure dependencies. To demonstrate the computational viability and structural consistency of this formalism, we explicitly constructed the superstrip algebra for two cases: the fermionic Fibonacci category and theories with a q-type $\mathbb Z_2$ symmetry.

Furthermore, we showcased the physical efficacy of our framework by directly applying it to the integrable deformations of $\mathcal N$=1 and $\mathcal N$=2 minimal models. Through this application, we successfully derived a systematic categorical structure of the BPS spectra, which matches perfectly with the integrability literature. Finally, we embedded our purely algebraic construction within a broader topological paradigm by recovering the representation theory of the superstrip algebra from a boundary SymTFT. This crucial step not only provides a powerful holographic perspective on fermionic non-invertible symmetries, but also solidifies the deep interplay between category theory and topological quantum field theory. This may pave the way for the future explorations and higher-dimensional generalizations discussed below.

A possible direction for future research is investigating how fermionic non-invertible symmetries constrain the exact $S$-matrices of integrable theories. It was recently realized \cite{Copetti:2024rqj} that the presence of non-invertible topological defect lines forces a modification of the standard crossing symmetry equations in bosonic systems. Generalizing this paradigm to the fermionic realm is not merely a technical exercise, but crucial for a complete understanding of integrable scattering equipped with topological symmetries. It would be interesting to explore how superfusion symmetries would alter the algebraic constraints governing the $S$-matrix. Our framework provides the necessary categorical data, such as explicit superfusion rules and module $F$-symbols, to systematically derive these modified crossing relations. By explicitly constructing these ingredients, our work establishes the foundation for a new bootstrap program tailored to fermionic integrable models, potentially leading to the discovery of novel $S$-matrices and their corresponding renormalization group flows.

Significantly, the applicability of our framework extends well beyond the restricted domains of supersymmetry and exact integrability, offering powerful new algebraic tools for strongly coupled dynamics. For example, a possible candidate for such applications is 2D massless adjoint QCD. While recent theoretical breakthroughs \cite{Komargodski:2020mxz, Cordova:2024goh, Cordova:2024nux} have successfully leveraged bosonized duals to constrain the deconfinement phase transition and the infrared spectrum of this theory, the native symmetries governing the dynamics are intrinsically fermionic. Operating solely in the bosonized framework can sometimes obscure these subtle fermionic features, complicating the analysis of the full state space. Applying our superstrip algebra directly to these native fermionic categories provides a more natural and robust approach. We anticipate that this direct fermionic treatment will clarify existing constraints on the IR physics, such as capturing subtle non-invertible 't Hooft anomalies that might otherwise be missed. Ultimately, this approach promises to shed new light on the strong-coupling dynamics and the intricate phase structure of IR theories.

Finally, we highlight a tantalizing connection to four-dimensional physics via the dynamics of superconformal vortex strings \cite{Tong:2006pa}. This well-established 2D/4D correspondence intricately relates the Coulomb branch operators of strongly coupled 4D $\mathcal{N}$=2 Argyres-Douglas (AD) theories to the chiral ring operators of 2D $\mathcal{N}$=2 minimal models residing on the vortex string worldsheet. By leveraging this dictionary, one can identify special loci within the minimal chamber of the 4D AD theory where the BPS spectrum exhibits a remarkable degeneracy, consisting of $k$ mutually non-local monopoles of exactly equal mass. This 4D spectrum precisely matches the BPS solitons of the corresponding 2D minimal model when deformed by its least relevant operator \cite{Shapere:1999xr}. Physically, this exact matching is elegantly realized by viewing each 2D soliton as a 4D monopole physically confined within the core of the vortex string \cite{Tong:2006pa}. We thus suspect that a (hidden) non-invertible symmetry in four dimensions underlies this degenerate BPS spectrum. A more detailed study of this hidden symmetry and its implications is left for future work.

\section*{Acknowledgments}
We thank Hiroshi Itoyama, Hikaru Kawai, Zohar Komargodski, Kantaro Ohmori, and Yunqin Zheng for helpful discussions. J.~Chen is supported by the Fujian Provincial Natural Science Foundation of China (No.2025J01004), and the National Natural Science Foundation of China (Grants No.12247103). Z.~Duan is supported by a startup grant from SIMIS. Z.~Duan was previously supported by the Swiss National Science Foundation (SNSF) through a Postdoctoral Fellowship (TMPFP2\_234009). QJ is supported by National Research Foundation of Korea (NRF) Grant No. RS-2024-00405629 and Jang Young-Sil Fellow Program at the Korea Advanced Institute of Science and Technology. S.~Lee is supported by KIAS Grants PG056502. 

\newpage
\appendix

\section{More on superstrip algebra}
\label{app:hopf-algebra}
\subsection{\texorpdfstring{Definition of $C^\ast$-Weak Hopf superalgebra}{Definition of C*-Weak Hopf superalgebra}}
In this section, we give the definition of the $C^\ast$-weak Hopf superalgebra based on \cite{bohm1999weak,andruskiewitsch2001triangular,aissaoui2014classification,Inamura:2022lun}. The $C^\ast$ structure is easy to understand. A $C^\ast$-superalgebra $A$ is equipped with a norm $||\cdot||$ and an antilinear involution $\ast$. This means $(a^\ast)^\ast = a$, $(a_1 a_2)^\ast = a_2^\ast a_1^\ast$ and the involution is compatible with the norm $||a^\ast a|| = ||a||^2$. On the other hand, the weak Hopf superalgebra is more complicated. It is a weak Hopf algebra whose algebraic objects are super vector space $\textbf{sVec}$ over $\mathbb{C}$. More explicitly, the weak Hopf superalgebra is a sextuple $(A,\mu,\eta,\Delta,\epsilon,S)$ satisfying the following axioms:

\begin{itemize}
    \item $A$ is a finite dimensional associative superalgebra over $\mathbb{C}$, which is $\mathbb{Z}_2$ graded
        \begin{equation}
            A= A_0 \oplus A_1\,.
        \end{equation}
    
    The multiplication $\mu : A\otimes A\rightarrow A$ and unit $\eta : \mathbb{C}\rightarrow A$ are both even $\mathbb{C}$-linear map, preserving the graded structure of $A$
    \begin{equation}
        \mu (A^i\otimes A^j) \subseteq A^{[i+j]}\,, \quad \eta (1) \in A^0\,,
    \end{equation}
where $[i+j] \in \mathbb{Z}_2$ means $i+j$ mod 2, and they satisfy
    \begin{equation}
    \begin{split}
        &\textrm{Associativity}: \quad \mu \circ (\mu \otimes \textrm{id}) = \mu \circ (\textrm{id} \otimes \mu)\,,\\
        &\textrm{Unit property}: \quad \mu \circ (\eta \otimes \textrm{id})=\mu \circ (\textrm{id}\otimes \eta)=\textrm{id}\,.
    \end{split}
    \end{equation}
We will suppress $\mu$ and $\eta$ and write $xy$ for $\mu(x,y)$ and use the unit element $1:=\eta(1)$ instead of $\eta$. Moreover, the multiplication over the tensor product of two superalgebra $A\otimes B$ satisfies
    \begin{equation}
        (a_1\otimes b_1)(a_2\otimes b_2) = (-1)^{s(a_2)s(b_1)} (a_1 a_2) \otimes (b_1 b_2)\,,
    \end{equation}
for $a_1,a_2\in A$ and $b_1,b_2\in B$, where we introduce the signature $s(a)$ so that
    \begin{equation}
        s(a) = \left\{\begin{array}{c}
            0\,, \quad a \in A_0\,,\\
            1\,, \quad a \in A_1\,.
        \end{array} \right.
    \end{equation}

\item $A$ is also a coassociative supercoalgebra over $\mathbb{C}$, with even $\mathbb{C}$-linear comultiplication $\Delta:A\rightarrow A\otimes A$ and counit $\epsilon:A\rightarrow \mathbb{C}$
    \begin{equation}
        \Delta (A^i) \subseteq \oplus_{j=0,1} A^j \otimes A^{[i+j]}\,, \quad \epsilon(A^1) = 0\,,
    \end{equation}
satisfying
    \begin{align}
        &\textrm{Coassociativity}: \quad (\Delta \otimes \textrm{id}) \circ \Delta = (\textrm{id} \otimes \Delta) \circ \Delta\,,\\
        &\textrm{Counit property}: \quad  (\epsilon \otimes \textrm{id})\circ \Delta =(\textrm{id}\otimes \epsilon)\circ \Delta=\textrm{id}\,.
    \end{align}
\item For compatibility of the superalgebra and co-superalgebra structure, we require
    \begin{align}
        &\textrm{Multiplicativity of the coproduct} : \quad \Delta(xy) = \Delta(x) \Delta(y)\,, \quad \forall x,y\in A\,.\\
        &\textrm{Weak Multiplicativity of the counit}:\nonumber\\
        &\qquad\begin{aligned}
        \epsilon(xyz)&=\sum_y\epsilon(xy^{(1)})\epsilon(y^{(2)}z)\\
        &=\sum_y (-1)^{s(y^{(1)})s(y^{(2)})}\epsilon(xy^{(2)})\epsilon(y^{(1)}z)\,,
        \quad \forall x,y,z\in A\,.
        \end{aligned}\\
        &\textrm{Weak Multiplicativity of the unit}:\nonumber\\
        &\qquad\begin{aligned}
        (\Delta\otimes \textrm{id})\Delta(1)
        &=(\Delta(1)\otimes 1)(1\otimes \Delta(1))\\
        &=(1\otimes \Delta(1))(\Delta(1)\otimes 1)\,.
        \end{aligned}
    \end{align}
Here the comultiplication $\Delta : A\rightarrow A\otimes A$ is expanded as
    \begin{equation}
        \Delta(x) = \sum_{x^{(1)},x^{(2)}}  x^{(1)}\otimes x^{(2)} \equiv \sum_{i,j} x^{(1)}_i \otimes x^{(2)}_j\,,
    \end{equation}
and it satisfies the relation
    \begin{equation}
        \Delta(xy) = \sum_{x^{(1)},x^{(2)},y^{(1)},y^{(2)}}  (-1)^{s(x^{(2)}) s(y^{(1)})}(x^{(1)} y^{(1)}) \otimes (x^{(2)} y^{(2)})\,.
    \end{equation}
\item There exists an even $\mathbb{C}$-linear map $S:A \rightarrow A$ called the \emph{antipode}, satisfying the following relations
    \begin{align}
        &\sum_{x^{(1)},x^{(2)}} x^{(1)}S(x^{(2)})=\sum_{1^{(1)},1^{(2)}}(-1)^{s(1^{(2)})s(x)}\epsilon(1^{(1)}x)1^{(2)}\,,\\
        &\sum_{x^{(1)},x^{(2)}} S(x^{(1)})x^{(2)}=\sum_{1^{(1)},1^{(2)}} (-1)^{s(x)s(1^{(1)})}1^{(1)}\epsilon(x1^{(2)})\,,\\
        &\sum_{x^{(1)},x^{(2)},x^{(3)}} S(x^{(1)}) x^{(2)}S(x^{(3)})=S(x)\,,
    \end{align}    
where $x^{(1)},x^{(2)},x^{(3)}$ in the last line are given by
    \begin{equation}
        (\Delta \otimes \text{id}) \Delta(x)=\sum_{x^{(1)},x^{(2)},x^{(3)}} x^{(1)}\otimes x^{(2)}\otimes x^{(3)}\,.
    \end{equation}
\end{itemize}

Finally, the $C^*$-structure is compatible with that of a weak Hopf superalgebra. The involution on the graded tensor product is defined by
\begin{equation}
    (x \otimes y)^* = (-1)^{s(x)s(y)} (x^* \otimes y^*)\,.
\end{equation}
This means the comultiplication is a $\ast$-algebra homomorphism,
\begin{equation}
 \Delta(h^\ast) = \Delta(h)^\ast = \sum_{h^{(1)},h^{(2)}} (-1)^{s(h^{(1)})s(h^{(2)})} (h^{(1)})^*\otimes (h^{(2)})^*\,,
\end{equation}
with the action of $\ast$ on complex numbers being complex conjugate. We also have
\begin{equation}
    (A^i)^* = A^i,\quad \epsilon(h^\ast) = \overline{\epsilon(h)},\quad S(h^*)^* = S^{-1}(h)\,.
\end{equation}

\subsection{\texorpdfstring{Realizing the superstrip algebra as $C^\ast$-Weak Hopf superalgebra}{Realizing the superstrip algebra as C*-Weak Hopf superalgebra}}
\label{App:multiplication}
In this section, we explain how to realize the superstrip algebra as a $C^\ast$-Weak Hopf superalgebra. First of all, it is easy to identify the $C^\ast$ structure. Physically, the antilinear involution $\ast$ corresponds to the Hermitian conjugate of operators, and is represented by the $\dagger$ operation,
\begin{equation}\label{app:fig:superstrip}
\dagger: \begin{gathered}
        \begin{tikzpicture}
            \draw[ultra thick,MidnightBlue,->-=.55] (0,1)--(0,0);
            \draw[ultra thick,MidnightBlue,->-=.55] (0,0)--(0,-1);
            \draw[ultra thick,MidnightBlue,->-=.55] (1,0)--(1,1);
            \draw[ultra thick,MidnightBlue,->-=.55] (1,-1)--(1,0);
            \draw[densely dashed,ultra thick,BurntOrange,-<-=.55] (0,0)--(1,0.15);
            \node at (0,1.25) {$i$};
            \node at (1,1.25) {$j$};
            \node at (0,-1.25) {$k$};
            \node at (1,-1.25) {$l$};
            \node at (-0.25,0.25) {$\alpha$};
            \node at (1.25,0.3) {$\beta$};
            \node at (0.5,0.3) {$a$};
            \filldraw[black] (0,0) circle (1.5pt);
            \filldraw[black] (1,0.15) circle (1.5pt);
        \end{tikzpicture}
    \end{gathered} \quad \mapsto \quad (-1)^{s(\alpha)s(\beta)}\sqrt{\frac{d_i d_j}{d_k d_l}} \begin{gathered}
        \begin{tikzpicture}
            \draw[ultra thick,MidnightBlue,->-=.55] (0,1)--(0,0);
            \draw[ultra thick,MidnightBlue,->-=.55] (0,0)--(0,-1);
            \draw[ultra thick,MidnightBlue,->-=.55] (1,0)--(1,1);
            \draw[ultra thick,MidnightBlue,->-=.55] (1,-1)--(1,0);
            \draw[densely dashed,ultra thick,BurntOrange,->-=.55] (0,0)--(1,0.15);
            \node at (0,1.25) {$k$};
            \node at (1,1.25) {$l$};
            \node at (0,-1.25) {$i$};
            \node at (1,-1.25) {$j$};
            \node at (-0.25,0.25) {$\alpha^\dagger$};
            \node at (1.3,0.3) {$\beta^\dagger$};
            \node at (0.5,0.3) {$a$};
            \filldraw[black] (0,0) circle (1.5pt);
            \filldraw[black] (1,0.15) circle (1.5pt);
        \end{tikzpicture}
    \end{gathered}\,.
\end{equation}
Here $d_i$ denotes the module quantum dimension of the simple object $i \in \mathcal M$. It ensures the compatibility of $C^\ast$-structure with the weak Hopf superalgebra maps.

Next, we will derive the multiplication $\mu$, comultiplication $\Delta$, counit $\epsilon$ and antipode $S$ of the superstrip algebra $\mathbf{sStr}_{\mathscr{C}}(\mathcal{M})$ to endow it with the structure of a weak Hopf superalgebra. 

Recall that the module $F$-move coefficients of the superfusion category are defined as

    \begin{equation}\label{app:Ftilde-move}
        \begin{gathered}

where $i,j,k$ label the vacua, $a,b$ label the simple TDL and $\beta,\beta',\gamma,\mu$ are components of the junction vector spaces.
The dual module $F$-move coefficients $\widetilde{F}^\vee$ are similarly given by

    \begin{equation}\label{app:dualFtilde-move}
        \begin{gathered}
     %
    
        \end{gathered}\,.
    \end{equation}

\paragraph{Multiplication}
Begin with the LHS of \eqref{eq_multiplication}. To merge the two dashed lines, we move the two dashed lines closer to each other
    \begin{equation}
        \begin{gathered}
            %
        \end{gathered}
    \end{equation}
and there could be a sign factor $(-1)^{s(\alpha) s(\delta)}$ if both $\alpha$ and $\delta$ are fermionic. Then we apply module $F$-move and dual module $F$-move on both sides and get

    \begin{equation}
(-1)^{s_{\alpha} s_{\delta}}\quad \sum_{c,\alpha',\gamma'} \sum_{d,\delta',\beta'} \widetilde{F}^{i a k,\alpha \gamma}_{b m c,\alpha' \gamma'} (\widetilde{F}^\vee)^{\bar{a}\bar{b}d,\delta' \beta'}_{njl,\delta \beta}
        \begin{gathered}
            %
        \end{gathered}\,.
    \end{equation}
To shrink the bubble, we apply the $\widetilde{O}$-move introduced in \cite{Zhou:2021ulc} as
    \begin{equation}
        \begin{gathered}
            %
        \end{gathered}
    \end{equation}
where $\rho$ is given by
    \begin{equation}
        \rho = \left\{\begin{array}{l}
        \gamma\,,\quad \textrm{if } s(\alpha) + s(\beta) = 0 \ \textrm{mod}\ 2\\
        \gamma \times f\,, \quad \textrm{if } s(\alpha) + s(\beta) = 1 \ \textrm{mod }2,\ \textrm{and } n\ \textrm{is q-type}
        \end{array} \right.
    \end{equation}
where $\gamma \times f$ means it fuses the leftover Majorana fermion after shrinking the bubble. 
Moreover, if $n$ is m-type and $s(\alpha) + s(\beta) = 1\ \textrm{mod }2$, the $\widetilde{O}$-coefficient is zero, since the m-type TDL cannot support the Majorana fermion. Therefore we get
\begin{equation}
\begin{aligned}
   \mu: &\quad \begin{gathered}
        %
    \end{gathered}\,,
    \end{aligned}
        \end{equation}

\paragraph{Comultiplication} To derive the comultiplication, we begin with the left-top diagram of Figure.~\ref{Fig-comultiplication}, and we slide the $\alpha$ and $\beta$ nodes down while keeping their relative position. Then we make use of an $F$-move and move one of the $\gamma$ nodes to the left as shown in the left-bottom in Figure.~\ref{Fig-comultiplication}, which may generate a non-trivial phase. Finally, we rebalance the position of the two middle lines to get the right-bottom Figure.~\ref{Fig-comultiplication}. Moving one of the $\gamma$ junction from right to left gives a phase factor $\Lambda$ as explained in \cite{Aasen:2017ubm}.\footnote{It satisfies $\Lambda^2 = -1$, so $\Lambda = \pm i$ depending on the choice of gauge of the $F$-moves. For the $F$-moves in Appendix~\ref{App:B}, we can choose $\Lambda = -i$.} Then adjusting the vertical positions of each junctions produces the sign factors in \eqref{eq_comultiplication}.
\begin{figure}
    \centering
\includegraphics[width=0.82\linewidth,height=0.66\textheight,keepaspectratio]{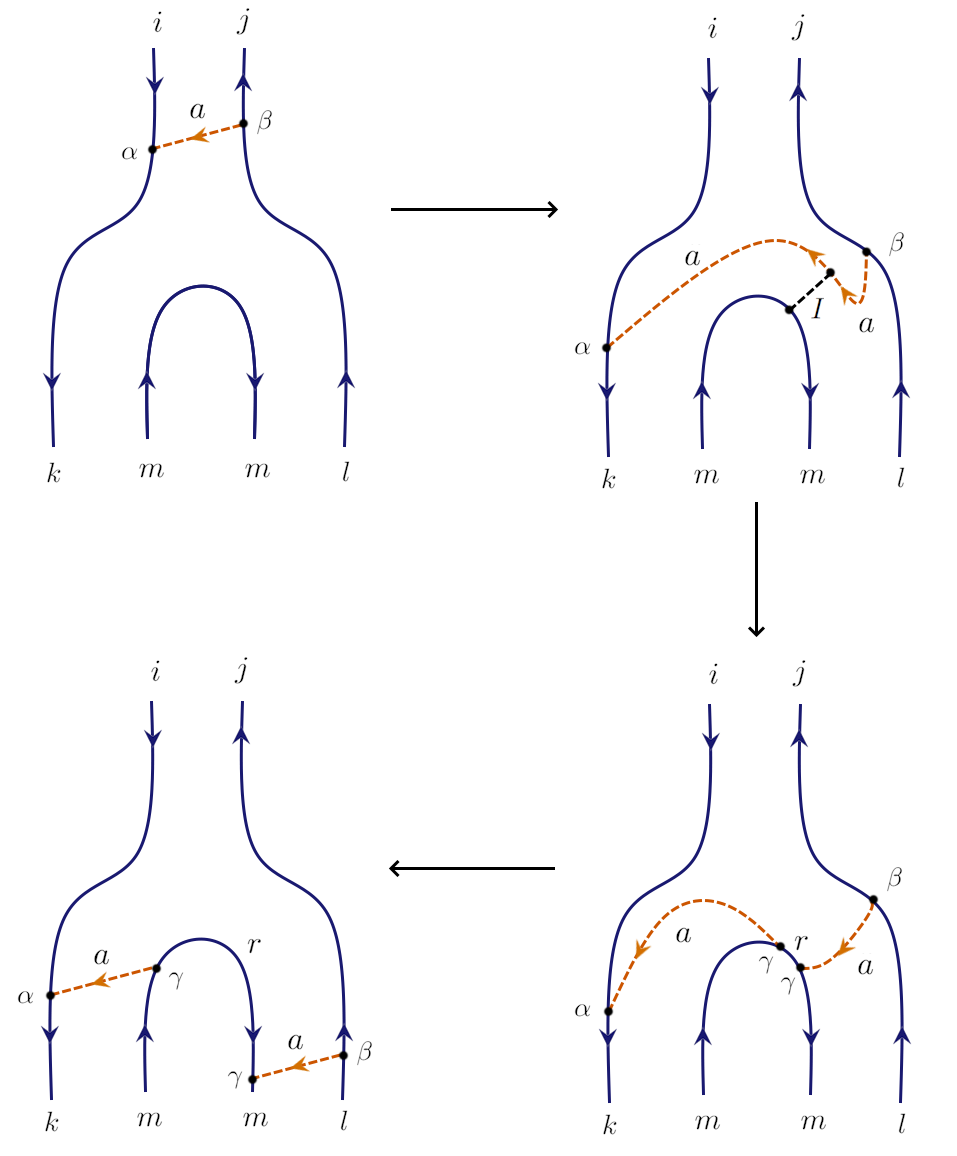}
\caption{The derivation of comultiplication.}
\label{Fig-comultiplication}
 \end{figure}
Collecting all factors, we arrive at \eqref{eq_comultiplication}.
\begin{equation}
    \Delta: \begin{gathered}

\end{gathered}\,.
\end{equation}   

\paragraph{Counit} The counit can be motivated from capping the superstrip algebra from above
\begin{align}
    %
\end{align}
where $\Lambda$ is a possible phase of moving the right junction $\beta$ to the left across the bridge, and we have used the dual $F$-move in the second equality. This means the counit is given by
\begin{equation}
    \epsilon:\quad
\begin{gathered}
        %
    \end{gathered}\quad \rightarrow \quad \Lambda^{s(\beta)}\delta_{i,j} \delta_{k,l} \delta_{\alpha\beta} \widetilde{F}_{akI,\bb\bb}^{k\bar{a}i,\alpha\alpha} \frac{d_a}{d_i}\,.
\end{equation}

\paragraph{Antipode}
The antipode map comes from the following manipulation,
\begin{align}
%
  \node[anchor=center] at (4.0, 1.0) {$= \Lambda^{ s(\alpha) + s(\beta) }$};
\begin{scope}[shift={(7.5,0)}]
        \coordinate (q) at (-R, 0);
        \coordinate (p) at (-r, 0);
        \coordinate (m) at (r, 0);
        \coordinate (n) at (R, 0);
        \draw[ultra thick,MidnightBlue, mid arrow] (n) arc (0:180:R);
        \draw[ultra thick,MidnightBlue, mid arrow] (p) arc (180:0:r);
        \coordinate (start_node) at ({R*cos(135)}, {R*sin(135)});
        \coordinate (end_node) at ({r*cos(120)}, {r*sin(120)});         
        \draw[ultra thick, BurntOrange, densely dashed, mid arrow] (start_node) -- (end_node) 
            node[midway, above right, text=black] {\large $a$};
        \node[dot] at (start_node) {};
        \node[dot] at (end_node) {};
        \node[below] at (q) {\large $j$};
        \node[below] at (p) {\large $i$};
        \node[below] at (m) {\large $k$};
        \node[below] at (n) {\large $l$};
        \node at (-0.2,0.4) {$\alpha$};
        \node at (-1.7,1.7) {$\beta$};
   \end{scope}
\end{tikzpicture}
\end{align}
This means the antipode $S$ is given by
\begin{equation}
    S:\quad \begin{gathered}

    \end{gathered}\,,
\end{equation}
where the extra $(-1)^{s_{\alpha} s_{\beta}}$ comes from exchanging the relative position of $\alpha$ and $\beta$.

\section{Explicit data for some superfusion categories}\label{App:B}
In this appendix, we summarize some explicit data for the two examples we considered in Section~\ref{sec_example_of_superstrip}. Because the module category $\tilde{\mathscr C}$ is obtained from $\mathscr C$ by quotienting by $(-1)^F$, the simple objects in $\tilde{\mathscr C}$ are given by TDLs. Therefore, the module $F$-move, and it dual coincide with the $F$-move and dual $F$-move in the superfusion category $\mathscr C$. We also provide explicit $O$- and $\widetilde{O}$-moves.

\subsection{Data for fermionic Fibonacci}

In this case, we have a trivial TDL $I$ and an m-type TDL $W$ with the fusion rules
\begin{equation}
    I^2 = \mathbb{C}^{1|0}I\,,\quad W  I = I W = \mathbb{C}^{1|0} W\,, \quad W^2 = \mathbb{C}^{1|0} I + \mathbb{C}^{1|1}W\,.
\end{equation}
We will use black and green colors for $I$ and $W$ separately. All the $F$-move coefficients are worked out in \cite{Zhou:2021ulc} under certain gauge choice, and we summarize the non-trivial ones in the following (we omit the blue dot for better visualization)
\begin{equation}
\,
\end{equation}
with $w = 1+\sqrt{2}$, and we omitted the arrows along the lines. The $O$-coefficients are represented as the bubble diagrams
    \begin{equation}
        \begin{gathered}
            %
    \end{gathered}\quad\,,
    \end{equation}
which reads $O^{WW,0}_I = \omega,O^{WW,0}_{W}=O^{WW,1}_W=\sqrt{\omega}$. The dual $F$-move coefficients are 
\begin{equation}
%
\,
\end{equation}
The non-zero $\widetilde{O}$-coefficients are the same as the $O$-coefficients given above.

\subsection{\texorpdfstring{Data for q-type $\mathbb{Z}_2$ symmetries}{Data for q-type Z2 symmetries}}
In this case, we have a trivial TDL $I$ and a q-type TDL $W$ with the fusion rules
\begin{equation}
    I^2 = \mathbb{C}^{1|0}I\,,\quad I W = W I = \mathbb{C}^{1|0}W,\quad W^2 = \mathbb{C}^{1|1}I\,,
\end{equation}
and we still use black and green colors for $I$ and $W$. The $F$-move symbols are given in \cite{Zhou:2021ulc} under certain gauge choice, and are summarized as below (we omit the blue dot for better visualization)

\begin{equation}
\,.
\end{equation}
The non-trivial $O$-coefficients are represented as the bubble diagrams
    \begin{equation}
        \begin{gathered}
            %
    \end{gathered}\quad \,,
    \end{equation}
which reads $O^{NN,0}_I = O^{NN,1}_I=\sqrt{2}$. The dual $F$-move coefficients are

\begin{equation}%
\,.
\end{equation}
The non-trivial $\widetilde{O}$-coefficients are the same as the $O$-coefficients given above.

\section{Modules of Frobenius algebra}
\label{App:algebra}
In this appendix, we discuss anyon condensation with respect to a Frobenius
algebra in a topological order described by a modular tensor category (MTC)
$\mathscr{C}$. Our first goal is to start with the bosonized minimal models
\begin{equation}
    \mathfrak B\left[\mathcal {SM}^{\mathcal{N}=1}_{p,p+2}\right]
    =
    \frac{\mathrm{SU}(2)_{p-2}\times \mathrm{SU}(2)_2}
    {\mathrm{SU}(2)_p}\,,
    \quad
    \mathfrak B\left[\mathcal {SM}^{\mathcal{N}=2}_{p,p+2}\right]
    =
    \frac{\mathrm{SU}(2)_{p-2} \times \mathrm{U}(1)_2}
    {\mathrm{U}(1)_{p}}\,,
\end{equation}
and fermionize them to obtain the $A$-type superconformal minimal models. This is
equivalent to performing fermionic anyon condensation with respect to a
commutative\footnote{In \cite{Lou:2020gfq}, the fermionic Frobenius algebra $\mathbf{1}\oplus f$ is described as supercommutative because $f$ has spin $1/2$. Here, following the convention of \cite{Bhardwaj:2024ydc}, we dress it with the transparent fermion $\pi$, so that $\pi f$ has integer spin and $\mathcal{A}_f$ becomes commutative.} Frobenius algebra
\begin{equation}
    \mathcal{A}_{f} = \mathbf 1\oplus \pi f\, .
\end{equation}
Here $f$ generates a $\mathbb{Z}_2$ one-form symmetry in the bosonized
coset model and has spin $1/2$, while $\pi$ is the transparent fermionic
identity introduced in \eqref{eq:pi-identity-iso}. For simplicity, we will
focus only on the sectors of TDLs preserving the relevant deformation. The
SSB vacua are described by left $\mathcal{A}_{f}$-supermodules, which we
collect into the module category $\mathcal{M}_{A}$, where the subscript $A$ means $A$-type. Since $\mathcal{A}_{f}$ is
supercommutative, left $\mathcal{A}_{f}$-supermodules can be canonically
regarded as bimodules, giving the
superfusion category $\tilde{\mathscr{C}}_{A}=\mathcal{M}_{A}$ preserved after
fermionization. We then further consider condensing the $D/E$-type Frobenius algebra
$\mathcal{A}_{D/E}$ in $\widetilde{\mathscr{C}}_{A}$ to obtain the deformed
$D/E$-type superminimal models. The corresponding SSB vacua are again
described by left $\mathcal{A}_{D/E}$-modules. The technique of anyon condensation has been discussed in many
places, for example in~\cite{Fuchs:2001qc,Fuchs:2002cm,kirillov2002q}, and
its fermionic generalization is discussed in~\cite{Lou:2020gfq}.

We first present some general results of anyon condensation with respect to an algebra $\mathcal{A}$. We will mainly focus on the left modules of $\mathcal{A}$, which are identified with the gapped vacua of the theory.
Given a bosonic MTC $\mathscr{C}$, after condensing $\mathcal{A}$, a simple
(indecomposable) left $\mathcal{A}$-module can be represented, after forgetting
the module structure, as a collection of anyons $W_i$ in $\mathscr{C}$
\begin{equation}\label{eq:module-forget}
    M=\bigoplus_i n_{i M} W_i\,.
\end{equation}
We use $M_x$ to label the simple left modules. Let us denote
\begin{equation}
    \langle W,W'\rangle_{\mathscr{C}}
    =
    \text{sdim} \operatorname{Hom}_{\mathscr{C}}(W,W')\,,
\end{equation}
where the subscript $\mathscr{C}$ emphasizes that the Hom space is evaluated
in the original category $\mathscr{C}$. Here we introduce the super dimension of the graded vector space such that $\text{sdim}(\mathbb{C}^{m|n})=m+n\pi$, where for later purpose we keep the fermionic identity $\pi$ to emphasize that the odd homomorphism is being retained. For simple objects in $\mathscr{C}$,
we have
\begin{equation}
    \langle W_i,W_j\rangle_{\mathscr{C}}=\delta_{i,j},
\end{equation}
where we assume that $\mathscr{C}$ is bosonic so there is no odd homomorphism.

A useful theorem is the Frobenius reciprocity (see, for example, Proposition 4.12 in \cite{Fuchs:2001qc}) 
\begin{equation}\label{eq:reciprocity_theorem}
    \langle W_i , M\rangle_{\mathscr{C}}
    =
    \langle \operatorname{Ind}(W_i) , M\rangle_{\mathcal{A}},
    \qquad
    \langle M,W_i\rangle_{\mathscr{C}}
    =
    \langle M,\operatorname{Ind}(W_i)\rangle_{\mathcal{A}}\,.
\end{equation}
Here
\begin{equation}
    \operatorname{Ind}(W_i)\equiv \mathcal{A}\otimes W_i
\end{equation}
is the induced module of $W_i$. As an object of $\mathscr{C}$, it is usually reducible and decomposes into simple
modules $M_x$. We also denote
\begin{equation}
    \langle M,N\rangle_{\mathcal{A}}
    =\textrm{sdim} 
    \operatorname{Hom}_{\mathcal{A}}(M,N).
\end{equation}
The subscript $\mathcal{A}$ means that the Hom space is computed in the
$\mathcal{A}$-module category, not in $\mathscr{C}$. Thus the same symbol
$M$ should be interpreted differently depending on the bracket: in
$\langle M,\cdot\rangle_{\mathscr{C}}$, it means the underlying object
\begin{equation}
    M=\bigoplus_i n_{iM}W_i
\end{equation}
of $\mathscr C$, while in $\langle M,\cdot\rangle_{\mathcal A}$, it means an
object of the $\mathcal A$-module category, decomposed as
\begin{equation}
    M=\bigoplus_x \widetilde n_{Mx}M_x\,.
\end{equation}
In the fermionic case
this Hom space is a super vector space, and for simple $M_x$ we have
\begin{equation}
    \langle M_x,M_x\rangle_{\mathcal{A}}=1
\end{equation}
for an m-type simple module, while
\begin{equation}
    \langle M_x,M_x\rangle_{\mathcal{A}}=1+\pi
\end{equation}
for a q-type simple module. 

The reciprocity theorem \eqref{eq:reciprocity_theorem}, together with the
induced module, is very useful for determining the resulting
$\mathcal{A}$-module category. Decompose the induced module as 
\begin{equation}\label{eq:module-induce}
    \operatorname{Ind}(W_i)
    =
    \bigoplus_x \widetilde{n}_{ix} M_x\, .
\end{equation}
Applying Frobenius reciprocity to a simple module $M_x$ gives the relation between $\widetilde{n}_{iM_x}$ and $n_{ix}$ in \eqref{eq:module-forget}:
\begin{equation}
    n_{iM_x}
    =
    \widetilde{n}_{ix}\,
    \langle M_x,M_x\rangle_{\mathcal{A}}\, .
\end{equation}
Setting
\begin{equation}
    M=\operatorname{Ind}(W_j)=\mathcal{A}\otimes W_j
\end{equation}
in \eqref{eq:reciprocity_theorem}, we obtain the useful relation
\begin{equation}\label{eq:reciprocity_induced-module}
    \langle W_i,\mathcal{A}\otimes W_j\rangle_{\mathscr{C}}
    =
    \langle \operatorname{Ind}(W_i),\operatorname{Ind}(W_j)\rangle_{\mathcal{A}}
    =
    \sum_x
    \widetilde{n}_{ix}\widetilde{n}_{jx}
    \langle M_x,M_x\rangle_{\mathcal{A}}\,.
\end{equation}
Moreover, if $\mathcal{A}$ is a commutative algebra, the left module can be lifted to a bimodule, and we can define the tensor product in the module category. The $\text{Ind}$ operation then becomes a tensor functor satisfying (see Proposition 5.11 in \cite{Fuchs:2001qc}) 
    \begin{equation}
        \text{Ind}(W_i\otimes W_j) = \text{Ind}(W_i) \otimes_{\mathcal{A}} \text{Ind}(W_j)\,,
    \end{equation}
where $\otimes_{\mathcal{A}}$ is the tensor product in the module category. We then further have
    \begin{equation}\label{eq:reciprocity_induced-module_fusion}
        \langle W_i,\mathcal{A}\otimes W_j\otimes W_k\rangle_{\mathscr{C}}
    =
    \langle \operatorname{Ind}(W_i),\operatorname{Ind}(W_j)\otimes_{\mathcal{A}}\operatorname{Ind}(W_k)  \rangle_{\mathcal{A}}\,.
    \end{equation}
These are the main practical tools for determining the simple
$\mathcal A$-modules from the original fusion category $\mathscr{C}$, as we shall see in the following examples.

\paragraph{$\mathcal{N}=1$ $A$-type minimal model}

Let us begin with the bosonized $N=1$ superconformal minimal model
\begin{equation}
    \mathfrak{B}[\mathcal {SM}_{k+2,k+4}]
    =
    \frac{SU(2)_{k}\times SU(2)_2}{SU(2)_{k+2}}\,.
\end{equation}
where the primaries are labelled by the quartet $(a,b,c,d)$. As discussed in Section~\ref{sec:N_1_minimal_model}, the TDLs
that preserve this deformation are
\begin{equation}
    \mathscr{C}_{\mathfrak{B}}
    =
    \{\cL_{(a,0,0,0)},\cL_{(a,2,0,0)}\mid a\ \mathrm{even}\}
    \cup
    \{\cL_{(a,1,0,0)}\mid a\ \mathrm{odd}\}\,,
\end{equation}
with $a=0,\ldots,k$. The parity constraint simply comes from the coset
selection rule with $c=0$.

The $\mathbb{Z}_2$ anyon used for fermionization is
\begin{equation}
    f=\cL_{(0,2,0,0)}\,,
\end{equation}
which has topological spin -1. It acts on the TDLs in $\mathscr{C}_{\mathfrak{B}}$ as
\begin{equation}
    f\cdot \cL_{(a,b,0,0)}
    =
    \cL_{(a,2-b,0,0)}.
\end{equation}
Thus, for even $a$, it exchanges
\begin{equation}
    \cL_{(a,0,0,0)}
    \longleftrightarrow
    \cL_{(a,2,0,0)},
\end{equation}
whereas for odd $a$ it fixes
\begin{equation}
    \cL_{(a,1,0,0)}.
\end{equation}

Let us condense the supercommutative Frobenius algebra
\begin{equation}
    \mathcal{A}_f=\mathbf 1\oplus \pi f\,,
\end{equation}
and focus on the induced left modules
    \begin{equation}
    \begin{split}
        \text{Ind}(\mathcal{L}_{(a,0,0,0)}) &= \mathcal{A}_f\otimes \mathcal{L}_{(a,0,0,0)} = \mathcal{L}_{(a,0,0,0)}\oplus \pi \mathcal{L}_{(a,2,0,0)} \,,\\
        \text{Ind}(\mathcal{L}_{(a,1,0,0)}) &= \mathcal{A}_f\otimes \mathcal{L}_{(a,1,0,0)} = (1 \oplus \pi) \mathcal{L}_{(a,1,0,0)}\,,
    \end{split}
    \end{equation}
for even and odd $a$, respectively. By Frobenius reciprocity \eqref{eq:reciprocity_theorem}, one finds
    \begin{equation}
        \langle \text{Ind}(\mathcal{L}_{(a,0,0,0)})\,,\text{Ind}(\mathcal{L}_{(a,0,0,0)})\rangle_{\mathcal{A}_f} = 1\,, \quad \langle \text{Ind}(\mathcal{L}_{(a,1,0,0)})\,, \text{Ind}(\mathcal{L}_{(a,1,0,0)})\rangle_{\mathcal{A}_f} = 1+\pi\,,
    \end{equation}
which implies that $\text{Ind}(\mathcal{L}_{(a,0,0,0)})$ are m-type simple modules and $\text{Ind}(\mathcal{L}_{(a,1,0,0)})$ are q-type simple modules. We label these simple modules by $\mathcal{L}_a$:
    \begin{equation}
        \mathcal{L}_a \equiv \left\{\begin{array}{l}
            \text{Ind}(\mathcal{L}_{(a,0,0,0)})\,, \quad \text{for $a$ even}\,,\\
            \text{Ind}(\mathcal{L}_{(a,1,0,0)})\,, \quad \text{for $a$ odd}\,,
        \end{array} \right.
    \end{equation}
and we denote the resulting module category by
    \begin{equation}
        \mathcal{M}_{A_k} = \left\{ \mathcal{L}_a|a=0,\dots,k\right\}\,.
    \end{equation}

Since $\mathcal{A}_f$ is a commutative algebra, we can also work out the fusion $\otimes_{\mathcal{A}_f}$ from \eqref{eq:reciprocity_induced-module_fusion}
    \begin{equation}
        \langle \mathcal{L}_c,\mathcal{L}_a\otimes_{\mathcal{A}_f} \mathcal{L}_b\rangle_{\mathcal{A}_f} = \langle \mathcal{L}_{(c,[c],0,0)} , \mathcal{A}_f\otimes \mathcal{L}_{(a,[a],0,0)}\otimes \mathcal{L}_{(b,[b],0,0)}\rangle_{\mathscr{C}}\,.
    \end{equation}
where $[a]=a \ \text{mod}\ 2$. Let us introduce the $su(2)_k$ fusion range $F_{ab}^{(k)}$:
\begin{equation}
    F^{(k)}_{ab}
    =
    \left\{
    c:
    |a-b|\le c\le \min(a+b,2k-a-b),
    \quad
    a+b+c\in 2\mathbb Z
    \right\}.
\end{equation}
When both $a,b$ are even, $c$ is also even and we simply have
    \begin{equation}
        \mathcal{L}_a \otimes_{\mathcal{A}_f} \mathcal{L}_b = \bigoplus_{c \in F^{(k)}_{ab}} \mathcal{L}_c\,.\quad (a,b \ \text{are even})
    \end{equation}
On the other hand, when both $a,b$ are odd, $c$ is even and one has the fusion rule
    \begin{equation}
        \mathcal{L}_{(a,1,0,0)} \otimes \mathcal{L}_{(b,1,0,0)} = \bigoplus_{c\in F_{ab}^{(p)}} \left(\mathcal{L}_{(c,0,0,0)}\oplus \mathcal{L}_{(c,2,0,0)}\right)\,.
    \end{equation}
Then $\mathcal{A}_f$ acts as
    \begin{equation}
        \mathcal{A}_f \otimes \left(\mathcal{L}_{(c,0,0,0)}\oplus \mathcal{L}_{(c,2,0,0)}\right) = (1\oplus \pi) \left(\mathcal{L}_{(c,0,0,0)}\oplus \mathcal{L}_{(c,2,0,0)}\right)\,,
    \end{equation}
and for any $c\in F_{ab}^{(k)}$ one has
    \begin{equation}
        \langle \mathcal{L}_c , \mathcal{L}_a \otimes_{\mathcal{A}_f} \mathcal{L}_b\rangle =  1+\pi\,.
    \end{equation}
Since $\mathcal{L}_c$ is m-type, one should have
    \begin{equation}
        \mathcal{L}_a \otimes_{\mathcal{A}_f} \mathcal{L}_b = \bigoplus_{c \in F^{(k)}_{ab}} \mathbb{C}^{1|1} \mathcal{L}_c\,.\quad (a,b \ \text{are odd})
    \end{equation}
Lastly, if one of $a,b$ is even and the other is odd, then $c$ is also odd so that $\mathcal{A}_f \otimes \mathcal{L}_{(c,1,0,0)} = (1\oplus \pi)\otimes \mathcal{L}_{(c,1,0,0)}$. Then for any $c\in F_{ab}^{(p)}$ one also has
    \begin{equation}
        \langle \mathcal{L}_c , \mathcal{L}_a \otimes_{\mathcal{A}_f} \mathcal{L}_b\rangle =  1+\pi\,.
    \end{equation}
This is consistent with $\mathcal{L}_c$ being a q-type object, and we have the fusion product
    \begin{equation}
        \mathcal{L}_a \otimes_{\mathcal{A}} \mathcal{L}_b = \bigoplus_{c \in F^{(k)}_{ab}} \mathcal{L}_c\,.\quad (a+b \ \text{is odd})
    \end{equation}
In summary, we have the fusion algebra
    \begin{equation}\label{eq:fusion_rule_N1}
        \mathcal{L}_a \otimes_{\mathcal{A}} \mathcal{L}_b = \left\{\begin{array}{l}
            \bigoplus_{c \in F^{(k)}_{ab}} \mathcal{L}_c \quad (ab\ \text{is even})\\
            \bigoplus_{c \in F^{(k)}_{ab}} \mathbb{C}^{1|1} \mathcal{L}_c \quad (ab\ \text{is odd})
        \end{array} \right.
    \end{equation}

The above procedure can be depicted as follows:
    \begin{equation}
        \begin{gathered}
     \begin{tikzpicture}[scale=1]
        \draw[ultra thick,MidnightBlue,-<-=.55] (-2.5,2)--(-2.5,0);
        \node at (-1.5,1) {$\mathscr{C}_{\mathfrak{B}}$};
        \node at (-3.5,1) {$\mathscr{C}_{A_k}$};
        \node at (-2.5,-0.5) {$\mathcal{M}_{A_k}$};
    \end{tikzpicture} 
\end{gathered}
\end{equation}
where we fermionize the theory on the left half-space by condensing $\mathcal{A}_f$. The interface is described by the left module $\mathcal{M}_{A_k}$, and the dual symmetry is therefore $\mathscr{C}_{A_k}=(-1)^F \otimes \widetilde{\mathscr{C}}_{A_k}$ with $\widetilde{\mathscr{C}}_{A_k}=\mathcal{M}_{A_k}$. Furthermore, the left module $\mathcal{M}_{A_k}$ is naturally a $(\mathscr{C}_{A_k}, \mathscr{C}_{\mathfrak{B}})$ bimodule since it is the interface separating the two regions, so that $\mathscr{C}_{\mathfrak{B}}$ is understood as the dual category of $\mathscr{C}_{A_k}$ with respect to the module $\mathcal{M}_{A_k}$.

Recall that in the 2D particle/soliton prescription, the module category $\mathcal{M}_{A_k}$ characterizes the vacuum structure of the theory, and the simple objects in the dual category $\mathscr{C}_{\mathfrak{B}}$ are in one-to-one correspondence with the irreducible representations of the particle/soliton multiplet. We will mainly focus on the representation associated to $\mathcal{L}_{(1,1,0,0)}\in \mathscr{C}_{\mathfrak{B}}$, depicted in the following diagram:
    \begin{equation}
        \begin{gathered}
 
\end{gathered}
\end{equation}
The junction vector $\mathcal{O}_{ab}$ can be understood as follows. The representation line $\mathcal{L}_{(1,1,0,0)}$ becomes the induced line $\text{Ind}(\mathcal{L}_{(1,1,0,0)}) = \mathcal{L}_1$ after crossing the interface. Then we can bend the $\mathcal{L}_1$ line to merge with the upper $\mathcal{L}_a$, as shown in the figure:
\begin{equation}
        \begin{gathered}
            %
        \end{gathered}
\end{equation}
Therefore the junction vector $\mathcal{O}_{ab}$ takes value in 
    \begin{equation}
        \mathcal{O}_{ab} \in \text{Hom}_{\mathcal{A}_f} (\mathcal{L}_b ,\mathcal{L}_1 \otimes_{\mathcal{A}_f} \mathcal{L}_a)\,.
    \end{equation}
From the fusion rule \eqref{eq:fusion_rule_N1} one has
    \begin{equation}
    \begin{gathered}
        \mathcal{O}_{a,a\pm 1} \in \mathbb{C}^{1|1}\,,\quad (a\neq 0,k)\\
        \mathcal{O}_{0,1},\mathcal{O}_{k,k-1} \in \mathbb{C}^{1|1}\,.
    \end{gathered}
    \end{equation}
This is summarized in the following quiver diagram, where blue and red arrows stand for the vector spaces $\mathbb{C}^{1|0}$ and $\mathbb{C}^{0|1}$, respectively.

\begin{equation*}
\begin{gathered}

\end{gathered}
\end{equation*}

\paragraph{$\mathcal{N}=1$ $D$-type minimal model}

We then consider condensing the Frobenius algebra $\mathcal{A}_{D_k}$
    \begin{equation}
        \mathcal{A}_{D_k}= \mathcal{L}_0 \oplus \mathcal{L}_{2k}\,,
    \end{equation}
in $\widetilde{\mathscr{C}}_{A_{2k}}$ to obtain the $\mathcal{N}=1$ $D$-type minimal model. The line $\mathcal{L}_{2k}$ is an m-type object and generates a $\mathbb{Z}_2$ symmetry in $\widetilde{\mathscr{C}}_{A_{2k}}$
    \begin{equation}
        \mathcal{L}_{2k} \cdot \mathcal{L}_{a} = \mathcal{L}_{2k-a}\,,
    \end{equation}
There is a single fixed point, $\mathcal{L}_k$, which is m-type for odd $k$ and q-type for even $k$. 

Let us consider the induced left modules $\text{Ind}(\mathcal{L}_a) = \mathcal{A}_{D_k} \otimes_{\mathcal{A}_f} \mathcal{L}_a$. Using Frobenius reciprocity,
    \begin{equation}
        \langle \text{Ind}(\mathcal{L}_a) , \text{Ind}(\mathcal{L}_b)\rangle_{\mathcal{A}_{D_k}} = \langle \mathcal{L}_a , \mathcal{A}_{D_k} \otimes_{\mathcal{A}_f} \mathcal{L}_b\rangle_{\mathcal{A}_{f}}\,,
    \end{equation}
we find
    \begin{equation}
    \begin{gathered}
        \langle \text{Ind}(\mathcal{L}_a) , \text{Ind}(\mathcal{L}_a)\rangle_{\mathcal{A}_{D_k}}=\langle \mathcal{L}_a,\mathcal{L}_a\rangle_{\mathcal{A}_{f}}\,, \quad (a=0,1,\dots,k-1)\\
        \langle \text{Ind}(\mathcal{L}_k) , \text{Ind}(\mathcal{L}_k)\rangle_{\mathcal{A}_{D_k}}=2\langle \mathcal{L}_k,\mathcal{L}_k\rangle_{\mathcal{A}_{f}}\,.
    \end{gathered}
    \end{equation}
Therefore, for $a=0,\dots,k-1$, the induced left modules are simple modules, and they are of $m/q$-type for even/odd $a$, respectively. We define
    \begin{equation}
        \mathcal{D}_{a} \equiv \text{Ind}(\mathcal{L}_a)\,.\quad (a=0,\dots,k-1)
    \end{equation}
For $a=k$, based on the discussion of the Witten index, we assume that
    \begin{itemize}
        \item For even $k$, $\text{Ind}(\mathcal{L}_k)$ splits into two m-type lines $\mathcal{D}_k^{\pm}$ according to $\text{Ind}(\mathcal{L}_k)=\mathcal{D}_k^{+}+\mathcal{D}_k^{-}$.
        \item For odd $k$, $\text{Ind}(\mathcal{L}_k)$ does not split and becomes an m-type line $\mathcal{D}_k$ according to $\text{Ind}(\mathcal{L}_k) = (1\oplus \pi) \mathcal{D}_k$.
    \end{itemize}

To find the particle/soliton multiplet in the $D$-type model, notice that in the $A$-type model we have
    \begin{equation}
        \mathcal{O}_{ab} \in \text{Hom}_{\mathcal{A}_f} (\mathcal{L}_b ,\mathcal{L}_1 \otimes_{\mathcal{A}_f} \mathcal{L}_a)\,,
    \end{equation}
after condensing $\mathcal{A}_{D_k}$ one has
    \begin{equation}
        \begin{gathered}
     \begin{tikzpicture}[scale=1]
        \draw[ultra thick,MidnightBlue,-<-=.55] (-2.5,2)--(-2.5,1);
        \draw[ultra thick,MidnightBlue,-<-=.55] (-2.5,1)--(-2.5,0);
        \draw[densely dashed,ultra thick,ForestGreen,->-=.6] (-1,1)--(-2.5,1);
        \node at (-3,1) {$\mathcal{O}_{ab}$};
         \filldraw[black] (-2.5,1) circle (2pt);
        \node at (-2.9,0.3) {$\mathcal{L}_a$};
        \node at (-2.9,1.7) {$\mathcal{L}_b$};
        \node at (-0.25,1) {$\mathcal{L}_{(1,1,0,0)}$};
    \end{tikzpicture} 
\end{gathered}\qquad \rightarrow \quad
        \begin{gathered}
     \begin{tikzpicture}[scale=1]
        \draw[ultra thick,MidnightBlue,-<-=.55] (-2.5,2)--(-2.5,1);
        \draw[ultra thick,MidnightBlue,-<-=.55] (-2.5,1)--(-2.5,0);
        \draw[densely dashed,ultra thick,ForestGreen,->-=.6] (-1,1)--(-2.5,1);
        \node at (-3,1) {$\widetilde{\mathcal{O}}_{ab}$};
         \filldraw[black] (-2.5,1) circle (2pt);
        \node at (-3.4,0.3) {$\text{Ind}(\mathcal{L}_a)$};
        \node at (-3.4,1.7) {$\text{Ind}(\mathcal{L}_b)$};
        \node at (-0.25,1) {$\mathcal{L}_{(1,1,0,0)}$};
    \end{tikzpicture} 
\end{gathered}
\end{equation}
where one should replace all lines in $\widetilde{\mathscr{C}}_{A_{2k}}$ with the corresponding induced modules in $\mathcal{M}_{\mathcal{A}_{D_{k}}}$. Thus the junction vector is mapped to $\widetilde{\mathcal{O}}_{ab}$, which takes values in
    \begin{equation}
        \widetilde{\mathcal{O}}_{ab} \in \text{Hom}_{\mathcal{A}_{D_k}}(\text{Ind}(\mathcal{L}_b) , \text{Ind}(\mathcal{L}_{1}\otimes_{\mathcal{A}_f}\mathcal{L}_a )) = \text{Hom}_{\mathcal{A}_{f}}(\mathcal{L}_b , \mathcal{A}_{D_k}\otimes_{\mathcal{A}_f} \mathcal{L}_{1}\otimes_{\mathcal{A}_f}\mathcal{L}_a)\,.
    \end{equation}
Then we decompose the induced module into the simple modules $\mathcal{D}_a$ and work out the corresponding junction vector space.

When $a,b=0,\dots,k-1$, it is easy to see that $\widetilde{\mathcal{O}}_{ab} \in \mathbb{C}^{1|1}$ when $|a-b|=1$. We then focus on the case where one of $a,b$ is $k$; for example, choose $b=k-1$ and $a=k$. We have
    \begin{equation}
        \widetilde{\mathcal{O}}_{k-1,k} \in \mathbb{C}^{2|2} = 2(1\oplus \pi)\,,
    \end{equation}
regardless of $k$. When $k$ is even, $\text{Ind}(\mathcal{L}_k)$ splits into two simple modules according to $\text{Ind}(\mathcal{L}_k)=\mathcal{D}_k^{+}+\mathcal{D}_k^{-}$, so that the junction vector connecting $\mathcal{D}_{k-1}$ to either $\mathcal{D}_k^{+}$ or $\mathcal{D}_k^{-}$ is $\mathbb{C}^{1|1}$. On the other hand, when $k$ is odd, $\text{Ind}(\mathcal{L}_k)= \mathbb{C}^{1|1} \mathcal{D}_k$ with $\mathcal{D}_k$ m-type. After modding out this coefficient, the junction vector connecting $\mathcal{D}_{k-1}$ to $\mathcal{D}_k$ is also $\mathbb{C}^{1|1}$.


\paragraph{$\mathcal{N}=1$ $E$-type minimal model}

We then consider the $E$-type Frobenius algebra in $\mathscr{C}_{\mathcal{A}_f}$ given by
    \begin{equation}
        \begin{gathered}
            \mathcal{A}_{E_6} = \mathcal{L}_0 \oplus \mathcal{L}_6\,,\\
            \mathcal{A}_{E_7} = \mathcal{L}_0 \oplus \mathcal{L}_8 \oplus \mathcal{L}_{16}\,,\\
            \mathcal{A}_{E_8}=\mathcal{L}_0 \oplus \mathcal{L}_{10} \oplus \mathcal{L}_{18} \oplus \mathcal{L}_{28}\,,
        \end{gathered}
    \end{equation}
for $k=10,16,28$, respectively. We work out the $E_6$ case explicitly as an illustration, and then summarize the results for $E_7$ and $E_8$.

Let us begin with $k=10$ and condense $\mathcal{A}_{E_6}$. Using Frobenius reciprocity, one can work out the homomorphism matrix for induced modules as
\begin{equation}
    \langle\text{Ind}(\mathcal{L}_a),\text{Ind}(\mathcal{L}_b) \rangle_{\mathcal{A}_{E_6}} =\left(
\begin{array}{ccccccccccc}
 1 & 0 & 0 & 0 & 0 & 0 & 1 & 0 & 0 & 0 & 0 \\
 0 & 1+\pi & 0 & 0 & 0 & 1+\pi & 0 & 1+\pi & 0 & 0 & 0 \\
 0 & 0 & 1 & 0 & 1 & 0 & 1 & 0 & 1 & 0 & 0 \\
 0 & 0 & 0 & 2(1+\pi) & 0 & 1+\pi & 0 & 1+\pi & 0 & 1+\pi & 0 \\
 0 & 0 & 1 & 0 & 2 & 0 & 1 & 0 & 1 & 0 & 1 \\
 0 & 1+\pi & 0 & 1+\pi & 0 & 2(1+\pi) & 0 & 1+\pi & 0 & 1+\pi & 0 \\
 1 & 0 & 1 & 0 & 1 & 0 & 2 & 0 & 1 & 0 & 0 \\
 0 & 1+\pi & 0 & 1+\pi & 0 & 1+\pi & 0 & 2(1+\pi) & 0 & 0 & 0 \\
 0 & 0 & 1 & 0 & 1 & 0 & 1 & 0 & 1 & 0 & 0 \\
 0 & 0 & 0 & 1+\pi & 0 & 1+\pi & 0 & 0 & 0 & 1+\pi & 0 \\
 0 & 0 & 0 & 0 & 1 & 0 & 0 & 0 & 0 & 0 & 1 \\
\end{array}
\right)
\end{equation}
One finds that $\text{Ind}(\mathcal{L}_0),\text{Ind}(\mathcal{L}_1),\text{Ind}(\mathcal{L}_2),\text{Ind}(\mathcal{L}_8),\text{Ind}(\mathcal{L}_9),\text{Ind}(\mathcal{L}_{10})$ are simple left modules. Moreover, since $\langle \text{Ind}(\mathcal{L}_2),\text{Ind}(\mathcal{L}_8) \rangle=1$, they give the same simple left module. Therefore we identify
    \begin{equation}\label{eq:E6_ind}
          \text{Ind}(\mathcal{L}_0)=\mathcal{E}_0\,,\quad \text{Ind}(\mathcal{L}_1)=\mathcal{E}_1\,,\quad \text{Ind}(\mathcal{L}_2)=\text{Ind}(\mathcal{L}_8)=\mathcal{E}_2\,,\quad \text{Ind}(\mathcal{L}_9)=\mathcal{E}_3\,,\quad  \text{Ind}(\mathcal{L}_{10})=\mathcal{E}_4\,,
    \end{equation}
with $\mathcal{E}_0,\mathcal{E}_2,\mathcal{E}_4$ of m-type and $\mathcal{E}_1,\mathcal{E}_3$ of q-type. The induced module $\text{Ind}(\mathcal{L}_3)$ is not simple, but in the matrix above it only has a non-trivial homomorphism with $\text{Ind}(\mathcal{L}_9)=\mathcal{E}_3$ among the simple modules in \eqref{eq:E6_ind}. Therefore it should also contain another q-type simple module which is not listed above. We denote it by $\mathcal{E}_5$ and write
    \begin{equation}
        \text{Ind}(\mathcal{L}_3) = \mathcal{E}_3\oplus \mathcal{E}_5\,,
    \end{equation}
and we can effectively represent $\mathcal{E}_5 = \text{Ind}(\mathcal{L}_3) -\text{Ind}(\mathcal{L}_9)$. Similarly, we can deduce the decompositions for the other induced modules:
    \begin{equation}
        \text{Ind}(\mathcal{L}_4) = \mathcal{E}_2\oplus \mathcal{E}_4\,,\quad \text{Ind}(\mathcal{L}_5) = \mathcal{E}_1\oplus \mathcal{E}_3\,,\quad \text{Ind}(\mathcal{L}_6) = \mathcal{E}_0\oplus \mathcal{E}_2\,,\quad \text{Ind}(\mathcal{L}_7) = \mathcal{E}_1\oplus \mathcal{E}_5\,.
    \end{equation}

To determine the particle/soliton structure, we then analyze
    \begin{equation}
        \widetilde{\mathcal{O}}_{ab} \in \text{Hom}_{\mathcal{A}_{E_6}}(\text{Ind}(\mathcal{L}_b) , \text{Ind}(\mathcal{L}_{1}\otimes_{\mathcal{A}_f}\mathcal{L}_a )) = \text{Hom}_{\mathcal{A}_{f}}(\mathcal{L}_b , \mathcal{A}_{E_6}\otimes_{\mathcal{A}_f} \mathcal{L}_{1}\otimes_{\mathcal{A}_f}\mathcal{L}_a)\,.
    \end{equation}
In terms of the basis $\mathcal{E}_{a}$, one obtains
    \begin{equation}
        \langle \widetilde{\mathcal{O}}(\mathcal{E}_a \rightarrow \mathcal{E}_b)\rangle = \left(
\begin{array}{cccccc}
 0 & 1+\pi & 0 & 0 & 0 & 0 \\
 1+\pi & 0 & 1+\pi & 0 & 0 & 0 \\
 0 & 1+\pi & 0 & 1+\pi & 0 & 1+\pi \\
 0 & 0 & 1+\pi & 0 & 1+\pi & 0 \\
 0 & 0 & 0 & 1+\pi & 0 & 0 \\
 0 & 0 & 1+\pi & 0 & 0 & 0 \\
\end{array}
\right)\,.
    \end{equation}

We then briefly summarize the results for $E_7$ and $E_8$. For $E_7$, we identify the simple modules as
    \begin{equation}
    \begin{gathered}
        \mathcal{E}_0=\text{Ind}(\mathcal{L}_0)\,, \quad\mathcal{E}_1=\text{Ind}(\mathcal{L}_1)\,,\quad \mathcal{E}_2=\text{Ind}(\mathcal{L}_2)\,,\quad \mathcal{E}_3=\text{Ind}(\mathcal{L}_3)\\
        \mathcal{E}_4=\text{Ind}(\mathcal{L}_6)-\text{Ind}(\mathcal{L}_2)\,,\quad \mathcal{E}_5=\text{Ind}(\mathcal{L}_5)-\text{Ind}(\mathcal{L}_3)\,,\quad \mathcal{E}_6=\text{Ind}(\mathcal{L}_2)+\text{Ind}(\mathcal{L}_4)-\text{Ind}(\mathcal{L}_6)\,,
    \end{gathered}
    \end{equation}
where $\mathcal{E}_0,\mathcal{E}_2,\mathcal{E}_4,\mathcal{E}_6$ are m-type, and $\mathcal{E}_1,\mathcal{E}_3,\mathcal{E}_5$ are q-type. The particle/soliton structure is encoded in the matrix
    \begin{equation}
        \langle \widetilde{\mathcal{O}}(\mathcal{E}_a \rightarrow \mathcal{E}_b)\rangle = \left(
\begin{array}{ccccccc}
 0 & 1+\pi & 0 & 0 & 0 & 0 & 0 \\
 1+\pi & 0 & 1+\pi & 0 & 0 & 0 & 0 \\
 0 & 1+\pi & 0 & 1+\pi & 0 & 0 & 0 \\
 0 & 0 & 1+\pi & 0 & 1+\pi & 0 & 1+\pi \\
 0 & 0 & 0 & 1+\pi & 0 & 1+\pi & 0 \\
 0 & 0 & 0 & 0 & 1+\pi & 0 & 0 \\
 0 & 0 & 0 & 1+\pi & 0 & 0 & 0 \\
\end{array}
\right)\,.
    \end{equation}
And for $E_8$, we have
    \begin{equation}
    \begin{gathered}
        \mathcal{E}_0=\text{Ind}(\mathcal{L}_0)\,, \quad\mathcal{E}_1=\text{Ind}(\mathcal{L}_1)\,,\quad \mathcal{E}_2=\text{Ind}(\mathcal{L}_2)\,,\quad \mathcal{E}_3=\text{Ind}(\mathcal{L}_3)\,,\quad \mathcal{E}_4=\text{Ind}(\mathcal{L}_4)\,,\\
        \mathcal{E}_5=\text{Ind}(\mathcal{L}_7)-\text{Ind}(\mathcal{L}_3)\,,\quad \mathcal{E}_6=\text{Ind}(\mathcal{L}_6)-\text{Ind}(\mathcal{L}_4)\,,\quad \mathcal{E}_7=\text{Ind}(\mathcal{L}_5)+\text{Ind}(\mathcal{L}_3)-\text{Ind}(\mathcal{L}_7)\,,
    \end{gathered}
    \end{equation}
where $\mathcal{E}_0,\mathcal{E}_2,\mathcal{E}_4,\mathcal{E}_6$ are m-type, and $\mathcal{E}_1,\mathcal{E}_3,\mathcal{E}_5,\mathcal{E}_7$ are q-type. The particle/soliton structure is encoded in the matrix
\begin{equation}
    \langle \widetilde{\mathcal{O}}(\mathcal{E}_a \rightarrow \mathcal{E}_b)\rangle = \left(
\begin{array}{cccccccc}
 0 & 1+\pi & 0 & 0 & 0 & 0 & 0 & 0 \\
 1+\pi & 0 & 1+\pi & 0 & 0 & 0 & 0 & 0 \\
 0 & 1+\pi & 0 & 1+\pi & 0 & 0 & 0 & 0 \\
 0 & 0 & 1+\pi & 0 & 1+\pi & 0 & 0 & 0 \\
 0 & 0 & 0 & 1+\pi & 0 & 1+\pi & 0 & 1+\pi \\
 0 & 0 & 0 & 0 & 1+\pi & 0 & 1+\pi & 0 \\
 0 & 0 & 0 & 0 & 0 & 1+\pi & 0 & 0 \\
 0 & 0 & 0 & 0 & 1+\pi & 0 & 0 & 0 \\
\end{array}
\right)\,.
\end{equation}
The quiver diagrams for $E$-type theories can be drawn according to the matrices. They can be found in Figure~\ref{fig:E8-type_N=1}. 


\paragraph{$\mathcal{N}=2$ $A$-type minimal model}

We then move to the bosonized $\mathcal N=2$ $A$-type minimal model
\begin{equation}
    \mathfrak{B}[\mathcal {SM}_{k+2,k+4}]
    =
    \frac{SU(2)_k\times U(1)_2}{U(1)_{k+2}}\,,
\end{equation}
where the primaries are labeled by the pair $(a,c)$. The TDLs preserving the relevant deformation are
\begin{equation}
    \mathscr{C}^{\mathcal N=2}_{\mathfrak{B}}
    =
    \{\cL_{(a,0)},\cL_{(a,k+2)}
    \mid a=0,\ldots,k\}.
\end{equation}

The anyon used for fermion condensation is
\begin{equation}
    f=\cL_{(k,k+2)}\,,
\end{equation}
which has topological spin $-1$. It acts on the TDLs in $\mathscr{C}_{\mathfrak{B}}$ as
\begin{equation}
    f\cdot \cL_{(a,c)}
    =
    \cL_{(k-a,c+k+2)}\,.
\end{equation}
Therefore it exchanges
\begin{equation}
    \cL_{(a,0)}
    \longleftrightarrow
    \cL_{(k-a,k+2)}\,,
\end{equation}
and there is no fixed point.

Let us condense the supercommutative Frobenius algebra
    \begin{equation}
        \mathcal{A}_f = \mathbf{1} \oplus \pi f\,,
    \end{equation}
and focus on the induced left modules
    \begin{equation}
        \begin{split}
            \text{Ind}(\mathcal{L}_{(a,0)}) = \mathcal{A}_f \otimes \mathcal{L}_{(a,0)} = \mathcal{L}_{(a,0)} \oplus \pi \mathcal{L}_{(k-a,k+2)}\,.
        \end{split}
    \end{equation}
By Frobenius reciprocity \eqref{eq:reciprocity_induced-module}, one finds
    \begin{equation}
        \langle \text{Ind}(\mathcal{L}_{(a,0)}),\text{Ind}(\mathcal{L}_{(a,0)})\rangle_{\mathcal{A}_f}=1\,,
    \end{equation}
which implies that $\text{Ind}(\mathcal{L}_{(a,0)})$ are all m-type simple modules. We label them by
    \begin{equation}
        \mathcal{L}_a \equiv \text{Ind}(\mathcal{L}_{(a,0)})\,.
    \end{equation}

Since $\mathcal{A}_f$ is a commutative algebra, we can also work out the fusion $\otimes_{\mathcal{A}_f}$ from \eqref{eq:reciprocity_induced-module_fusion}:
    \begin{equation}
        \langle \mathcal{L}_c,\mathcal{L}_a\otimes_{\mathcal{A}_f} \mathcal{L}_b\rangle_{\mathcal{A}_f} = \langle \mathcal{L}_{(c,0)} , \mathcal{A}_f\otimes \mathcal{L}_{(a,0)}\otimes \mathcal{L}_{(b,0)}\rangle_{\mathscr{C}}\,.
    \end{equation}
When $ab$ is even, we simply have
    \begin{equation}
        \mathcal{L}_a \otimes_{\mathcal{A}_f} \mathcal{L}_b = \bigoplus_{c\in F_{ab}^{(k)}} \mathcal{L}_c\,. \quad (ab \text{ is even})
    \end{equation}
On the other hand, when $ab$ is odd, we have
    \begin{equation}
        \mathcal{L}_{(a,0)} \otimes \mathcal{L}_{(b,0)} = \bigoplus_{c\in F_{ab}^{(k)}} \mathcal{L}_{(k-c,k+2)}\,, \quad (ab \text{ is odd})
    \end{equation}
and also
    \begin{equation}
        \mathcal{A}_f \otimes \mathcal{L}_{(k-c,k+2)} = \mathcal{L}_{(k-c,k+2)}\oplus \pi \mathcal{L}_{(c,0)}\,.
    \end{equation}
Therefore, for any $c\in F_{ab}^{(k)}$, we have
    \begin{equation}
        \langle \mathcal{L}_c, \mathcal{L}_a \otimes_{\mathcal{A}_f} \mathcal{L}_b\rangle_{\mathcal{A}_f} = \pi\,, \quad (ab \text{ is odd})
    \end{equation}
and thus
    \begin{equation}
        \mathcal{L}_a \otimes_{\mathcal{A}_f} \mathcal{L}_b = \bigoplus_{c\in F_{ab}^{(k)}} \mathbb{C}^{0|1}\mathcal{L}_c\,. \quad (ab \text{ is odd})
    \end{equation}
In summary, we have the fusion algebra
    \begin{equation}\label{eq:fusion_rule_N2}
        \mathcal{L}_a \otimes_{\mathcal{A}} \mathcal{L}_b = \left\{\begin{array}{l}
            \bigoplus_{c \in F^{(p)}_{ab}} \mathbb{C}^{1|0}\mathcal{L}_c \quad (ab\ \text{is even})\\
            \bigoplus_{c \in F^{(p)}_{ab}} \mathbb{C}^{0|1} \mathcal{L}_c \quad (ab\ \text{is odd})
        \end{array} \right.
    \end{equation}

Since $\mathcal{L}_1 = \mathcal{L}_{(1,0)} \oplus \pi \mathcal{L}_{(k-1,k+2)}$, there are two relevant representation lines in $\mathcal{C}_{\mathfrak{B}}$, unlike in the $\mathcal{N}=1$ case. They are depicted in the following two diagrams:
    \begin{equation}
        \begin{gathered}
     \begin{tikzpicture}[scale=1]
        \draw[ultra thick,MidnightBlue,-<-=.55] (-2.5,2)--(-2.5,1);
        \draw[ultra thick,MidnightBlue,-<-=.55] (-2.5,1)--(-2.5,0);
        \draw[densely dashed,ultra thick,ForestGreen,->-=.6] (-1,1)--(-2.5,1);
        \node at (-3,1) {$\mathcal{O}_{ab}$};
         \filldraw[black] (-2.5,1) circle (2pt);
        \node at (-2.9,0.3) {$\mathcal{L}_a$};
        \node at (-2.9,1.7) {$\mathcal{L}_b$};
        \node at (-0.25,1) {$\mathcal{L}_{(1,0)}$};
    \end{tikzpicture} 
\end{gathered}\,, \quad         \begin{gathered}
     \begin{tikzpicture}[scale=1]
        \draw[ultra thick,MidnightBlue,-<-=.55] (-2.5,2)--(-2.5,1);
        \draw[ultra thick,MidnightBlue,-<-=.55] (-2.5,1)--(-2.5,0);
        \draw[densely dashed,ultra thick,ForestGreen,->-=.6] (-1,1)--(-2.5,1);
        \node at (-3,1) {$\mathcal{O}'_{ab}$};
         \filldraw[black] (-2.5,1) circle (2pt);
        \node at (-2.9,0.3) {$\mathcal{L}_a$};
        \node at (-2.9,1.7) {$\mathcal{L}_b$};
        \node at (-0.05,1) {$\mathcal{L}_{(k-1,k+2)}$};
    \end{tikzpicture}
\end{gathered}\,.
\end{equation}
For the first one, since $\text{Ind}(\mathcal{L}_{(1,0)})=\mathcal{L}_1$, the junction vector $\mathcal{O}_{ab}$ takes value in 
    \begin{equation}
        \mathcal{O}_{ab} \in \text{Hom}_{\mathcal{A}_f} (\mathcal{L}_b ,\mathcal{L}_1 \otimes_{\mathcal{A}_f} \mathcal{L}_a)\,,
    \end{equation}
and from fusion rule \eqref{eq:fusion_rule_N2}, one has
    \begin{equation}
    \begin{gathered}
        \mathcal{O}_{a,a\pm 1} \in \mathbb{C}^{1|0}\,,\quad (a\text{ even and }a\neq 0,k)\,,\\
        \mathcal{O}_{a,a\pm 1} \in \mathbb{C}^{0|1}\,,\quad (a\text{ odd and }a\neq 0,k)\,,\\
        \mathcal{O}_{0,1} \in \mathbb{C}^{1|0}\,,\quad \mathcal{O}_{k,k-1}\in \left\{
\begin{array}{l}
            \mathbb{C}^{1|0}\,,\quad (k\text{ even})\\
            \mathbb{C}^{0|1}\,,\quad (k\text{ odd})
        \end{array} \right.
    \end{gathered}
    \end{equation}
which is summarized in the following diagram
\begin{align}
    \begin{gathered}
         %
    \end{gathered}\nonumber
\end{align}
On the other hand, $\text{Ind}(\mathcal{L}_{(k-1,k+2)}) = \pi\otimes \mathcal{L}_1$. Therefore the junction vector $\mathcal{O}'_{ab}$ is essentially the same as $\mathcal{O}_{ab}$, with $\mathbb{C}^{0|1}$ and $\mathbb{C}^{1|0}$ swapped:
\begin{align}
    \begin{gathered}
         %
    \end{gathered}\nonumber
\end{align}

\paragraph{$\mathcal{N}=2$ $D$-type minimal model}

We then consider condensing the Frobenius algebra $\mathcal{A}_{D_k}$, and the discussion is parallel to the $\mathcal{N}=1$ case. For $a=0,\dots,k-1$, the induced left modules are simple modules, and they are all of m-type. We define
    \begin{equation}
        \mathcal{D}_{a} \equiv \text{Ind}(\mathcal{L}_a)\,.\quad (a=0,\dots,k-1)
    \end{equation}
For $a=k$, it splits into two m-type lines $\mathcal{D}_k^{\pm}$ according to $\text{Ind}(\mathcal{L}_k)=\mathcal{D}_k^{+}+\mathcal{D}_k^{-}$.

For the junction vector $\widetilde{\mathcal{O}}_{ab}$, 
when $a,b=0,\dots,k-1$, one has $\widetilde{\mathcal{O}}_{ab} \in \mathbb{C}^{1|0}$ when $|a-b|=1$ and $a$ is even, and $\widetilde{\mathcal{O}}_{ab} \in \mathbb{C}^{1|0}$ when $|a-b|=1$ and $a$ is odd. When one of $a,b$ is $k$, we have
    \begin{equation}
        \widetilde{\mathcal{O}}_{k-1,k} \in \mathbb{C}^{2|0}\,,\quad \widetilde{\mathcal{O}}_{k,k-1} \in \mathbb{C}^{0|2}\,,
    \end{equation}
when $k$ is odd, with $\mathbb{C}^{0|2}$ and $\mathbb{C}^{2|0}$ swapped when $k$ is even. Since $\text{Ind}(\mathcal{L}_k)$ splits into two simple modules according to $\text{Ind}(\mathcal{L}_k)=\mathcal{D}_k^{+}+\mathcal{D}_k^{-}$, the junction vector connecting $\mathcal{D}_{k-1}$ to either $\mathcal{D}_k^{+}$ or $\mathcal{D}_k^{-}$ is $\mathbb{C}^{1|0}$ for odd $k$ and $\mathbb{C}^{0|1}$ for even $k$. Similarly, the junction vector connecting either $\mathcal{D}_k^{+}$ or $\mathcal{D}_k^{-}$ to $\mathcal{D}_{k-1}$ is $\mathbb{C}^{0|1}$ for odd $k$ and $\mathbb{C}^{1|0}$ for even $k$. As a result, the quiver diagram is
\begin{equation}

\end{equation}
for the representation associated to $\mathcal{L}_{(1,0)}$. For $\mathcal{L}_{(k-1,k+2)}$ one similarly swaps the red and blue lines as
\begin{equation*}
    %
\end{equation*}

\paragraph{$\mathcal{N}=2$ $E$-type minimal model}

Finally, we move to the $E$-type Frobenius algebra. The analysis is parallel to the previous case, so we only summarize the results here.

For $E_6$, the simple left modules are
    \begin{equation}
    \begin{gathered}
        \mathcal{E}_0=\text{Ind}(\mathcal{L}_0)\,,\quad \mathcal{E}_1=\text{Ind}(\mathcal{L}_1)\,,\quad \mathcal{E}_2=\text{Ind}(\mathcal{L}_2)\,,\quad \mathcal{E}_3=\text{Ind}(\mathcal{L}_9)\,,\\
        \mathcal{E}_4=\text{Ind}(\mathcal{L}_{10})\,,\quad \mathcal{E}_{5}=\text{Ind}(\mathcal{L}_3)-\text{Ind}(\mathcal{L}_9)\,,
    \end{gathered}
    \end{equation}
and all of them are m-type. The particle/soliton structure for $\mathcal{R}_{\mathcal{L}_{(1,0)}}$ is given by the matrix
\begin{equation}
    \langle \widetilde{\mathcal{O}}(\mathcal{E}_a \rightarrow \mathcal{E}_b)\rangle = \left(
\begin{array}{cccccc}
 0 & 1 & 0 & 0 & 0 & 0 \\
 \pi & 0 & \pi & 0 & 0 & 0 \\
 0 & 1 & 0 & 1 & 0 & 1 \\
 0 & 0 & \pi & 0 & \pi & 0 \\
 0 & 0 & 0 & 1 & 0 & 0 \\
 0 & 0 & \pi & 0 & 0 & 0 \\
\end{array}
\right)\,.
\end{equation}
For $E_7$, we identify the simple modules as
    \begin{equation}
    \begin{gathered}
        \mathcal{E}_0=\text{Ind}(\mathcal{L}_0)\,, \quad\mathcal{E}_1=\text{Ind}(\mathcal{L}_1)\,,\quad \mathcal{E}_2=\text{Ind}(\mathcal{L}_2)\,,\quad \mathcal{E}_3=\text{Ind}(\mathcal{L}_3)\\
        \mathcal{E}_4=\text{Ind}(\mathcal{L}_6)-\text{Ind}(\mathcal{L}_2)\,,\quad \mathcal{E}_5=\text{Ind}(\mathcal{L}_5)-\text{Ind}(\mathcal{L}_3)\,,\quad \mathcal{E}_6=\text{Ind}(\mathcal{L}_2)+\text{Ind}(\mathcal{L}_4)-\text{Ind}(\mathcal{L}_6)\,,
    \end{gathered}
    \end{equation}
and all of them are m-type. The particle/soliton structure for $\mathcal{R}_{\mathcal{L}_{(1,0)}}$ is encoded in the matrix
    \begin{equation} \label{eq:N1_quiver_matrix_E7}
        \langle \widetilde{\mathcal{O}}(\mathcal{E}_a \rightarrow \mathcal{E}_b)\rangle = \left(
\begin{array}{ccccccc}
 0 & 1 & 0 & 0 & 0 & 0 & 0 \\
 \pi & 0 & \pi & 0 & 0 & 0 & 0 \\
 0 & 1 & 0 & 1 & 0 & 0 & 0 \\
 0 & 0 & \pi & 0 & \pi & 0 & \pi \\
 0 & 0 & 0 & 1 & 0 & 1 & 0 \\
 0 & 0 & 0 & 0 & \pi & 0 & 0 \\
 0 & 0 & 0 & 1 & 0 & 0 & 0 \\
\end{array}
\right)\,.
    \end{equation}
For $E_8$, we have
    \begin{equation}
    \begin{gathered}
        \mathcal{E}_0=\text{Ind}(\mathcal{L}_0)\,, \quad\mathcal{E}_1=\text{Ind}(\mathcal{L}_1)\,,\quad \mathcal{E}_2=\text{Ind}(\mathcal{L}_2)\,,\quad \mathcal{E}_3=\text{Ind}(\mathcal{L}_3)\,,\quad \mathcal{E}_4=\text{Ind}(\mathcal{L}_4)\,,\\
        \mathcal{E}_5=\text{Ind}(\mathcal{L}_7)-\text{Ind}(\mathcal{L}_3)\,,\quad \mathcal{E}_6=\text{Ind}(\mathcal{L}_6)-\text{Ind}(\mathcal{L}_4)\,,\quad \mathcal{E}_7=\text{Ind}(\mathcal{L}_5)+\text{Ind}(\mathcal{L}_3)-\text{Ind}(\mathcal{L}_7)\,,
    \end{gathered}
    \end{equation}
and all of them are m-type. The particle/soliton structure for $\mathcal{R}_{\mathcal{L}_{(1,0)}}$ is encoded in the matrix
\begin{equation}
    \langle \widetilde{\mathcal{O}}(\mathcal{E}_a \rightarrow \mathcal{E}_b)\rangle = \left(
\begin{array}{cccccccc}
 0 & 1 & 0 & 0 & 0 & 0 & 0 & 0 \\
 \pi & 0 & \pi & 0 & 0 & 0 & 0 & 0 \\
 0 & 1 & 0 & 1 & 0 & 0 & 0 & 0 \\
 0 & 0 & \pi & 0 & \pi & 0 & 0 & 0 \\
 0 & 0 & 0 & 1 & 0 & 1 & 0 & 1 \\
 0 & 0 & 0 & 0 & \pi & 0 & \pi & 0 \\
 0 & 0 & 0 & 0 & 0 & 1 & 0 & 0 \\
 0 & 0 & 0 & 0 & \pi & 0 & 0 & 0 \\
\end{array}
\right)\,.
\end{equation}
As an example, the quiver diagram associated to $E_7$ theory is drawn according to the matrix \eqref{eq:N1_quiver_matrix_E7} as
\begin{equation}
\centering
    %
\end{equation}
for representation $\mathcal{R}_{\mathcal{L}_{(k-1,k+2)}}$ where $k = 16$.

\bibliographystyle{JHEP}
\bibliography{main.bib}

\providecommand{\href}[2]{#2}\begingroup\raggedright\begin{thebibliography}{10}

\bibitem{Chang:2018iay}
C.-M.~Chang, Y.-H.~Lin, S.-H.~Shao, Y.~Wang and X.~Yin, \emph{{Topological
  Defect Lines and Renormalization Group Flows in Two Dimensions}},
  \href{https://doi.org/10.1007/JHEP01(2019)026}{\emph{JHEP} {\bfseries 01}
  (2019) 026} [\href{https://arxiv.org/abs/1802.04445}{{\ttfamily
  1802.04445}}].

\bibitem{Thorngren:2019iar}
R.~Thorngren and Y.~Wang, \emph{{Fusion category symmetry. Part I. Anomaly
  in-flow and gapped phases}},
  \href{https://doi.org/10.1007/JHEP04(2024)132}{\emph{JHEP} {\bfseries 04}
  (2024) 132} [\href{https://arxiv.org/abs/1912.02817}{{\ttfamily
  1912.02817}}].

\bibitem{Komargodski:2020mxz}
Z.~Komargodski, K.~Ohmori, K.~Roumpedakis and S.~Seifnashri, \emph{{Symmetries
  and strings of adjoint QCD$_{2}$}},
  \href{https://doi.org/10.1007/JHEP03(2021)103}{\emph{JHEP} {\bfseries 03}
  (2021) 103} [\href{https://arxiv.org/abs/2008.07567}{{\ttfamily
  2008.07567}}].

\bibitem{Thorngren:2021yso}
R.~Thorngren and Y.~Wang, \emph{{Fusion category symmetry. Part II.
  Categoriosities at c = 1 and beyond}},
  \href{https://doi.org/10.1007/JHEP07(2024)051}{\emph{JHEP} {\bfseries 07}
  (2024) 051} [\href{https://arxiv.org/abs/2106.12577}{{\ttfamily
  2106.12577}}].

\bibitem{Cordova:2022ieu}
C.~Cordova and K.~Ohmori, \emph{{Noninvertible Chiral Symmetry and Exponential
  Hierarchies}}, \href{https://doi.org/10.1103/PhysRevX.13.011034}{\emph{Phys.
  Rev. X} {\bfseries 13} (2023) 011034}
  [\href{https://arxiv.org/abs/2205.06243}{{\ttfamily 2205.06243}}].

\bibitem{Apte:2022xtu}
A.~Apte, C.~Cordova and H.T.~Lam, \emph{{Obstructions to gapped phases from
  noninvertible symmetries}},
  \href{https://doi.org/10.1103/PhysRevB.108.045134}{\emph{Phys. Rev. B}
  {\bfseries 108} (2023) 045134}
  [\href{https://arxiv.org/abs/2212.14605}{{\ttfamily 2212.14605}}].

\bibitem{Lin:2023uvm}
Y.-H.~Lin and S.-H.~Shao, \emph{{Bootstrapping noninvertible symmetries}},
  \href{https://doi.org/10.1103/PhysRevD.107.125025}{\emph{Phys. Rev. D}
  {\bfseries 107} (2023) 125025}
  [\href{https://arxiv.org/abs/2302.13900}{{\ttfamily 2302.13900}}].

\bibitem{Kaidi:2023maf}
J.~Kaidi, E.~Nardoni, G.~Zafrir and Y.~Zheng, \emph{{Symmetry TFTs and
  anomalies of non-invertible symmetries}},
  \href{https://doi.org/10.1007/JHEP10(2023)053}{\emph{JHEP} {\bfseries 10}
  (2023) 053} [\href{https://arxiv.org/abs/2301.07112}{{\ttfamily
  2301.07112}}].

\bibitem{Zhang:2023wlu}
C.~Zhang and C.~C{\'o}rdova, \emph{{Anomalies of (1+1)-dimensional categorical
  symmetries}}, \href{https://doi.org/10.1103/PhysRevB.110.035155}{\emph{Phys.
  Rev. B} {\bfseries 110} (2024) 035155}
  [\href{https://arxiv.org/abs/2304.01262}{{\ttfamily 2304.01262}}].

\bibitem{Damia:2023ses}
J.A.~Damia, R.~Argurio, F.~Benini, S.~Benvenuti, C.~Copetti and L.~Tizzano,
  \emph{{Non-invertible symmetries along 4d RG flows}},
  \href{https://doi.org/10.1007/JHEP02(2024)084}{\emph{JHEP} {\bfseries 02}
  (2024) 084} [\href{https://arxiv.org/abs/2305.17084}{{\ttfamily
  2305.17084}}].

\bibitem{Cordova:2023bja}
C.~Cordova, P.-S.~Hsin and C.~Zhang, \emph{{Anomalies of non-invertible
  symmetries in (3+1)d}},
  \href{https://doi.org/10.21468/SciPostPhys.17.5.131}{\emph{SciPost Phys.}
  {\bfseries 17} (2024) 131}
  [\href{https://arxiv.org/abs/2308.11706}{{\ttfamily 2308.11706}}].

\bibitem{Choi:2023pdp}
Y.~Choi, M.~Forslund, H.T.~Lam and S.-H.~Shao, \emph{{Quantization of
  Axion-Gauge Couplings and Noninvertible Higher Symmetries}},
  \href{https://doi.org/10.1103/PhysRevLett.132.121601}{\emph{Phys. Rev. Lett.}
  {\bfseries 132} (2024) 121601}
  [\href{https://arxiv.org/abs/2309.03937}{{\ttfamily 2309.03937}}].

\bibitem{Antinucci:2023ezl}
A.~Antinucci, F.~Benini, C.~Copetti, G.~Galati and G.~Rizi, \emph{{Anomalies of
  non-invertible self-duality symmetries: fractionalization and gauging}},
  \href{https://arxiv.org/abs/2308.11707}{{\ttfamily 2308.11707}}.

\bibitem{Bhardwaj:2024kvy}
L.~Bhardwaj, L.E.~Bottini, S.~Schafer-Nameki and A.~Tiwari, \emph{{Lattice
  Models for Phases and Transitions with Non-Invertible Symmetries}},
  \href{https://doi.org/10.21468/SciPostPhys.20.5.134}{\emph{SciPost Phys.}
  {\bfseries 20} (2026) 134}
  [\href{https://arxiv.org/abs/2405.05964}{{\ttfamily 2405.05964}}].

\bibitem{Bhardwaj:2024qiv}
L.~Bhardwaj, D.~Pajer, S.~Schafer-Nameki, A.~Tiwari, A.~Warman and J.~Wu,
  \emph{{Gapped phases in (2+1)d with non-invertible symmetries: Part I}},
  \href{https://doi.org/10.21468/SciPostPhys.19.2.056}{\emph{SciPost Phys.}
  {\bfseries 19} (2025) 056}
  [\href{https://arxiv.org/abs/2408.05266}{{\ttfamily 2408.05266}}].

\bibitem{Cordova:2024ypu}
C.~Cordova, S.~Hong and S.~Koren, \emph{{Noninvertible Peccei-Quinn Symmetry
  and the Massless Quark Solution to the Strong CP Problem}},
  \href{https://doi.org/10.1103/PhysRevX.15.031011}{\emph{Phys. Rev. X}
  {\bfseries 15} (2025) 031011}
  [\href{https://arxiv.org/abs/2402.12453}{{\ttfamily 2402.12453}}].

\bibitem{Copetti:2024rqj}
C.~Copetti, L.~Cordova and S.~Komatsu, \emph{{Noninvertible Symmetries,
  Anomalies, and Scattering Amplitudes}},
  \href{https://doi.org/10.1103/PhysRevLett.133.181601}{\emph{Phys. Rev. Lett.}
  {\bfseries 133} (2024) 181601}
  [\href{https://arxiv.org/abs/2403.04835}{{\ttfamily 2403.04835}}].

\bibitem{DelZotto:2024arv}
M.~Del~Zotto, S.N.~Meynet, D.~Migliorati and K.~Ohmori, \emph{{Emergent
  Non-Invertible Symmetries Bridging UV and IR Phases -- The Adjoint QCD
  Example}},  \href{https://arxiv.org/abs/2408.07123}{{\ttfamily 2408.07123}}.

\bibitem{Nakayama:2024msv}
Y.~Nakayama and T.~Tanaka, \emph{{Infinitely many new renormalization group
  flows between Virasoro minimal models from non-invertible symmetries}},
  \href{https://doi.org/10.1007/JHEP11(2024)137}{\emph{JHEP} {\bfseries 11}
  (2024) 137} [\href{https://arxiv.org/abs/2407.21353}{{\ttfamily
  2407.21353}}].

\bibitem{Cordova:2024goh}
C.~Cordova and D.~Garc{\'\i}a-Sep{\'u}lveda, \emph{{Topological Cosets via
  Anyon Condensation and Applications to Gapped $\mathrm{\bf{QCD_{2}}}$}},
  \href{https://arxiv.org/abs/2412.01877}{{\ttfamily 2412.01877}}.

\bibitem{Cordova:2024nux}
C.~Cordova, D.~Garc{\'\i}a-Sep{\'u}lveda and N.~Holfester,
  \emph{{Particle-Soliton Degeneracy in 2D Quantum Chromodynamics}},
  \href{https://arxiv.org/abs/2412.21153}{{\ttfamily 2412.21153}}.

\bibitem{Chen:2023qnv}
J.~Chen, W.~Cui, B.~Haghighat and Y.-N.~Wang, \emph{{SymTFTs and Duality
  Defects from 6d SCFTs on 4-manifolds}},
  \href{https://arxiv.org/abs/2305.09734}{{\ttfamily 2305.09734}}.

\bibitem{Bhardwaj:2025piv}
L.~Bhardwaj, S.~Schafer-Nameki, A.~Tiwari and A.~Warman, \emph{{Gapped Phases
  in (2+1)d with Non-Invertible Symmetries: Part II}},
  \href{https://arxiv.org/abs/2502.20440}{{\ttfamily 2502.20440}}.

\bibitem{Bhardwaj:2025jtf}
L.~Bhardwaj, Y.~Gai, S.-J.~Huang, K.~Inamura, S.~Schafer-Nameki, A.~Tiwari
  et~al., \emph{{Gapless Phases in (2+1)d with Non-Invertible Symmetries}},
  \href{https://arxiv.org/abs/2503.12699}{{\ttfamily 2503.12699}}.

\bibitem{Seiberg:2025bqy}
N.~Seiberg and S.~Seifnashri, \emph{{Symmetry transmutation and anomaly
  matching}}, \href{https://doi.org/10.1007/JHEP09(2025)014}{\emph{JHEP}
  {\bfseries 09} (2025) 014}
  [\href{https://arxiv.org/abs/2505.08618}{{\ttfamily 2505.08618}}].

\bibitem{Antinucci:2025fjp}
A.~Antinucci, C.~Copetti, Y.~Gai and S.~Schafer-Nameki, \emph{{Categorical
  Anomaly Matching}},  \href{https://arxiv.org/abs/2508.00982}{{\ttfamily
  2508.00982}}.

\bibitem{Zhang:2026gqp}
J.-R.~Zhang, J.-H.~Jin, T.-K.~Chen and J.~Chen, \emph{{Defect Conformal
  Manifolds along RG Domain Walls between $\mathbb Z_N$-Parafermions and
  Minimal Models}},  \href{https://arxiv.org/abs/2605.24978}{{\ttfamily
  2605.24978}}.

\bibitem{Bhardwaj:2017xup}
L.~Bhardwaj and Y.~Tachikawa, \emph{{On finite symmetries and their gauging in
  two dimensions}}, \href{https://doi.org/10.1007/JHEP03(2018)189}{\emph{JHEP}
  {\bfseries 03} (2018) 189}
  [\href{https://arxiv.org/abs/1704.02330}{{\ttfamily 1704.02330}}].

\bibitem{Aasen:2017ubm}
D.~Aasen, E.~Lake and K.~Walker, \emph{{Fermion condensation and super pivotal
  categories}}, \href{https://doi.org/10.1063/1.5045669}{\emph{J. Math. Phys.}
  {\bfseries 60} (2019) 121901}
  [\href{https://arxiv.org/abs/1709.01941}{{\ttfamily 1709.01941}}].

\bibitem{Chang:2022hud}
C.-M.~Chang, J.~Chen and F.~Xu, \emph{{Topological defect lines in two
  dimensional fermionic CFTs}},
  \href{https://doi.org/10.21468/SciPostPhys.15.5.216}{\emph{SciPost Phys.}
  {\bfseries 15} (2023) 216}
  [\href{https://arxiv.org/abs/2208.02757}{{\ttfamily 2208.02757}}].

\bibitem{Cordova:2024vsq}
C.~Cordova, D.~Garc{\'\i}a-Sep{\'u}lveda and N.~Holfester,
  \emph{{Particle-soliton degeneracies from spontaneously broken non-invertible
  symmetry}}, \href{https://doi.org/10.1007/JHEP07(2024)154}{\emph{JHEP}
  {\bfseries 07} (2024) 154}
  [\href{https://arxiv.org/abs/2403.08883}{{\ttfamily 2403.08883}}].

\bibitem{Cordova:2024iti}
C.~Cordova, N.~Holfester and K.~Ohmori, \emph{{Representation theory of
  solitons}}, \href{https://doi.org/10.1007/JHEP06(2025)001}{\emph{JHEP}
  {\bfseries 06} (2025) 001}
  [\href{https://arxiv.org/abs/2408.11045}{{\ttfamily 2408.11045}}].

\bibitem{Wan:2016php}
Y.~Wan and C.~Wang, \emph{{Fermion Condensation and Gapped Domain Walls in
  Topological Orders}},
  \href{https://doi.org/10.1007/JHEP03(2017)172}{\emph{JHEP} {\bfseries 03}
  (2017) 172} [\href{https://arxiv.org/abs/1607.01388}{{\ttfamily
  1607.01388}}].

\bibitem{Lou:2020gfq}
J.~Lou, C.~Shen, C.~Chen and L.-Y.~Hung, \emph{{A (dummy\textquoteright{}s)
  guide to working with gapped boundaries via (fermion) condensation}},
  \href{https://doi.org/10.1007/JHEP02(2021)171}{\emph{JHEP} {\bfseries 02}
  (2021) 171} [\href{https://arxiv.org/abs/2007.10562}{{\ttfamily
  2007.10562}}].

\bibitem{Hu:2021qhm}
Y.~Hu, Z.~Huang, L.-y.~Hung and Y.~Wan, \emph{{Anyon condensation: coherent
  states, symmetry enriched topological phases, Goldstone theorem, and
  dynamical rearrangement of symmetry}},
  \href{https://doi.org/10.1007/JHEP03(2022)026}{\emph{JHEP} {\bfseries 03}
  (2022) 026} [\href{https://arxiv.org/abs/2109.06145}{{\ttfamily
  2109.06145}}].

\bibitem{Wang:2023iqt}
Y.-N.~Wang and Y.~Zhang, \emph{{Fermionic higher-form symmetries}},
  \href{https://doi.org/10.21468/SciPostPhys.15.4.142}{\emph{SciPost Phys.}
  {\bfseries 15} (2023) 142}
  [\href{https://arxiv.org/abs/2303.12633}{{\ttfamily 2303.12633}}].

\bibitem{Ambrosino:2024ggh}
F.~Ambrosino, R.~Luo, Y.-N.~Wang and Y.~Zhang, \emph{{Understanding fermionic
  generalized symmetries}},
  \href{https://doi.org/10.1103/PhysRevD.110.105020}{\emph{Phys. Rev. D}
  {\bfseries 110} (2024) 105020}
  [\href{https://arxiv.org/abs/2404.12301}{{\ttfamily 2404.12301}}].

\bibitem{Bhardwaj:2024ydc}
L.~Bhardwaj, K.~Inamura and A.~Tiwari, \emph{{Fermionic non-invertible
  symmetries in (1+1)d: Gapped and gapless phases, transitions, and symmetry
  TFTs}}, \href{https://doi.org/10.21468/SciPostPhys.18.6.194}{\emph{SciPost
  Phys.} {\bfseries 18} (2025) 194}
  [\href{https://arxiv.org/abs/2405.09754}{{\ttfamily 2405.09754}}].

\bibitem{Balasubramanian:2024nei}
M.~Balasubramanian, M.~Buican and R.~Radhakrishnan, \emph{{On the
  Classification of Bosonic and Fermionic One-Form Symmetries in $2+1$d and
  {\textquoteright}t Hooft Anomaly Matching}},
  \href{https://doi.org/10.1007/s00220-025-05494-0}{\emph{Commun. Math. Phys.}
  {\bfseries 406} (2025) 319}
  [\href{https://arxiv.org/abs/2408.00866}{{\ttfamily 2408.00866}}].

\bibitem{Cappelli:1987xt}
A.~Cappelli, C.~Itzykson and J.B.~Zuber, \emph{{The ADE Classification of
  Minimal and A1(1) Conformal Invariant Theories}},
  \href{https://doi.org/10.1007/BF01221394}{\emph{Commun. Math. Phys.}
  {\bfseries 113} (1987) 1}.

\bibitem{Fuchs:2002cm}
J.~Fuchs, I.~Runkel and C.~Schweigert, \emph{{TFT construction of RCFT
  correlators 1. Partition functions}},
  \href{https://doi.org/10.1016/S0550-3213(02)00744-7}{\emph{Nucl. Phys. B}
  {\bfseries 646} (2002) 353}
  [\href{https://arxiv.org/abs/hep-th/0204148}{{\ttfamily hep-th/0204148}}].

\bibitem{Carqueville:2012dk}
N.~Carqueville and I.~Runkel, \emph{{Orbifold completion of defect
  bicategories}}, \href{https://doi.org/10.4171/qt/76}{\emph{Quantum Topol.}
  {\bfseries 7} (2016) 203} [\href{https://arxiv.org/abs/1210.6363}{{\ttfamily
  1210.6363}}].

\bibitem{Diatlyk:2023fwf}
O.~Diatlyk, C.~Luo, Y.~Wang and Q.~Weller, \emph{{Gauging non-invertible
  symmetries: topological interfaces and generalized orbifold groupoid in 2d
  QFT}}, \href{https://doi.org/10.1007/JHEP03(2024)127}{\emph{JHEP} {\bfseries
  03} (2024) 127} [\href{https://arxiv.org/abs/2311.17044}{{\ttfamily
  2311.17044}}].

\bibitem{Bernard:1990ti}
D.~Bernard and A.~Leclair, \emph{{The Fractional supersymmetric Sine-Gordon
  models}}, \href{https://doi.org/10.1016/0370-2693(90)90901-H}{\emph{Phys.
  Lett. B} {\bfseries 247} (1990) 309}.

\bibitem{Fendley:1991ve}
P.~Fendley and K.A.~Intriligator, \emph{{Scattering and thermodynamics of
  fractionally charged supersymmetric solitons}},
  \href{https://doi.org/10.1016/0550-3213(92)90365-I}{\emph{Nucl. Phys. B}
  {\bfseries 372} (1992) 533}
  [\href{https://arxiv.org/abs/hep-th/9111014}{{\ttfamily hep-th/9111014}}].

\bibitem{Chen:2025qub}
J.~Chen, Z.~Duan, Q.~Jia and S.~Lee, \emph{{Fermionic Non-invertible Symmetry
  Behind Supersymmetric ADE Solitons}},
  \href{https://arxiv.org/abs/2511.22129}{{\ttfamily 2511.22129}}.

\bibitem{Baume:2023kkf}
F.~Baume, J.J.~Heckman, M.~H{\"u}bner, E.~Torres, A.P.~Turner and X.~Yu,
  \emph{{SymTrees and Multi-Sector QFTs}},
  \href{https://doi.org/10.1103/PhysRevD.109.106013}{\emph{Phys. Rev. D}
  {\bfseries 109} (2024) 106013}
  [\href{https://arxiv.org/abs/2310.12980}{{\ttfamily 2310.12980}}].

\bibitem{Bhardwaj:2023bbf}
L.~Bhardwaj, L.E.~Bottini, D.~Pajer and S.~Schafer-Nameki, \emph{{The club
  sandwich: Gapless phases and phase transitions with non-invertible
  symmetries}},
  \href{https://doi.org/10.21468/SciPostPhys.18.5.156}{\emph{SciPost Phys.}
  {\bfseries 18} (2025) 156}
  [\href{https://arxiv.org/abs/2312.17322}{{\ttfamily 2312.17322}}].

\bibitem{Braeger:2024jcj}
N.~Braeger, V.~Chakrabhavi, J.J.~Heckman and M.~H{\"u}bner, \emph{{Generalized
  symmetries of nonsupersymmetric orbifolds}},
  \href{https://doi.org/10.1103/PhysRevD.111.066015}{\emph{Phys. Rev. D}
  {\bfseries 111} (2025) 066015}
  [\href{https://arxiv.org/abs/2404.17639}{{\ttfamily 2404.17639}}].

\bibitem{Heckman:2024zdo}
J.J.~Heckman and M.~H{\"u}bner, \emph{{Celestial Topology, Symmetry Theories,
  and Evidence for a NonSUSY D3-Brane CFT}},
  \href{https://doi.org/10.1002/prop.202400270}{\emph{Fortsch. Phys.}
  {\bfseries 73} (2025) 2400270}
  [\href{https://arxiv.org/abs/2406.08485}{{\ttfamily 2406.08485}}].

\bibitem{Cvetic:2024dzu}
M.~Cveti\v{c}, R.~Donagi, J.J.~Heckman, M.~H\"ubner and E.~Torres,
  \emph{{Cornering Relative Symmetry Theories}},
  \href{https://arxiv.org/abs/2408.12600}{{\ttfamily 2408.12600}}.

\bibitem{Choi:2024tri}
Y.~Choi, B.C.~Rayhaun and Y.~Zheng, \emph{{Generalized Tube Algebras,
  Symmetry-Resolved Partition Functions, and Twisted Boundary States}},
  \href{https://doi.org/10.1007/s00220-025-05543-8}{\emph{Commun. Math. Phys.}
  {\bfseries 407} (2026) 62}
  [\href{https://arxiv.org/abs/2409.02159}{{\ttfamily 2409.02159}}].

\bibitem{Bhardwaj:2024igy}
L.~Bhardwaj, C.~Copetti, D.~Pajer and S.~Schafer-Nameki, \emph{{Boundary
  SymTFT}},  \href{https://arxiv.org/abs/2409.02166}{{\ttfamily 2409.02166}}.

\bibitem{etingof2015tensor}
P.~Etingof, S.~Gelaki, D.~Nikshych and V.~Ostrik, \emph{Tensor categories},
  vol.~205, American Mathematical Soc. (2015).

\bibitem{etingof2005fusion}
P.~Etingof, D.~Nikshych and V.~Ostrik, \emph{On fusion categories},
  {\emph{Annals of mathematics} (2005) 581}.

\bibitem{USHER2018453}
R.~Usher, \emph{Fermionic 6j-symbols in superfusion categories},
  \href{https://doi.org/https://doi.org/10.1016/j.jalgebra.2018.02.015}{\emph{Journal
  of Algebra} {\bfseries 503} (2018) 453}.

\bibitem{Fuchs:2012dt}
J.~Fuchs, C.~Schweigert and A.~Valentino, \emph{{Bicategories for boundary
  conditions and for surface defects in 3-d TFT}},
  \href{https://doi.org/10.1007/s00220-013-1723-0}{\emph{Commun. Math. Phys.}
  {\bfseries 321} (2013) 543}
  [\href{https://arxiv.org/abs/1203.4568}{{\ttfamily 1203.4568}}].

\bibitem{Blumenhagen:2009zz}
R.~Blumenhagen and E.~Plauschinn, \emph{{Introduction to conformal field
  theory}: {with applications to String theory}}, vol.~779 (2009),
  \href{https://doi.org/10.1007/978-3-642-00450-6}{10.1007/978-3-642-00450-6}.

\bibitem{Cappelli:1986ed}
A.~Cappelli, \emph{{Modular Invariant Partition Functions of Superconformal
  Theories}}, \href{https://doi.org/10.1016/0370-2693(87)91532-2}{\emph{Phys.
  Lett. B} {\bfseries 185} (1987) 82}.

\bibitem{Gepner:1986hr}
D.~Gepner and Z.-a.~Qiu, \emph{{Modular Invariant Partition Functions for
  Parafermionic Field Theories}},
  \href{https://doi.org/10.1016/0550-3213(87)90348-8}{\emph{Nucl. Phys. B}
  {\bfseries 285} (1987) 423}.

\bibitem{Gepner:1987qi}
D.~Gepner, \emph{{Space-Time Supersymmetry in Compactified String Theory and
  Superconformal Models}},
  \href{https://doi.org/10.1016/0550-3213(88)90397-5}{\emph{Nucl. Phys. B}
  {\bfseries 296} (1988) 757}.

\bibitem{Vafa:1988uu}
C.~Vafa and N.P.~Warner, \emph{{Catastrophes and the Classification of
  Conformal Theories}},
  \href{https://doi.org/10.1016/0370-2693(89)90473-5}{\emph{Phys. Lett. B}
  {\bfseries 218} (1989) 51}.

\bibitem{Cecotti:1992rm}
S.~Cecotti and C.~Vafa, \emph{{On classification of N=2 supersymmetric
  theories}}, \href{https://doi.org/10.1007/BF02096804}{\emph{Commun. Math.
  Phys.} {\bfseries 158} (1993) 569}
  [\href{https://arxiv.org/abs/hep-th/9211097}{{\ttfamily hep-th/9211097}}].

\bibitem{Ambrosino:2026kfu}
F.~Ambrosino, M.R.~Gaberdiel and Y.~Nakayama, \emph{{$\mathcal{N}=2$ RG flows,
  Non-Invertible Symmetries and Matrix Factorisations}},
  \href{https://arxiv.org/abs/2608.02717}{{\ttfamily 2608.02717}}.

\bibitem{Dijkgraaf:1990dj}
R.~Dijkgraaf, H.L.~Verlinde and E.P.~Verlinde, \emph{{Topological strings in d
  {\ensuremath{<}} 1}},
  \href{https://doi.org/10.1016/0550-3213(91)90129-L}{\emph{Nucl. Phys. B}
  {\bfseries 352} (1991) 59}.

\bibitem{Gray:2008je}
O.~Gray, \emph{{On the complete classification of the unitary N=2 minimal
  superconformal field theories}},
  \href{https://doi.org/10.1007/s00220-012-1478-z}{\emph{Commun. Math. Phys.}
  {\bfseries 312} (2012) 611}
  [\href{https://arxiv.org/abs/0812.1318}{{\ttfamily 0812.1318}}].

\bibitem{Cordova:2023qei}
C.~Cordova and G.~Rizi, \emph{{Non-invertible symmetry in Calabi-Yau conformal
  field theories}}, \href{https://doi.org/10.1007/JHEP01(2025)045}{\emph{JHEP}
  {\bfseries 01} (2025) 045}
  [\href{https://arxiv.org/abs/2312.17308}{{\ttfamily 2312.17308}}].

\bibitem{Friedan:1984rv}
D.~Friedan, Z.-a.~Qiu and S.H.~Shenker, \emph{{Superconformal Invariance in
  Two-Dimensions and the Tricritical Ising Model}},
  \href{https://doi.org/10.1016/0370-2693(85)90819-6}{\emph{Phys. Lett. B}
  {\bfseries 151} (1985) 37}.

\bibitem{EICHENHERR198526}
H.~Eichenherr, \emph{Minimal operator algebras in superconformal quantum field
  theory},
  \href{https://doi.org/https://doi.org/10.1016/0370-2693(85)90817-2}{\emph{Physics
  Letters B} {\bfseries 151} (1985) 26}.

\bibitem{Schellekens:1989am}
A.N.~Schellekens and S.~Yankielowicz, \emph{{Extended Chiral Algebras and
  Modular Invariant Partition Functions}},
  \href{https://doi.org/10.1016/0550-3213(89)90310-6}{\emph{Nucl. Phys. B}
  {\bfseries 327} (1989) 673}.

\bibitem{Gaberdiel:2026sfg}
M.R.~Gaberdiel and L.~Merkens, \emph{{Defects in $ \mathcal{N}=1 $ minimal
  models and RG flows}},
  \href{https://doi.org/10.1007/JHEP05(2026)018}{\emph{JHEP} {\bfseries 05}
  (2026) 018} [\href{https://arxiv.org/abs/2601.03879}{{\ttfamily
  2601.03879}}].

\bibitem{Zhou:2021ulc}
J.-R.~Zhou, Q.-R.~Wang and Z.-C.~Gu, \emph{{Towards a complete classification
  of nonchiral topological phases in two-dimensional fermion systems}},
  \href{https://doi.org/10.1103/PhysRevB.106.245120}{\emph{Phys. Rev. B}
  {\bfseries 106} (2022) 245120}
  [\href{https://arxiv.org/abs/2112.06124}{{\ttfamily 2112.06124}}].

\bibitem{Witten:2023snr}
E.~Witten, \emph{{Anomalies and Nonsupersymmetric D-Branes}},
  \href{https://arxiv.org/abs/2305.01012}{{\ttfamily 2305.01012}}.

\bibitem{Schoutens:1990vb}
K.~Schoutens, \emph{{Supersymmetry and Factorizable Scattering}},
  \href{https://doi.org/10.1016/0550-3213(90)90674-3}{\emph{Nucl. Phys. B}
  {\bfseries 344} (1990) 665}.

\bibitem{Ahn:1990gn}
C.~Ahn, D.~Bernard and A.~LeClair, \emph{{Fractional Supersymmetries in
  Perturbed Coset Cfts and Integrable Soliton Theory}},
  \href{https://doi.org/10.1016/0550-3213(90)90287-N}{\emph{Nucl. Phys. B}
  {\bfseries 346} (1990) 409}.

\bibitem{Goldstone:1981kk}
J.~Goldstone and F.~Wilczek, \emph{{Fractional Quantum Numbers on Solitons}},
  \href{https://doi.org/10.1103/PhysRevLett.47.986}{\emph{Phys. Rev. Lett.}
  {\bfseries 47} (1981) 986}.

\bibitem{Chang:2026pdu}
C.-M.~Chang, J.~Chen and F.~Xu, \emph{{Classification of 2D Fermionic Systems
  with a $\mathbb Z_2$ Flavor Symmetry}},
  \href{https://arxiv.org/abs/2604.09503}{{\ttfamily 2604.09503}}.

\bibitem{Copetti:2024onh}
C.~Copetti, \emph{{Defect Charges, Gapped Boundary Conditions, and the Symmetry
  TFT}},  \href{https://arxiv.org/abs/2408.01490}{{\ttfamily 2408.01490}}.

\bibitem{Izumi:2000qa}
M.~Izumi, \emph{{The structure of sectors associated with Longo-Rehren
  inclusions. I: General theory}},
  \href{https://doi.org/10.1007/s002200000234}{\emph{Commun. Math. Phys.}
  {\bfseries 213} (2000) 127}.

\bibitem{Evans:2010yr}
D.E.~Evans and T.~Gannon, \emph{{The exoticness and realisability of twisted
  Haagerup-Izumi modular data}},
  \href{https://doi.org/10.1007/s00220-011-1329-3}{\emph{Commun. Math. Phys.}
  {\bfseries 307} (2011) 463}
  [\href{https://arxiv.org/abs/1006.1326}{{\ttfamily 1006.1326}}].

\bibitem{MUGER2003159}
M.~Müger, \emph{From subfactors to categories and topology ii: The quantum
  double of tensor categories and subfactors},
  \href{https://doi.org/https://doi.org/10.1016/S0022-4049(02)00248-7}{\emph{Journal
  of Pure and Applied Algebra} {\bfseries 180} (2003) 159}.

\bibitem{Lin:2022dhv}
Y.-H.~Lin, M.~Okada, S.~Seifnashri and Y.~Tachikawa, \emph{{Asymptotic density
  of states in 2d CFTs with non-invertible symmetries}},
  \href{https://doi.org/10.1007/JHEP03(2023)094}{\emph{JHEP} {\bfseries 03}
  (2023) 094} [\href{https://arxiv.org/abs/2208.05495}{{\ttfamily
  2208.05495}}].

\bibitem{Lan:2014uaa}
T.~Lan, J.C.~Wang and X.-G.~Wen, \emph{{Gapped Domain Walls, Gapped Boundaries
  and Topological Degeneracy}},
  \href{https://doi.org/10.1103/PhysRevLett.114.076402}{\emph{Phys. Rev. Lett.}
  {\bfseries 114} (2015) 076402}
  [\href{https://arxiv.org/abs/1408.6514}{{\ttfamily 1408.6514}}].

\bibitem{Cong:2016ayp}
I.~Cong, M.~Cheng and Z.~Wang, \emph{{Topological Quantum Computation with
  Gapped Boundaries}},  \href{https://arxiv.org/abs/1609.02037}{{\ttfamily
  1609.02037}}.

\bibitem{Davydov:2010kfz}
A.~Davydov, M.~Mueger, D.~Nikshych and V.~Ostrik, \emph{{The Witt group of
  non-degenerate braided fusion categories}},
  \href{https://arxiv.org/abs/1009.2117}{{\ttfamily 1009.2117}}.

\bibitem{Kitaev:2000nmw}
A.~Kitaev, \emph{{Unpaired Majorana fermions in quantum wires}},
  \href{https://doi.org/10.1070/1063-7869/44/10S/S29}{\emph{Phys. Usp.}
  {\bfseries 44} (2001) 131}
  [\href{https://arxiv.org/abs/cond-mat/0010440}{{\ttfamily
  cond-mat/0010440}}].

\bibitem{Barkeshli:2014cna}
M.~Barkeshli, P.~Bonderson, M.~Cheng and Z.~Wang, \emph{{Symmetry
  Fractionalization, Defects, and Gauging of Topological Phases}},
  \href{https://doi.org/10.1103/PhysRevB.100.115147}{\emph{Phys. Rev. B}
  {\bfseries 100} (2019) 115147}
  [\href{https://arxiv.org/abs/1410.4540}{{\ttfamily 1410.4540}}].

\bibitem{Ji:2026yfj}
W.~Ji, R.A.~Lanzetta, Z.~Zhou and C.~Wang, \emph{{Self-dual Higgs transitions:
  Toric code and beyond}},  \href{https://arxiv.org/abs/2601.20945}{{\ttfamily
  2601.20945}}.

\bibitem{Tong:2006pa}
D.~Tong, \emph{{Superconformal vortex strings}},
  \href{https://doi.org/10.1088/1126-6708/2006/12/051}{\emph{JHEP} {\bfseries
  12} (2006) 051} [\href{https://arxiv.org/abs/hep-th/0610214}{{\ttfamily
  hep-th/0610214}}].

\bibitem{Shapere:1999xr}
A.D.~Shapere and C.~Vafa, \emph{{BPS structure of Argyres-Douglas
  superconformal theories}},
  \href{https://arxiv.org/abs/hep-th/9910182}{{\ttfamily hep-th/9910182}}.

\bibitem{bohm1999weak}
G.~B{\"o}hm, F.~Nill and K.~Szlach{\'a}nyi, \emph{Weak hopf algebras: I.
  integral theory and c-structure}, {\emph{Journal of Algebra} {\bfseries 221}
  (1999) 385}.

\bibitem{andruskiewitsch2001triangular}
N.~Andruskiewitsch, P.~Etingof and S.~Gelaki, \emph{Triangular hopf algebras
  with the chevalley property}, {\emph{Michigan Mathematical Journal}
  {\bfseries 49} (2001) 277}.

\bibitem{aissaoui2014classification}
S.~Aissaoui, A.~Makhlouf et~al., \emph{On classification of finite-dimensional
  superbialgebras and hopf superalgebras}, {\emph{SIGMA. Symmetry,
  Integrability and Geometry: Methods and Applications} {\bfseries 10} (2014)
  001}.

\bibitem{Inamura:2022lun}
K.~Inamura, \emph{{Fermionization of fusion category symmetries in 1+1
  dimensions}}, \href{https://doi.org/10.1007/JHEP10(2023)101}{\emph{JHEP}
  {\bfseries 10} (2023) 101}
  [\href{https://arxiv.org/abs/2206.13159}{{\ttfamily 2206.13159}}].

\bibitem{Fuchs:2001qc}
J.~Fuchs and C.~Schweigert, \emph{{Category theory for conformal boundary
  conditions}}, {\emph{Fields Inst. Commun.} {\bfseries 39} (2003) 25}
  [\href{https://arxiv.org/abs/math/0106050}{{\ttfamily math/0106050}}].

\bibitem{kirillov2002q}
A.~Kirillov~Jr and V.~Ostrik, \emph{On a q-analogue of the mckay correspondence
  and the ade classification of sl2 conformal field theories}, {\emph{Advances
  in Mathematics} {\bfseries 171} (2002) 183}.

\end{thebibliography}\endgroup
\end{document}